\documentclass[preprint,12pt]{JHEP3} 

\JHEPspecialurl{http://jhep.sissa.it/JOURNAL/JHEP3.tar.gz}

\usepackage{subfigure}
\usepackage{amsmath, amssymb, multicol}
\usepackage{graphicx}
\usepackage{epsfig,eepic}
\usepackage{rotating}
\usepackage{setspace}
\usepackage{cite}
\newcommand{\bc}{\begin{center}}
\newcommand{\ec}{\end{center}}
\newcommand{\beq}{\begin{equation}}         
\newcommand{\eeq}{\end{equation}}           %
\newcommand{\beqn}{\begin{eqnarray}}
\newcommand{\eeqn}{\end{eqnarray}}
\newcommand{\bpic}{\begin{picture}}            %
\newcommand{\epic}{\end{picture}}              

\def\Pom{I\!\!P}

\def\eq#1{{Eq.~(\ref{#1})}}

\numberwithin{figure}{section}\numberwithin{equation}{section}

\title{Saturation model in the non-Glauber approach}
\author{A.
Kormilitzin\thanks{Email: andreyk1@post.tau.ac.il}
\\
 Department of Particle Physics, School of Physics and Astronomy\\ Raymond and
Beverly Sackler Faculty of Exact Science\\  Tel Aviv University, Tel
Aviv, 69978, Israel}

\abstract{In this paper a new saturation model is presented. This
model is based on the theoretical solution for the generating
functional, and it is quite different and not more complicated than
the Glauber-like approach used before. The model describes the
structure function $F_{2}$ of the proton, as well as the diffractive
structure function $F_{2}^{D}$. We show the difference between our
model and the eikonal approach by calculating the multiplicity
distribution, using the AGK cutting rules strategy.}


\preprint{TAUP -2859-07\\
\today}

\begin{document}

\section{Introduction}

The very intriguing observation of the HERA experiments
is the rapid rise of the total $\gamma^{*}p$ cross section, with
energy in the deep inelastic scattering (DIS) region. The linear
evolution equations approach is not valid in the region of high
energies, since it predicts an increase of the parton density,
steeper than is allowed by the Froissart bound. The solution to this
problem is hidden in the effect of the parton saturation. At high
energy the density of partons increases, filling the whole
transverse area of the target. This stage is called saturation. At
small values of Bjorken-$x$, the interaction between partons should
be taken into account leading to the recombination of partons. This
process tames the growth of the parton density.

The main goal of this paper is to build a model which follows on
from the self consistent theoretical approach, and is able to
describe the matching between soft and hard processes. This approach
is based on the generating functional method. The model incorporates
the following
features :\\

$\;\;\;i$) Evolution of the parton density with energy\\

$\;ii$) The phenomena of parton saturation. The model takes into
  account the effect of parton recombination at high density and
  preserves unitarity.\\

$\;iii$) This is a simple model, different from the
model based on the eikonal approach.\\

$\;iv$) Good description of all recent experimental data on
inclusive, hard, soft, and diffractive processes.\\

In the next section, we start with a brief discussion of the main
properties of the generating functional \cite{Mueller:1993rr}. We
find the solution to the simplified evolution equation, assuming
that the dipoles do not change their transverse sizes, during the
interaction.

Based on this solution, we build our model. All details and
properties of the model are discussed in section \ref{desc_model}.
This model is quite different from the other saturation models,
which are based on the eikonal approach
\cite{Golec-Biernat:1998js,Golec-Biernat:2001mm,Bartels:2002cj,Kowalski:2003hm,Golec-Biernat:2006ba,Kowalski:2006hc}.
In section \ref{desc_HERA}, we show how our model describes the HERA
data using the recent data on the proton structure function $F_{2}$,
in a wide range of kinematics from different collaborations
\cite{Adloff:2000qk,Chekanov:2001qu,Breitweg:2000yn,Chekanov:2003rb,Adloff:2001zj,Adloff:1997mf,E665,Chekanov:2005vv,Adloff:2001rw},
we fit our model and find the free parameters. We compare the
developed model, to other saturation models in section
\ref{comparison_models}, where we discuss the multiplicity
distribution of the two models, and we present the differences
between them. In the next section using the parameters from the fit,
we describe the diffractive dissociation experimental data, the
charm quark structure function $F^{c\overline{c}}_{2}$, and the
slopes $dF_{2}/d(\ln Q^{2})$ and $d\ln F_{2}/d(\ln 1/x)$. Finally,
we summarize our results.

\section{Description of the Model}\label{desc_model}
\subsection{General properties of the generating functional}

This approach is based on the concept of color dipoles
\cite{Mueller:1993rr}, which in the large $N_c$ limit are correct
degrees of freedom at high energies \cite{MUDM}, since they
diagonalize the scattering matrix. At high energy, every parton
tends to emit more partons, which leads to the evolution of the
initial parton density. The dipole decay
 produces a parton cascade, originating from the
one parent dipole at initial high rapidity $Y$. We define by $P_{n}$
to be the probability to obtain $n$ dipoles from the parent dipole,
at some rapidity $y$.

Using the generating functional approach, one can construct an
evolution equation for the dipole probability $P_{n}$ at rapidity
$y$ \cite{Levin:2003nc}. The interesting property of this approach,
is the manifestation of the saturation phenomena, and the
conservation of unitarity. For the sake of simplicity, we consider
the generating functional approach in the toy-model
\cite{Levin:2003nc,MUUN}, where we neglect changes to the dipole
sizes during interaction. In this approach, we obtain the following
evolution equation for the probability $P_{n}$
\beqn\label{toy_evo_eq} -\frac{\partial P_{n}}{\partial
y}\,=\,-\omega_{0}nP_{n}\,+\,\omega_{0}(n-1)P_{n-1} \eeqn

where $\omega_{0}$ is the probability for one dipole to decay into
two. The generating functional, with our assumption degenerates into
the function \beqn\label{toy_gen_func}
Z(y,\,u)\,\,=\,\,\sum_n\,\,P_n(y)\,\,u^n\,, \eeqn

Equation \eq{toy_evo_eq}, for the probability $P_{n}$ can be
rewritten in terms of the generating functional \eq{toy_gen_func},
as \beqn\label{GFEQ} -\,\,\frac{\partial\,Z(y,\,u)}{\partial\,
y}\,\, =\,\,-\,\omega_0 \,u\,(1\,-\,u)
\,\,\frac{\partial\,Z(y,\,u)}{\partial\, u} \eeqn This is a
\emph{Liouville} equation which has the solution \beq\label{SOLZ}
Z(y_{0}-y,\,u)\,\,=\,\,\frac{u\,\,e^{\,-\,\omega_0\, (y -
y_{0})}}{1\,\,+\,\,u\,\, (e^{\,-\,\omega_0 \,(y\, -\,
y_{0})}\,-\,1)}\,. \eeq To find the interaction amplitude, we follow
the procedure suggested in Ref.\cite{K}, namely,\beqn\label{AMP}
N(y,\,b) \,\,=\,\,Im\,A^{el}(y,\,b) \,\,=\,\,-
\sum^{\infty}_{n=1}\,\,\frac{ 1}{n!}\,\, \frac{\partial^n
Z}{(\partial u)^n  }|_{u = 1} \,\,\gamma_n(b)\,\,=\,\,-
\sum^{\infty}_{n=1}\,\,\frac{ 1}{n!}\,\, \frac{\partial^n
Z}{(\partial u)^n  }|_{u = 1} \,\,\gamma^n_1(b)\eeqn Here,
$-\,\gamma_n(b)$ is the imaginary part of the amplitude for the
scattering of $n$ dipoles off the target, at fixed impact parameter
$b$, at low energies. The most important assumption, is the
independent interaction of $n$ dipoles with the target, which is
expressed by the factor $\gamma_n =\gamma^n_1$ factor in \eq{AMP}.

\subsection{The Model}\label{model}

Using the generating functional approach, we propose a model for the
interaction amplitude of the dipole with the target. The amplitude
$N(y,\,b)$ can be found from the following relation \beq
\label{AMP1} N(y;\,b,\,r) \,\,=\,\,1\,\,-\,\,Z(y,\,u(r))\,. \eeq
Inserting \eq{SOLZ} into \eq{AMP1}, one obtains the final expression
for the interaction amplitude \beqn\label{int_amp}
N(y;\,b,\,r)\,=\,\frac{(1-u)e^{\omega_{0}(y -
y_{0})}}{(1-u)e^{\omega_{0}(y - y_{0})}\,+\,u}\,=\,\frac{\gamma(r)
e^{\omega_{0}(y - y_{0})}}{\gamma(r) e^{\omega_{0}(y -
y_{0})}\,+\,(1 - \gamma(r))} \eeqn \eq{int_amp} describes the sum of
so-called"fan" diagrams. Introducing the new variable
\cite{Levin:2007yv} \beq
\gamma_{R}\;=\;\frac{\gamma}{1-\gamma}\eeq\label{new_gam_R} the
amplitude now has the simple form \beqn\label{fin_amp}
N(y;\,b,\,r)\,=\,\frac{\gamma_{R}(y_{0};\,b,\,r)e^{\omega_{0}(y-y_{0})}}{\gamma_{R}(y_{0};\,b,\,r)
e^{\omega_{0}(y-y_{0})}\,+\,1}\eeqn and satisfies the following
equation \beqn \frac{\partial N(y;\,b,\,r)}{\partial
y}\;=\;\omega_{0}\gamma_{R}\,\frac{\partial
N(y;\,b,\,r)}{\partial\gamma_{R}}\eeqn This means that in fact, the
generating functional describes the system of non-interacting hard
pomerons, but its interaction with the target should be renormilized
using Eq.(\ref{new_gam_R}). Our main assumption, is that we can
include the dependence on the size of the dipoles incorporates it in
the low energy amplitude $\gamma$ condition. \beq
\gamma_{R}(y_{0};\,b)\;\rightarrow\;\gamma_{R}(y_{0};\,b,\,r)\eeq
where $r$ is the transverse size of the dipole. At small values of
$r$, the amplitude of \eq{fin_amp} should match the expression
originating from the perturbative calculation of the BFKL pomeron
exchange, namely \beqn\label{conv_omega}
\Omega\,=\,\gamma_{R}(y_{0};\,b,\,r)e^{\omega_{0}(y -
y_{0})}\;\equiv\;\frac{\pi^{2}}{N_{c}}r^{2}\alpha_{s}(\mu^{2})G(y_{0},\mu^{2})G(y
- y_{0},\mu^{2})S(b) \eeqn where $G(y - y_{0},\mu^{2})$ is the gluon
density at some scale $\mu^{2}$, $y$ is the rapidity defined in
\eq{conv_omega} as $y\;=\;\ln(1/x)$, and S(b) stands for the proton
profile function in the form\footnote{This $S(b)$ is the Fourier
transform of the dipole formula for the electromagnetic form factor
of the proton. We can use this as the first approximation, having in
mind that the real $b$-dependance should be determined by the impact
parameter dependence of the low energy amplitude for dipole - proton
scattering, so-called "two gluon form factor".} \beqn\label{s_b}
S(b)\,=\,\frac{2}{\pi
R^{2}}\left(\frac{\sqrt{8}b}{R}\right)K_{1}\left(\frac{\sqrt{8}b}{R}\right)
\eeqn where $K_{1}$ is the McDonald function. The evolution of the
parton density is given by the DGLAP evolution equation, with the
initial parton density at some scale $Q^{2}_{0}$ \beq
G(x,Q^{2}_{0})\,=\,\frac{A}{x^{\omega_{0}}} \eeq where $A$ and
$\omega_{0}$ are to be determined from the data fit. The expression
for the gluon density $G(x, Q^{2})$, is obtained from the solution
to the DGLAP equation, and is given by the inverse Mellin transform
\cite{Ellis:1993rb} \beq\label{mastereqxG} G(y,t)=\frac{1}{2\pi
i}\int_{c-i\infty}^{c+i\infty} d\omega \exp (\frac{t}{\omega} - t
+\omega y) \frac{A}{(\omega-\omega_{0})}\eeq where the variable $t$
is defined as \beqn
t=\frac{4N_{c}}{b_{0}}\ln\frac{\ln(Q^{2}/\Lambda^{2})}{\ln(Q_{0}^{2}/\Lambda^{2})},
\,\,\,\ b_{0}=(11-\frac{2n_{f}}{3}),\;\;\;
\mbox{and}\,\,\,y\,=\,\ln(1/x)\eeqn where $Q^{2}_{0}$ is the initial
condition, and $n_{f}$ is the number of flavors. The analytical
solution to \eq{mastereqxG} reads as \beq\label{sol_diff_eq_BK}
G(y,t)\,=\,Ae^{-t+\omega_{0}y}\left[\int_{0}^{y}dy^{'}
e^{-\omega_{0}y^{'}}\sqrt{\frac{t}{y^{'}}}I_{1}(2\sqrt{ty'})\,+\,1\right]
\eeq Here, we introduce the hard scale $\mu^{2}$, which corresponds
to the transverse size of the dipoles in the following way
\footnote{Such a form of parametrization was used in
\cite{Golec-Biernat:1998js,Golec-Biernat:2001mm,Bartels:2002cj,Kowalski:2003hm,Golec-Biernat:2006ba,Kowalski:2006hc},
and we introduce the same parametrization, to make the comparison
easier
 between our model and other models.}
\beqn Q^{2}\rightarrow\mu^{2}\,=\,\frac{C}{r^{2}}\,+\,\mu^{2}_{0}
\eeqn where we rewrite the gluon density in terms of $\mu^{2}$,
rather than $Q^{2}$. The parameters $C$ and $\mu^{2}_{0}$, are
obtained from the fit to data. Finally, the expression for the
interaction amplitude takes the following form
\beqn\label{int_amp_fin}
N(r,b,x)\,=\,\frac{\Omega(r,b,x)}{\Omega(r,b,x) + 1} \eeqn with
\beqn
\Omega(r,b,x)\;=\;\frac{\pi^2}{N_{c}}\,r^{2}\,\alpha_{s}\left(\mu^{2}(r)\,\right)G\left(\ln(1/x_{0}),\mu^{2}(r)\right)\,G\left(\ln(x/x_{0}),\mu^{2}(r)\right)\,S(b)\eeqn
This is a new form for the interaction amplitude of the dipoles with
the target, which originates from the generating functional
approach.

\subsection{Saturation Scale}

\begin{figure}[htbp]
\centerline{\includegraphics[width=10 cm,height=7cm]{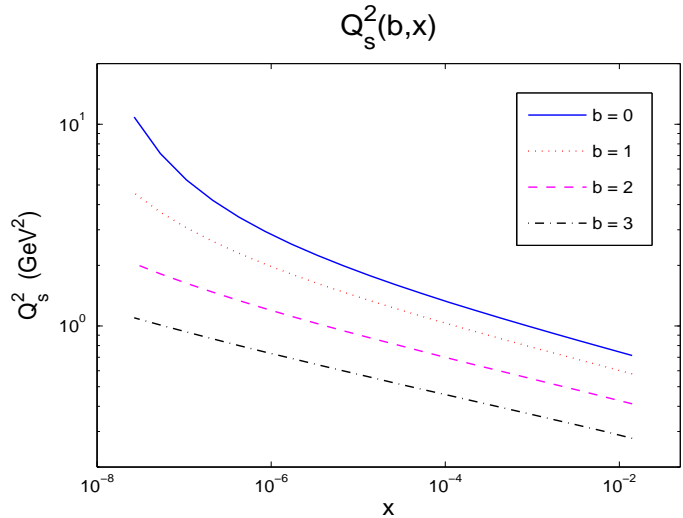}}
\caption{\it Saturation scale for various values of impact parameter
\emph{b}(fm) as a function of Bjorken-x.} \label{Qs_plot}
\end{figure}

At some energy value, partons start to populate densely, and this
leads to the effect of saturation. We define a new scale, which
separates the two regions, namely, the region of low parton density,
where we can apply perturbative methods, and the region of high
parton density, where we should take into account the recombination
effects, and non-liner corrections to the parton density. In order
to determine the saturation scale, we demand that the packing factor
of the partons \eq{p}, at some energy, be equal to 1. The packing
factor is defined as \beqn\label{p}
\kappa\;=\;\sigma_{0}\frac{N_{g}}{\pi R^{2}}\eeqn where $\sigma_{0}$
is the typical interaction cross section of the partons, which is
proportional to $Q^{2}$. $N_{g}$ corresponds to the number of
partons, and $\pi\,R^{2}$ denotes the area of the transverse slice
of the hadron. Putting everything together, one obtains
\beqn\label{sat_mom_xg}
1\,&=&\,\frac{\pi^{2}}{N_{c}}r^{2}_{s}\alpha_{s}(\mu_{s}^{2})G(y_{0},\mu_{s}^{2})G(y-y_{0},\mu_{s}^{2})S(b)
\nonumber
\\1\,&=&\,
\frac{4}{Q^{2}}\frac{\pi^{2}}{N_{c}}\alpha_{s}(\mu_{s}^{2})G(y_{0},\mu_{s}^{2})G(y-y_{0},\mu_{s}^{2})S(b)
\nonumber
\\Q^{2}_{s}(x)\,&=&\,4\frac{\pi^{2}}{N_{c}}\alpha_{s}(\mu_{s}^{2})G(y_{0},\mu_{s}^{2})G(y-y_{0},\mu_{s}^{2})S(b)
\eeqn where $\mu^{2}_{s}\,=\,\frac{CQ^{2}_{s}}{4}\,+\,\mu^{2}_{0}$.
The estimation of the saturation momentum is obtained from the
numerical solution of the \eq{sat_mom_xg} by iteration. The result
obtained for the saturation scale is plotted in Fig. \ref{Qs_plot}.

 Different graphs correspond to different values of the impact
parameter $b$. The value of the saturation scale, is calculated in
units of $GeV^{2}$.

\section{Description of DIS}\label{desc_HERA}
\begin{figure}[htbp]
\begin{center}
\begin{tabular}{c c}
\epsfig{file=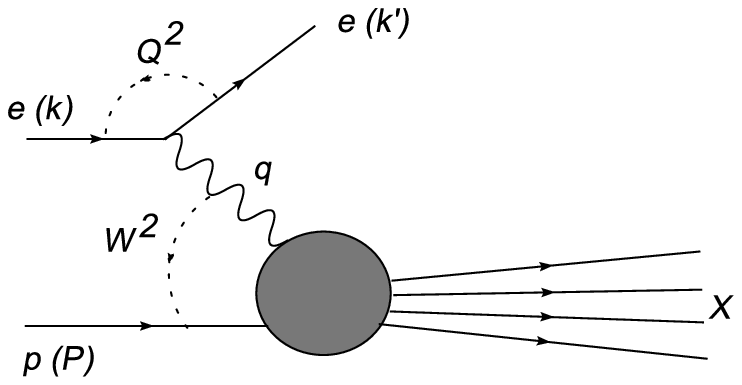,width=80mm, height=40mm}&
\epsfig{file=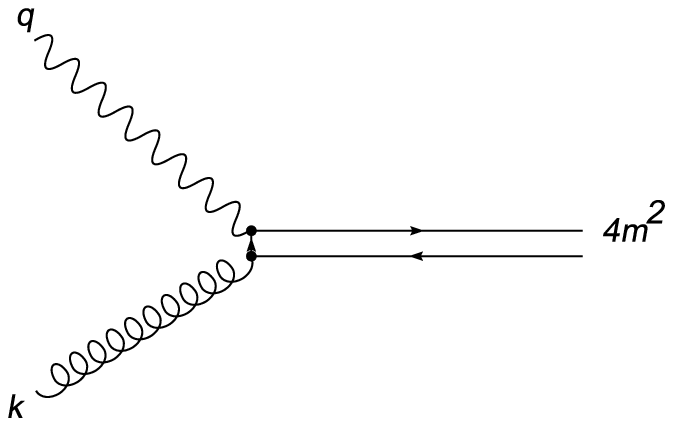,width=50mm, height=30mm}\\
(a) & (b)\\
\end{tabular}\caption{\it Kinematics
of deep inelastic scattering and photon-gluon fusion
diagram.}\label{dis_kin}
\end{center}
\end{figure}

We start by overviewing the main features of DIS.  The deep
inelastic scattering process is shown in Fig. \ref{dis_kin}$(a)$
 and the standard variables are defined as
\beqn\label{dis_kin_var}
\mbox{incoming proton momentum}&:&\;\;\; P \nonumber\\
\mbox{photon's virtuality}&:&\;\;\; Q^{2} = - q^{2} \nonumber\\
\mbox{fraction of electron energy transferred to the proton}&:&\;\;\; y\,=\,\frac{P\cdot q}{P\cdot k}\nonumber\\
\mbox{photon - proton system energy}&:&\;\;\; (q + P)^{2}\,=\,W^{2}\nonumber\\
\mbox{Bjorken-$x$}&:&\;\;\; x = \frac{Q^{2}}{2p\cdot q}\,=\,\frac{Q^{2}}{Q^{2} + W^{2}}\nonumber\\
\eeqn For small values of Bjorken-$x$, one can separate between the
photon wave function, which corresponds to the creation of the quark
- antiquark pair, and the interaction of the dipole with the target
\cite{Mueller:1989st,Barone:1993sy}. In the dipole picture of DIS,
the total photon-proton scattering cross section from transverse (T)
and longitudinal (L) polarized virtual photons, is given by the
convolution of the photon wave function, $\Psi_{T,L}$, and the
interaction amplitude $N$ of the dipoles with the target
\eq{int_amp_fin}. The proton structure function $F_{2}$, can be
written as the sum of two contributions \cite{Barone:1993sy}
\beq\label{F2}
F_{2}(x,Q^{2})\,=\,2\sum_{n=1}^{n_{f}}\frac{Q^{2}}{4\pi^{2} \alpha_{em}}\int d^{2}r\int d^{2}b\int_{0}^{1}dz \left\{|\Psi_{T}(r,z;m_{f},e_{f})|^{2}\,+\,|\Psi_{L}(r,z;m_{f},e_{f})|^{2}\right\}N(r,b,x)
\eeq where $f$ stands for the flavor of the quark-antiquark pair,
and $m_{f}$ and $e_{f}$, denote the mass and electric charge of the
quark with flavor $f$, respectively. The squared wave function of
the photon is given by \beqn\label{psi_T}
|\Psi_{T}(r,z;m_{f},e_{f})|^{2}\,=\,\frac{N_{c}e_{f}^{2}\alpha_{em}}{2\pi^{2}}\left\{[z^{2}
+ (1-z)^{2}]a^{2}K_{1}^{2}(ar) + m_{f}^{2}K_{0}^{2}(ar)\right\}
\eeqn and \beqn\label{psi}
|\Psi_{L}(r,z;m_{f},e_{f})|^{2}\,=\,\frac{N_{c}e_{f}^{2}\alpha_{em}}{2\pi^{2}}\left\{4Q^{2}z^{2}(1-z)^{2}K_{0}^{2}(ar)\right\}
\eeqn where \beqn\label{a2} a^{2}\,=\,z(1-z)Q^{2} + m^{2}_{f}\eeqn
and $K_{1}$, $K_{0}$ where $K_{1}$, $K_{0}$ are McDonald functions.
The significant contribution to the inclusive processes, appears
from the heavy quark production. In the present paper we investigate
the influence of the charm quark contribution to the proton
structure function $F_{2}$.

\subsection{Heavy quark}

The recent data from HERA, shows that the heavy quark contribution
is up to 30\%-40\% of the total value of the structure function
\cite{Chekanov:2003rb,Adloff:2001zj}, and can not be neglected.
Thus, we take into account the contribution of the charm quark and
confront it with recent data. The analysis does not include any new
parameters. The proton structure function may be written, as the
contribution from all possible participating quarks. Since the
contribution of each quark flavor is proportional to its
electromagnetic charge, and to the inverse mass, we may conclude
that the contribution of very heavy quarks like, \emph{top} and
\emph{bottom}, can be neglected compared with the light quarks (i.e.
\emph{up, down, strange}) and also the \emph{charm} quark. The light
quark masses are considered to be the same. Hence, the proton
structure function may be written as \beqn
F_{2}\;=\;F_{2}^{u\overline{u}}\;+\;F_{2}^{d\overline{d}}\;+\;F_{2}^{s\overline{s}}\;+\;F_{2}^{c\overline{c}}\eeqn
Each component of the structure function associated with different
quark flavors, is proportional to the particular quark wavefunction
of the corresponding flavor, \beqn
F_{2}^{f\overline{f}}(x,Q^{2})\,=\,2\frac{Q^{2}}{4\pi
\alpha_{em}}\int d^{2}b\int d^{2}r\int dz
|\Psi^{f\overline{f}}(r,z;m_{f},e_{f})|^{2}N(r,b,x^{f\overline{f}})\eeqn
In the case of the charm quark, its contribution to the total
structure function, has the following form \beq\label{F2_ch}
F_{2}^{c\overline{c}}(x,Q^{2})\,=\,2\frac{Q^{2}}{4\pi
\alpha_{em}}\int d^{2}b\int d^{2}r\int dz
|\Psi^{c\overline{c}}(r,z;m_{c},e_{c})|^{2}N(r,b,x^{c\overline{c}})
\eeq where the photon wavefunction, which decays into the
charm-anticharm pair is given by \beqn
|\Psi^{c\overline{c}}(r,z;m_{c},e_{c})|^{2}\,=\,\frac{N_{c}\alpha_{em}}{2\pi^{2}}\frac{4}{9}\{
[z^{2}&+&(1+z)^{2}]a^{2}K_{1}^{2}(ar)+m^{2}_{c}K_{0}^{2}(ar)\nonumber
\\ &+& 4Q^{2}z^{2}(1-z)^{2}K_{0}^{2}(ar)\}\eeqn where $a^{2}$ is given by \eq{a2}, and the numerical factor
$\frac{4}{9}$ is the squared value of the charge quantum number of
the charm, in units of $e$. It is important to stress, that every
quark flavor has an independent value of Bjorken-$x$, which depends
on the quark mass. This point is presented in the next section.

\subsection{Redefinition of Bjorken $x$}

The important ingredient of our approach, is the redefinition of
Bjorken-$x$. The motivation for this comes from the saturated
region, or the region which lies below the saturation scale
($Q^{2}<Q^{2}_{s}$). In this region, the transverse momentum of the
partons is proportional to the saturation scale, and this leads to
the new definition of Bjorken-$x$ \beqn\label{new_x}
4m^{2}\,&=&\,(q\,+\,k)^{2}\nonumber\\
4m^{2}\,&=&\,-Q^{2} + 2P\cdot q + k^{2}\,=\,-Q^{2} + 2P\cdot q - Q^{2}_{s}\nonumber\\ \nonumber\\
\widetilde{x}\,&=&\,\frac{Q^{2}\,+\,Q^{2}_{s}\,+\,4m^{2}}{Q^{2}\,+W^{2}}\eeqn
The kinematics is shown in Fig. \ref{dis_kin}$(b)$ We can easily
see, that this new definition of Bjorken-$x$ allows a smooth
transition from large values of $Q^{2}$, to the region of low values
of $Q^{2}$. At large $Q^{2}$, we get back the ordinary expression
for Bjorken-$x$, and at low values of photon virtuality, the main
contribution comes from the saturation scale.

\subsection{Description of Fit}

From the fit to the experimental data, we can find values of the
appropriate parameters for our model. For this purpose, we use all
the latest data for the proton structure function $F_{2}$ from
different collaborations in the region $x<0.01$ and
$0.045<Q^{2}<150\,\, GeV^{2}$. The small $x$ cut leads to an upper
limit on $Q^{2}$
\cite{Adloff:2000qk,Chekanov:2001qu,Breitweg:2000yn,Chekanov:2003rb,Adloff:2001zj,Adloff:1997mf,Chekanov:2005vv,E665,Adloff:2001rw}.
It was observed, that the set of data from \cite{Adloff:2000qk}
should be rescaled with the factor of $1.037$, in order to satisfy
the best fit. For the numerical evaluation, we have implemented the
formulae Eqs. (\ref{F2}) - (\ref{psi}). We have also included the
contribution from the heavy charm quark \eq{F2_ch}, and the new
definition of Bjorken-$x$ \eq{new_x}. The model contains free
parameters, whose values are to be determined from the fit to the
experimental data. The result of our fit, with four quark flavors is
represented in table 1.\\
\begin{table}[htbp]
\begin{center}
\begin{tabular}{|l|c|c|c|c|c|c|c|c|}
    \hline
$$ &  $A$    &   $\mu^{2}_{0}$  & $\omega_{0}$   &  $Q^{2}_{0}$ & $C$ & $m_{q}$ & $m_{c}$ & $\chi^{2}/d.o.f.$\\
    \hline\hline
four quarks (u,d,s,c)   & 0.785 & 1.294 & 0.060  & 1.236 &  \textbf{1.0} & \textbf{0.24} & \textbf{1.3} & 354/341\\
    \hline
\end{tabular}\label{table:tab}\caption{\it The parameters determined from the fit to the proton structure function $F_{2}$}
\end{center}
\end{table} \\ The parameters have the following physical meaning. $A$ with
$\omega_{0}$ and $C$ with $\mu^{2}_{0}$ and  $Q^{2}_{0}$, determine
the hard scale and the initial gluon density, respectively. We
observed, that the parameter $C$, has a strong correlation to other
parameters. For every value of $C$, it is possible to find a set of
four parameters, which satisfy almost the same $\chi^{2}$. Hence, it
was fixed with the arbitrary value $C\,=1.0$. The value of the light
quark mass, which corresponds to the best fit is
$m_{q}\,=\,0.24\,\,(GeV)$. For the charm quark, the mass was taken
to be $m_{c}\,=\,1.3\,\,(GeV)$. The reson for taking the same values
for the three light quarks ($u,d,s$) originates from the fact, that
the influence of the strange quark to the total fit is relatively
small, since its contribution in comparison with $u$ and $d$ is
proportional to the ratio of charges squared. The resulting fit is
plotted in Figs. \ref{DIS_1},~\ref{DIS_2}.

\begin{figure}[htbp]
\begin{center}
\begin{tabular}{ccccc ccccc ccccc ccccc}
\epsfig{file=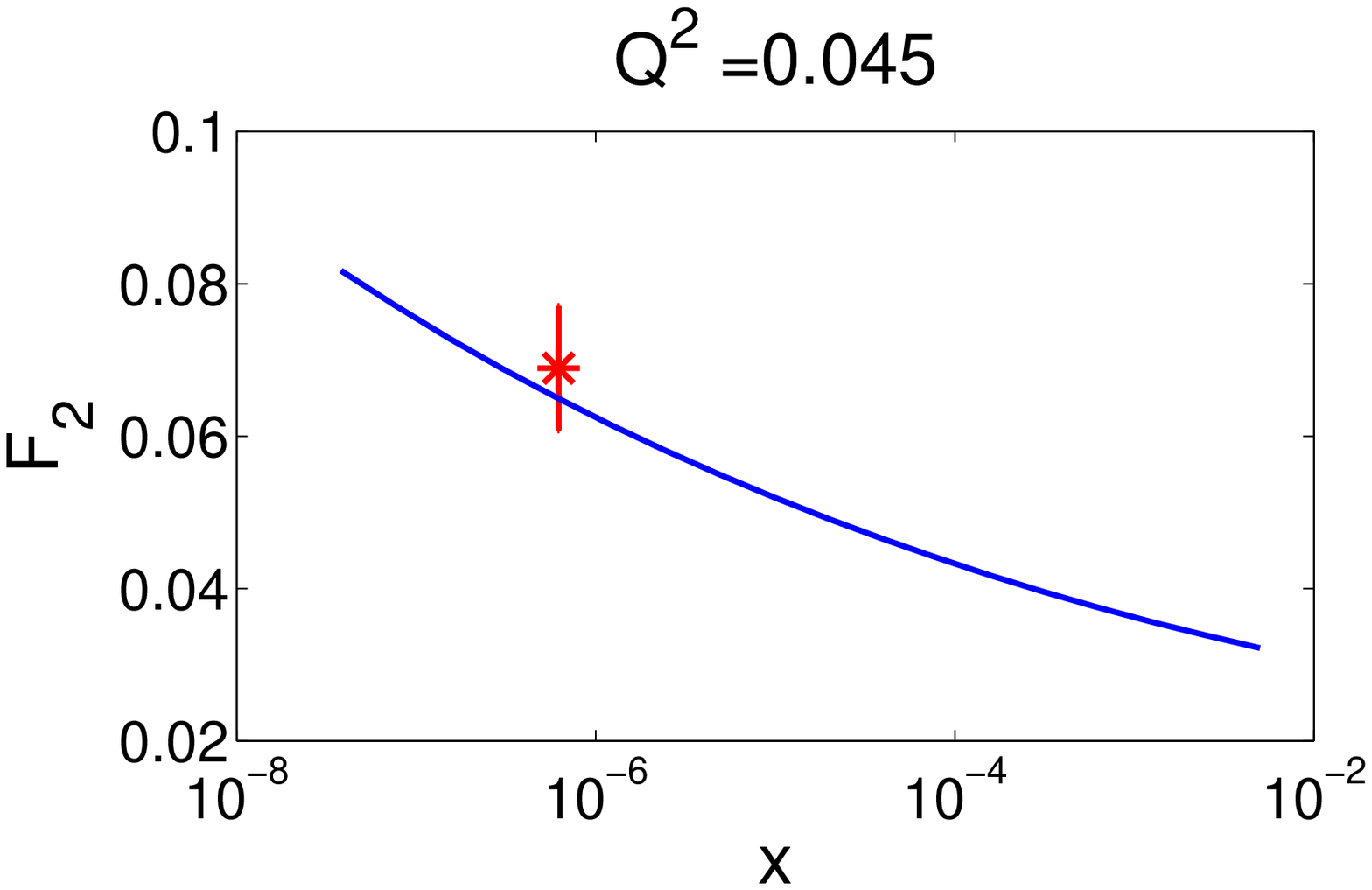,width=40mm, height=25mm}&
\epsfig{file=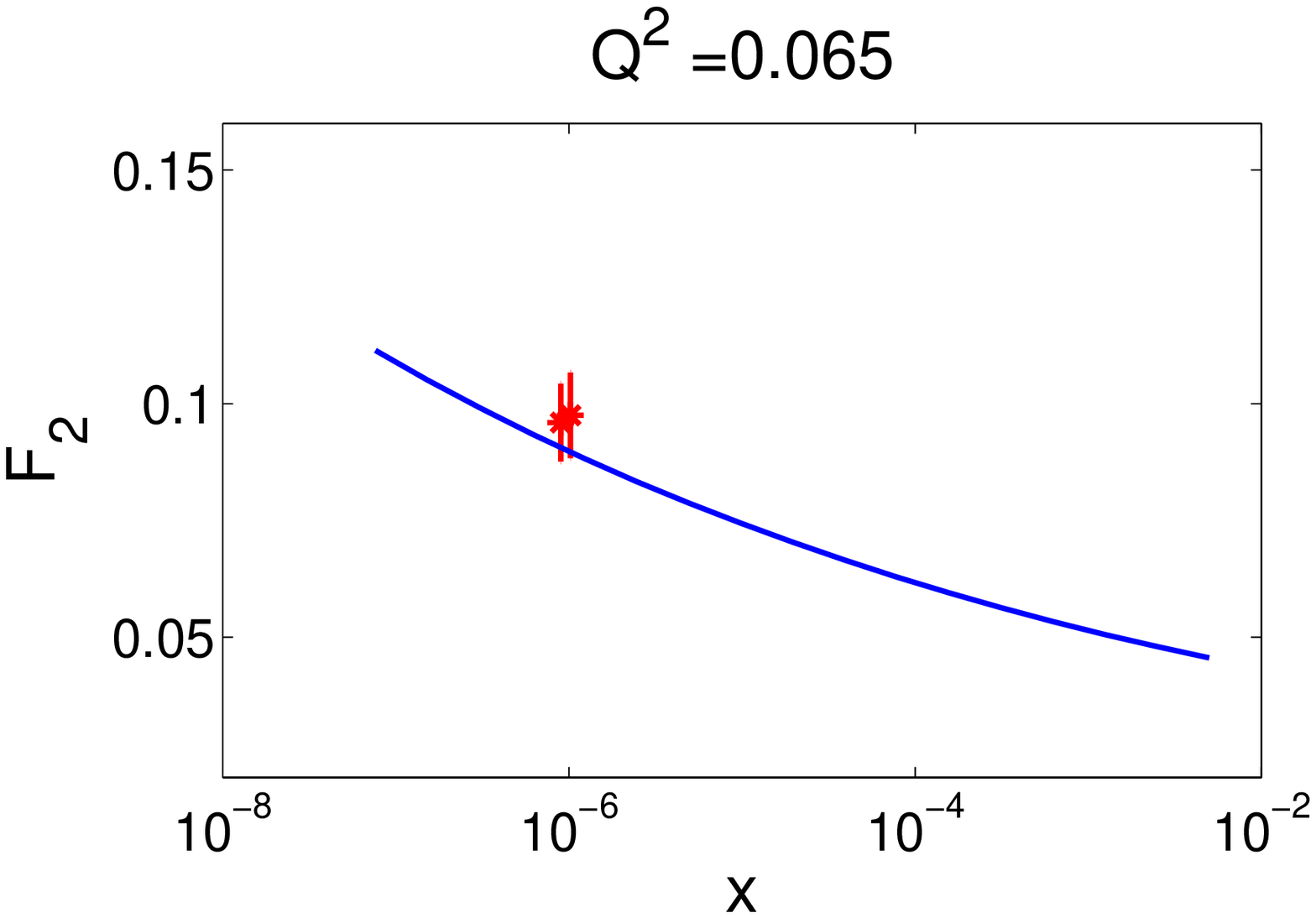,width=40mm, height=25mm}&
\epsfig{file=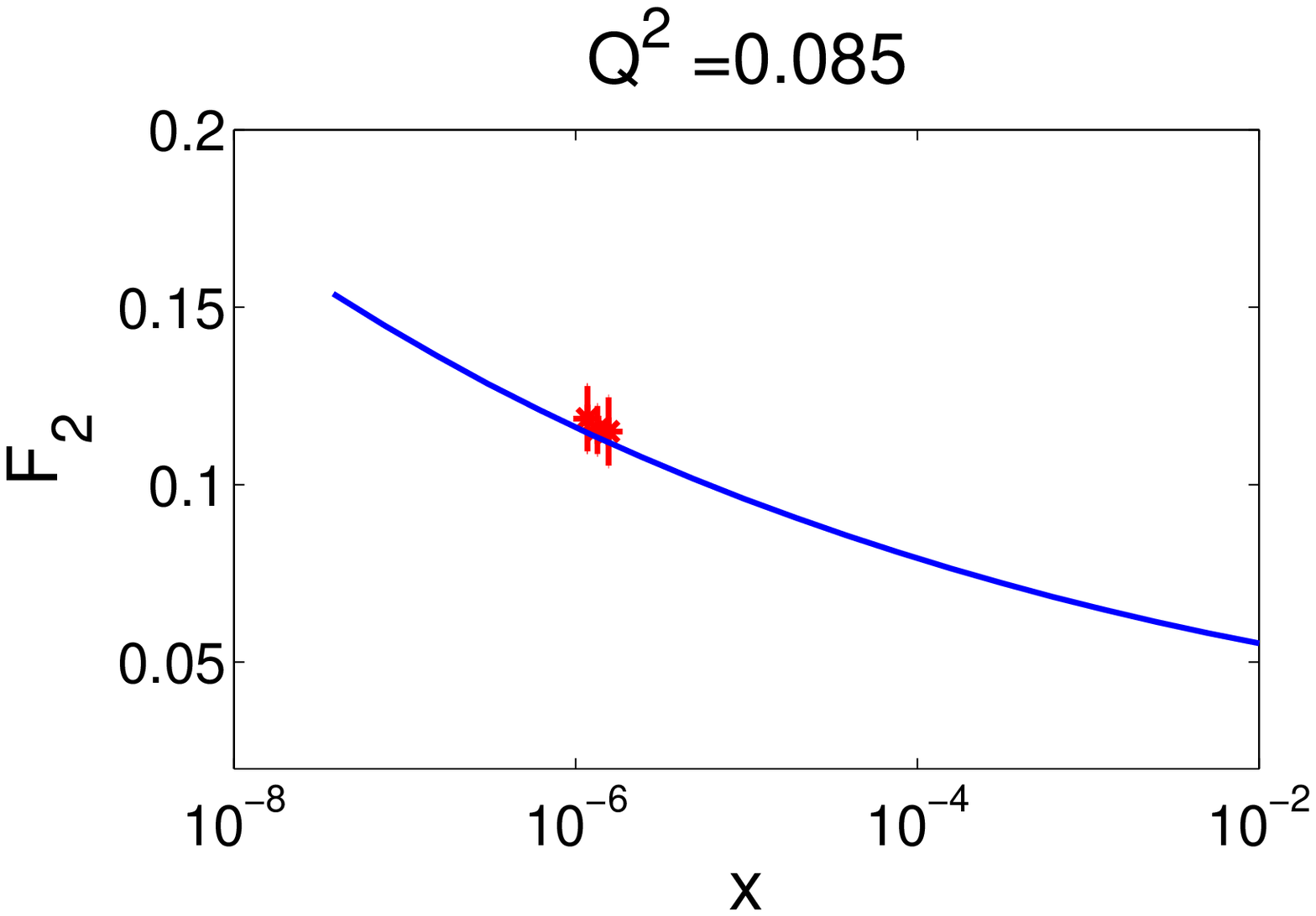,width=40mm, height=25mm}&
\epsfig{file=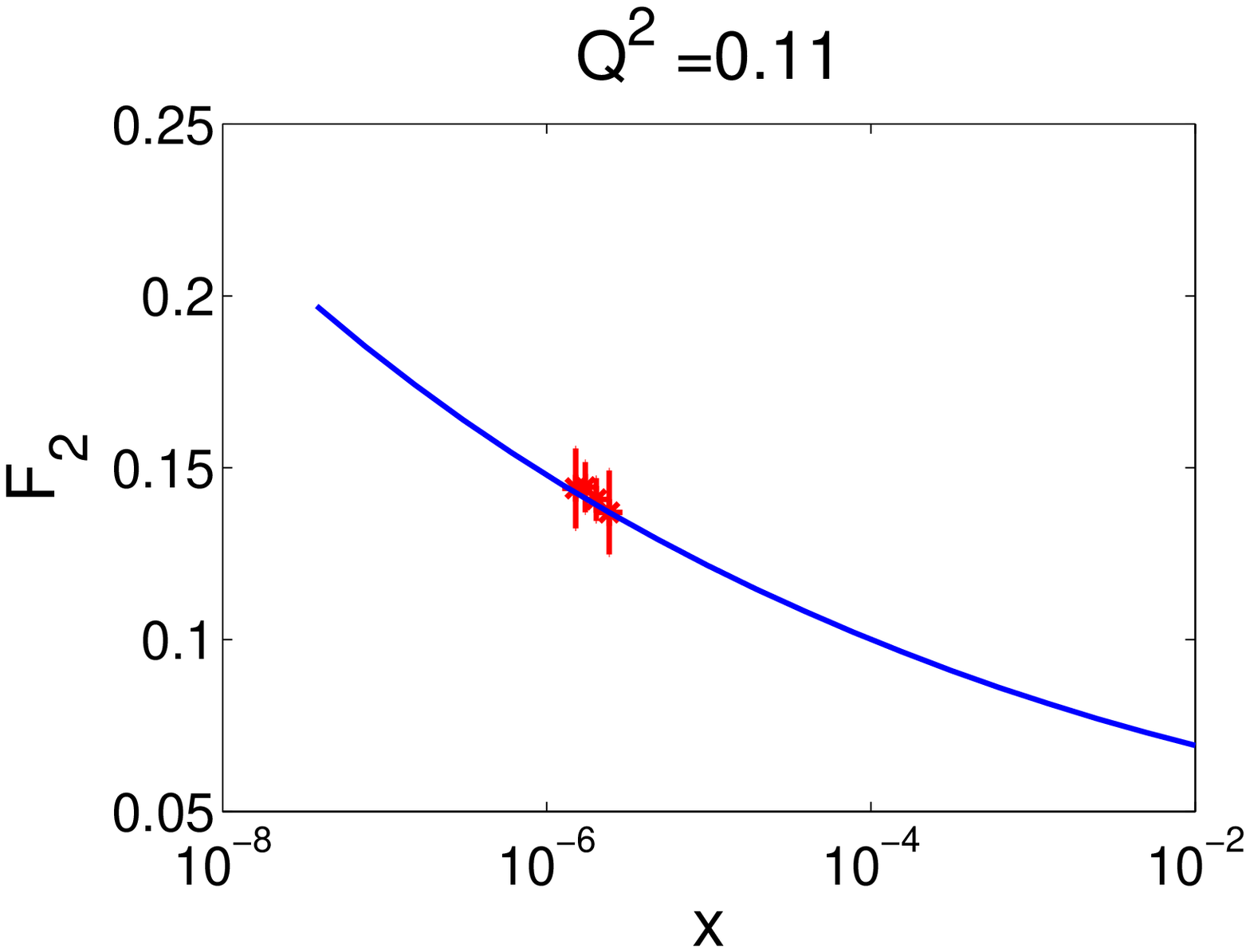,width=40mm, height=25mm}\\
\epsfig{file=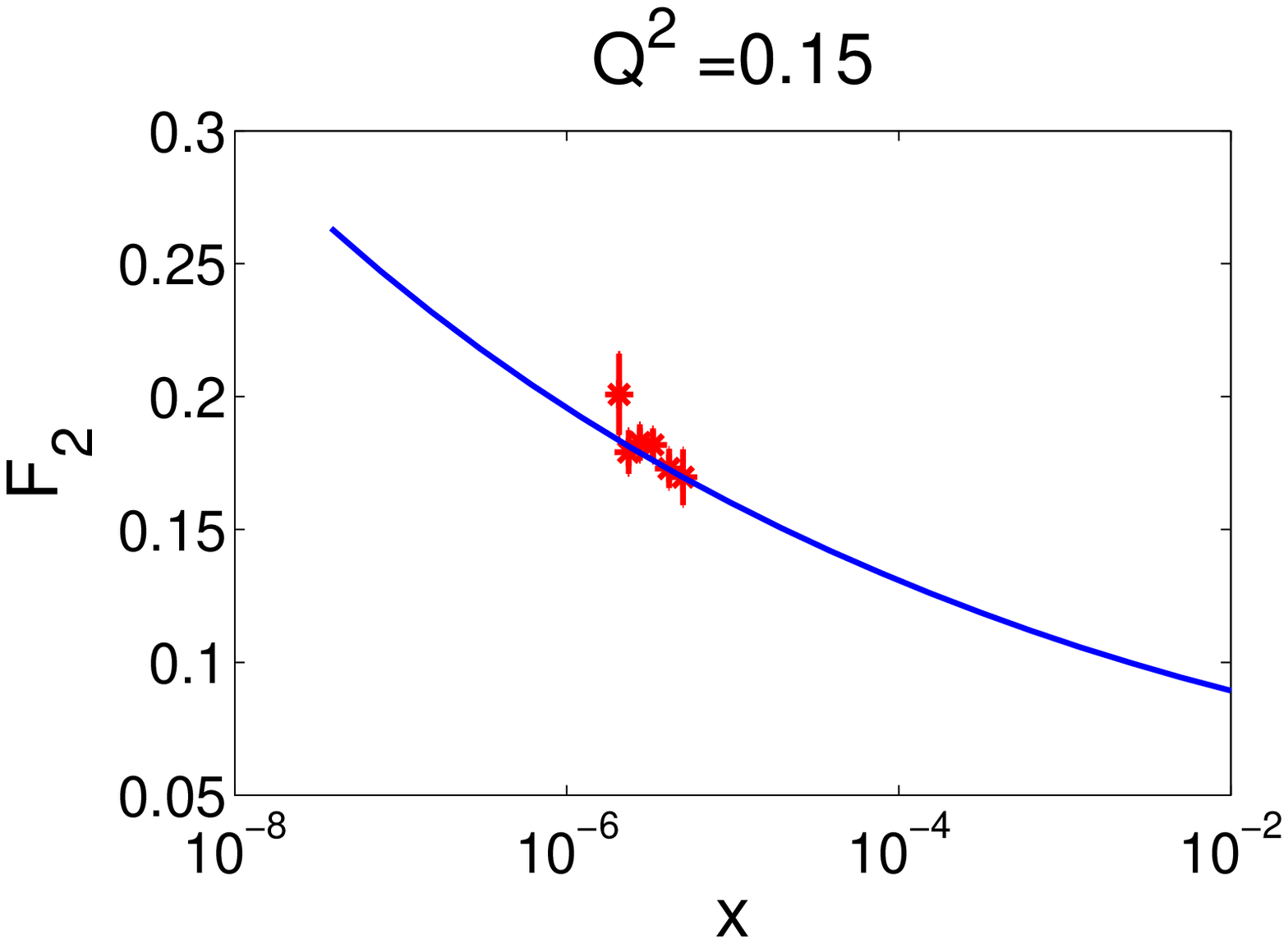,width=40mm, height=25mm}&
\epsfig{file=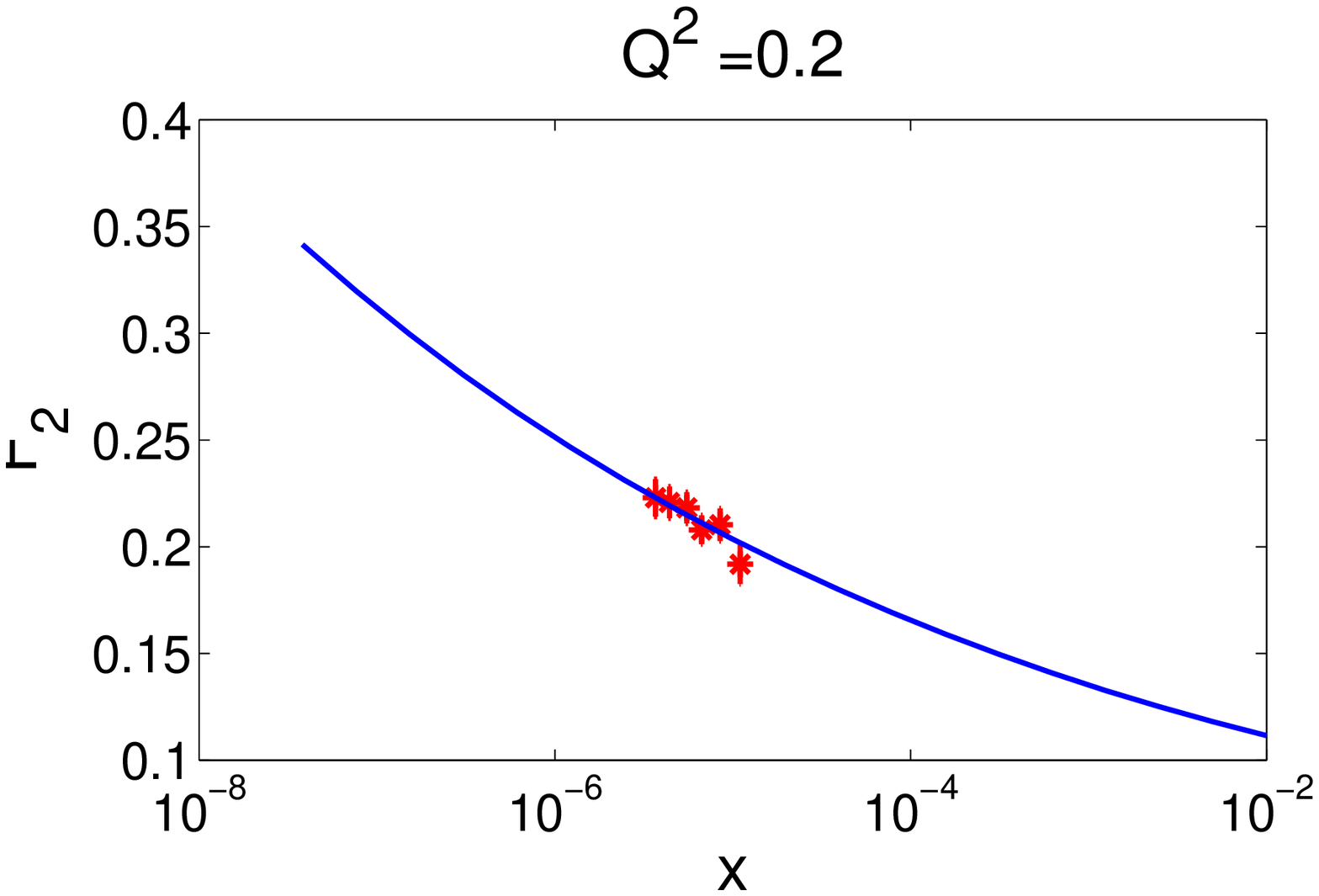,width=40mm, height=25mm}&
\epsfig{file=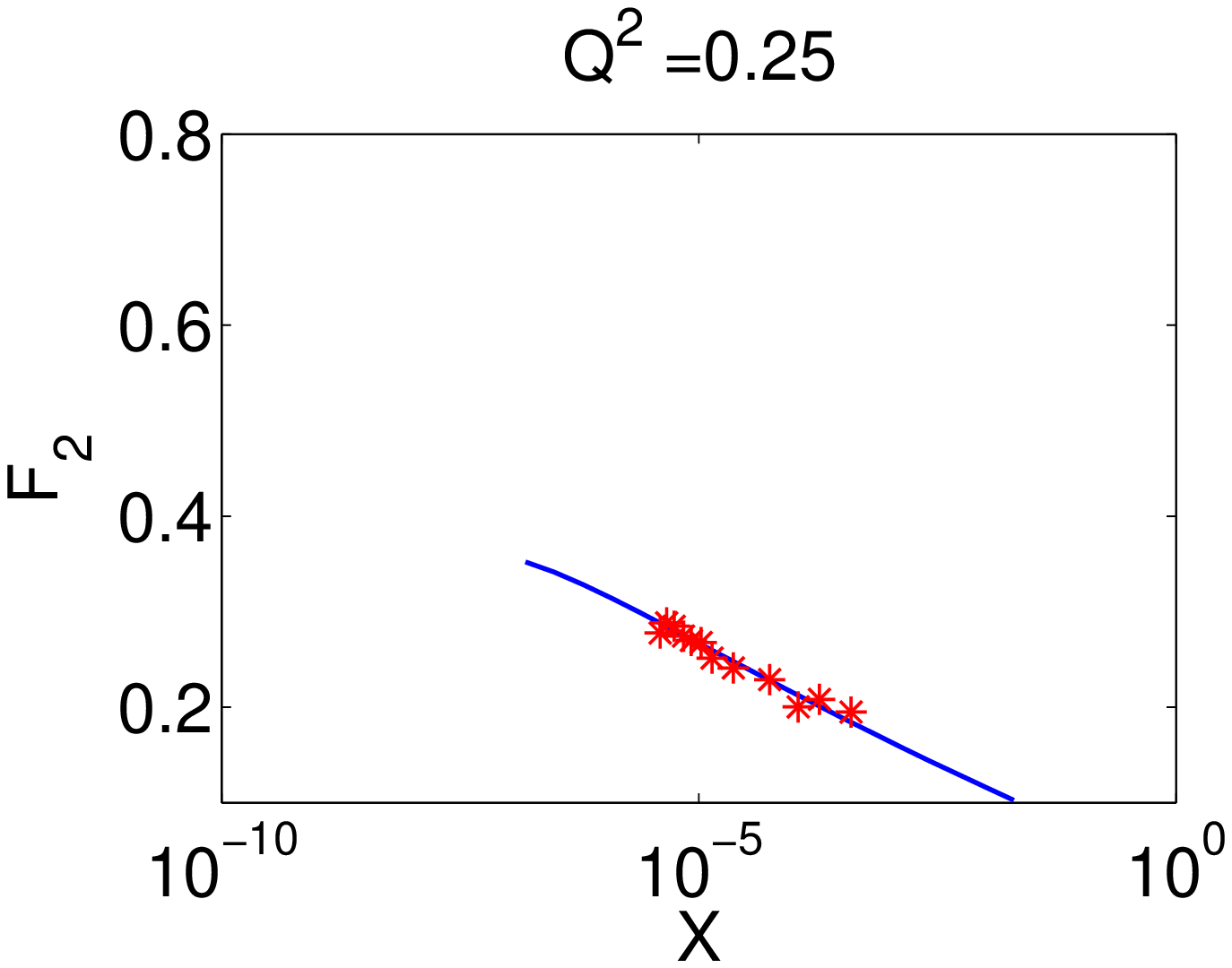,width=40mm, height=25mm}&
\epsfig{file=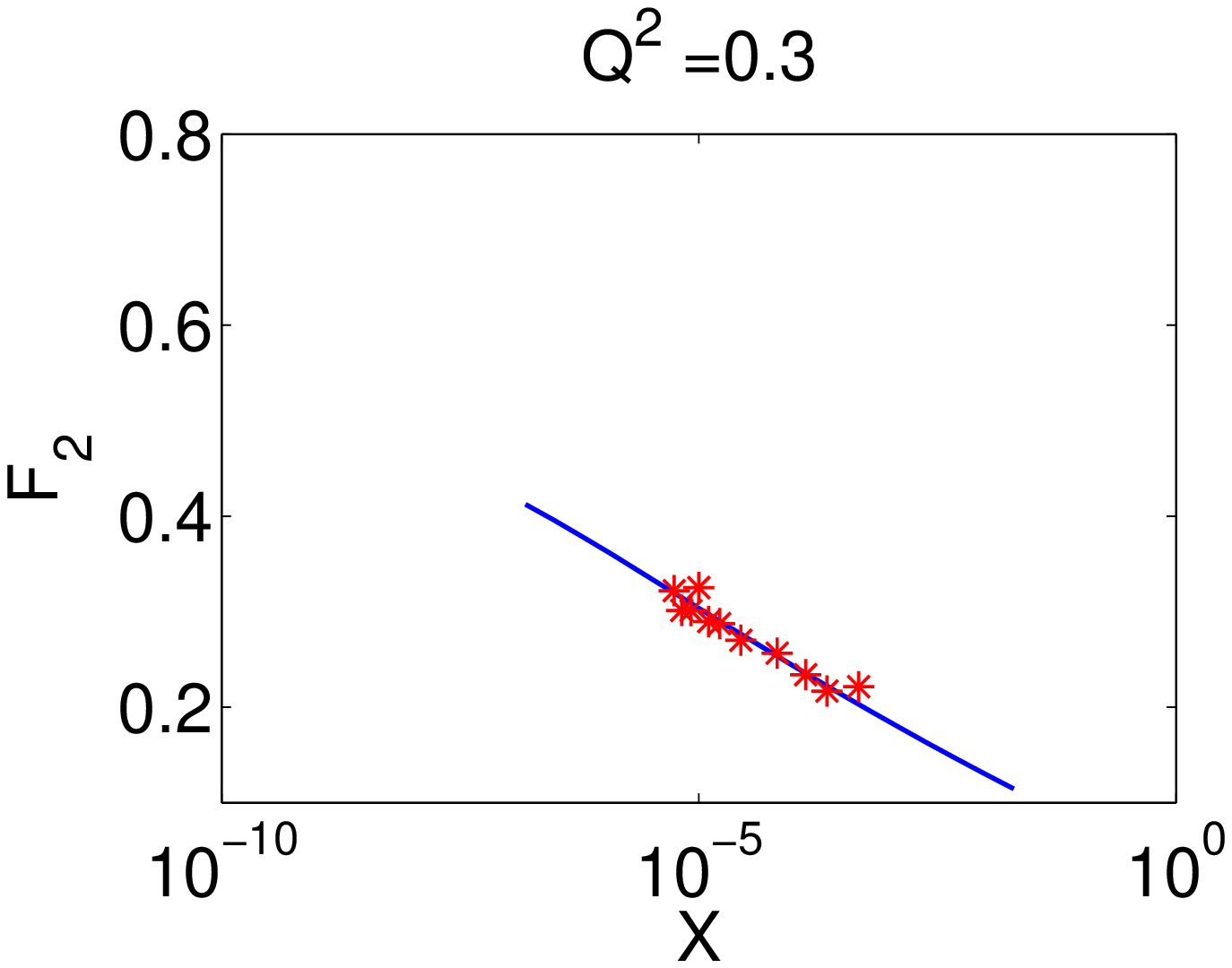,width=40mm, height=25mm}\\
\epsfig{file=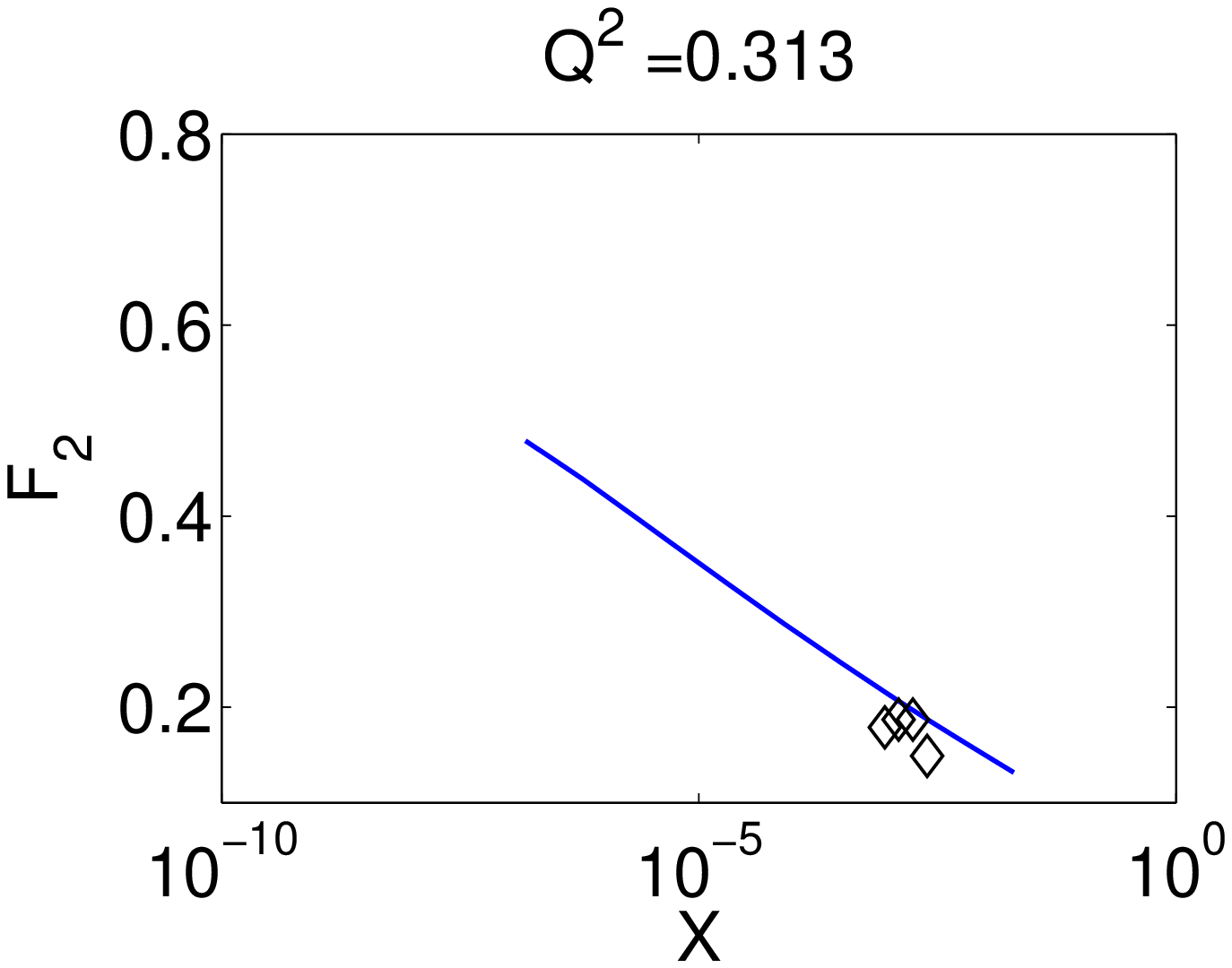,width=40mm, height=25mm}&
\epsfig{file=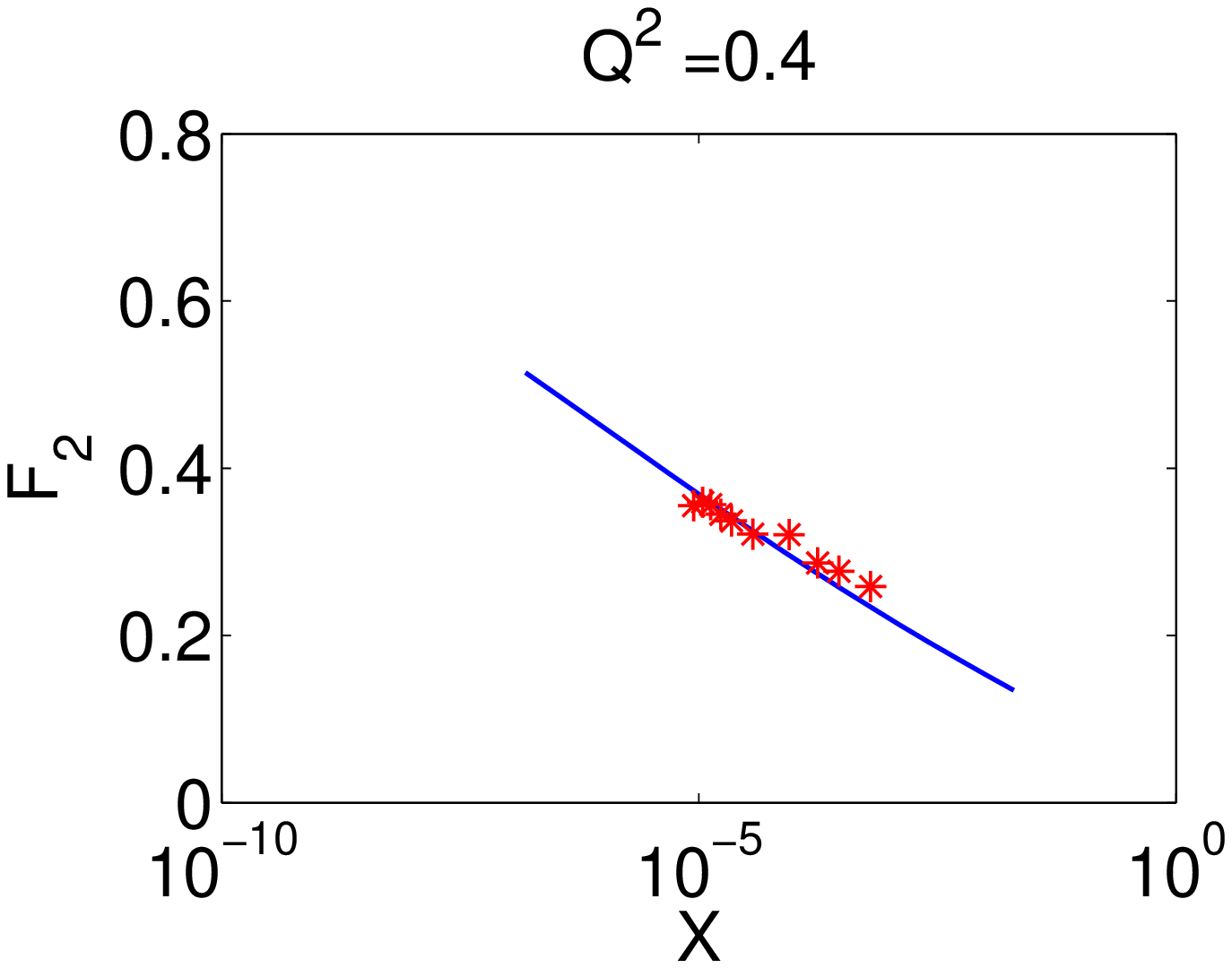,width=40mm, height=25mm}&
\epsfig{file=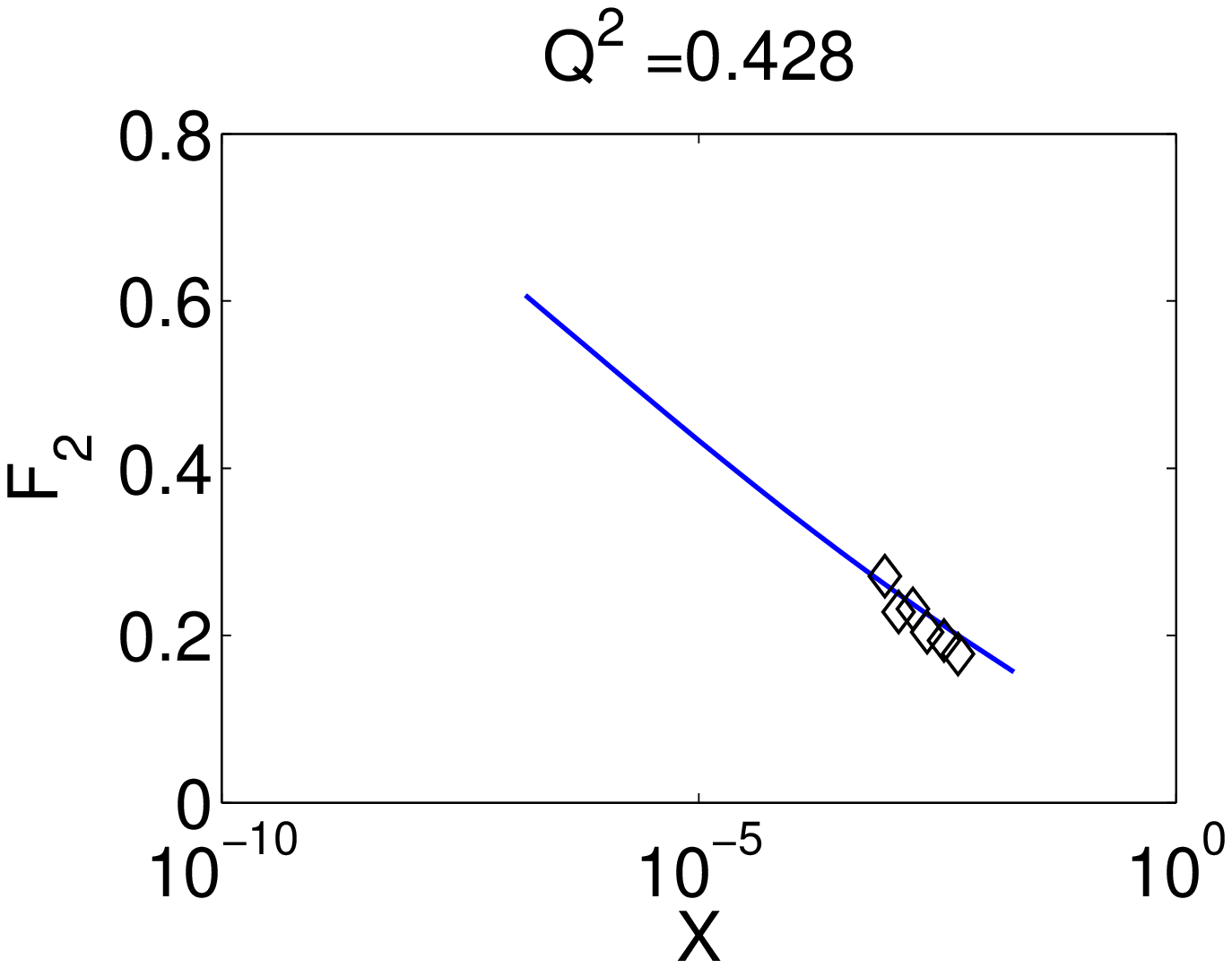,width=40mm, height=25mm}&
\epsfig{file=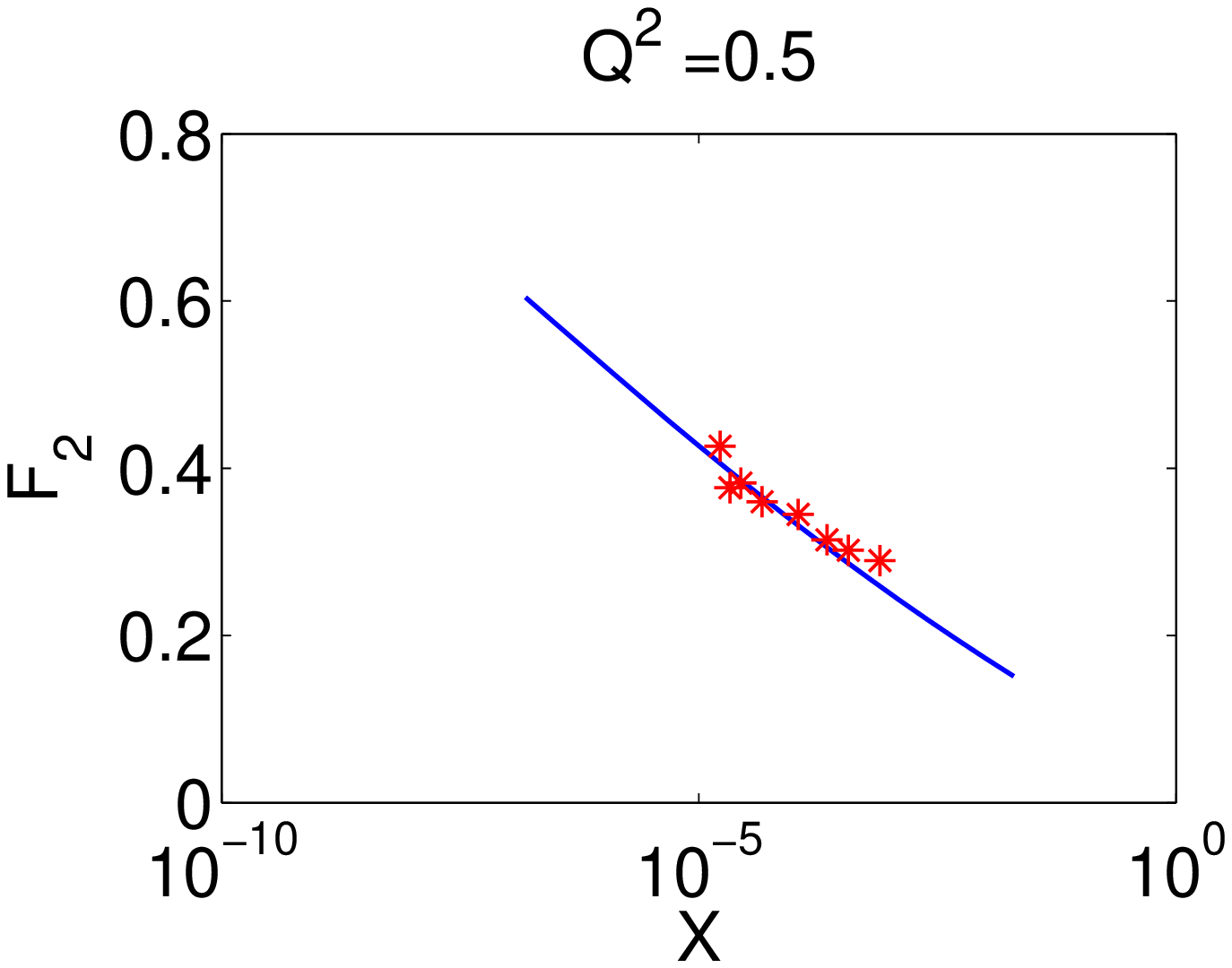,width=40mm, height=25mm}\\
\epsfig{file=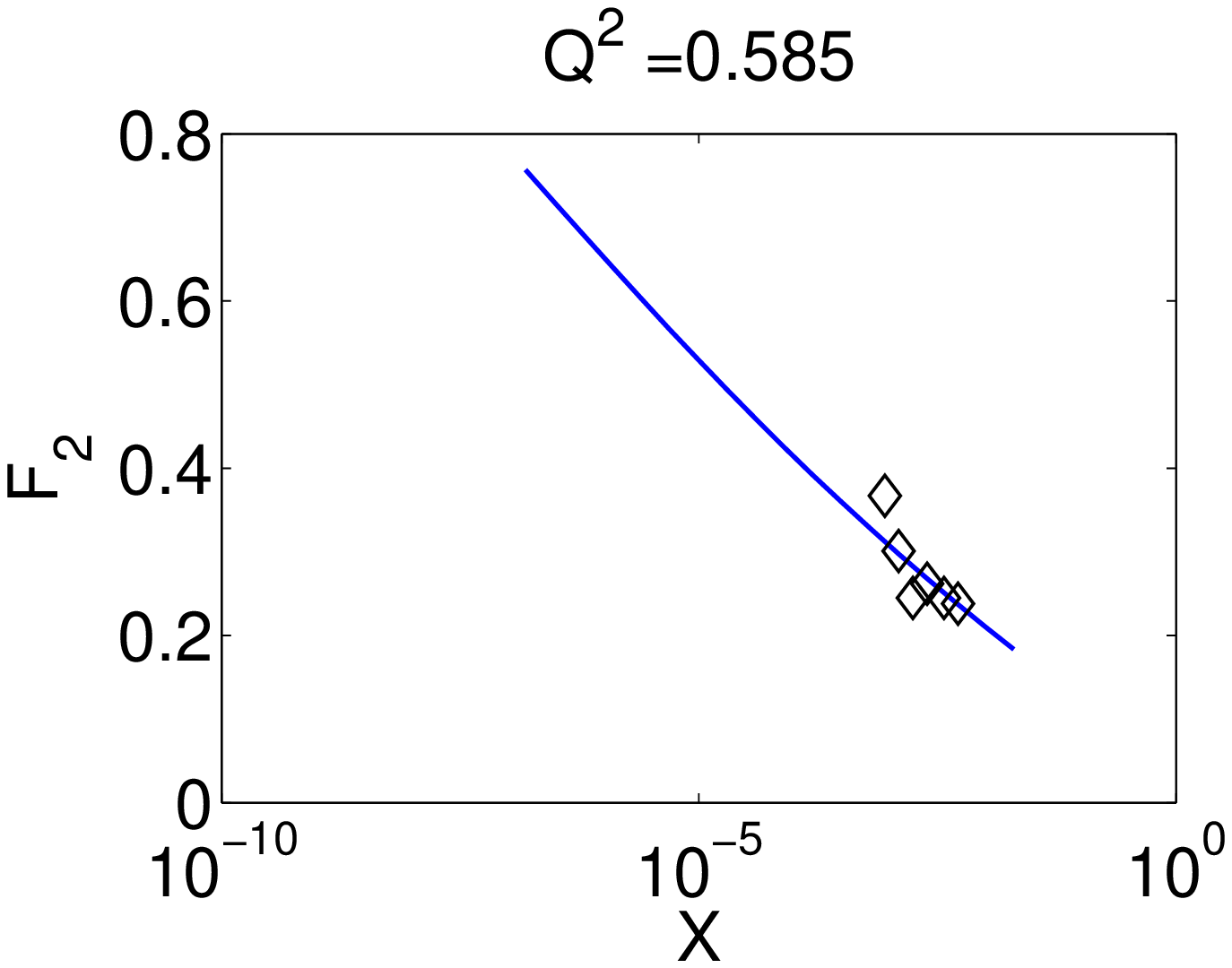,width=40mm, height=25mm}&
\epsfig{file=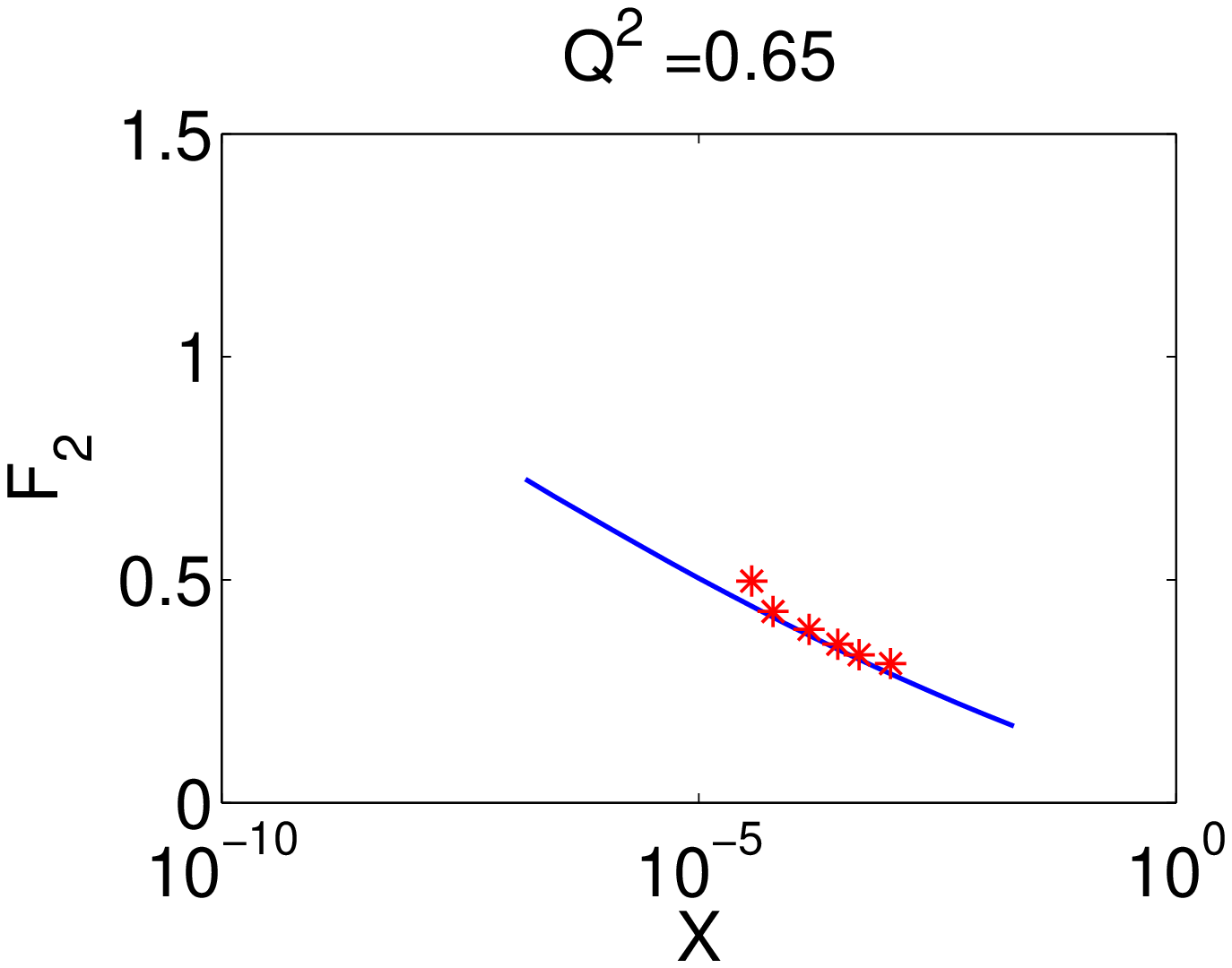,width=40mm, height=25mm}&
\epsfig{file=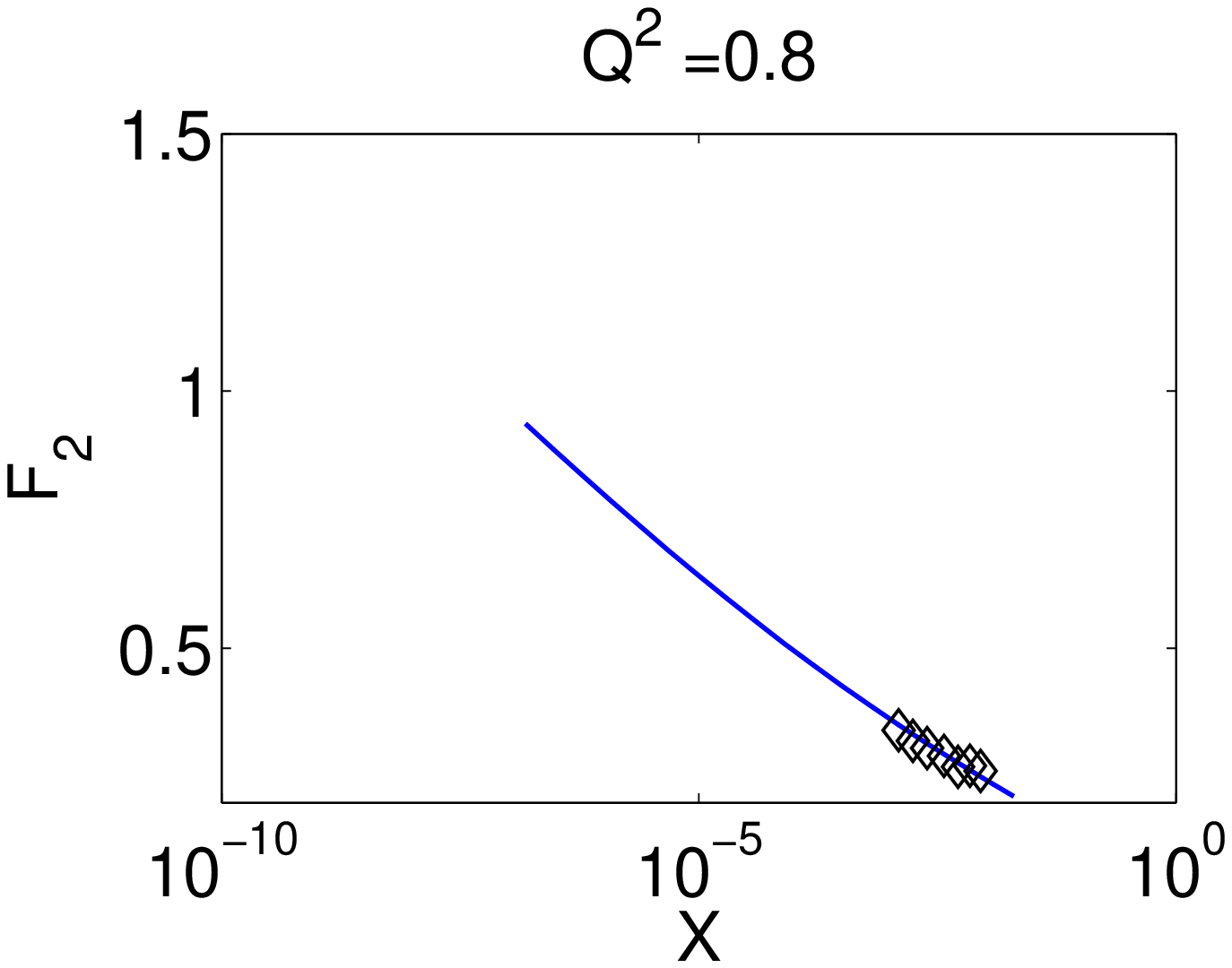,width=40mm, height=25mm}&
\epsfig{file=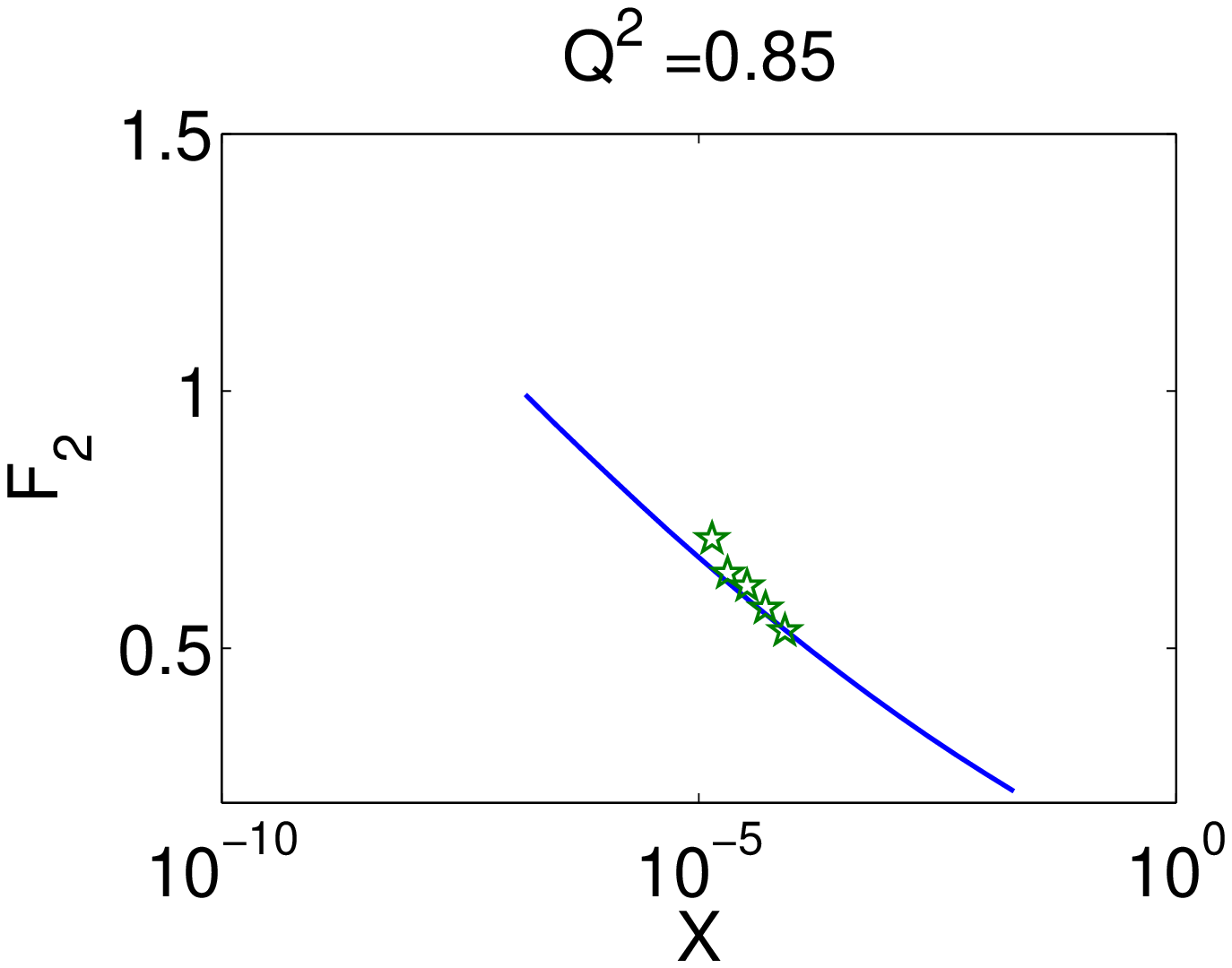,width=40mm, height=25mm}\\
\epsfig{file=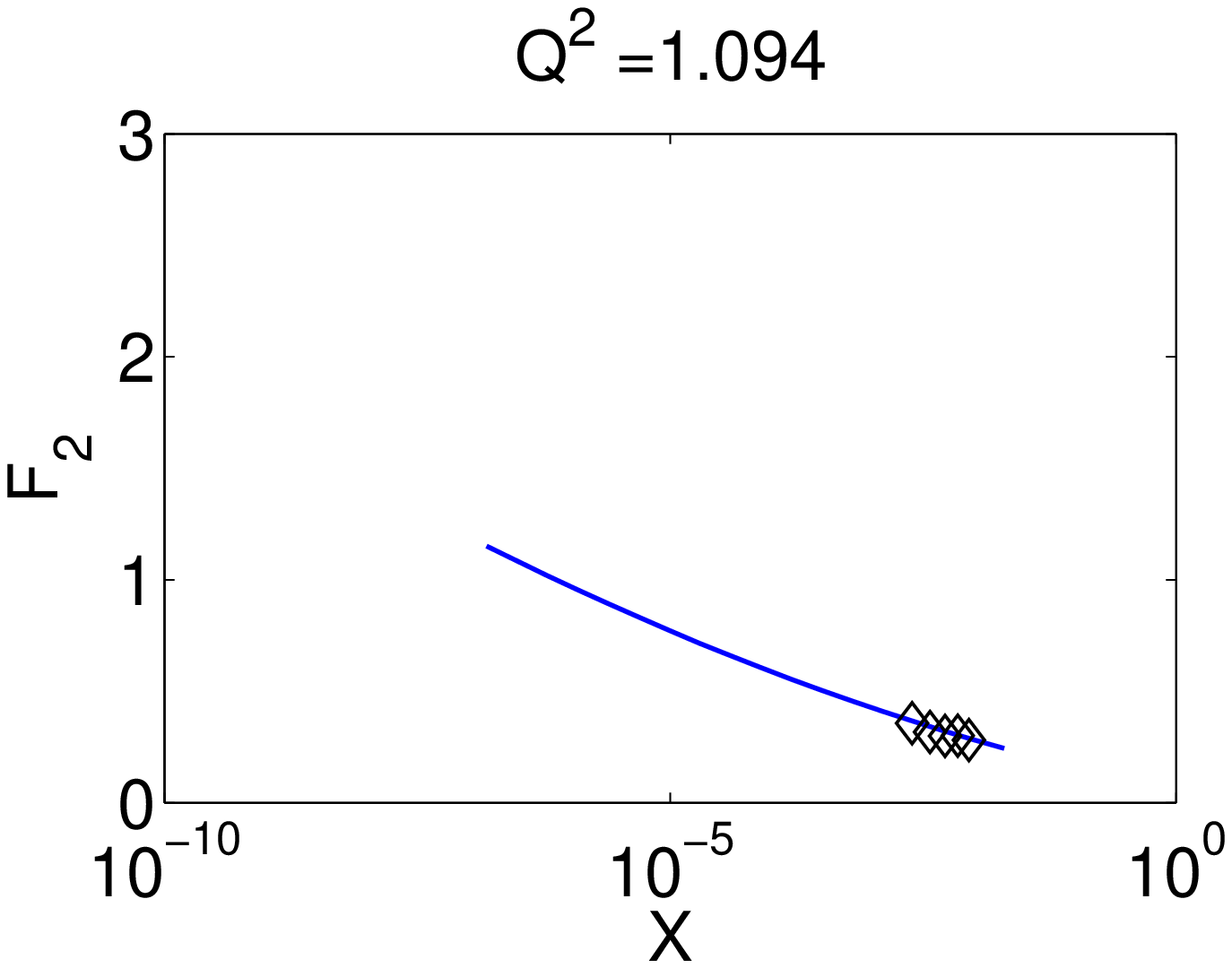,width=40mm, height=25mm}&
\epsfig{file=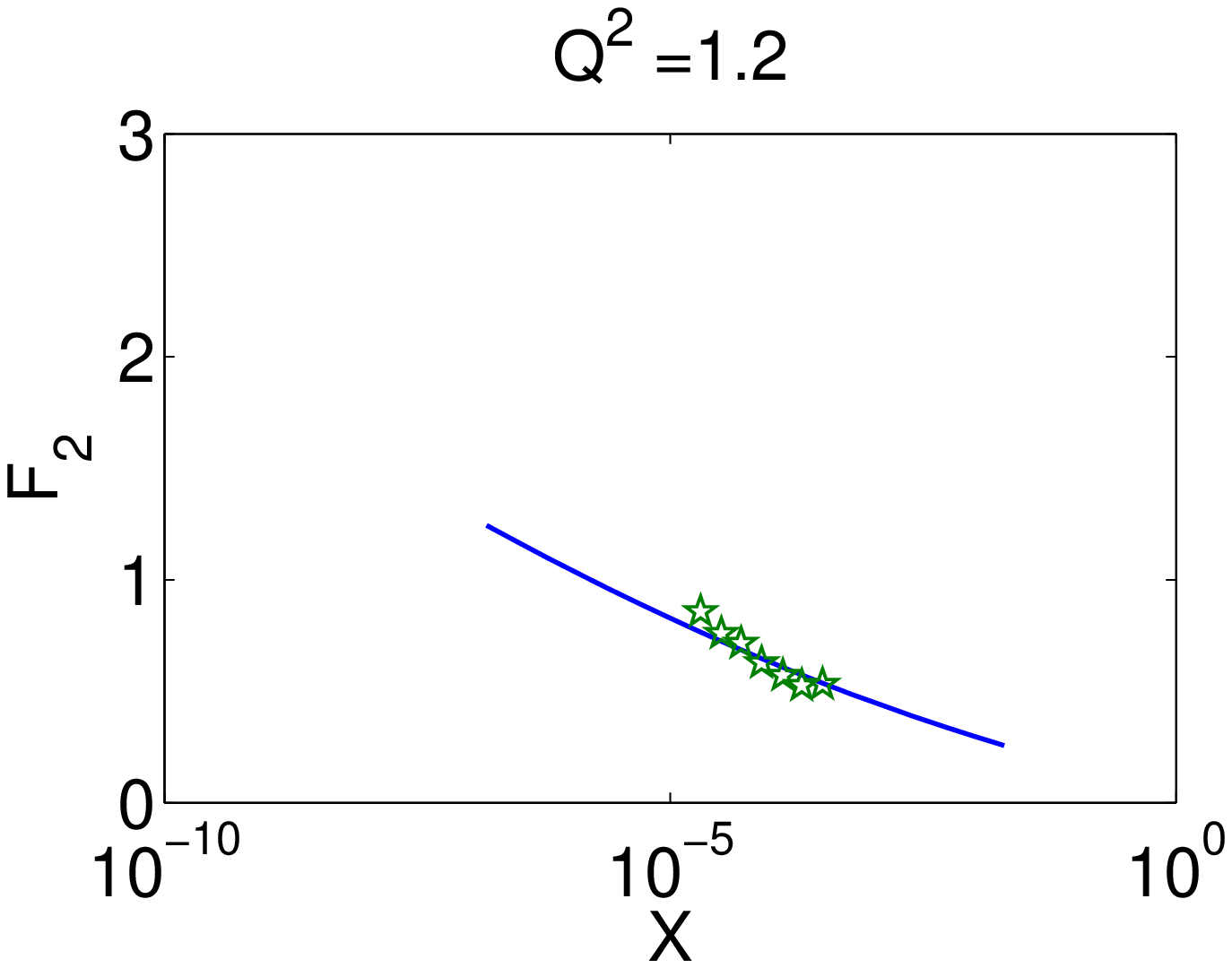,width=40mm, height=25mm}&
\epsfig{file=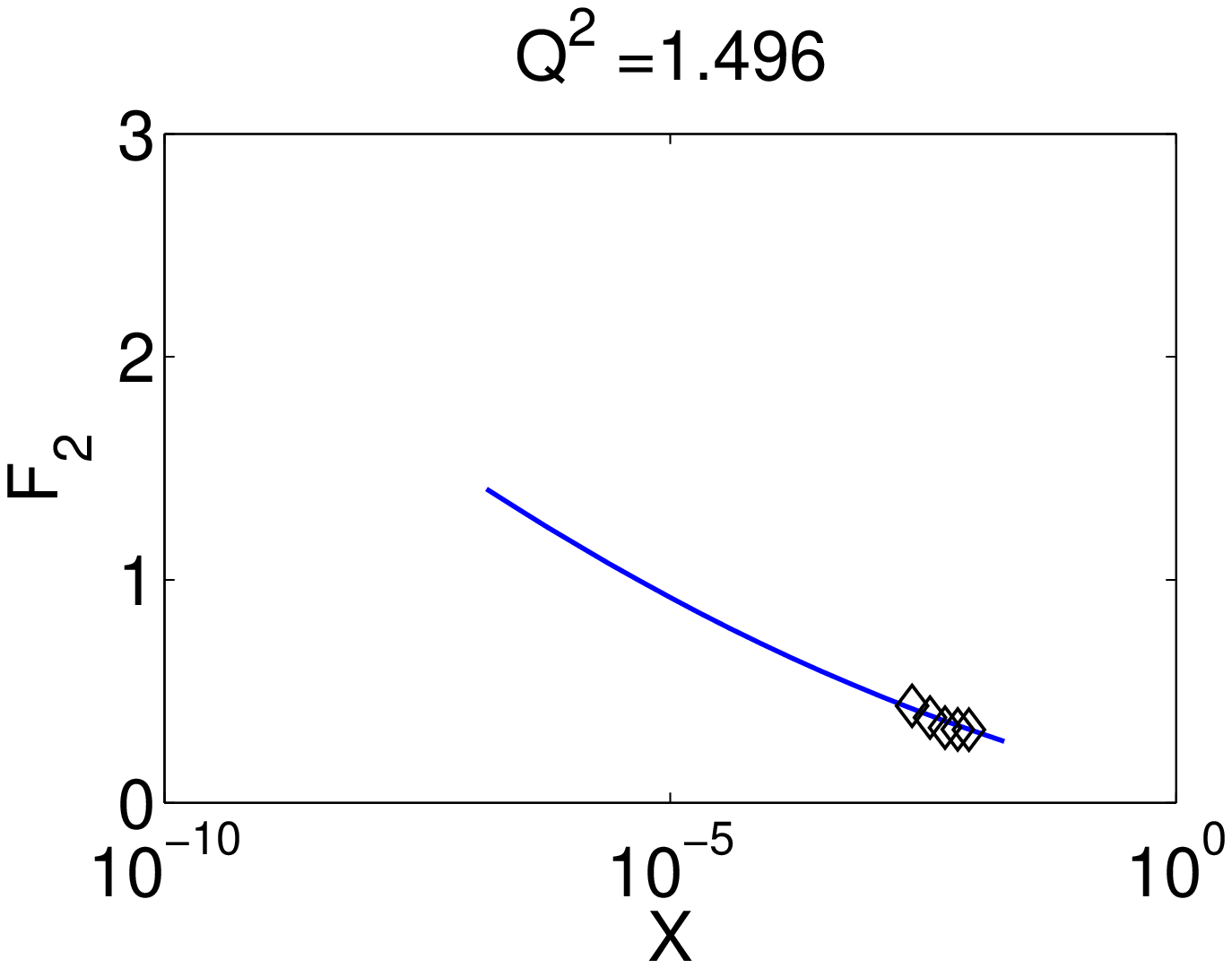,width=40mm, height=25mm}&
\epsfig{file=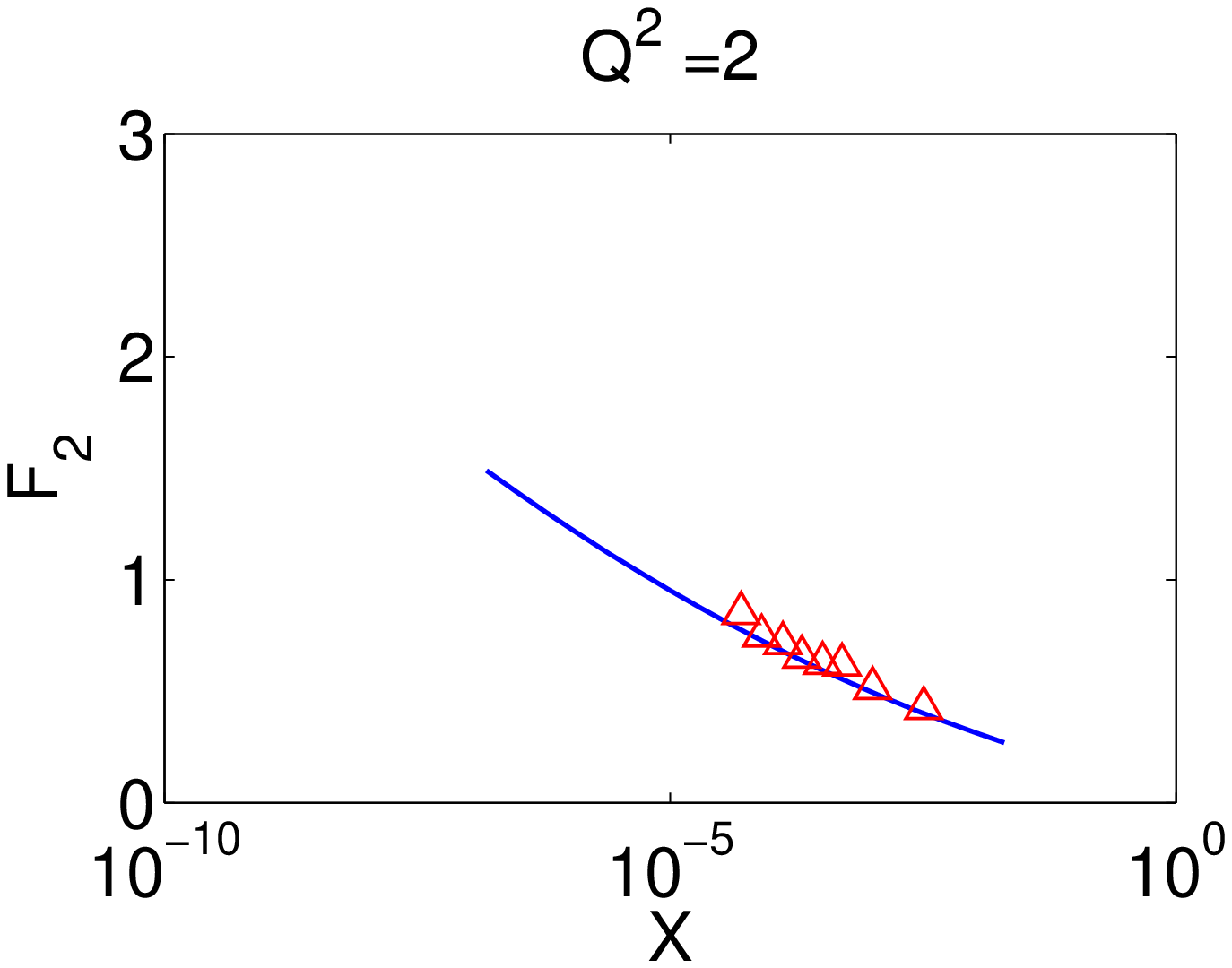,width=40mm, height=25mm}\\
\epsfig{file=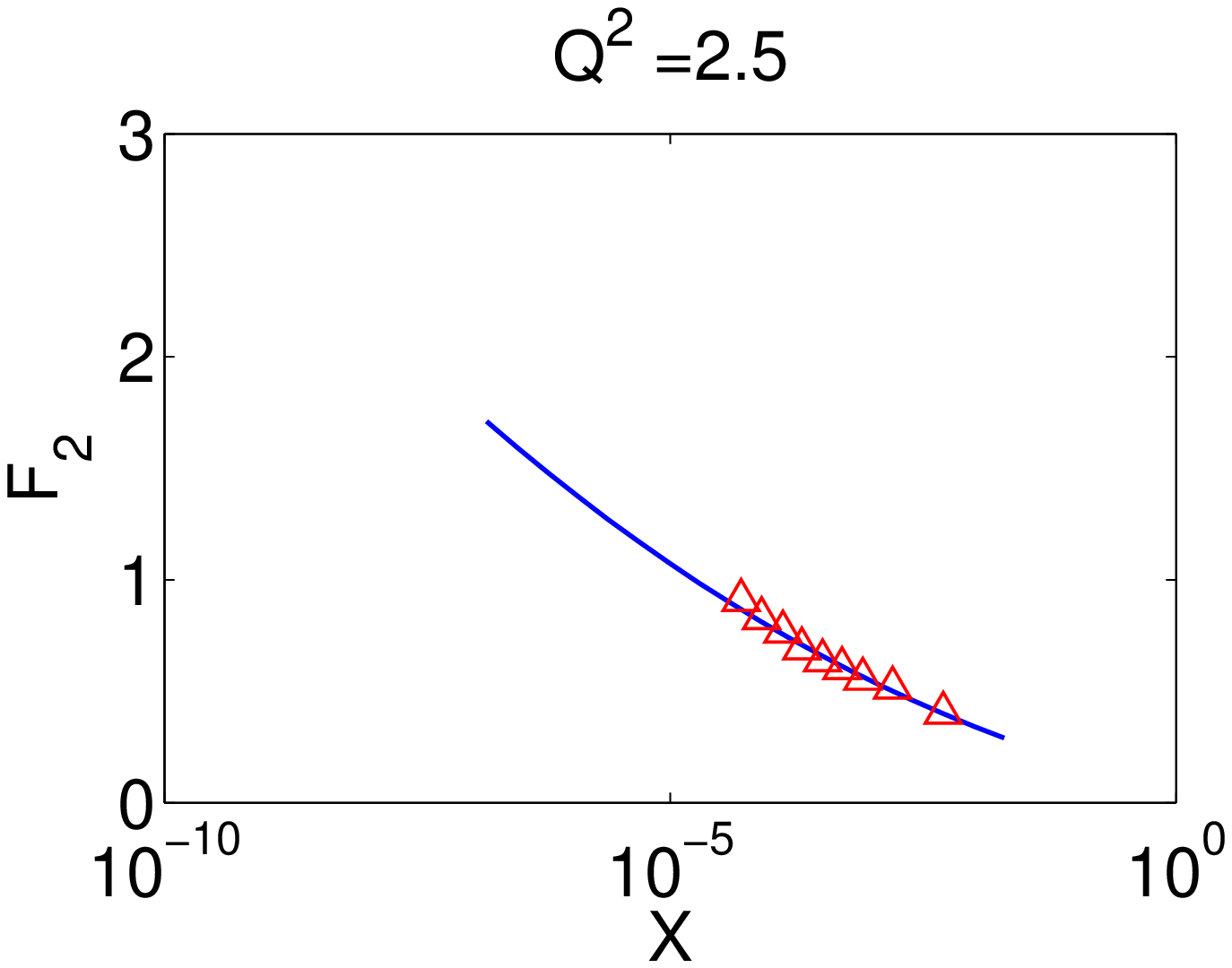,width=40mm, height=24mm}&
\epsfig{file=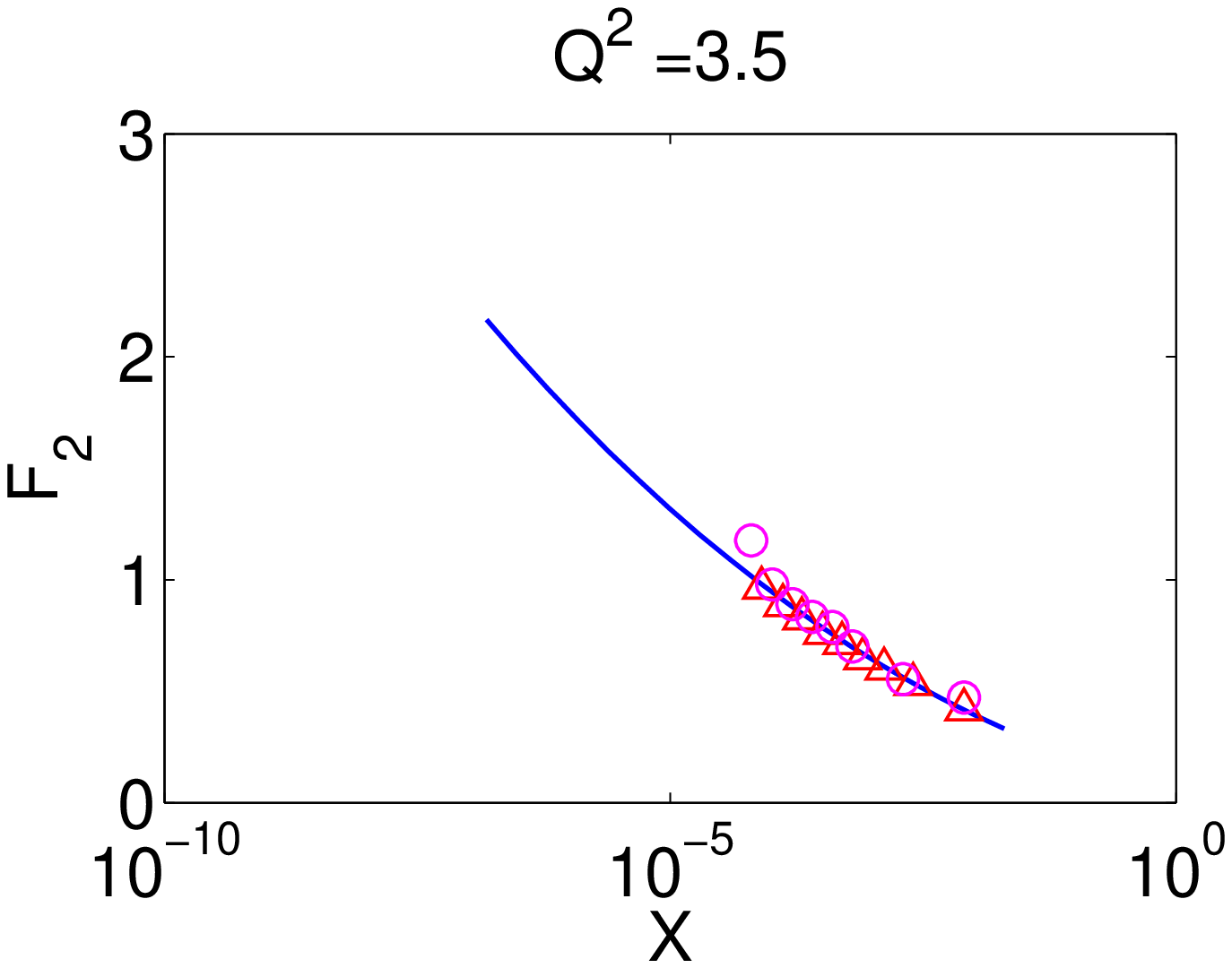,width=40mm, height=24mm}&
\epsfig{file=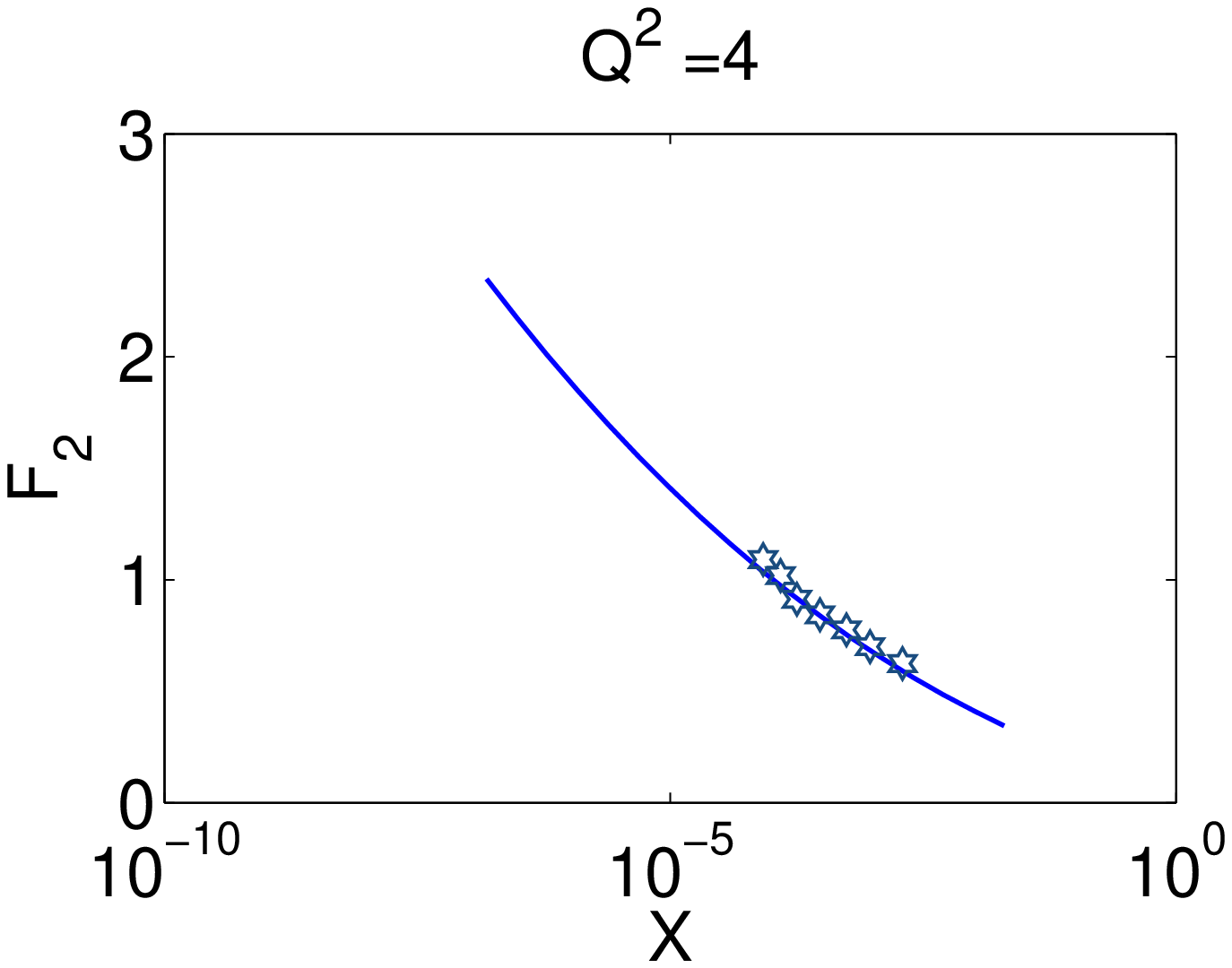,width=40mm, height=24mm}&
\epsfig{file=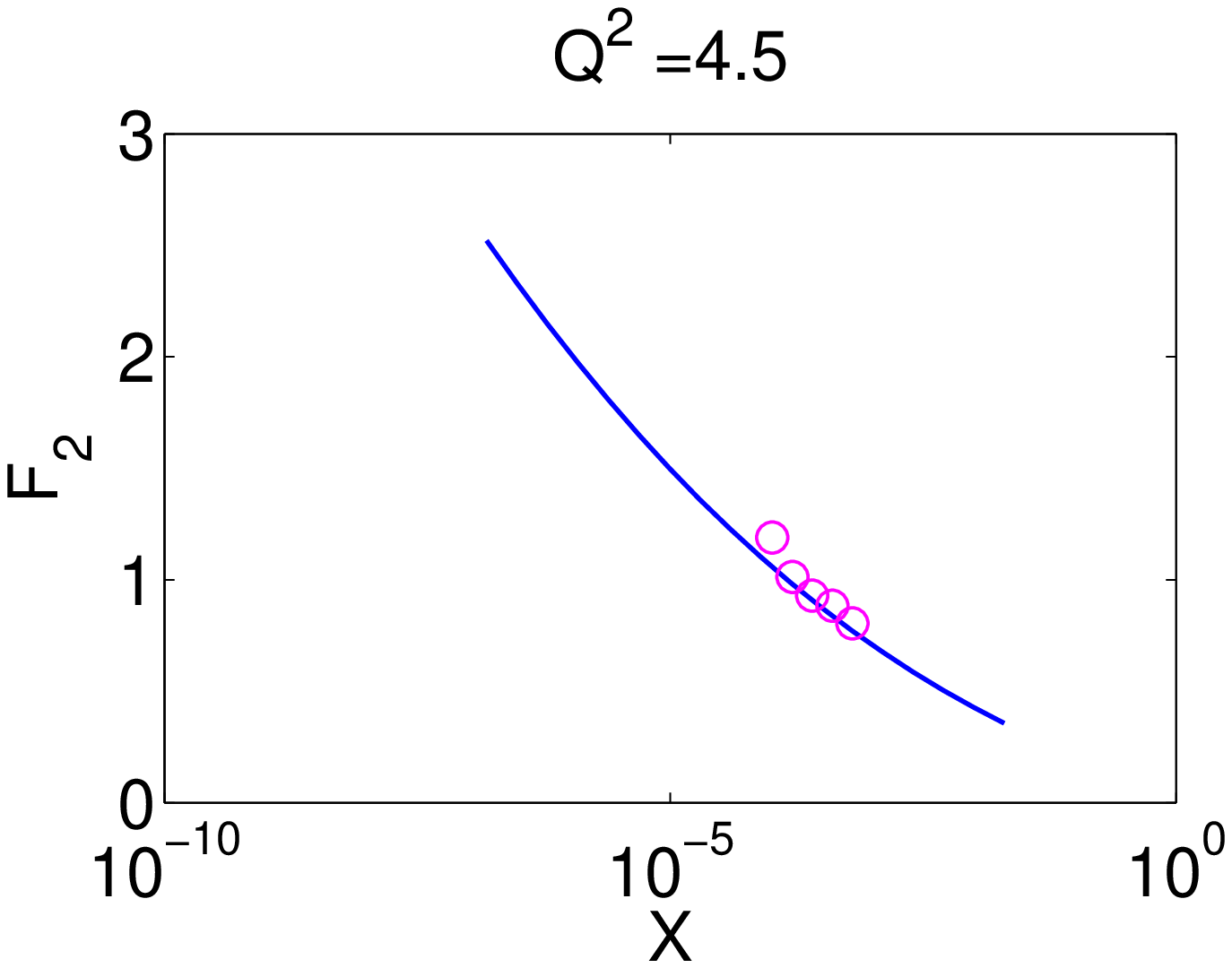,width=40mm, height=24mm}\\
\epsfig{file=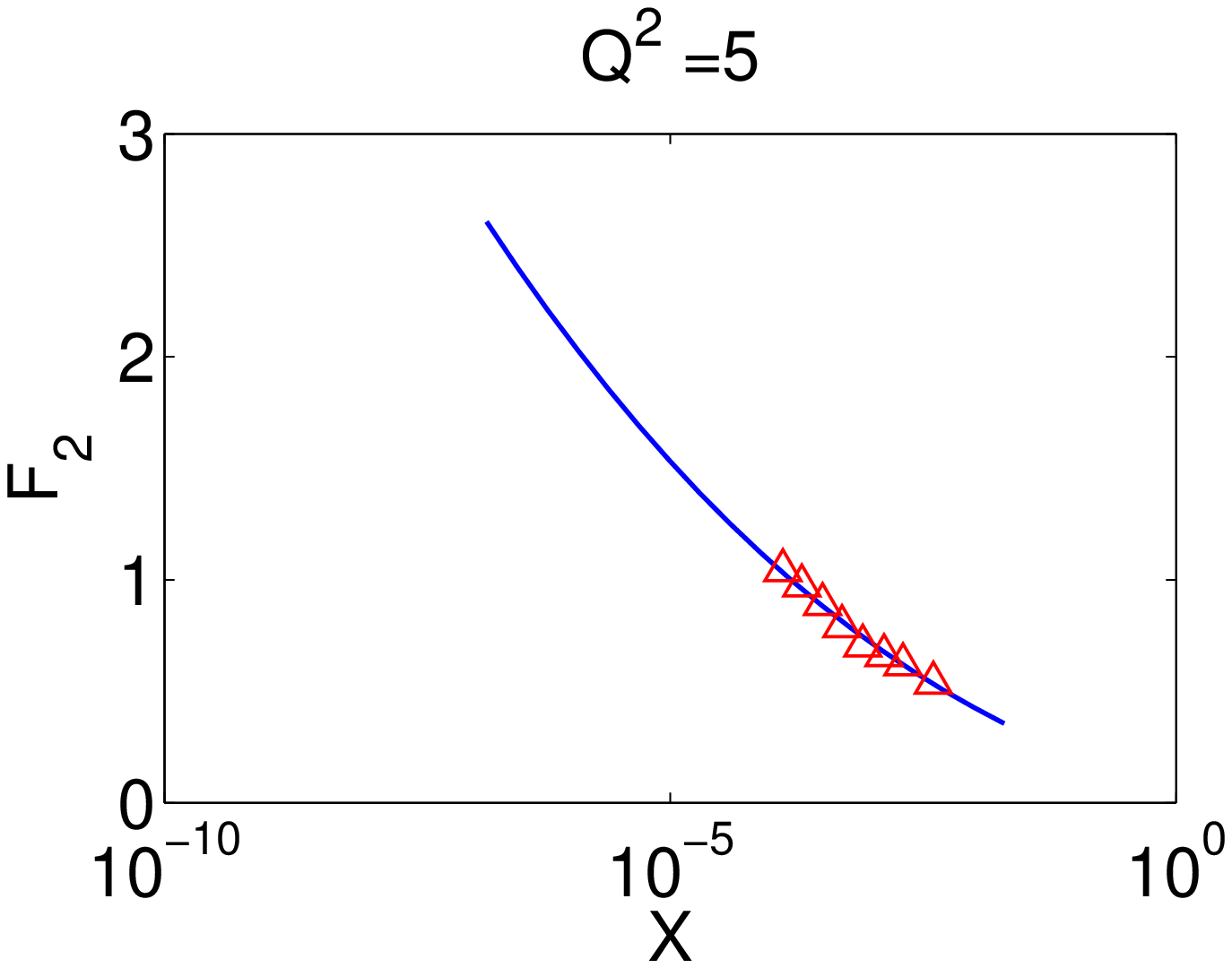,width=40mm, height=24mm}&
\epsfig{file=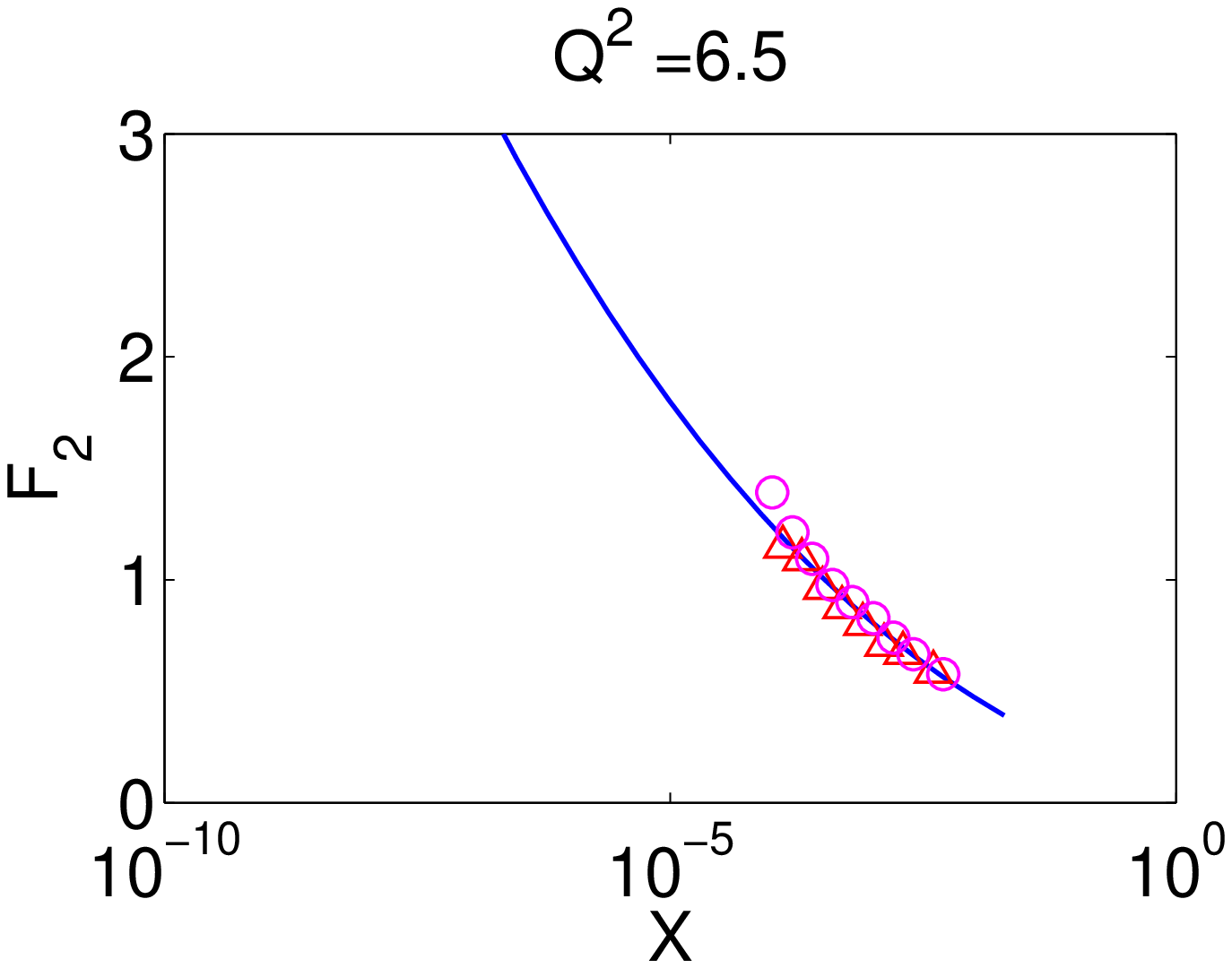,width=40mm, height=24mm}&
\epsfig{file=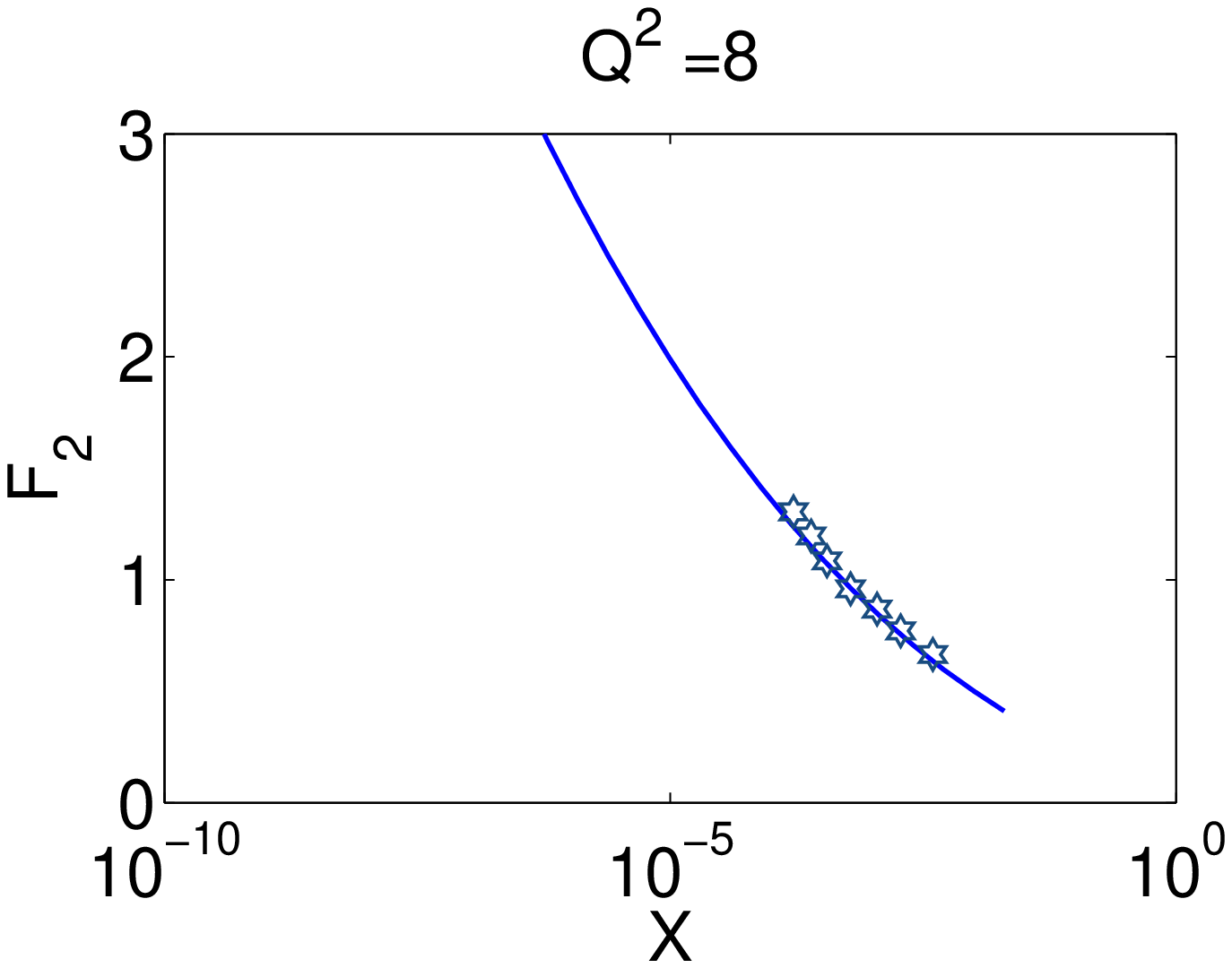,width=40mm, height=24mm}&
\epsfig{file=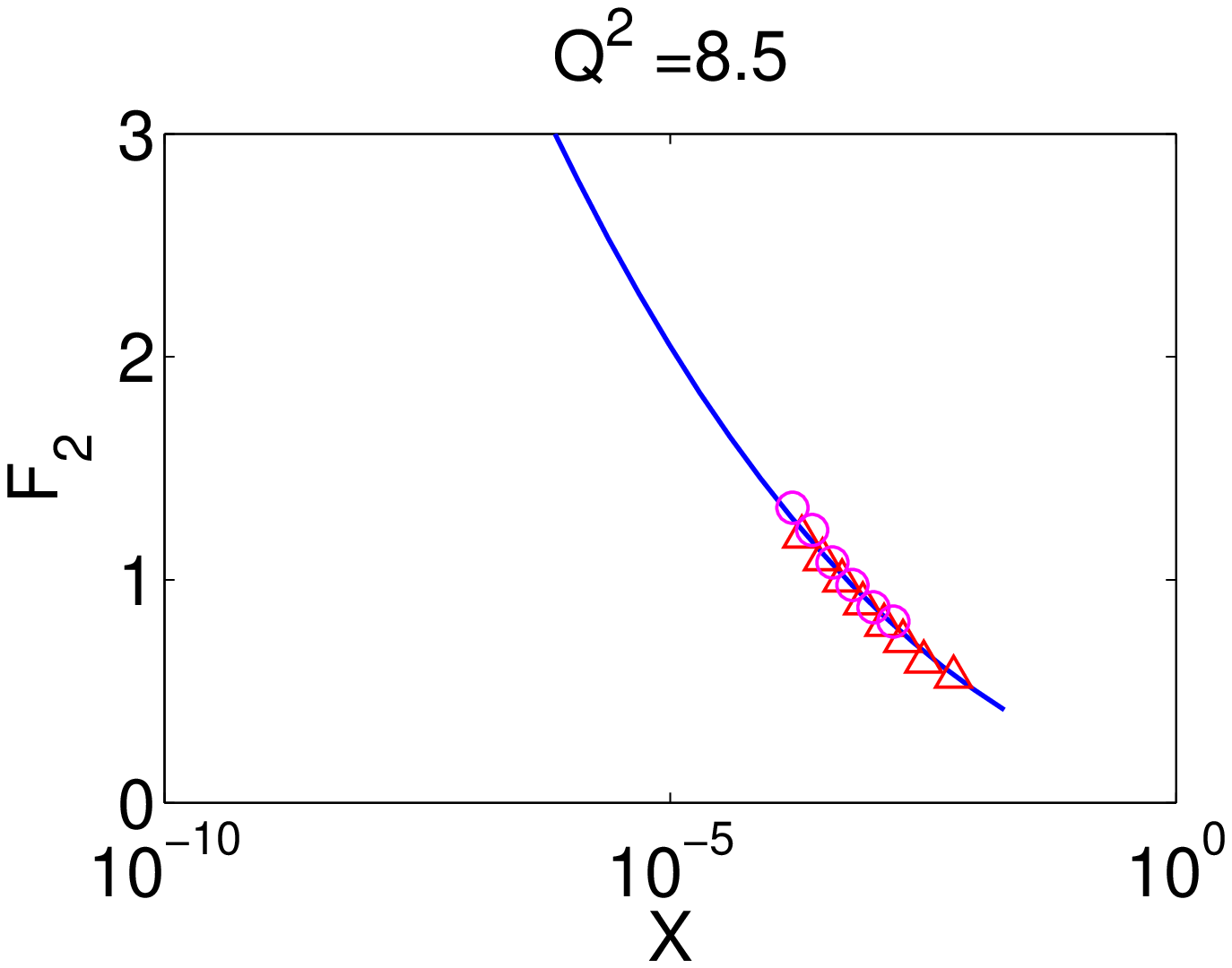,width=40mm, height=24mm}\\
\end{tabular}
\caption{\it $F_{2}(x,Q^{2})$ structure function of proton, as a
function of Bjorken $x$ for fixed value of photon virtuality
$Q^{2}$. This data was taken for small values of $x<0.01$. These
data points correspond to different collaborations. Asterisk
corresponds to \cite{Breitweg:2000yn}, diamond to \cite{E665}
collaboration, pentagram to \cite{Adloff:1997mf}, triangles
\cite{Adloff:2000qk}, circles correspond to \cite{Chekanov:2001qu}
and hexagons to \cite{Chekanov:2005vv}.}\label{DIS_1}
\end{center}
\end{figure}

\begin{figure}[htbp]
\begin{center}
\begin{tabular}{ccccc ccccc ccccc ccccc}
\epsfig{file=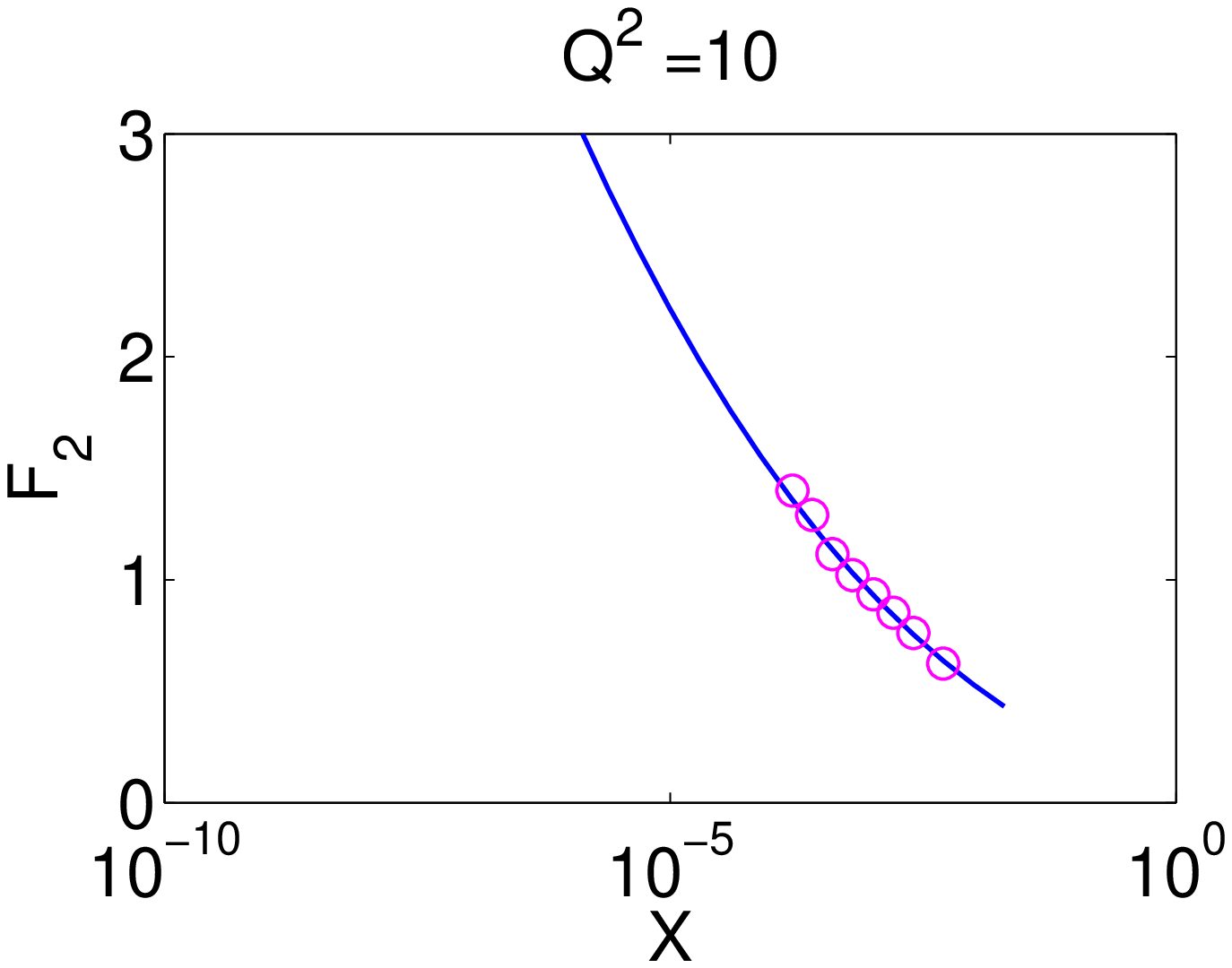,width=40mm, height=25mm}&
\epsfig{file=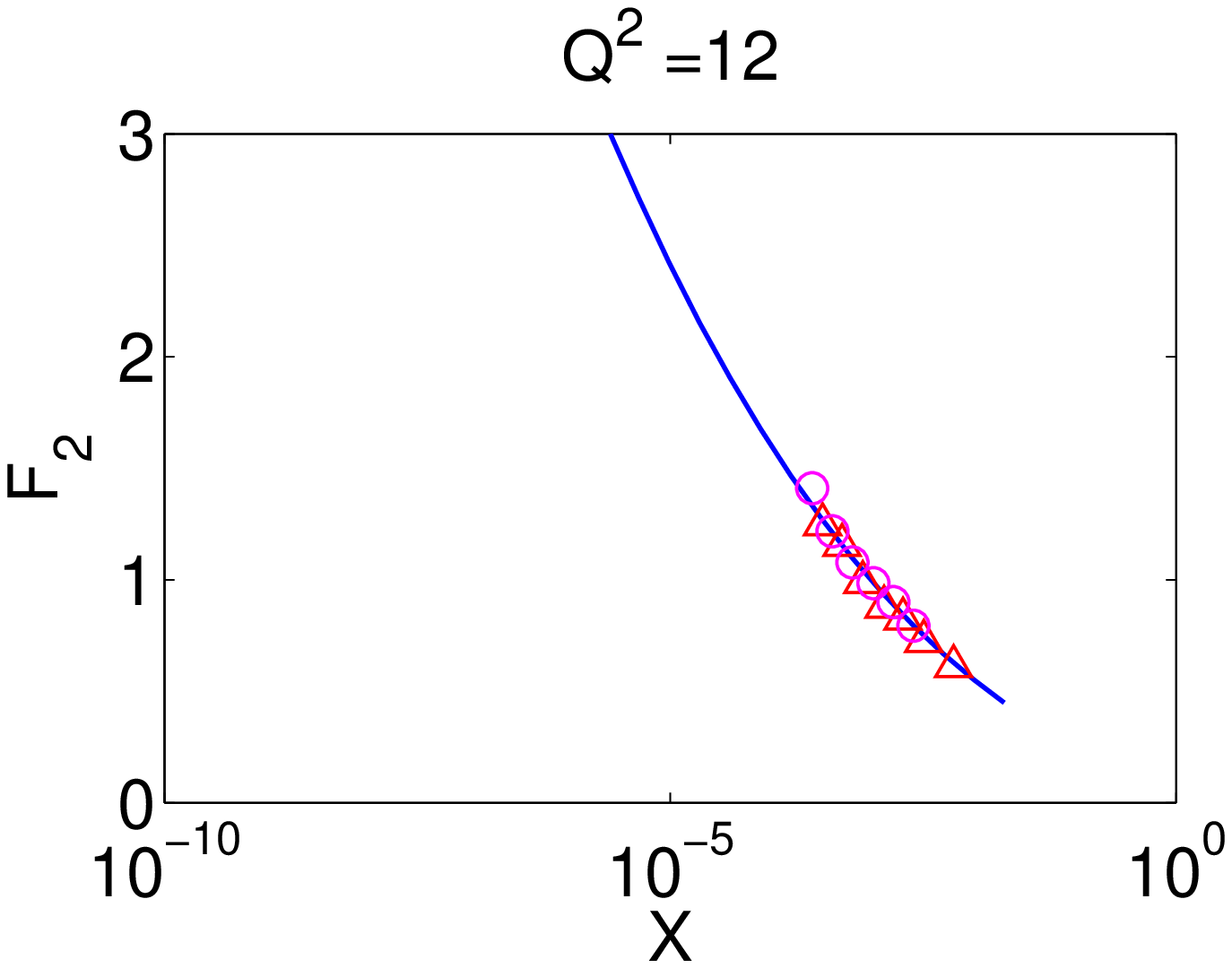,width=40mm, height=25mm}&
\epsfig{file=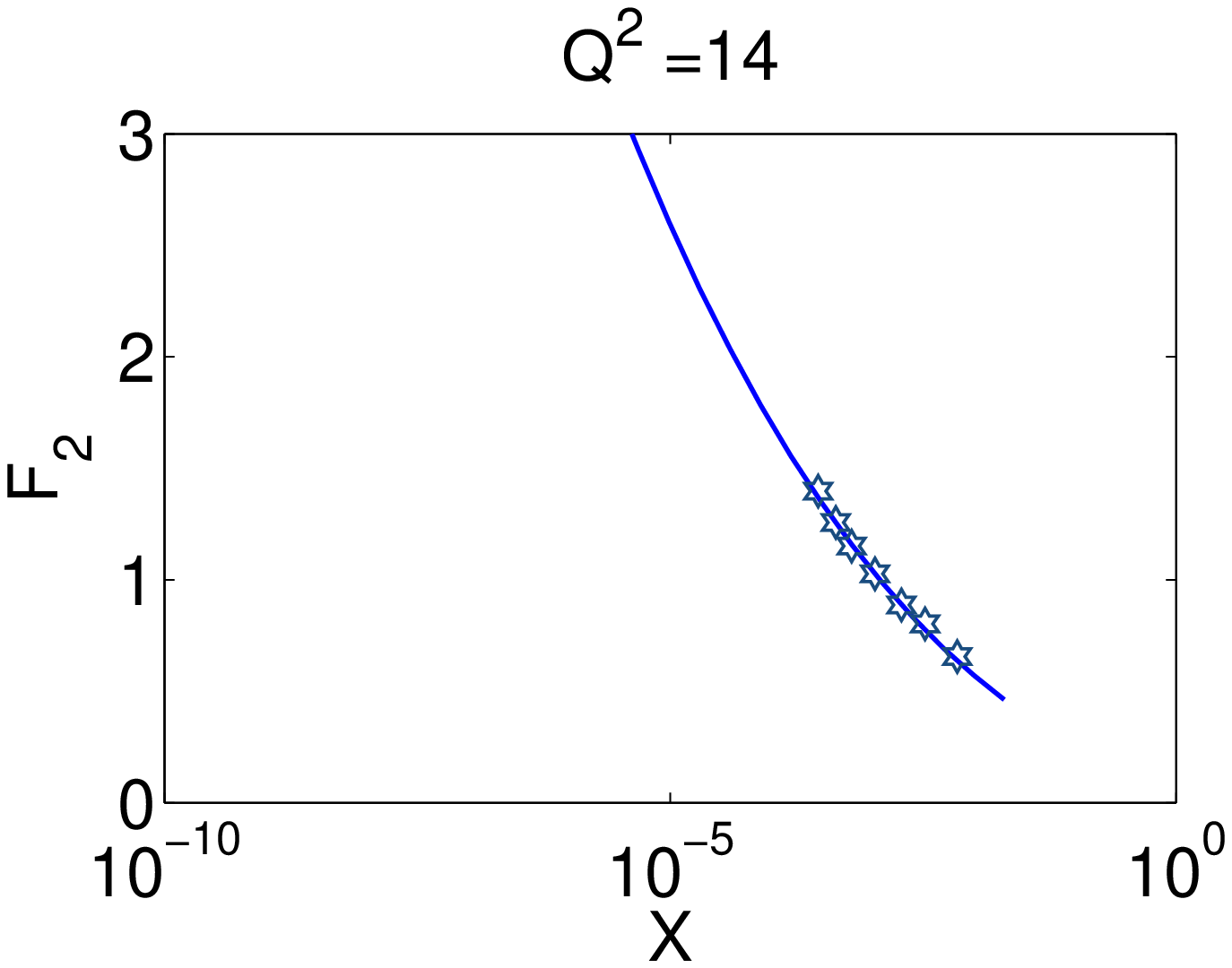,width=40mm, height=25mm}&
\epsfig{file=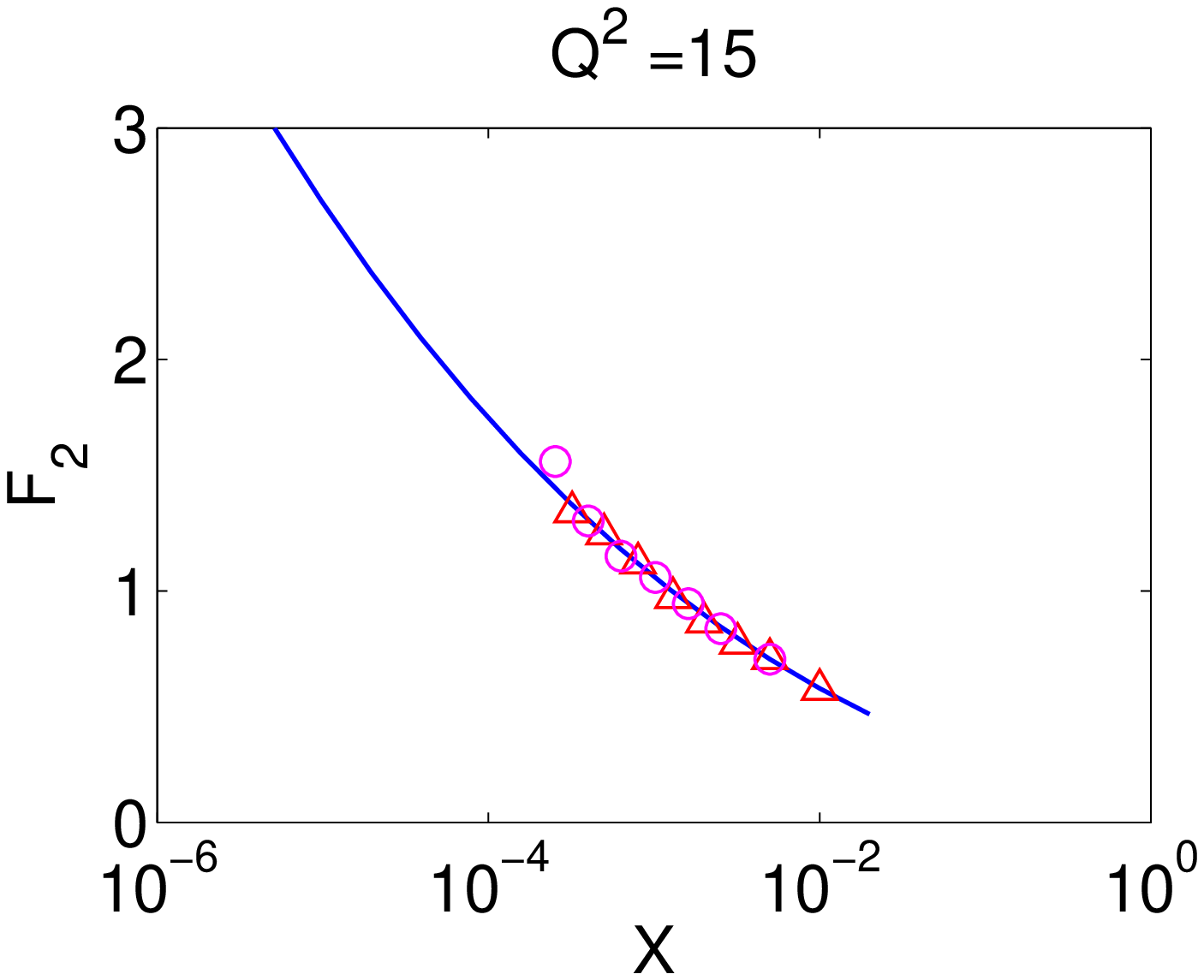,width=40mm, height=25mm}\\
\epsfig{file=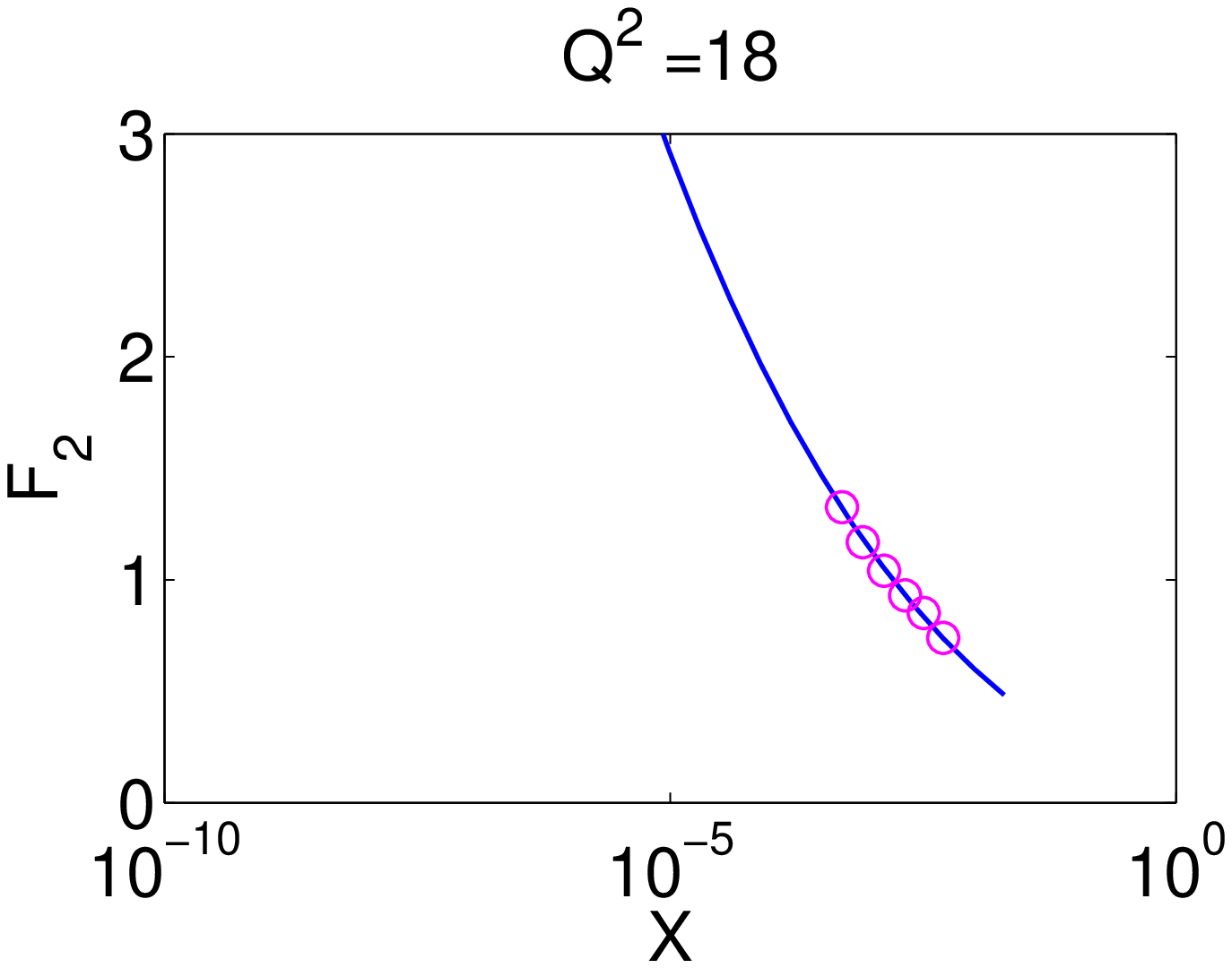,width=40mm, height=25mm}&
\epsfig{file=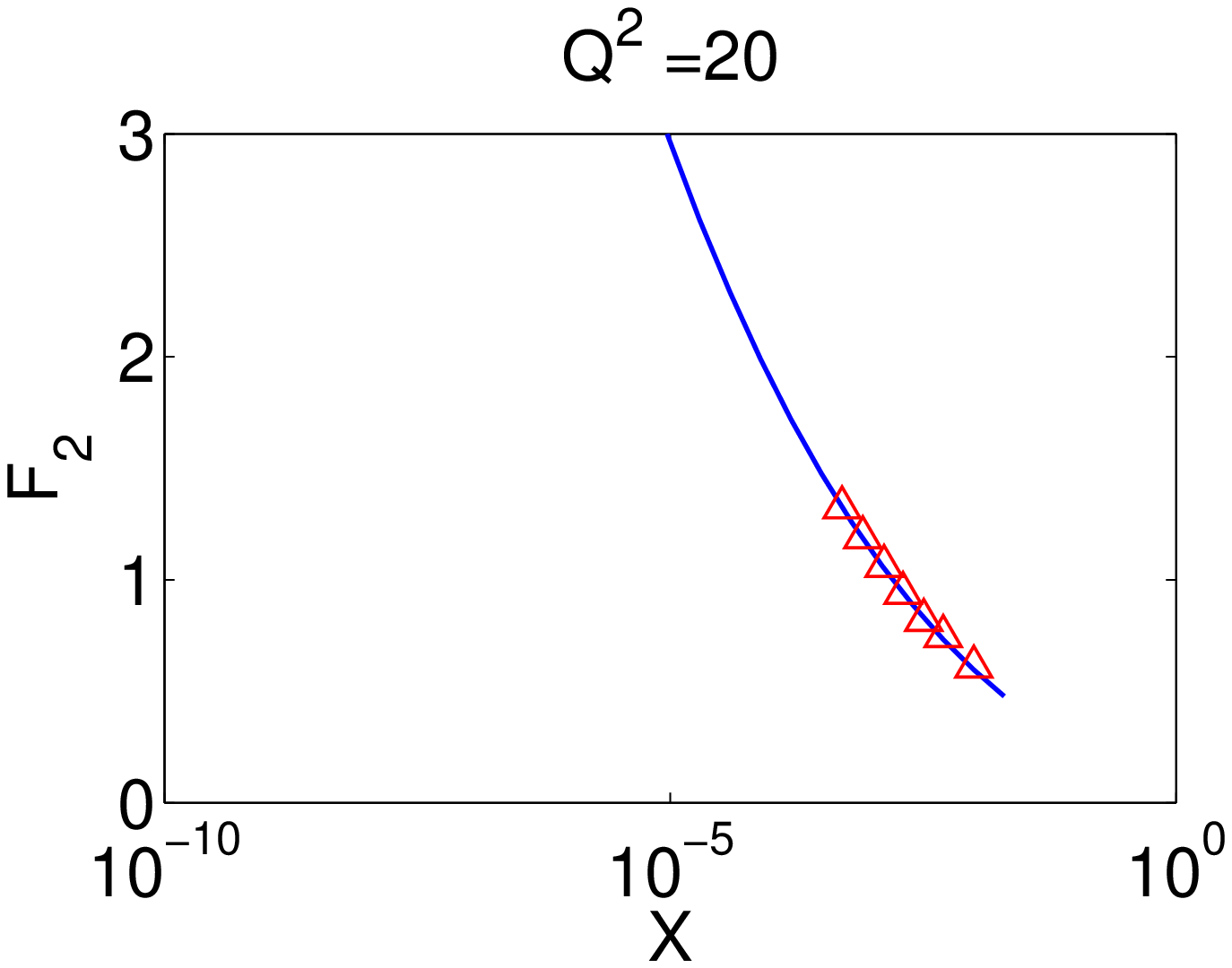,width=40mm, height=25mm}&
\epsfig{file=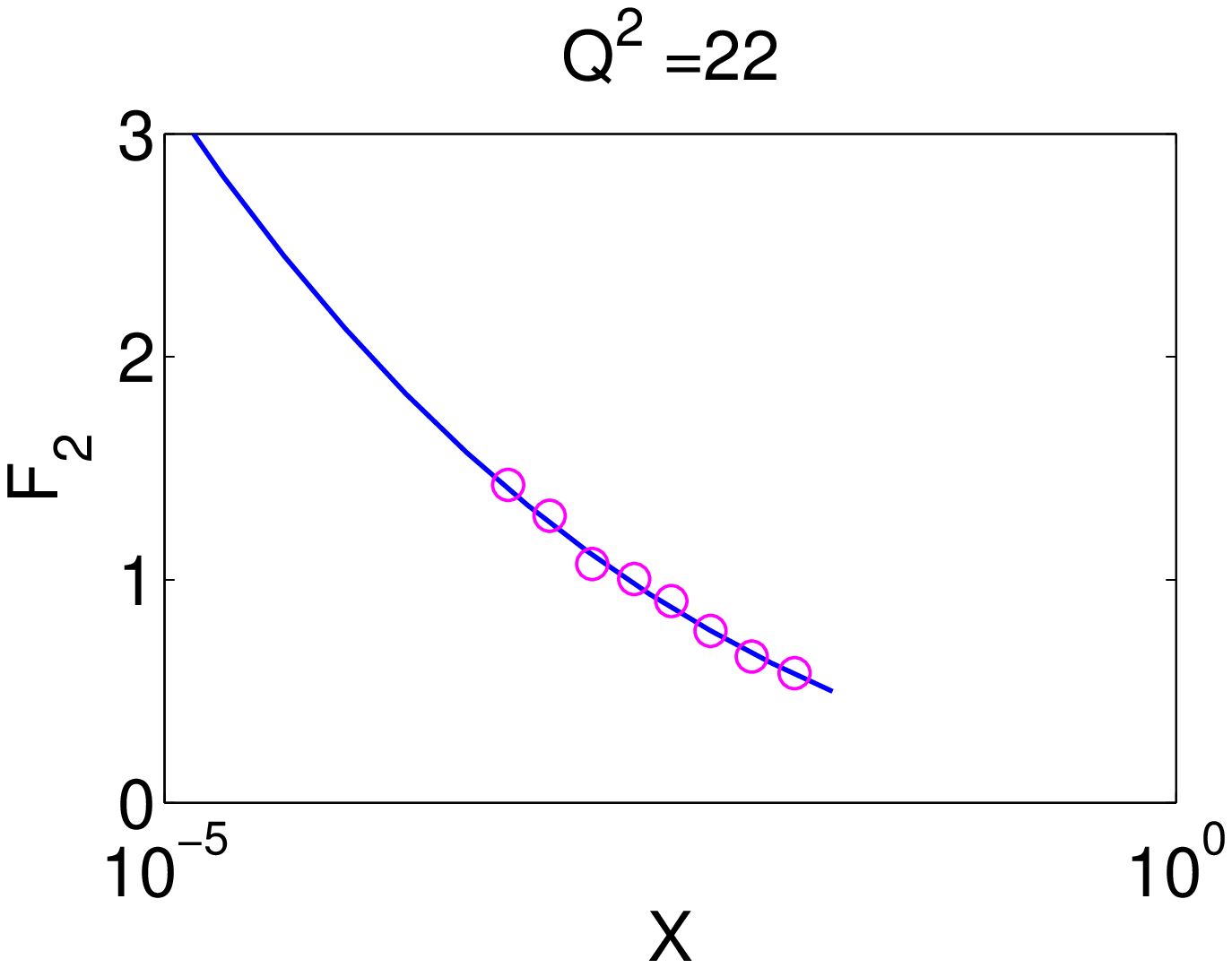,width=40mm, height=25mm}&
\epsfig{file=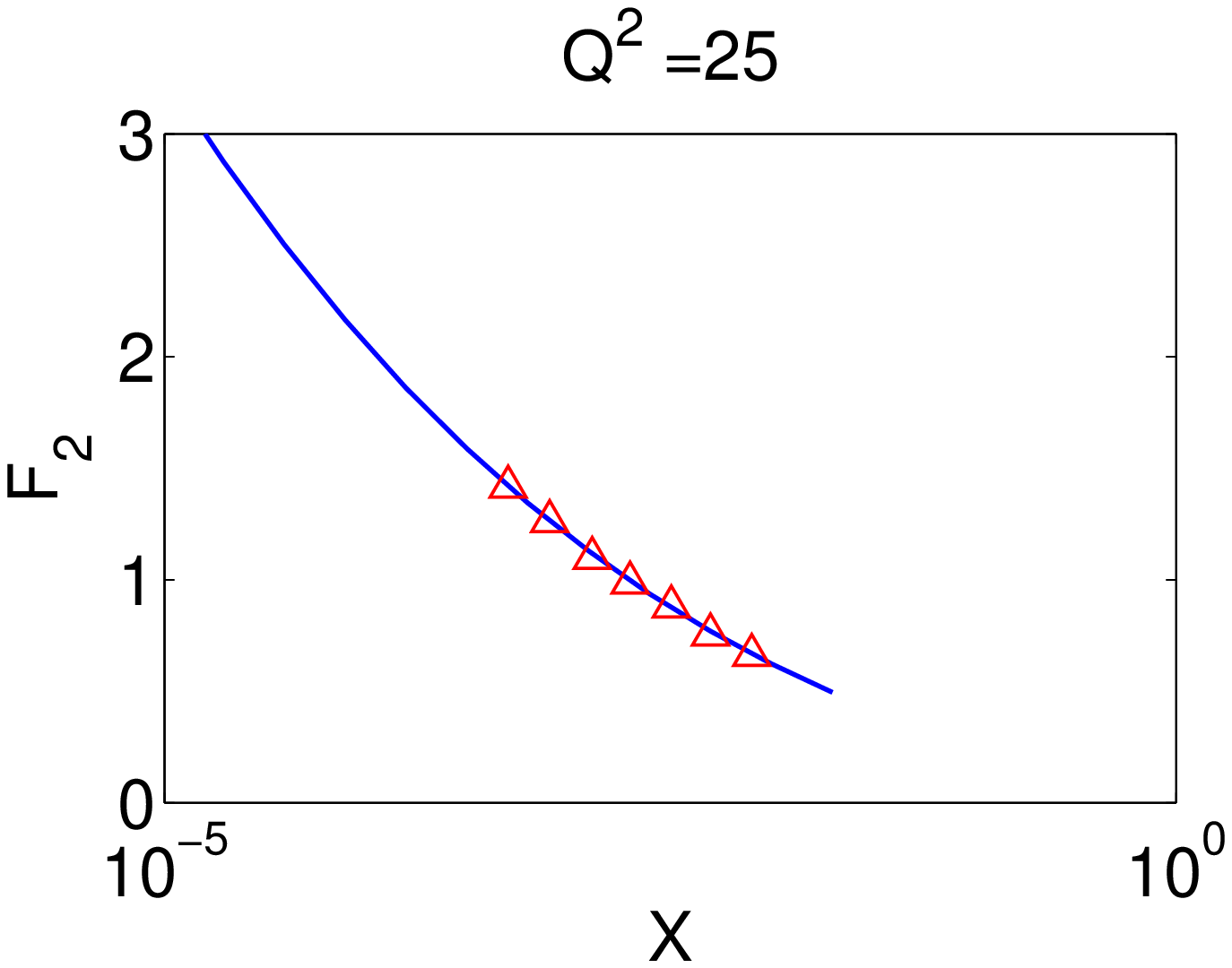,width=40mm, height=25mm}\\
\epsfig{file=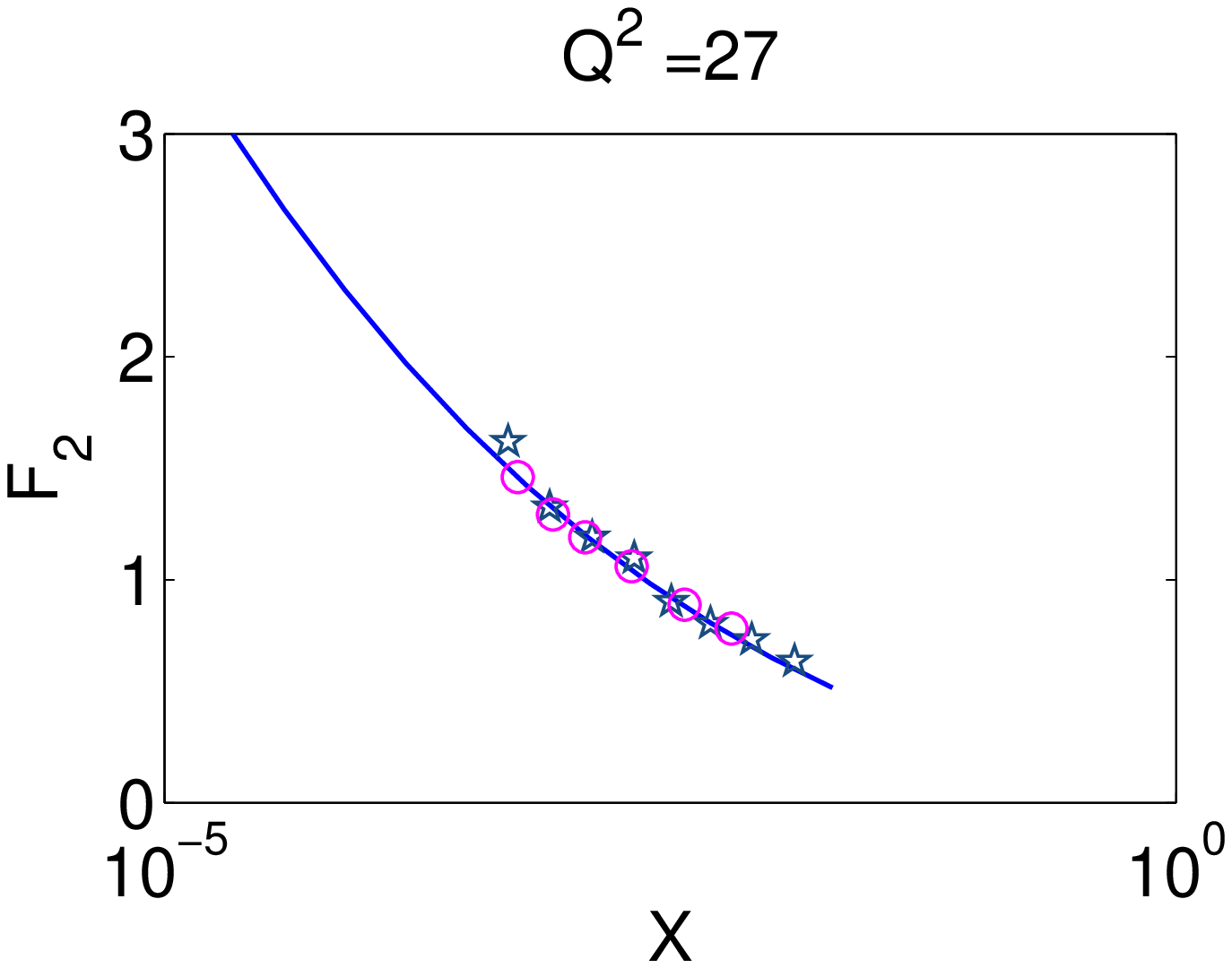,width=40mm, height=25mm}&
\epsfig{file=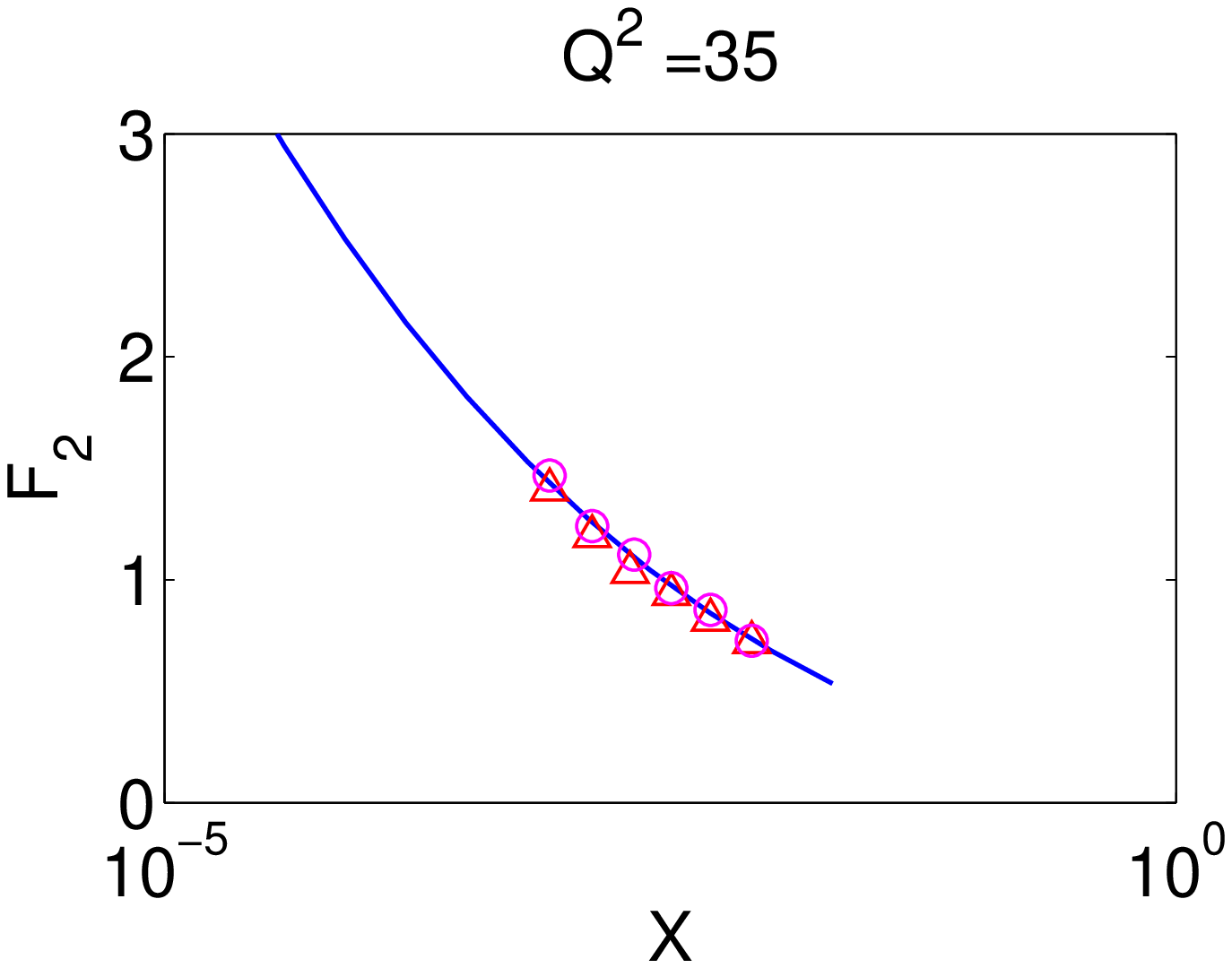,width=40mm, height=25mm}&
\epsfig{file=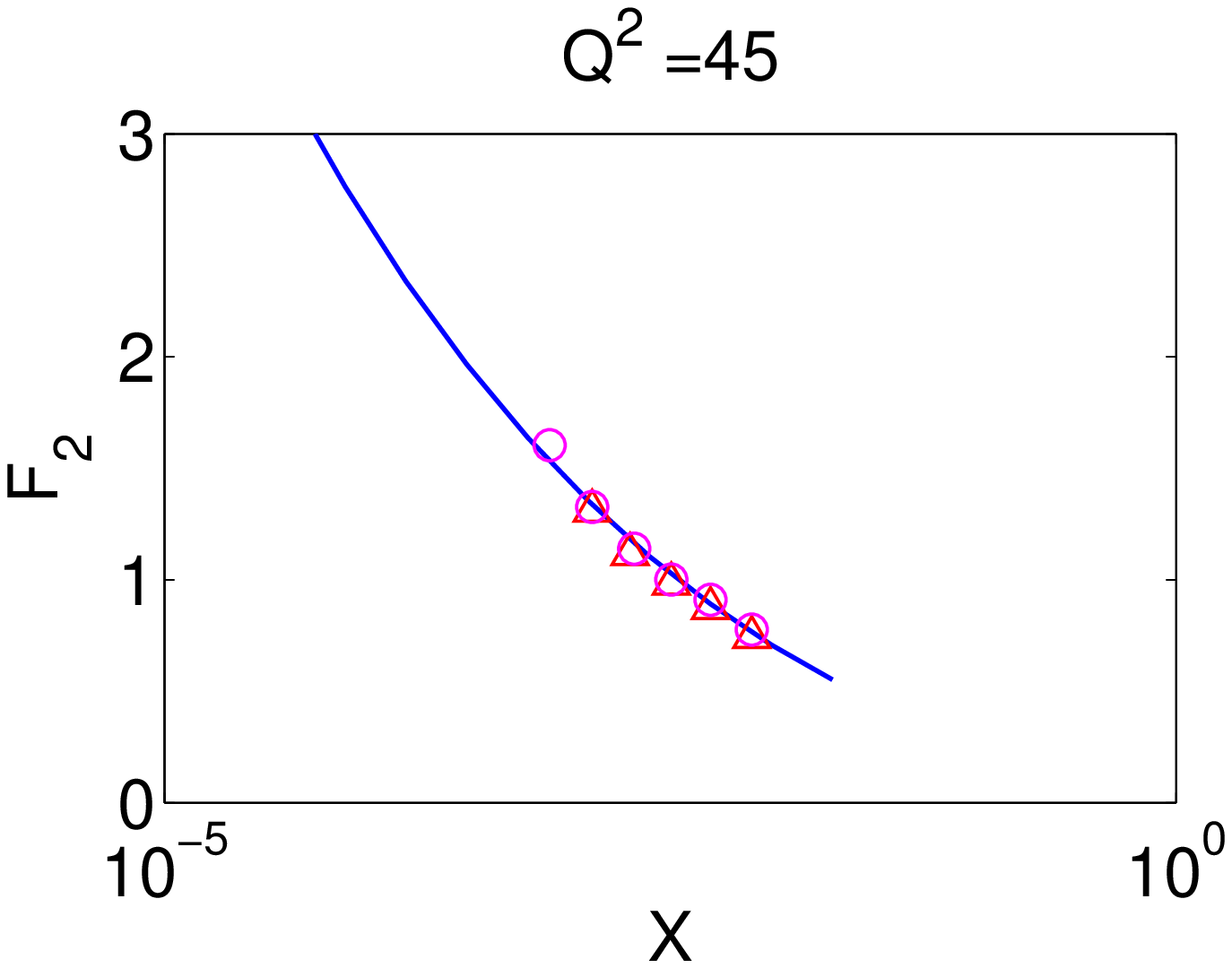,width=40mm, height=25mm}&
\epsfig{file=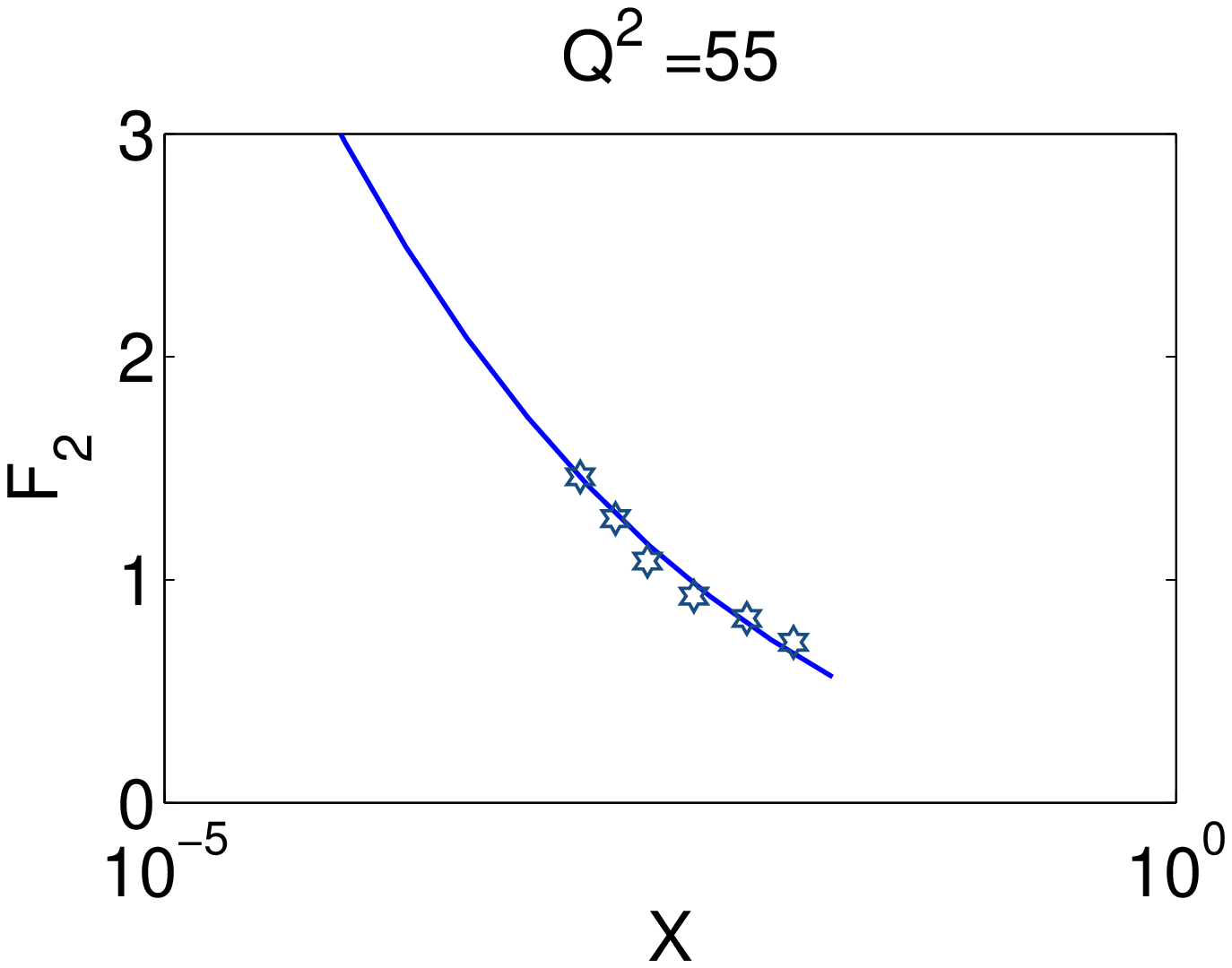,width=40mm, height=25mm}\\
\epsfig{file=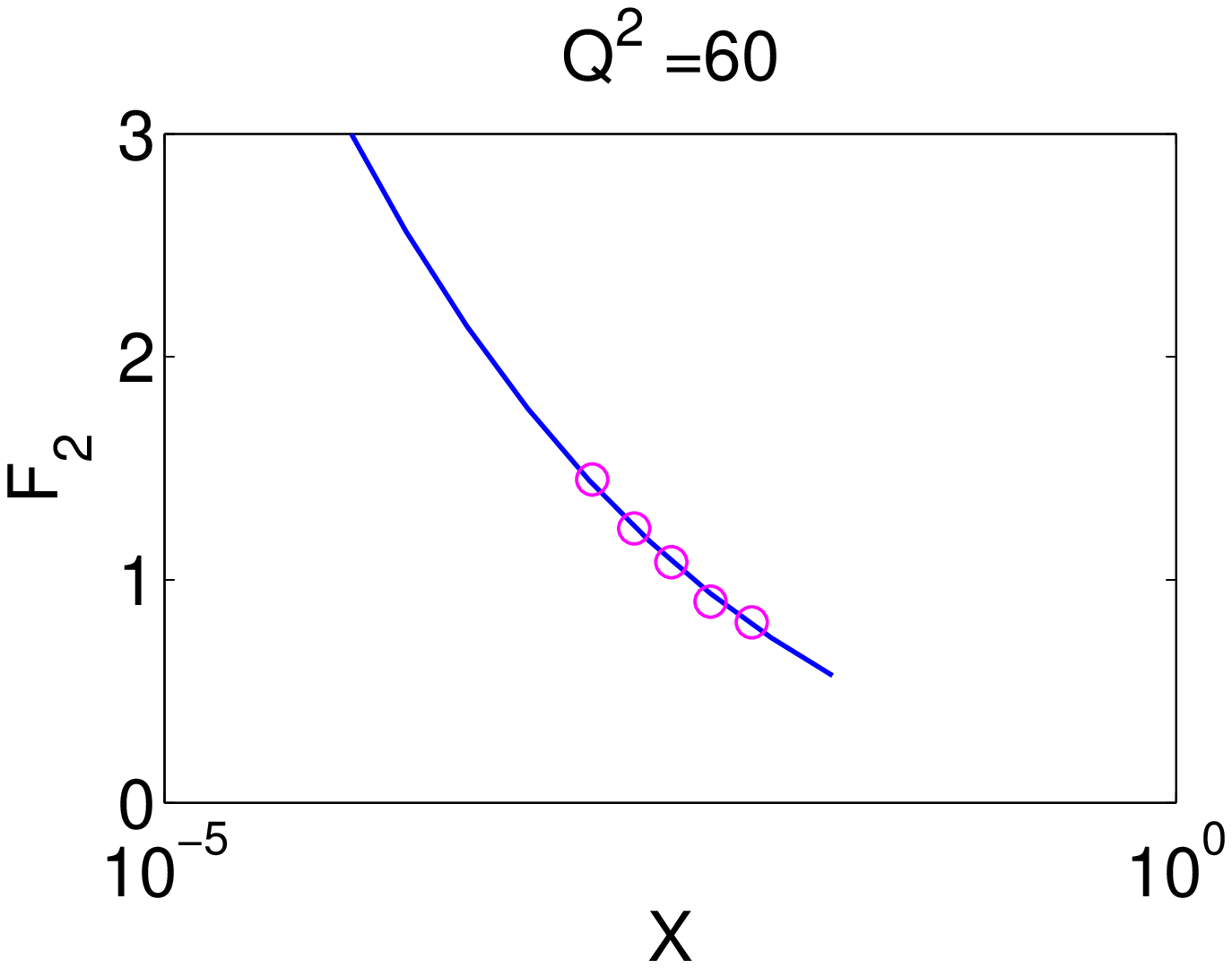,width=40mm, height=25mm}&
\epsfig{file=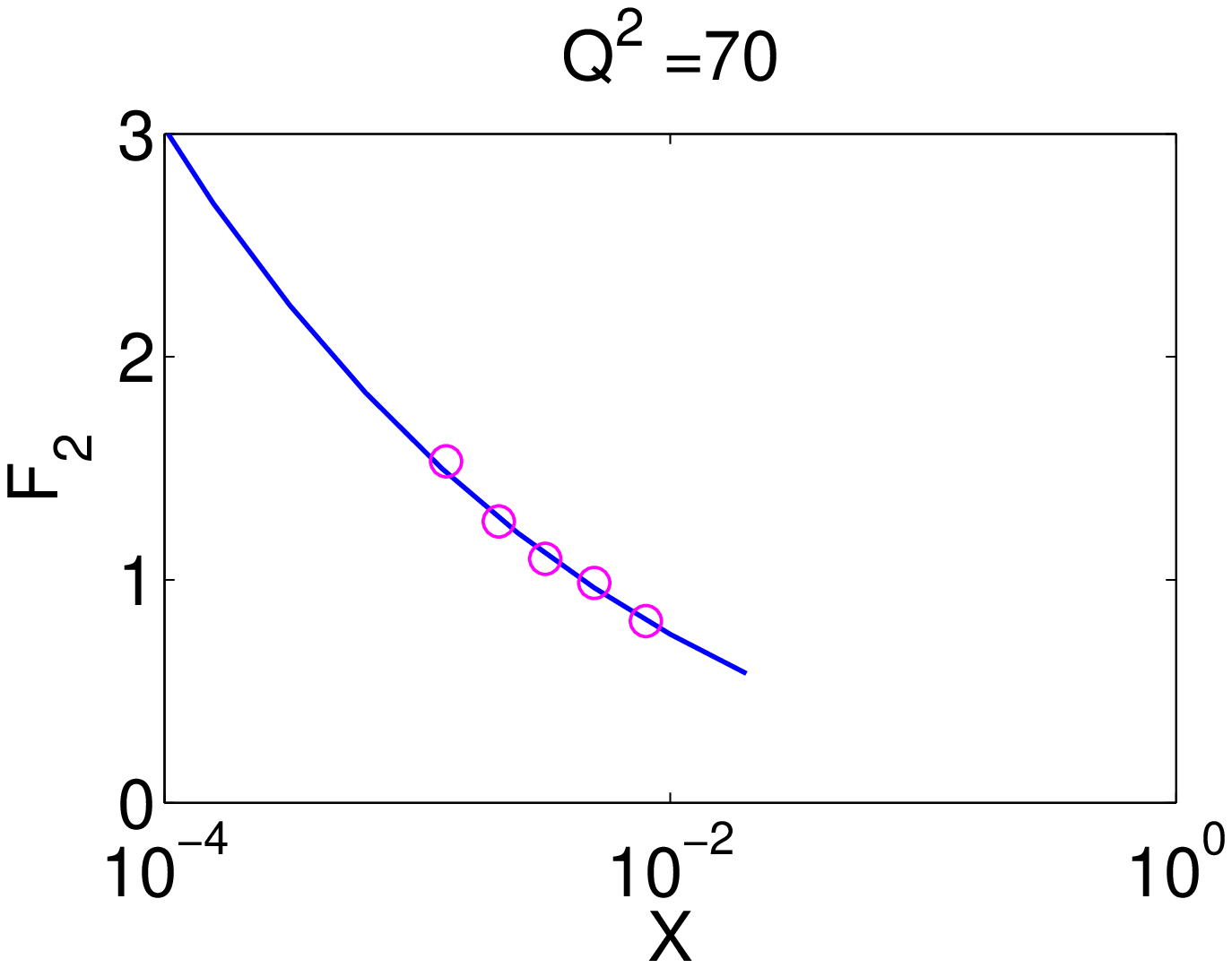,width=40mm, height=25mm}&
\epsfig{file=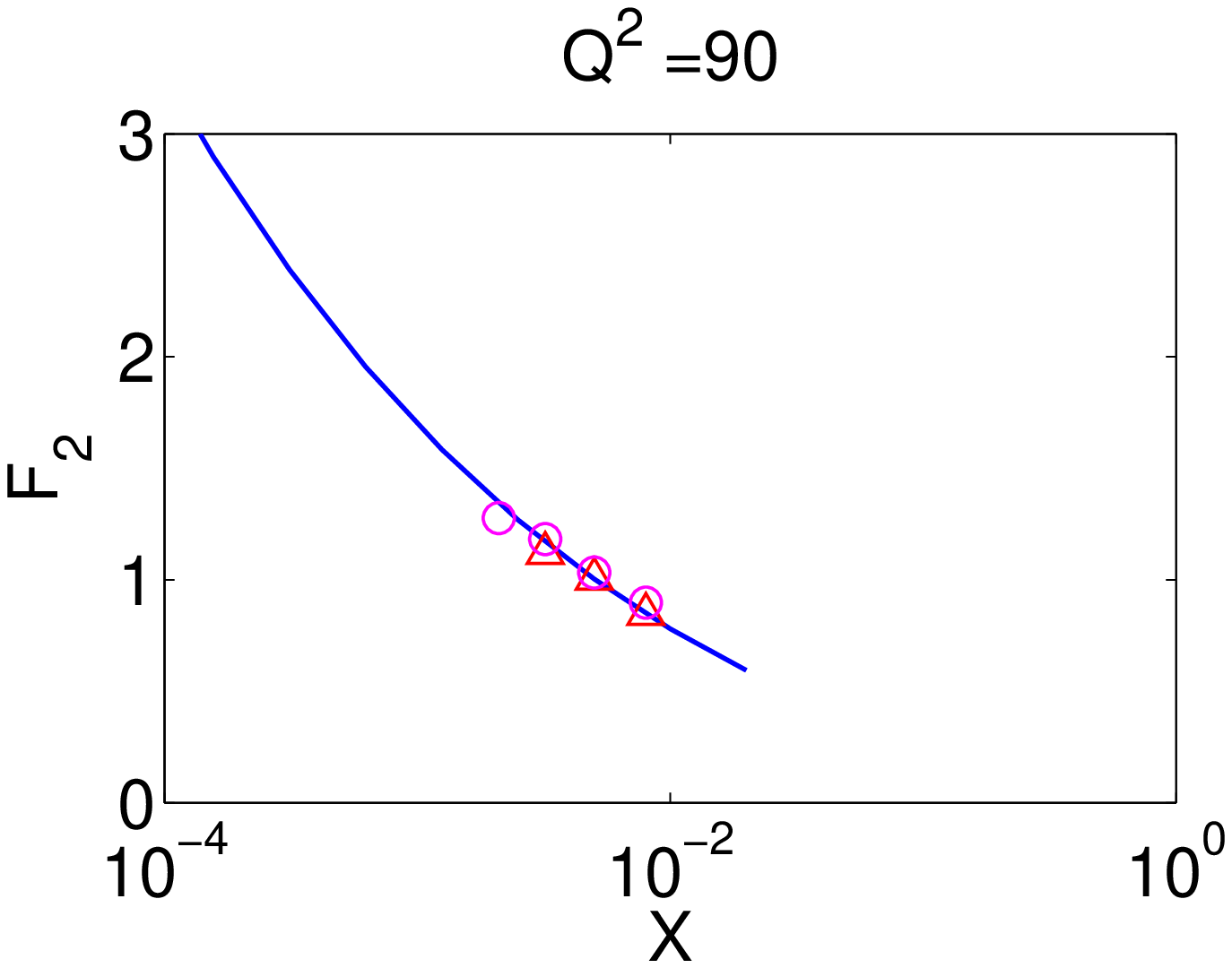,width=40mm, height=25mm}&
\epsfig{file=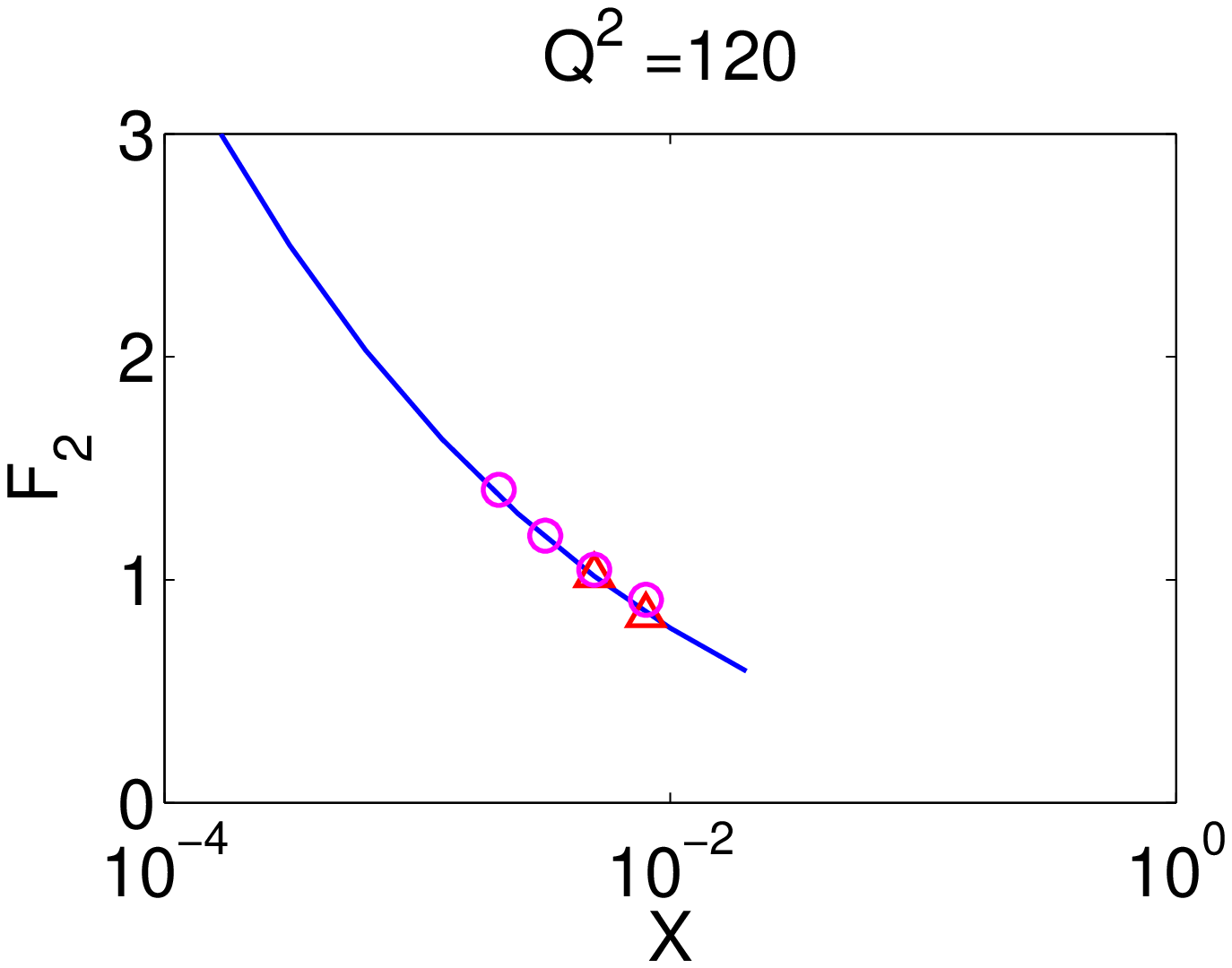,width=40mm, height=25mm}\\
\epsfig{file=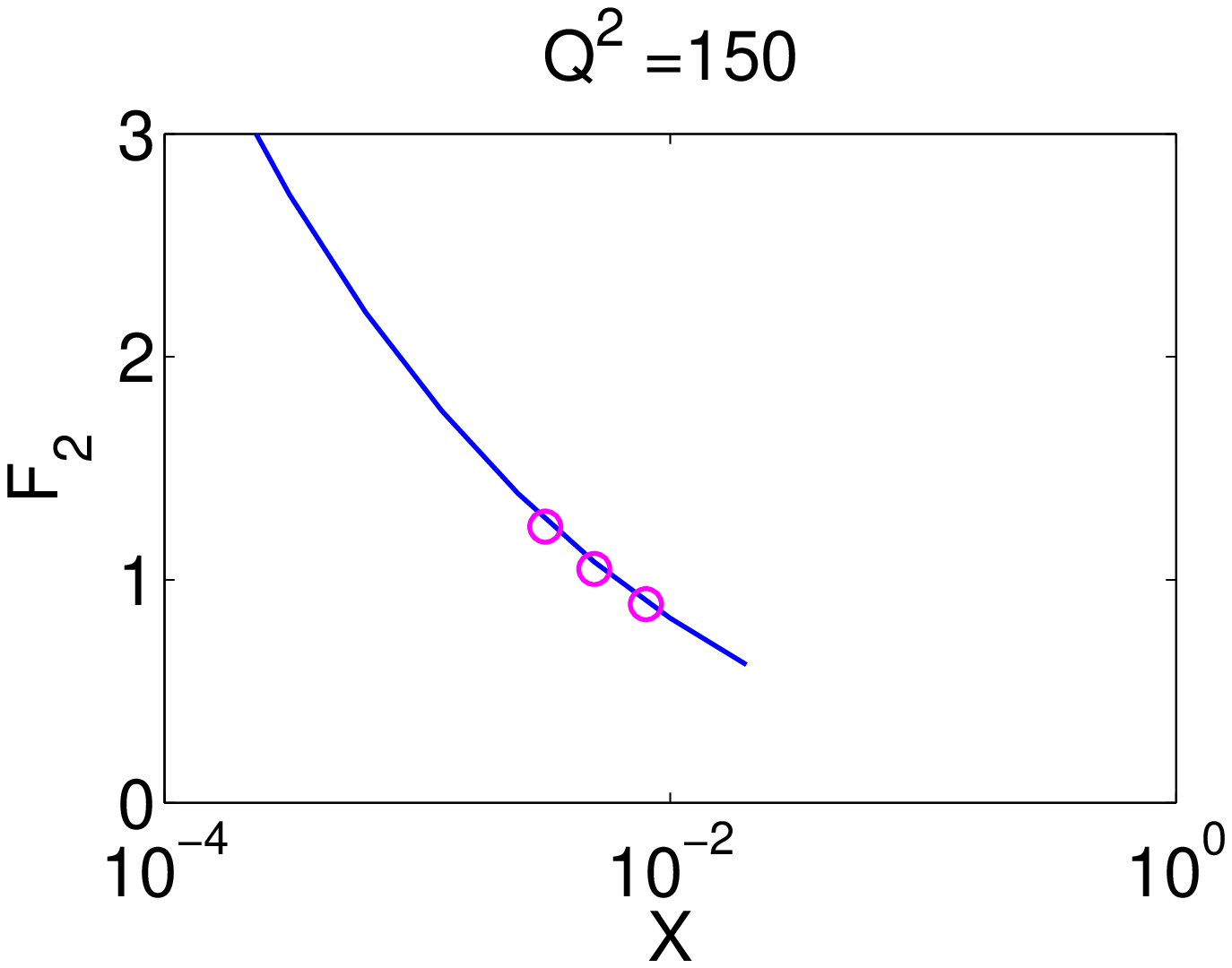,width=40mm, height=25mm}&
\end{tabular}
\caption{\it (continued) $F_{2}(x,Q^{2})$ structure function of
proton, as a function of Bjorken $x$ for fixed value of photon
virtuality $Q^{2}$. This data was taken for small values of $x<0.01$
These data points correspond to different collaborations. Asterisk
corresponds to \cite{Breitweg:2000yn}, triangles
\cite{Adloff:2000qk}, circles correspond to \cite{Chekanov:2001qu}
and hexagons to \cite{Chekanov:2005vv}.}\label{DIS_2}
\end{center}
\end{figure}

\section{Analysis of the model}\label{comparison_models}
\subsection{Comparison between the models}\label{com_mod}

\begin{figure}[htbp]
\centerline{\includegraphics[width=6cm,height=4.0cm]{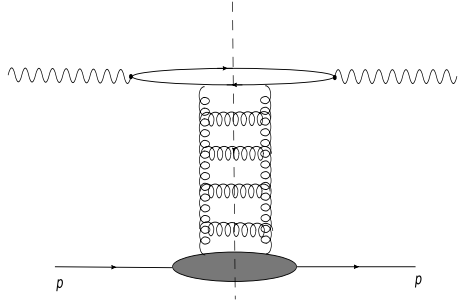}}
\caption{The single gluon-ladder contribution to the total
$\gamma^{*}$ P cross section} \label{lad} \end{figure}

The most widely spread saturation model, is based on the eikonal
approach and has the following form \beqn
\frac{d\sigma}{d^{2}b}\,\equiv\,N(r,b,x)\,=\,2\left(
1\,-\,e^{-\frac{\Omega(r,b,x)}{2}}\right) \eeqn where $\Omega$ is
defined in \eq{conv_omega}. Using the AGK cutting rules
\cite{Abramovsky:1973fm}, we can calculate the cross sections with
different multiplicities $k$. This is the content of the AGK rules,
and which relates the cross-section $\sigma_{k}$ for observing a
final state with $k$-cut Pomerons, with the amplitudes for the
exchange of $m$ Pomerons $F^{(m)}$ \beq\label{dsig_d2b}
\frac{d\sigma_{k}}{d^{2}b}\,=\,\sum_{m=k}^{\infty}(-1)^{m-k}2^{m}\frac{m!}{k!(m-k)!}F^{(m)}
\eeq We will apply this concept and calculate the predictions for
the $k$-cut cross-sections. We make a comparison between two models
\beqn\label{dif_N} N^{I}\,=\,2\frac{\Omega}{\Omega + 1}\;\;\;\;\;
\mbox{and}\;\;\;\;\;N^{I\!\!I}\,=\,2\left( 1 -
 e^{-\frac{\Omega}{2}}\right)\eeqn For small values of $\Omega$, the dipole cross sections in
\eq{dif_N} are equal to $\Omega$, and are proportional to the gluon
density. This allows one to identify the opacity, with the single
Pomeron exchange amplitude of Fig. \ref{lad}.
 Hence, the
multiple Pomeron amplitude is determined from the expansion of the
amplitudes in the form of a series expansion. For our model we
obtain \beq N^{I}\,=\,2\sum^{\infty}_{m=1}(-1)^{m-1} \Omega^{m}\eeq
with \beq\label{F_m_I} F^{(m)}_{I}\,=\,\Omega^{m} \eeq and for the
eikonal model, the same procedure leads to \beq
N^{I\!\!I}\,=\,2\sum^{\infty}_{m=1}(-1)^{m-1}
\left(\frac{\Omega}{2}\right)^{m}\frac{1}{m!}\eeq where
\beq\label{F_m_II}F^{(m)}_{I\!\!I}\,=\,\left(\frac{\Omega}{2}\right)^{m}\frac{1}{m!}
\eeq The dipole cross section can be rewritten \cite{Mueller:1996te}
in terms of $F^{(m)}$, as a sum over multi-Pomeron amplitudes \beq
N\,=\,2\sum_{m=1}^{\infty}(-1)^{m-1}F^{(m)} \eeq The expression for
the $k$ cut Pomeron cross section, is obtained from the AGK cutting
rules of Eqs.(\ref{dsig_d2b}), (\ref{F_m_I}) and (\ref{F_m_II})
\beqn
\frac{d\sigma_{k}^{I}}{d^{2}b}\,&=&\,\sum^{\infty}_{m=k}(-1)^{m-k}\,
2^{m}\frac{m!}{k!(m-k)!}\,\Omega^{m}\,=\,\frac{1}{1+2\Omega}\left(\frac{2\Omega}{1+2\Omega}\right)^{k}\eeqn
and \beqn
\frac{d\sigma_{k}^{I\!\!I}}{d^{2}b}\,&=&\,\sum^{\infty}_{m=k}(-1)^{m-k}\,
2^{m}\frac{m!}{k!(m-k)!}
\left(\frac{\Omega}{2}\right)^{m}\frac{1}{m!}\,=\,\frac{\Omega^{k}}{k!}e^{-\Omega}\eeqn

The diffractive cross-section, is given by the difference between
the total, and the sum over all cut cross sections
\beqn\label{sigma_diff}
\frac{d\sigma_{diff}}{d^{2}b}\,=\,\frac{d\sigma_{tot}}{d^{2}b}\,-\,\sum_{k=1}^{\infty}\frac{d\sigma_{k}}{d^{2}b}\eeqn
and for two models reads as follows \beqn
\frac{d\sigma_{diff}^{I}}{d^{2}b} \,=\,
2\left(\frac{\Omega}{1+\Omega}\right)\,-\,\frac{2\Omega}{1+\Omega}\,=\,\frac{2\Omega^{2}}{(2\Omega+1)(\Omega+1)}
\eeqn and \beqn \frac{d\sigma_{diff}^{I\!\!I}}{d^{2}b} \,=\,2\left(1
- e^{-\frac{\Omega}{2}}\right)\,-\,(1 - e^{-\Omega})\,=\,\left(1 -
e^{-\frac{\Omega}{2}}\right)^{2}\eeqn

Since we want to compare to different-valued functions, we need to
normalize them. Finally, we compare between the ratios \beqn
R_{k}^{I}\;=\;\frac{d\sigma_{k}^{I}/d^{2}b}{d\sigma_{el}^{I}/d^{2}b}\eeqn
and \beqn
R_{k}^{I\!\!I}\;=\;\frac{d\sigma_{k}^{I\!\!I}/d^{2}b}{d\sigma_{el}^{I\!I}/d^{2}b}\eeqn
where
\beqn\frac{d\sigma_{el}}{d^{2}b}\;=\;\sum_{k=1}^{\infty}\frac{d\sigma_{k}}{d^{2}b}\eeqn

Below we present the plots describing the partial normalized cross
sections $(d\sigma_{k}/d^{2}b)/(d\sigma_{el}/d^{2}b)$, which
correspond to $k$-cut Pomerons and the diffractive cross section
$(d\sigma_{diff}/d^{2}b)$. We compare the two models. Our approach,
uses a solution of the non-linear evolution equation in a
"toy-model" approach, and the Glauber approach. Our plots were
calculated for fixed values of the dipole size $r = 0.1 fm$, and the
value $x_{b}\;=\;0.001$. Since higher cuts ($k>1$) are strongly
suppressed, they were rescaled by a factor $10^{3k}$.
\begin{figure}[htbp]
\begin{center}
\begin{tabular}{c c c c}
\epsfig{file=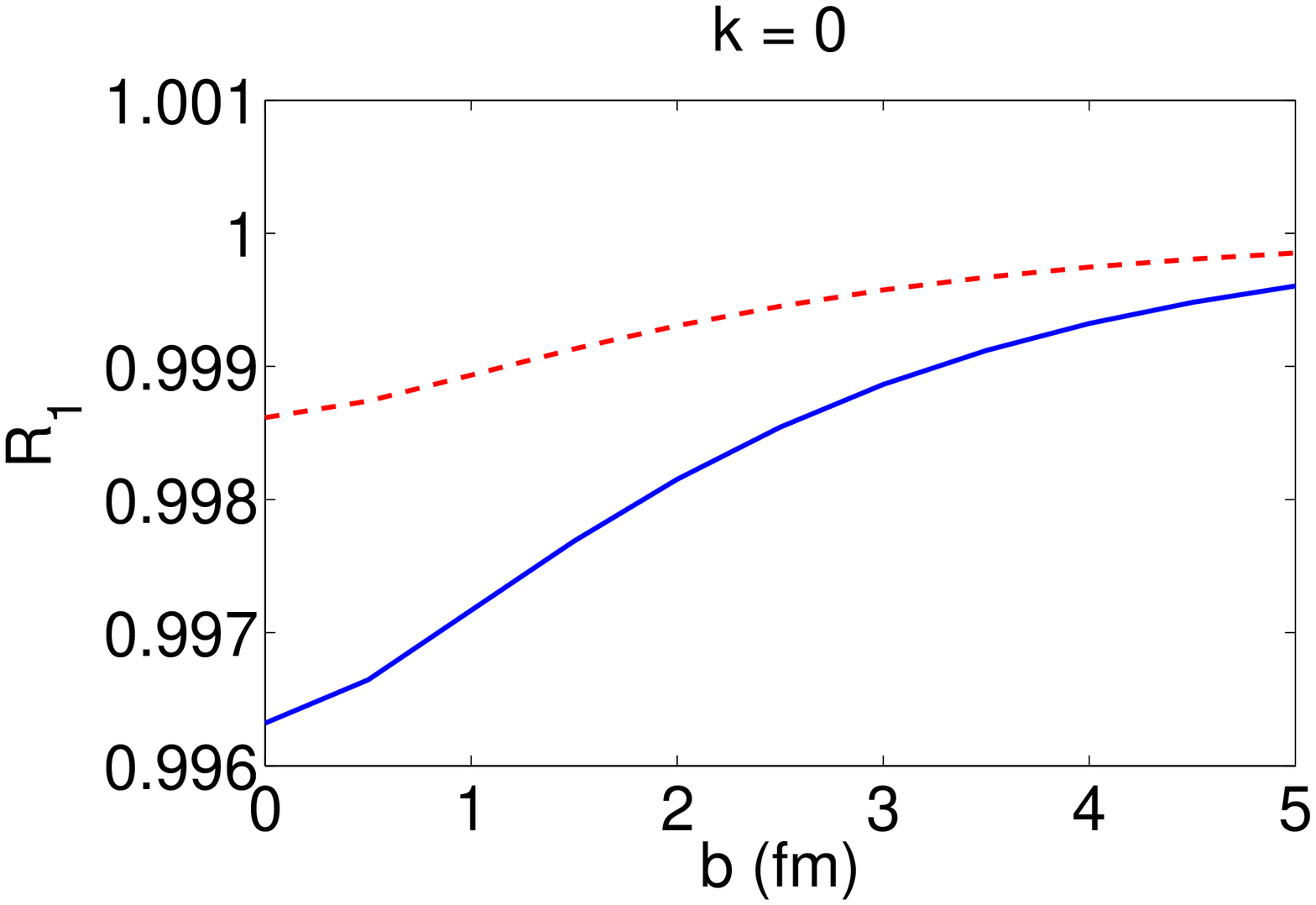,width=4cm, height=3cm} &
\epsfig{file=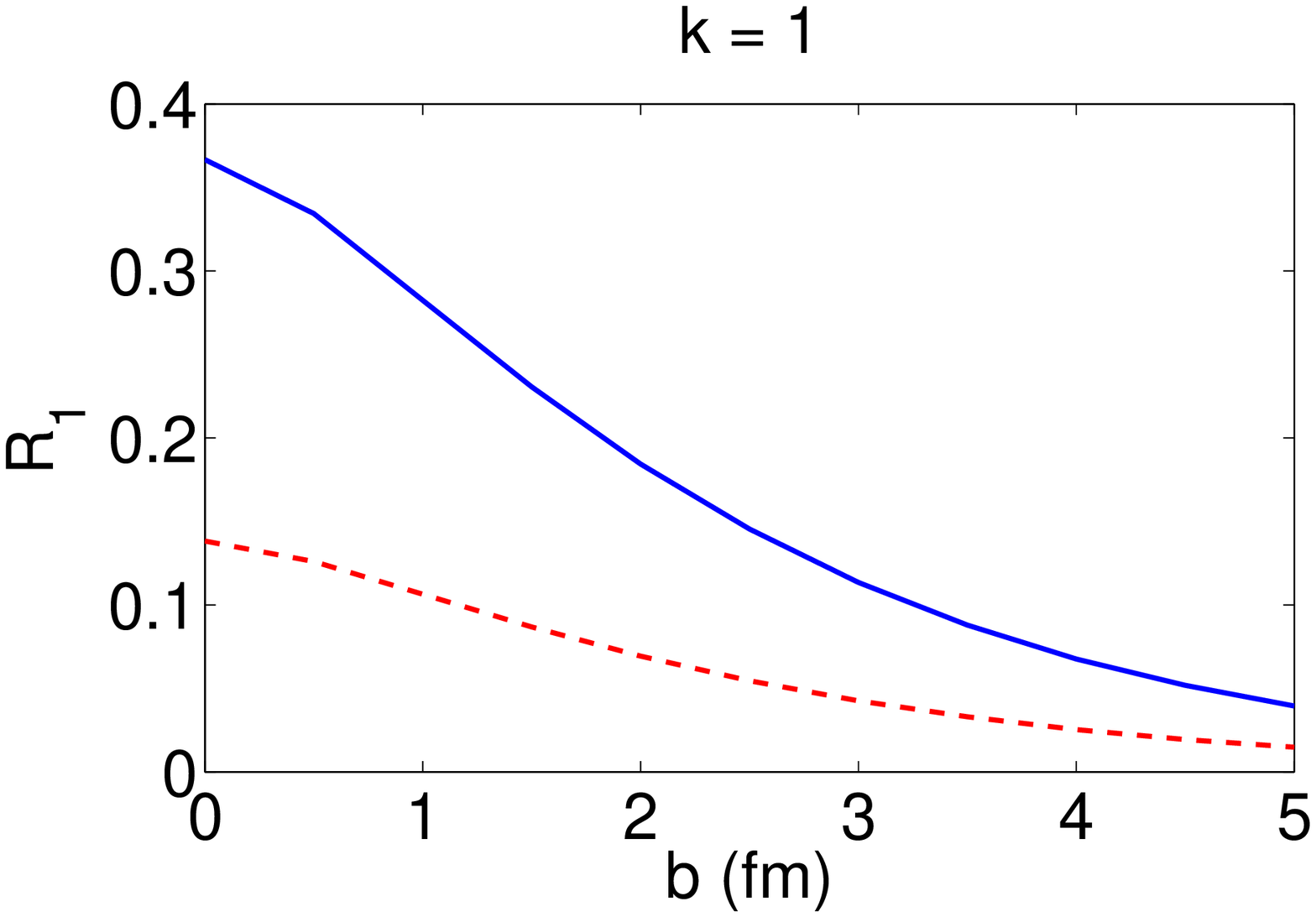,width=4cm, height=3cm} &
\epsfig{file=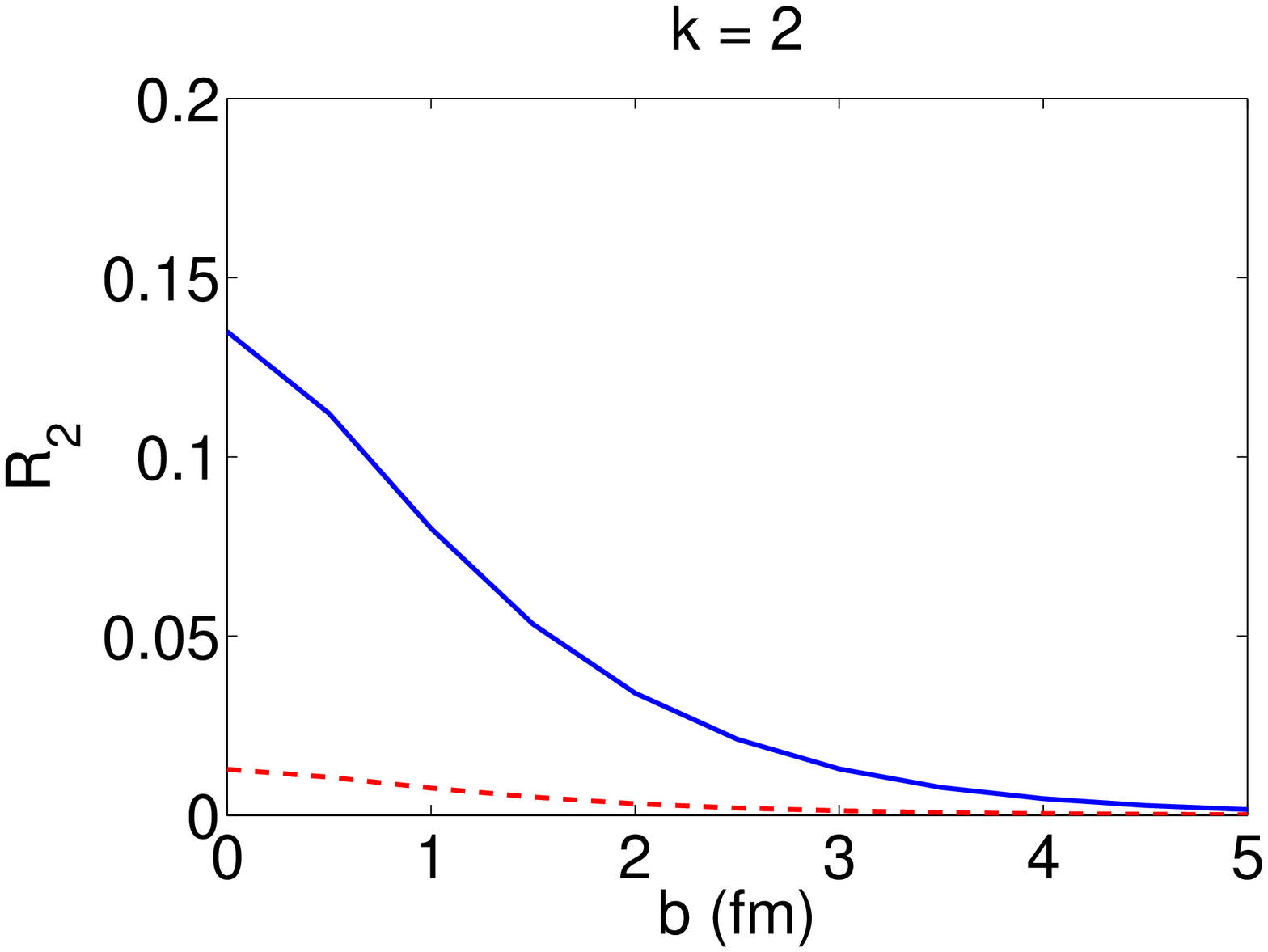,width=4cm, height=3cm} &
\epsfig{file=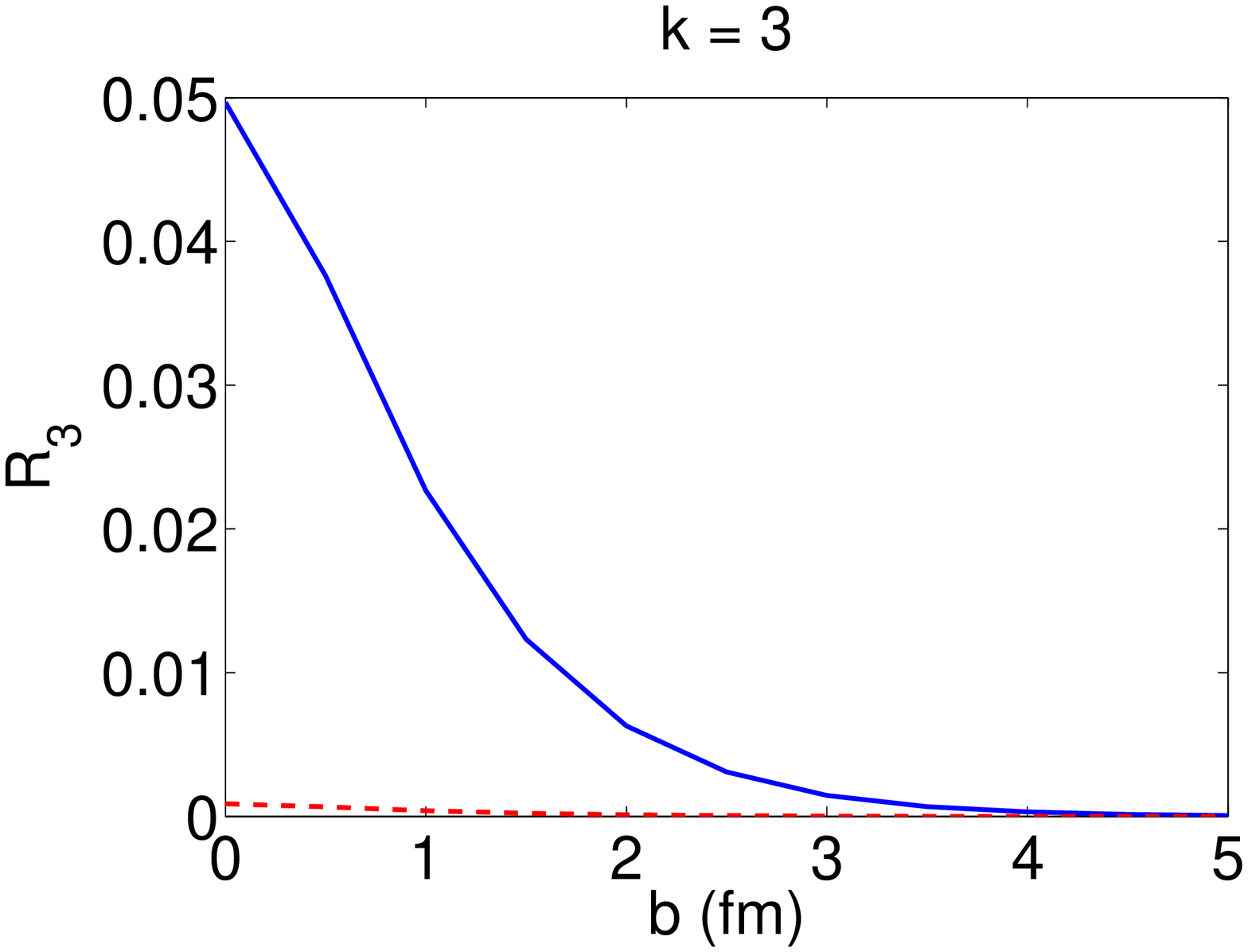,width=4cm, height=3cm} \\
\end{tabular}\caption{\it Cross sections with different multiplicities as a function of impact
parameter $b\;(fm)$. Solid line corresponds to our model and dashed
line to the eikonal approach.}\label{AGK_plots_lin}
\end{center}
\end{figure}

Although both models fit the DIS data quite well, there is a
difference in the value, and in the shape of the partial cross
sections, as shown in the plots.

\subsection{Shadowing corrections}

In this subsection, we want to discus the difference in the behavior
of the two models (our model and the Glauber-like model), from the
shadowing corrections point of view. Shadowing corrections (SC), are
defined as the next to leading order terms in the series expansion
of the amplitude near the amplitude zero point.
\begin{subequations}
\beqn N^{GF}\;&=&\;\frac{\Omega_{Gf}}{\Omega_{Gf} +
1}\;=\;\Omega_{Gf} - \Omega_{Gf}^{2} + \Omega_{Gf}^{3} - ...\eeqn
\beqn\label{amplitudes} N^{Gl}\;&=&\;1 -
e^{-\frac{\Omega_{Gl}}{2}}\;=\;\frac{\Omega_{Gl}}{2} -
\frac{\Omega_{Gl}^{2}}{4\cdot2!} + \frac{\Omega_{Gl}^{2}}{8\cdot3!}
- ...\eeqn
\end{subequations} We can easily conclude, that our proposed generating functional
$(GF)$ based model, and the Glauber-like $(Gl)$ function, are the
same at leading order, as was expected. Assuming this, we
immediately obtain the next condition
\beqn\label{cond_gl_gf}\Omega_{Gl}\;=\;2\Omega_{Gf}\eeqn Since we
want to check the difference in the behavior of the two different
parameterizations, we substitute the condition of \eq{cond_gl_gf}
into \eq{amplitudes}. Finally, we obtain, the fact that our model,
predicts larger SC than those proposed by the Glauber-like model.
\begin{subequations}
\beqn SC^{GF}\;\equiv\;N^{GF} - \Omega_{Gf}\;= - \Omega_{Gf}^{2} +
\Omega_{Gf}^{3} - ...\eeqn \beqn SC^{Gl}\;\equiv\;N^{Gl} -
\Omega_{Gl}\;= -\frac{\Omega_{Gf}^{2}}{2!} +
\frac{\Omega_{Gf}^{2}}{3!} - ...\eeqn
\end{subequations} It is obvious that $SC^{GF} > SC^{Gl}$. In spite of the fact that,
generically, the $SC^{GF}$ are larger, we can fit all the
experimental data, using the Glauber parametrization. It turns out
that $\Omega_{Gl}$, in such a parametrization, is larger than
$\Omega_{Gf}$ by $20-30\%$. The leading term is significant in the
low energy domain, and should be consistent with the perturbative
calculations of the amplitude. SC becomes significant at higher
energies, especially in the saturation domain. Our model predicts
the slower growth of the amplitude with energy, than that of the
Glauber model.

\section{Predictions and descriptions within our model}
\subsection{Charm quark contribution}

The developed dipole model, allows us to calculate a prediction for
the inclusive charm quark contribution, to $F_{2}$ of the proton.
Using the ansatz \eq{F2_ch}, we can easily obtain the contribution
to the structure function from the charm quark
$F_{2}^{c\overline{c}}$.

\begin{figure}[htbp]
\centering
\begin{tabular}{cccc cccc cccc ccc}
\epsfig{file=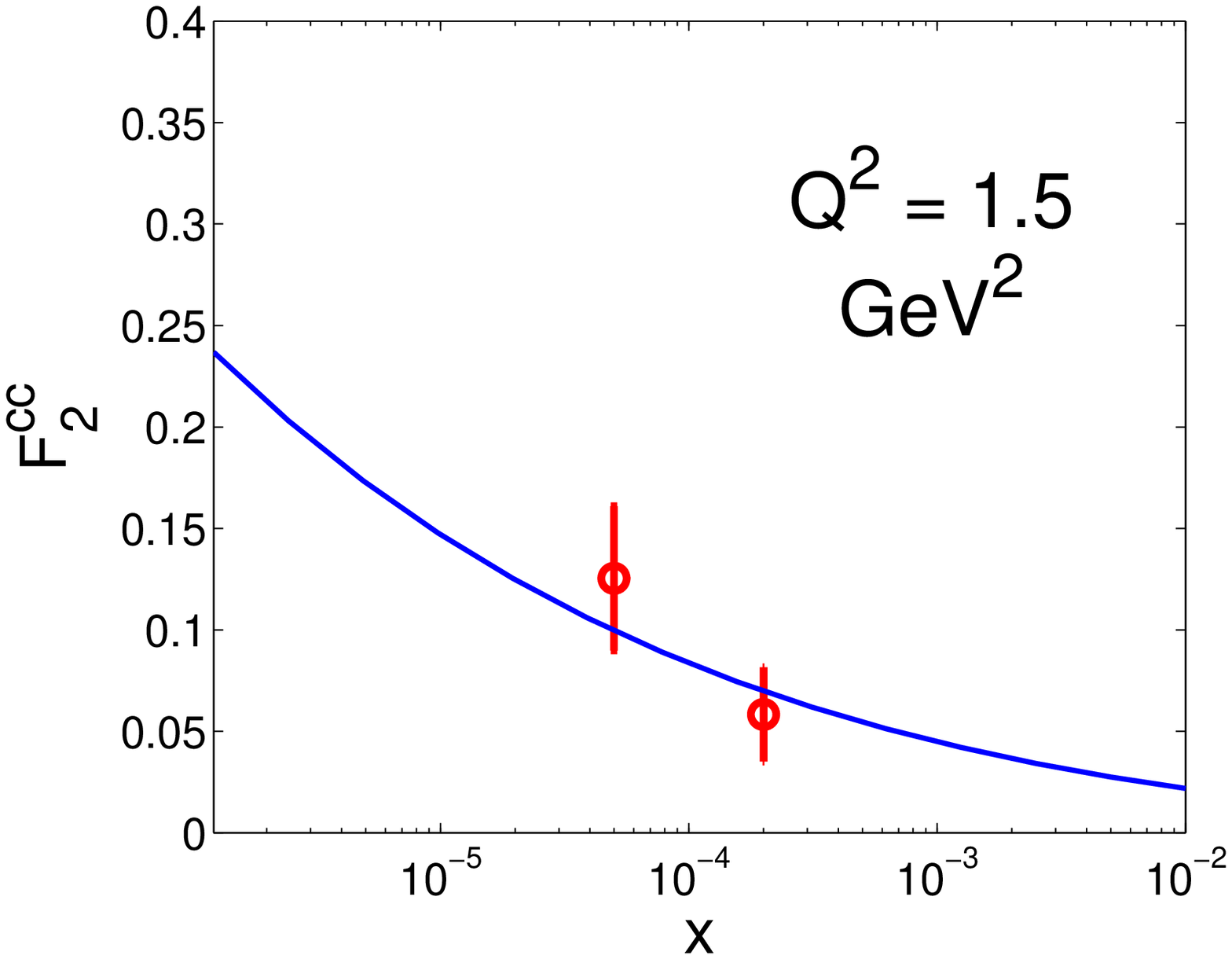,width=35mm, height=20mm}&
\epsfig{file=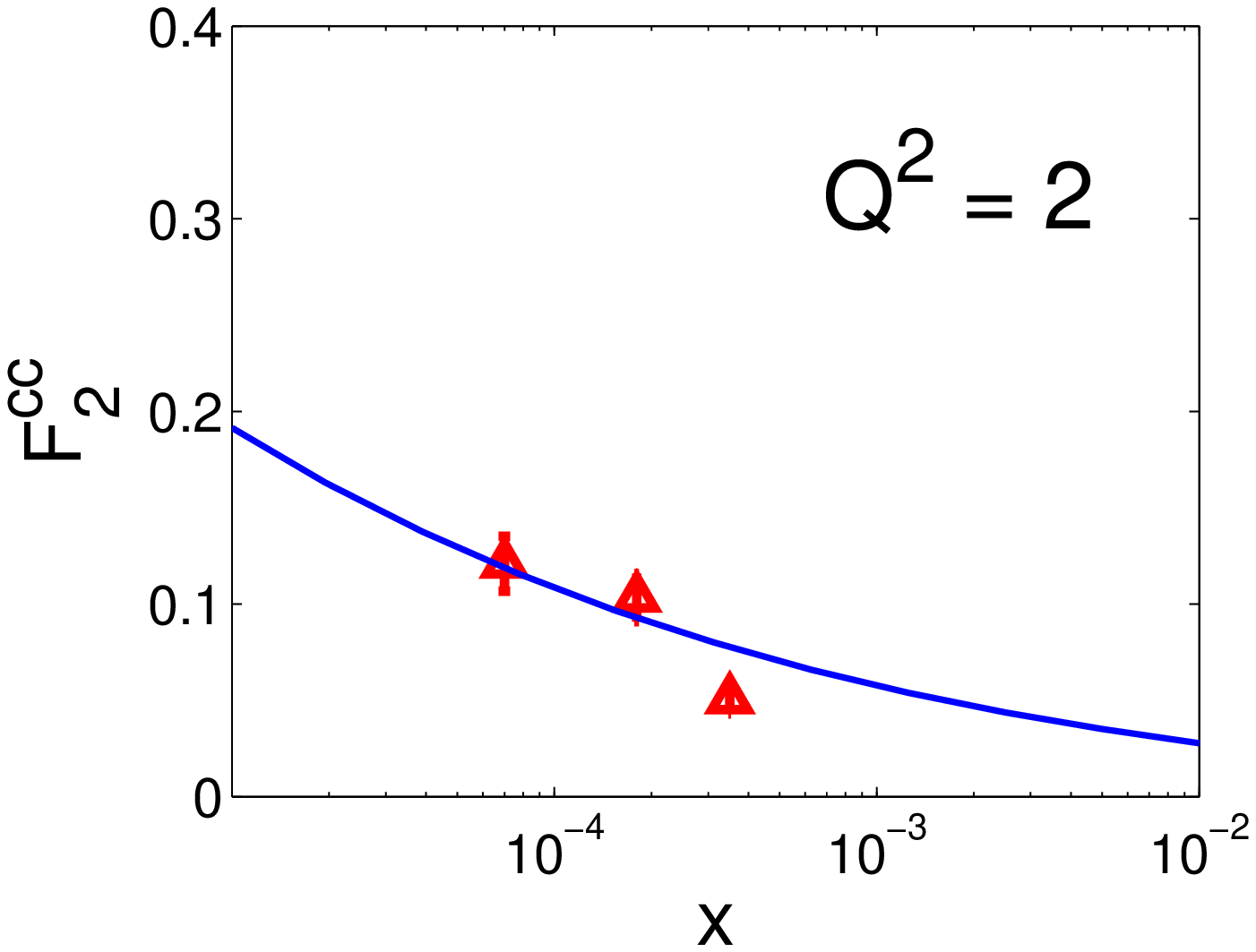,width=35mm, height=20mm}&
\epsfig{file=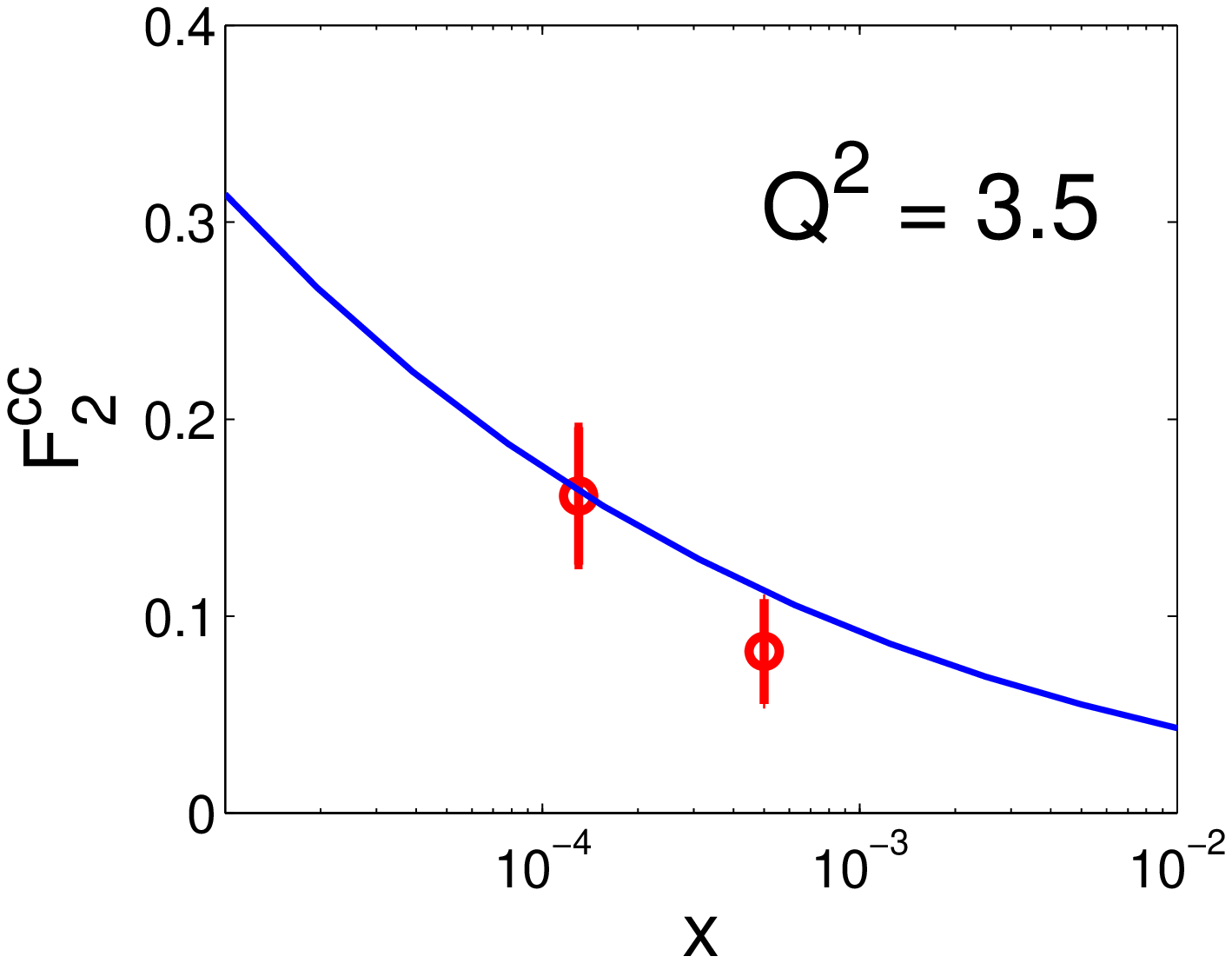,width=35mm, height=20mm}&
\epsfig{file=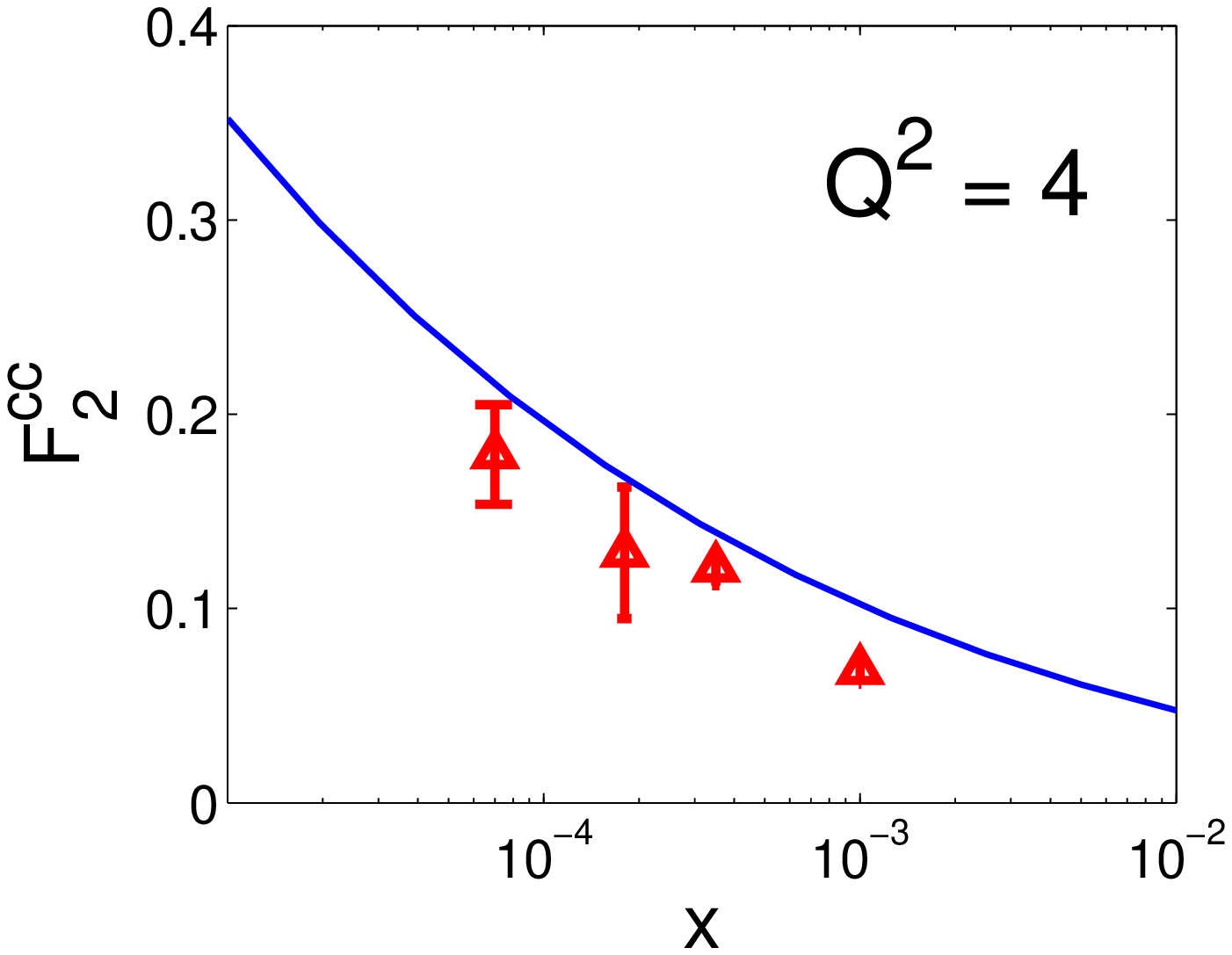,width=35mm, height=20mm}&\\
\epsfig{file=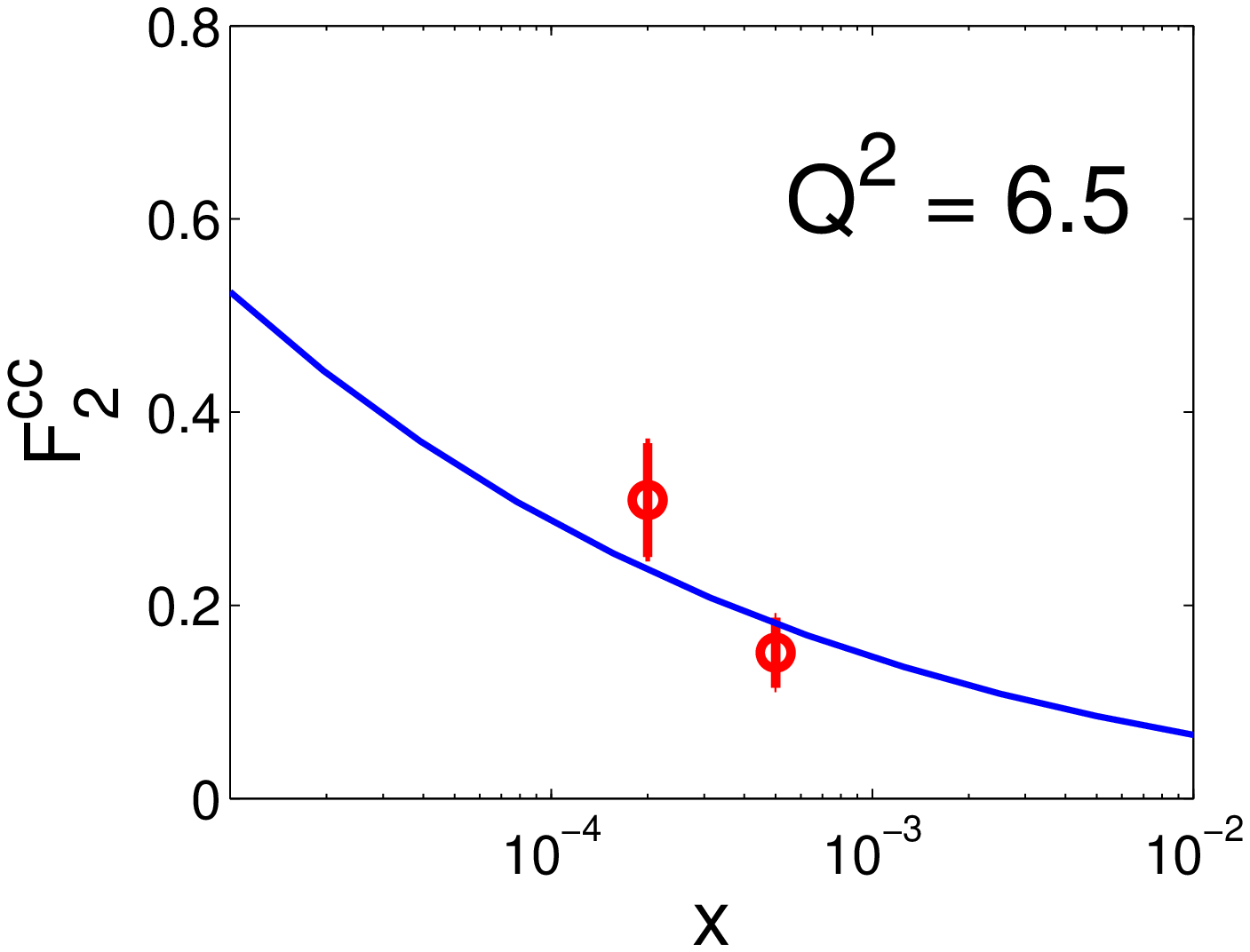,width=35mm, height=20mm}&
\epsfig{file=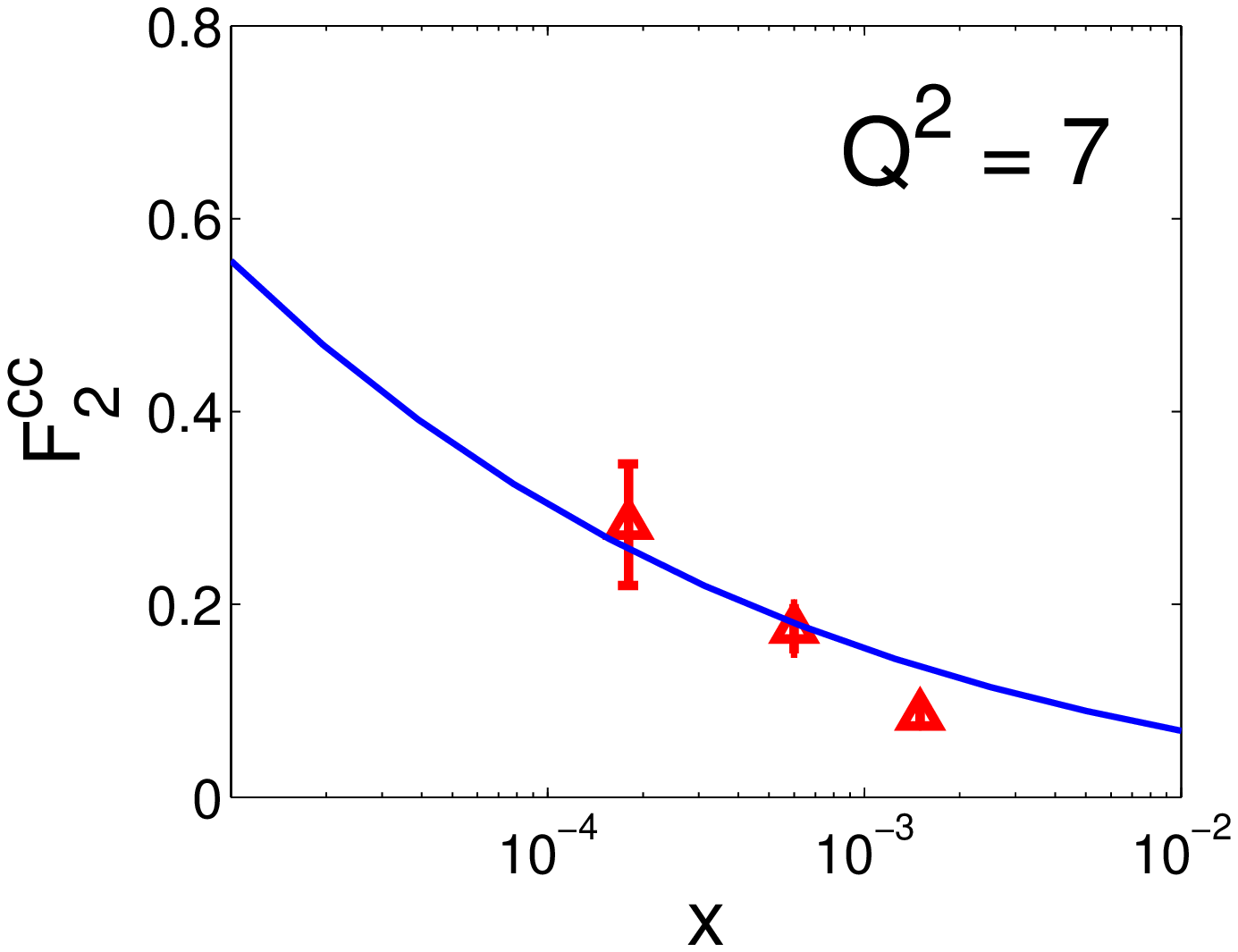,width=35mm, height=20mm}&
\epsfig{file=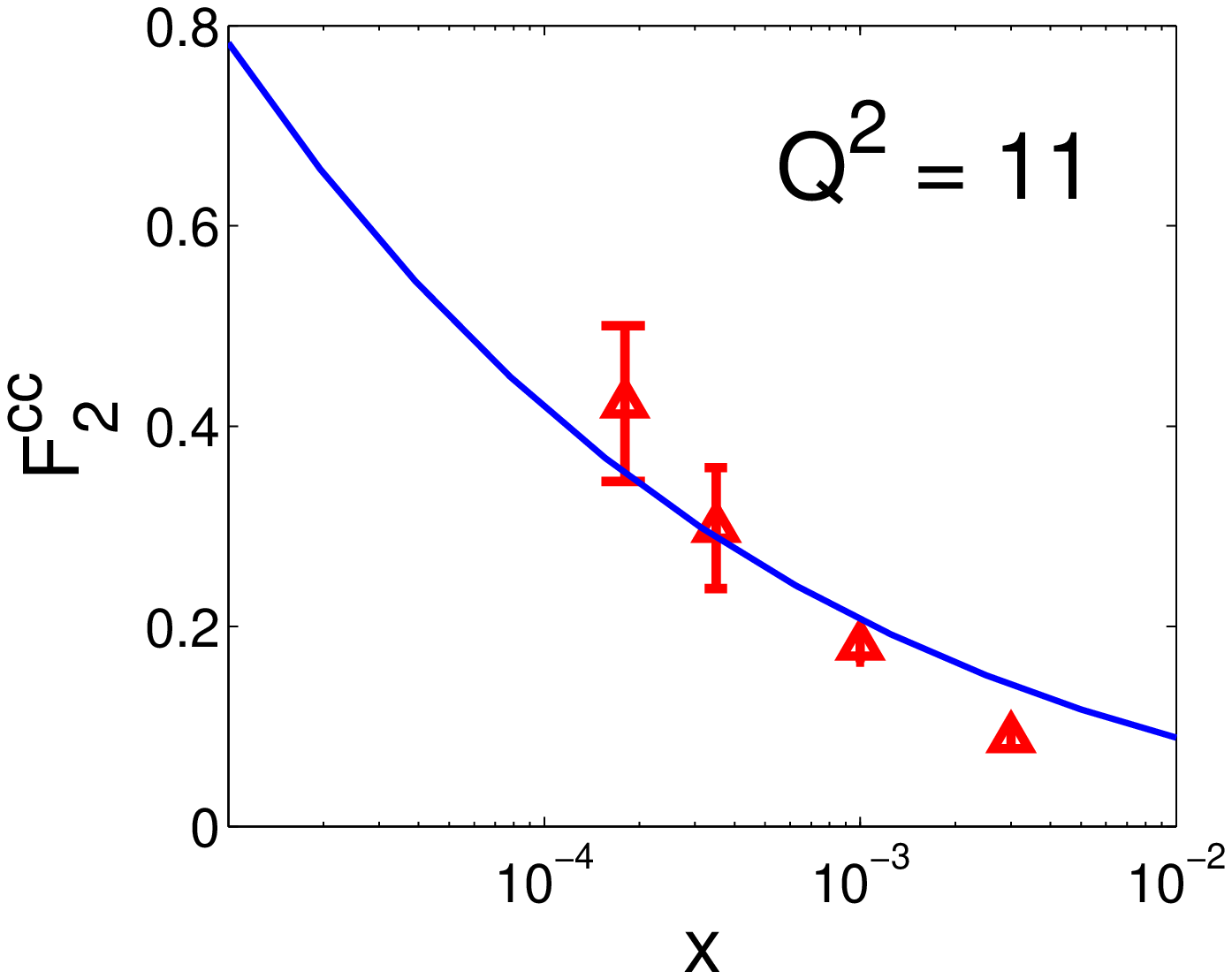,width=35mm, height=20mm}&
\epsfig{file=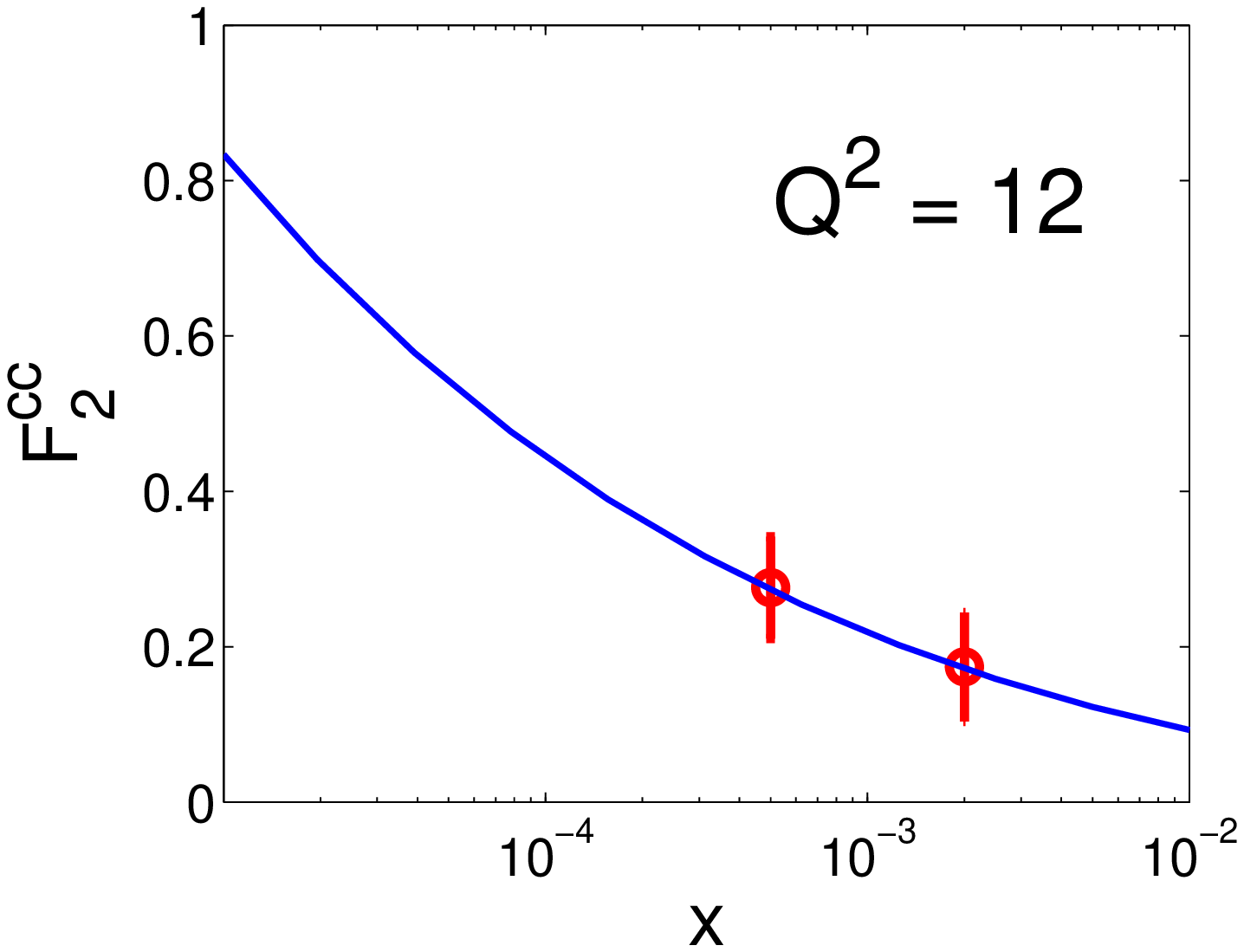,width=35mm, height=20mm}&\\
\epsfig{file=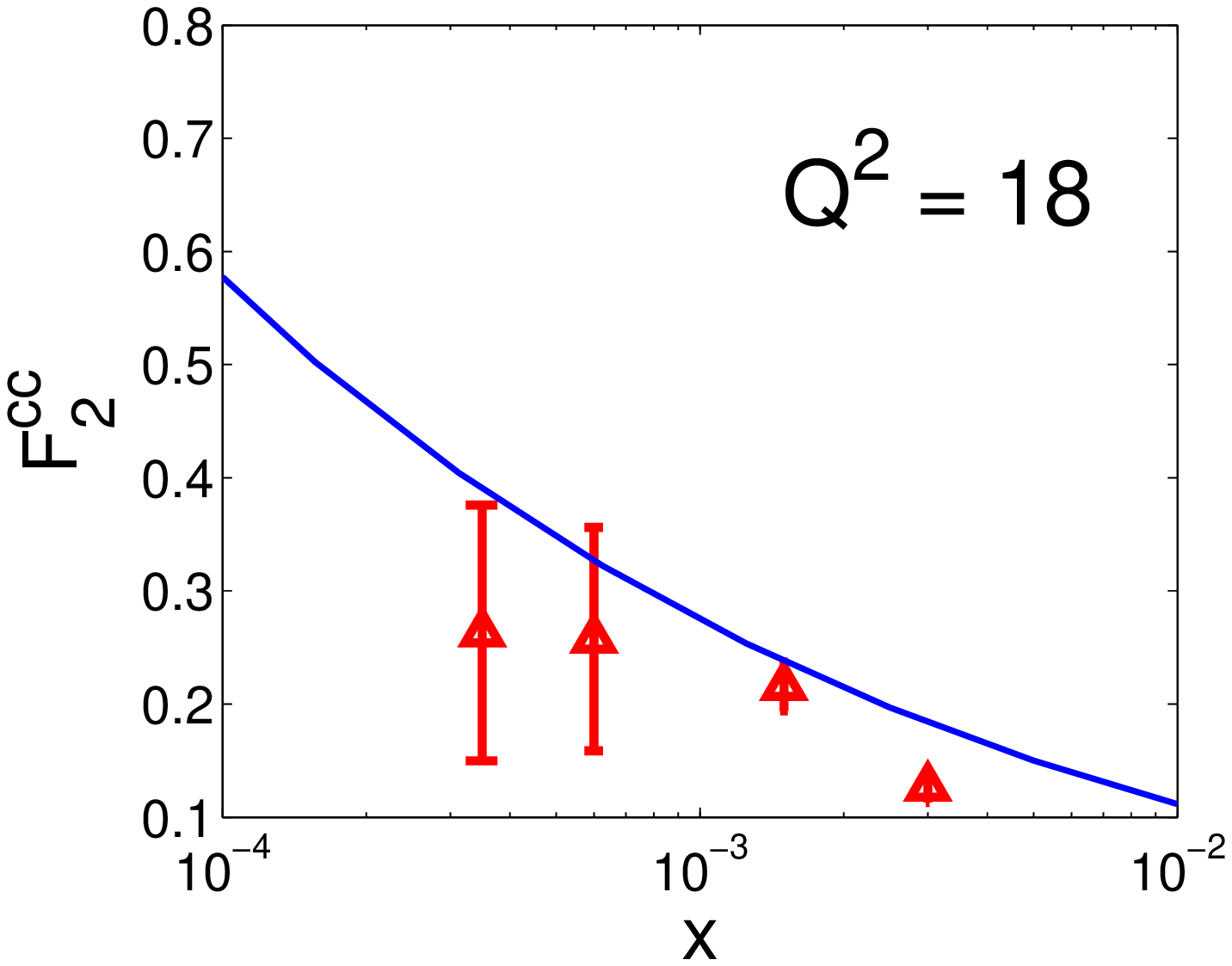,width=35mm, height=20mm}&
\epsfig{file=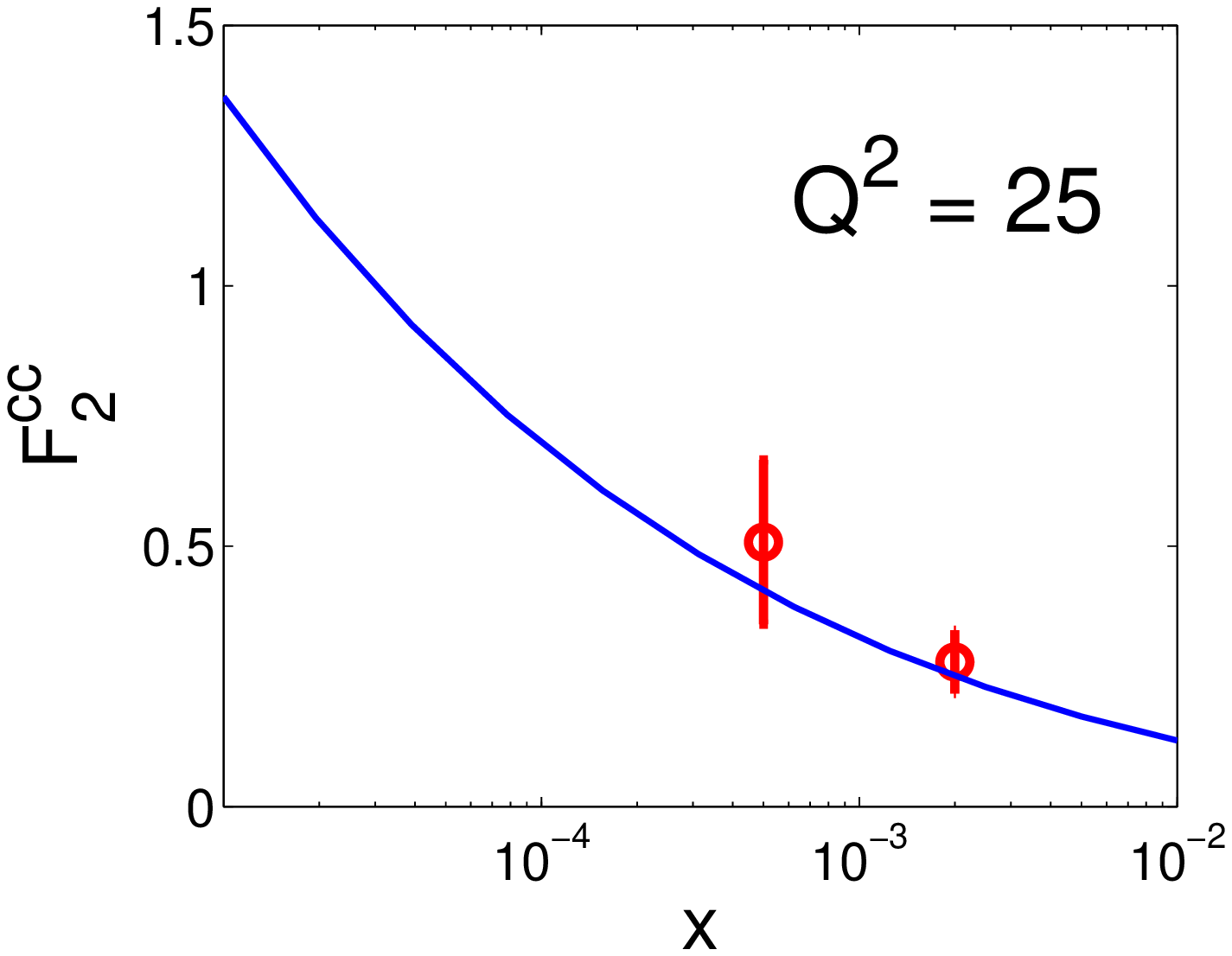,width=35mm, height=20mm}&
\epsfig{file=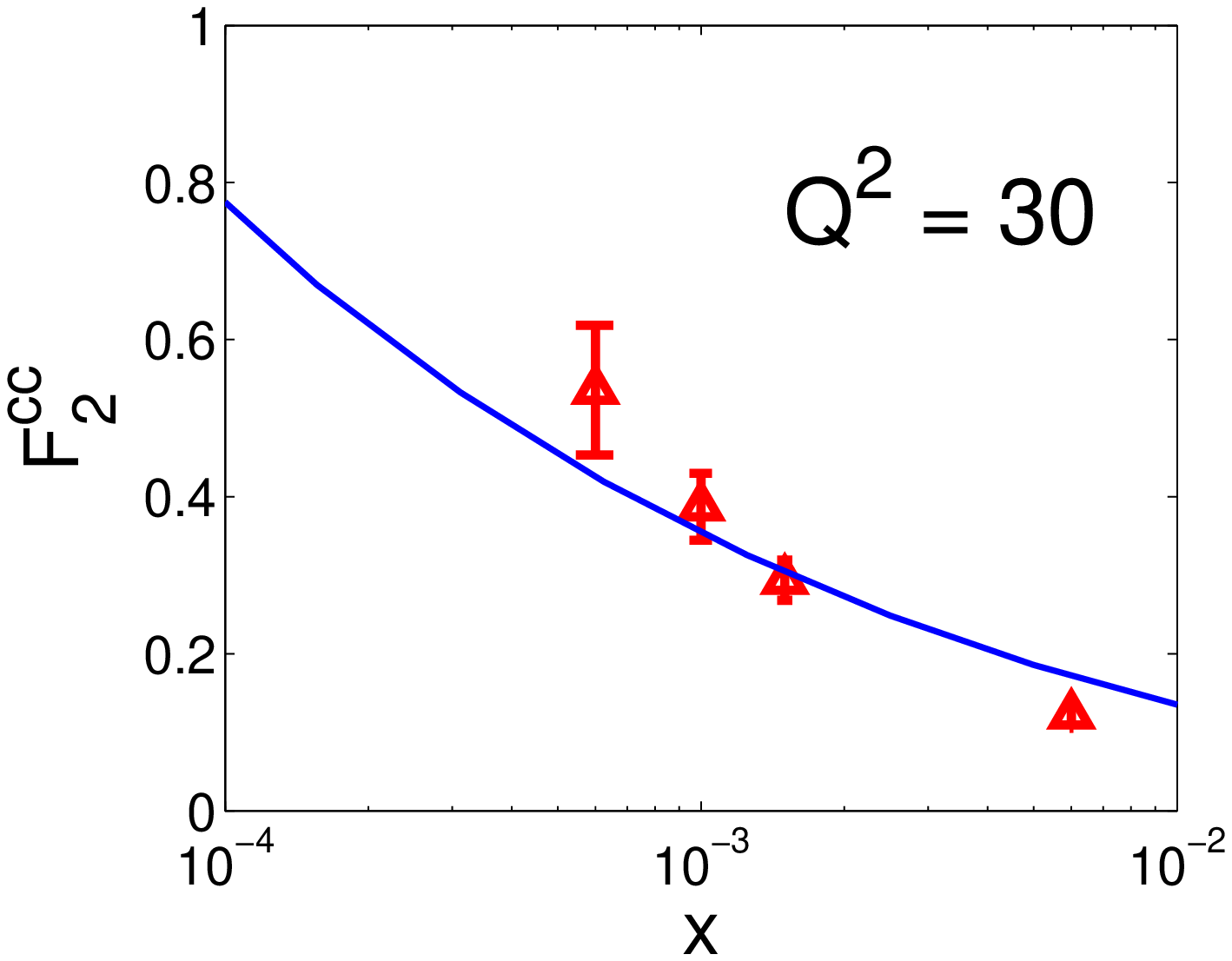,width=35mm, height=20mm}&
\epsfig{file=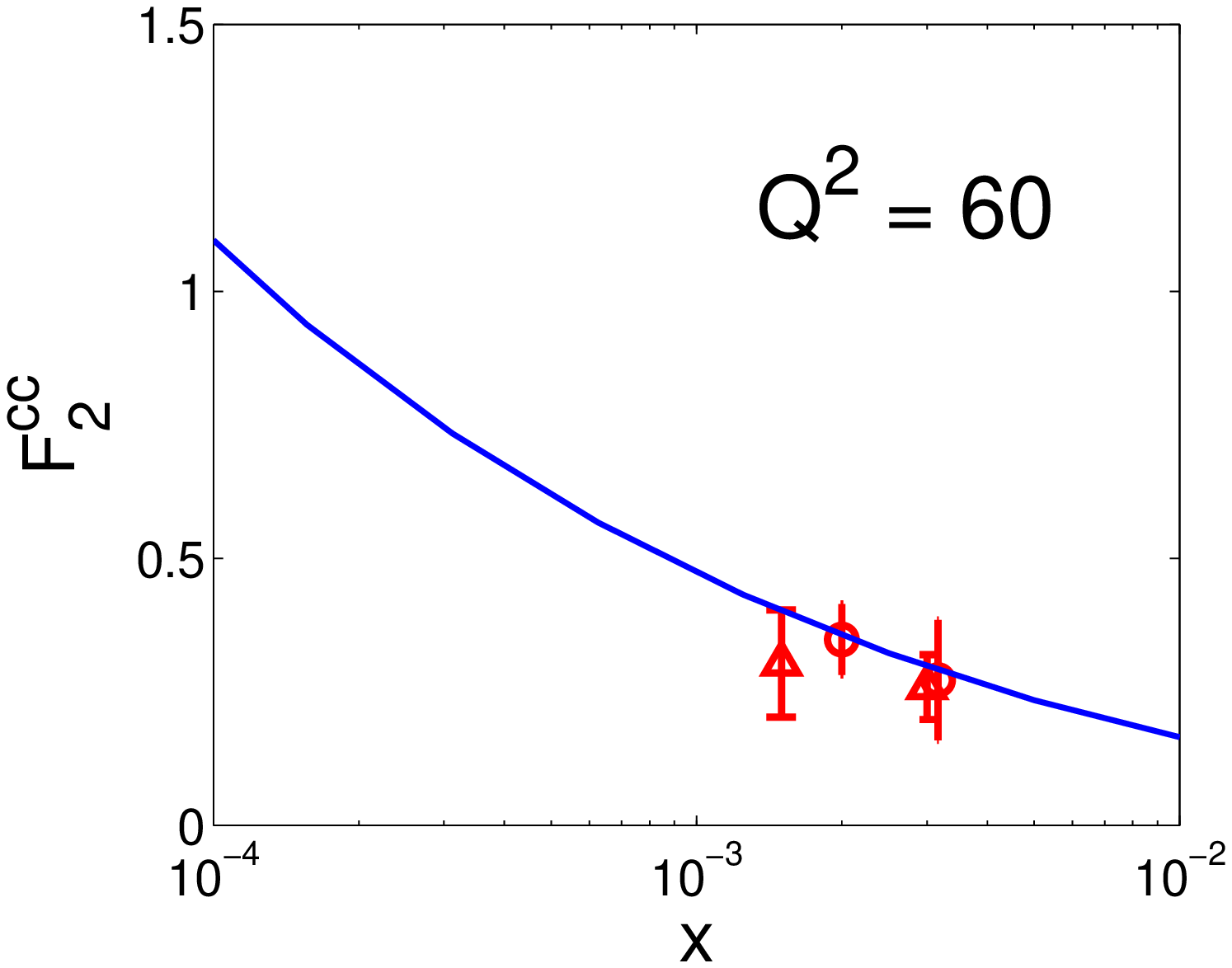,width=35mm, height=20mm}&\\
\epsfig{file=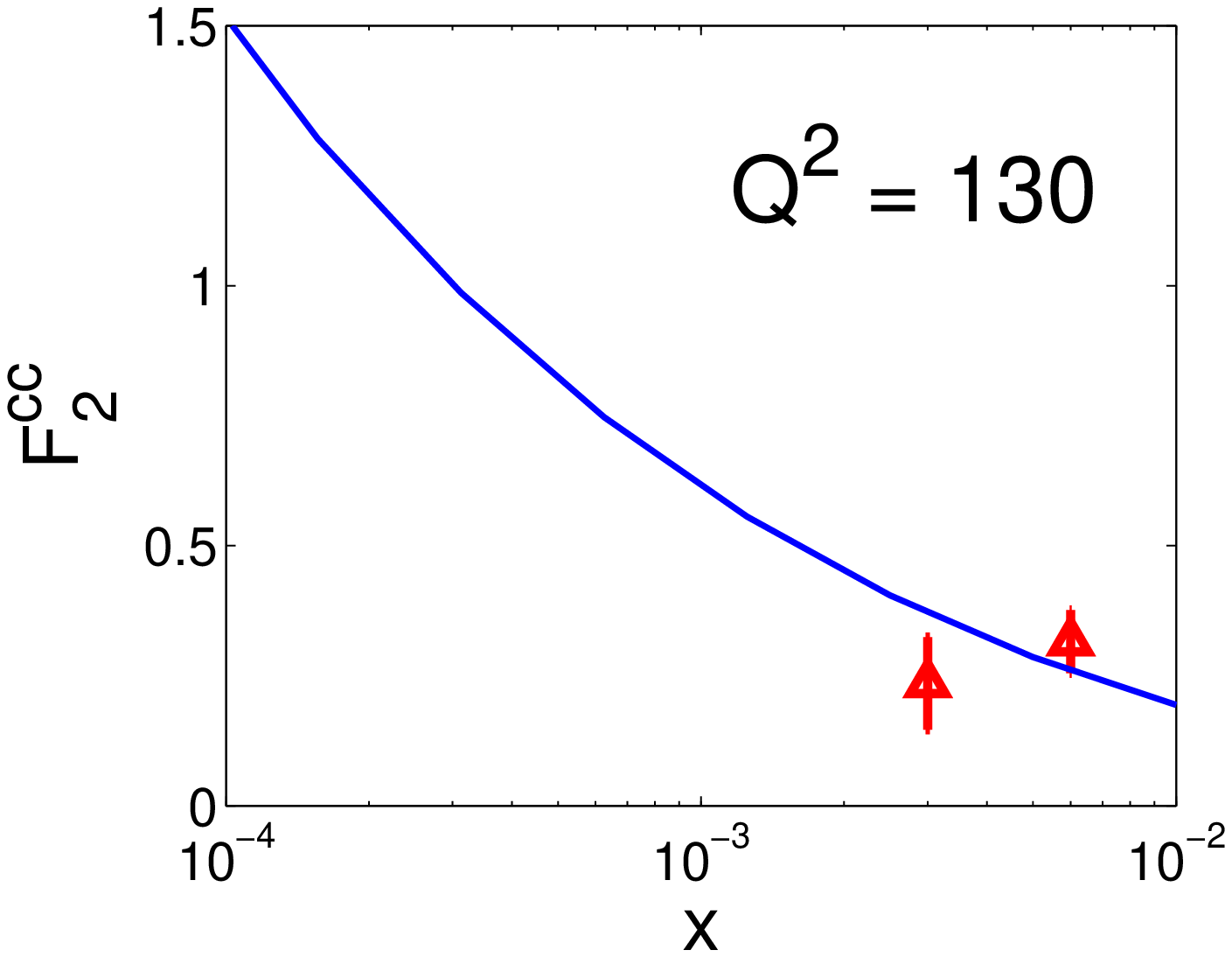,width=35mm, height=20mm}&
\end{tabular}
\label{plot_cc}\caption{\it Prediction of our model to
$F^{c\overline{c}}_{2}$. Solid line is our model, triangles
correspond to \emph{H1 96-97} and circles to \emph{ZEUS 99-00} data
\cite{Adloff:2001zj,{Chekanov:2003rb}} respectively. Mass of the
charm quark has the value: $m_{c}\;=\;1.3 \;\;GeV^{2}$}
\end{figure}

Comparing this to the recent experimental data
\cite{Adloff:2001zj,Chekanov:2003rb} we find a good agreement
between the calculated prediction, and the experimental result, as
it can be viewed in Fig. \ref{plot_cc}.

\subsection{Description of $\partial F_{2}/\partial(\ln Q^{2})$ in HERA and the LHC kinematic region}

\begin{figure}[htbp]
\centering
{\begin{rotate}{90}\vspace{2cm}$\partial F_{2}/\partial(\ln
Q^{2})$\end{rotate}}
\begin{tabular}{cccc cccc cccc ccc}\label{dF2_dLnQ_1}
\epsfig{file=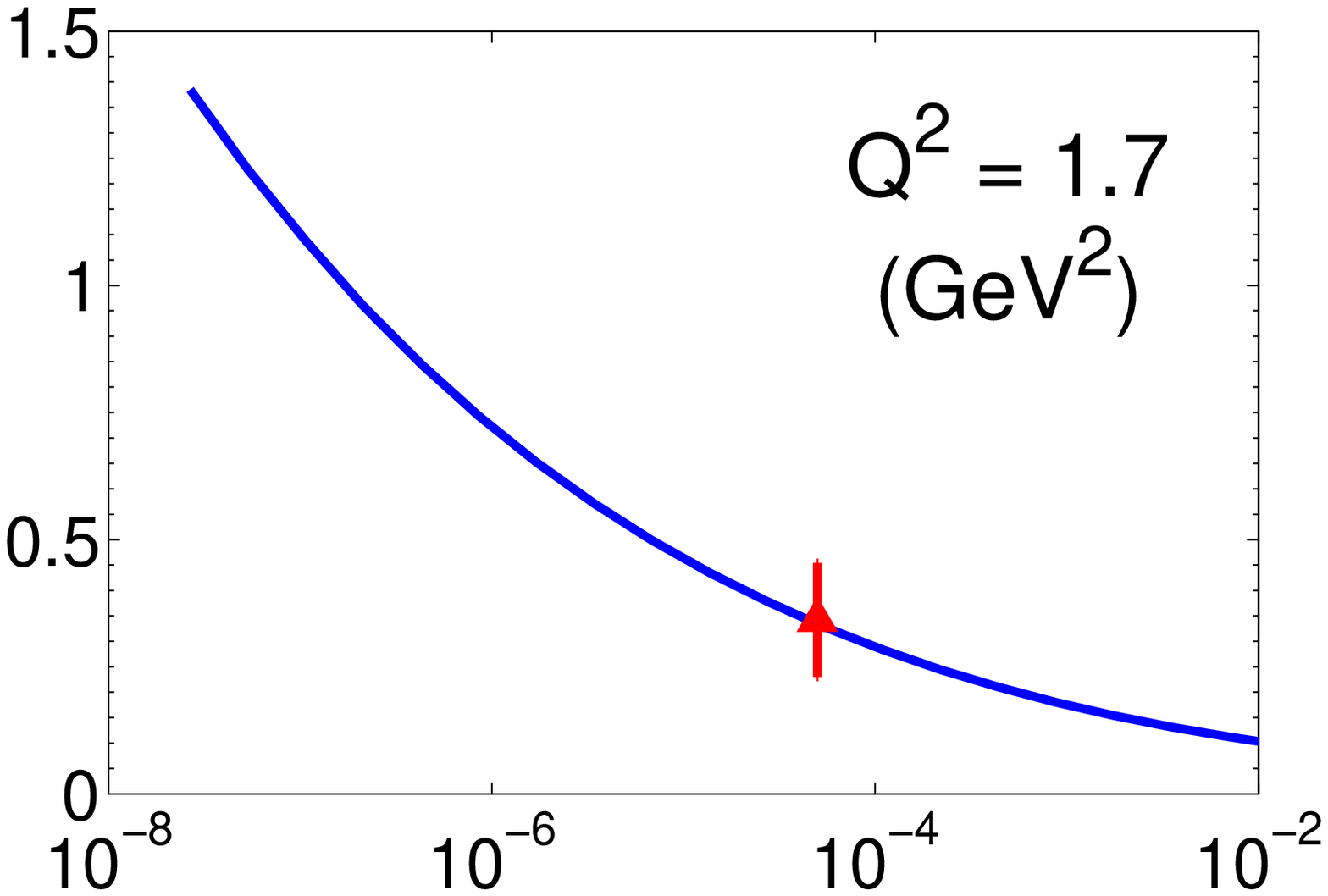,width=35mm, height=20mm}&
\epsfig{file=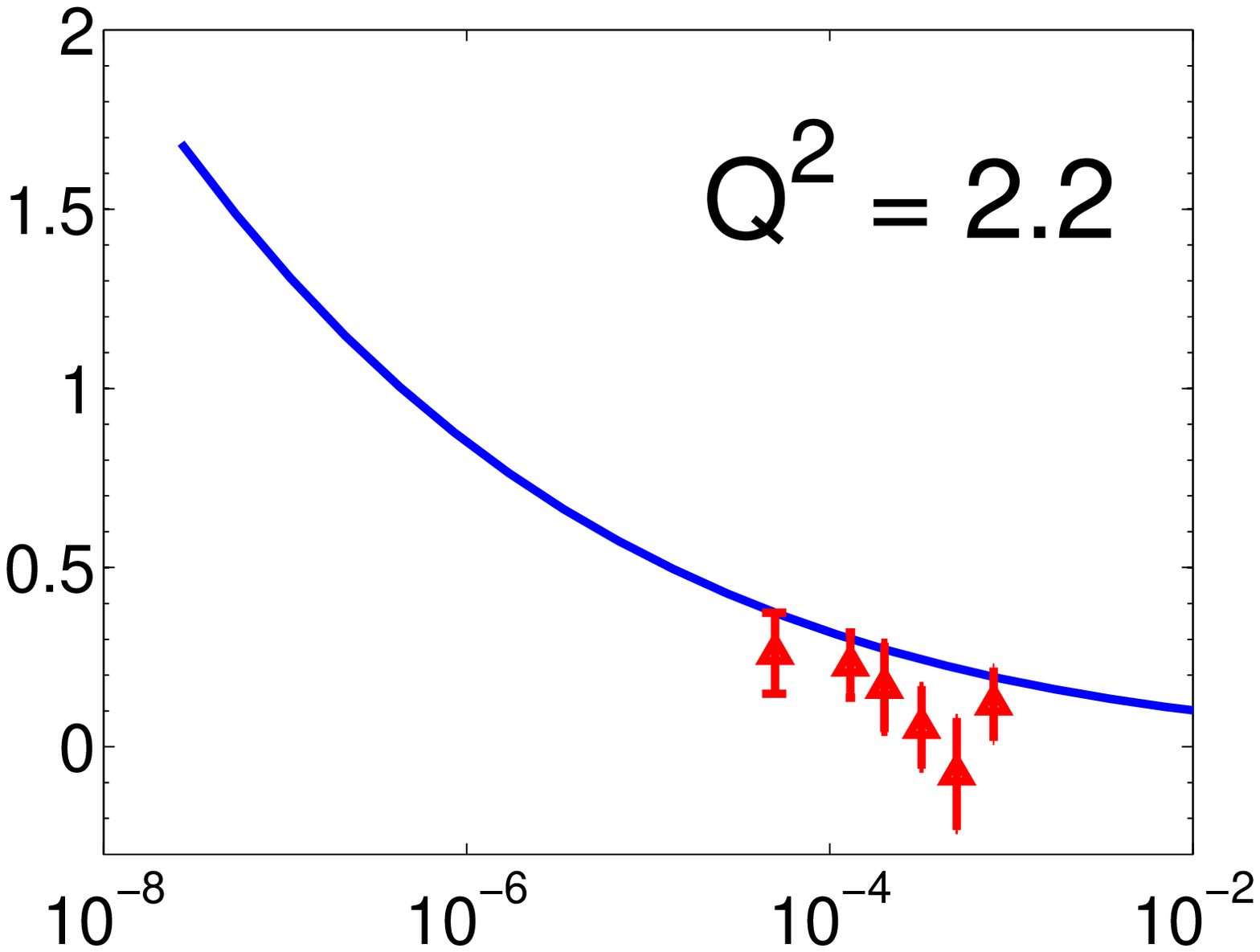,width=35mm, height=20mm}&
\epsfig{file=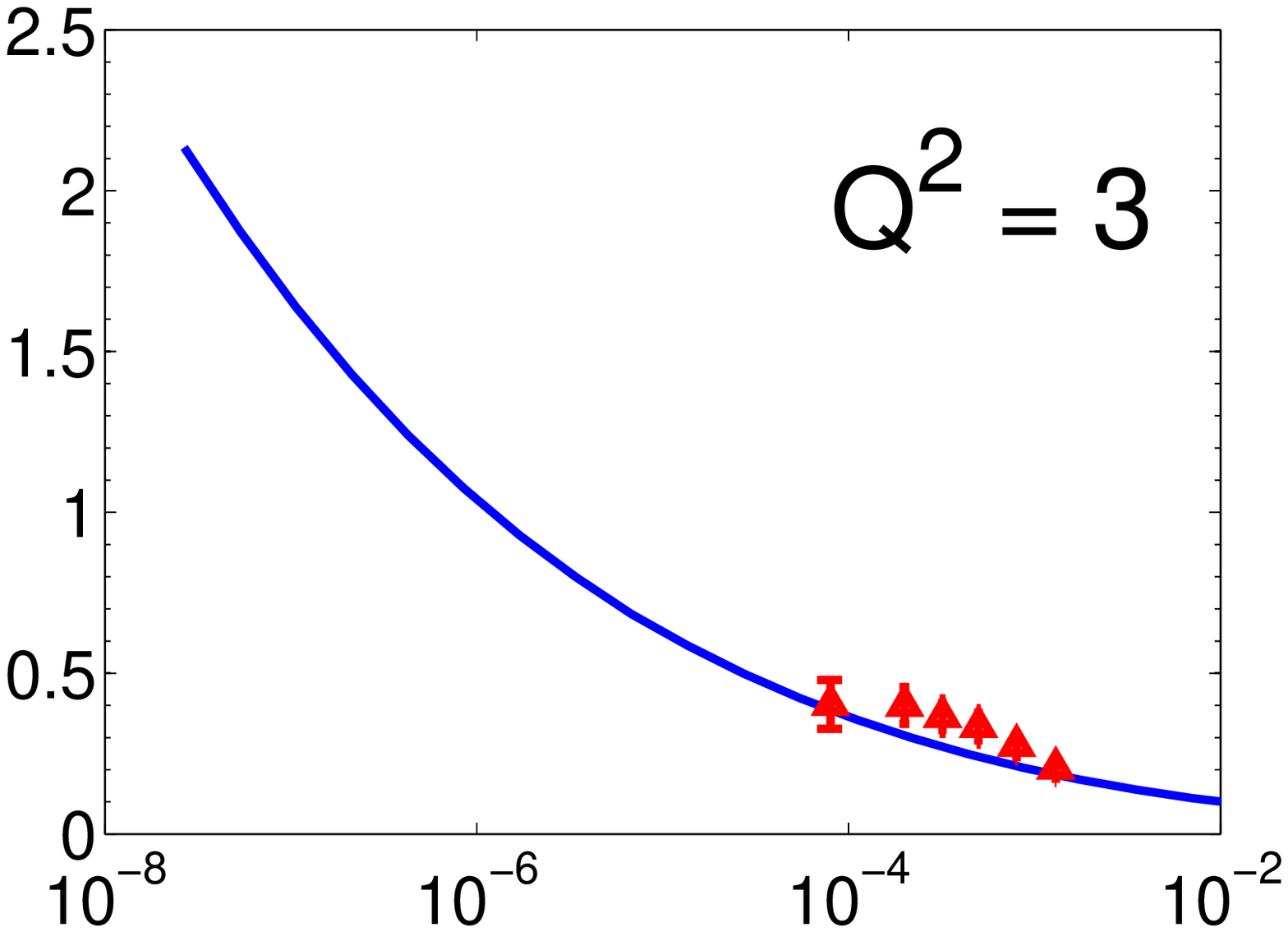,width=35mm, height=20mm}&
\epsfig{file=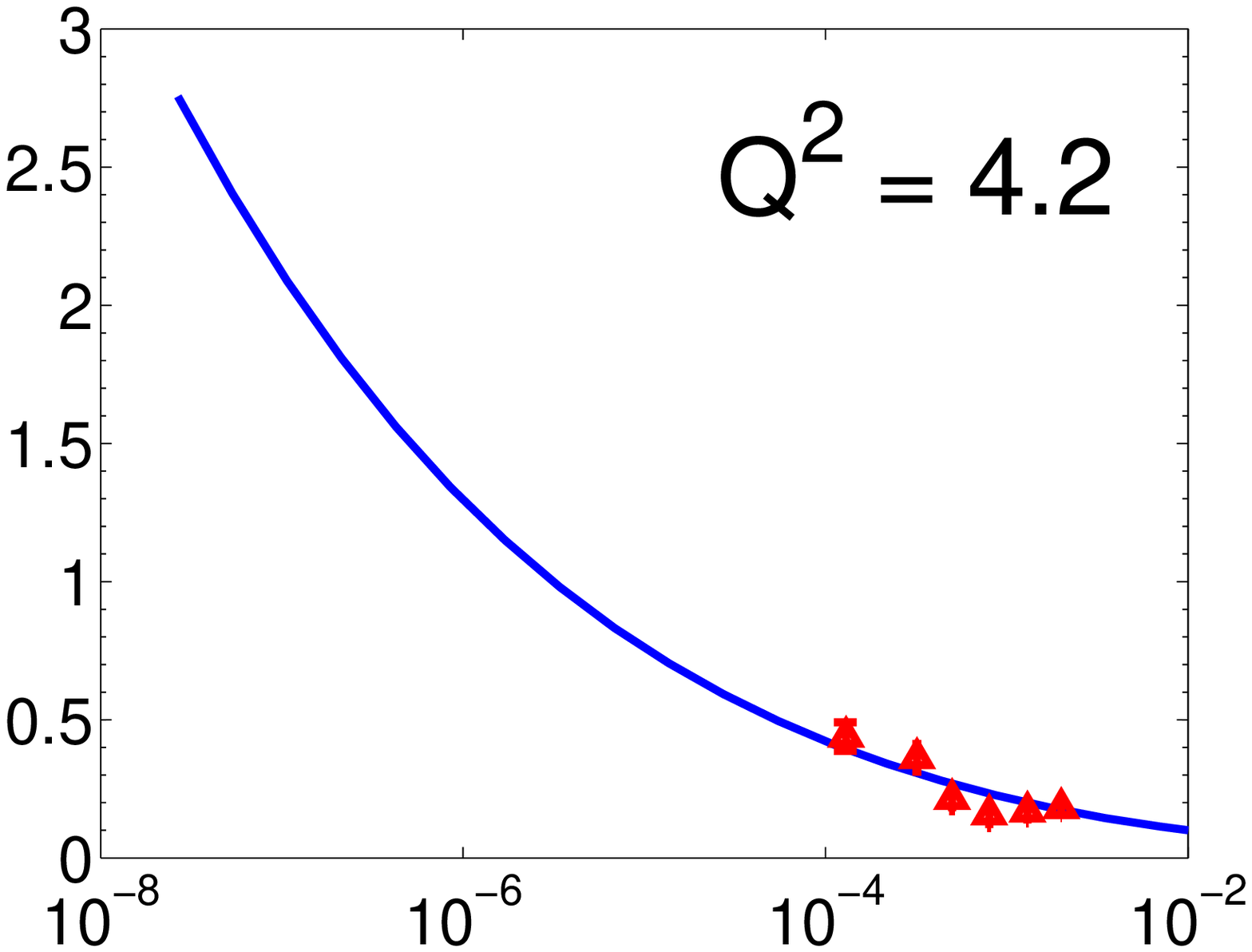,width=35mm, height=20mm}\\
\epsfig{file=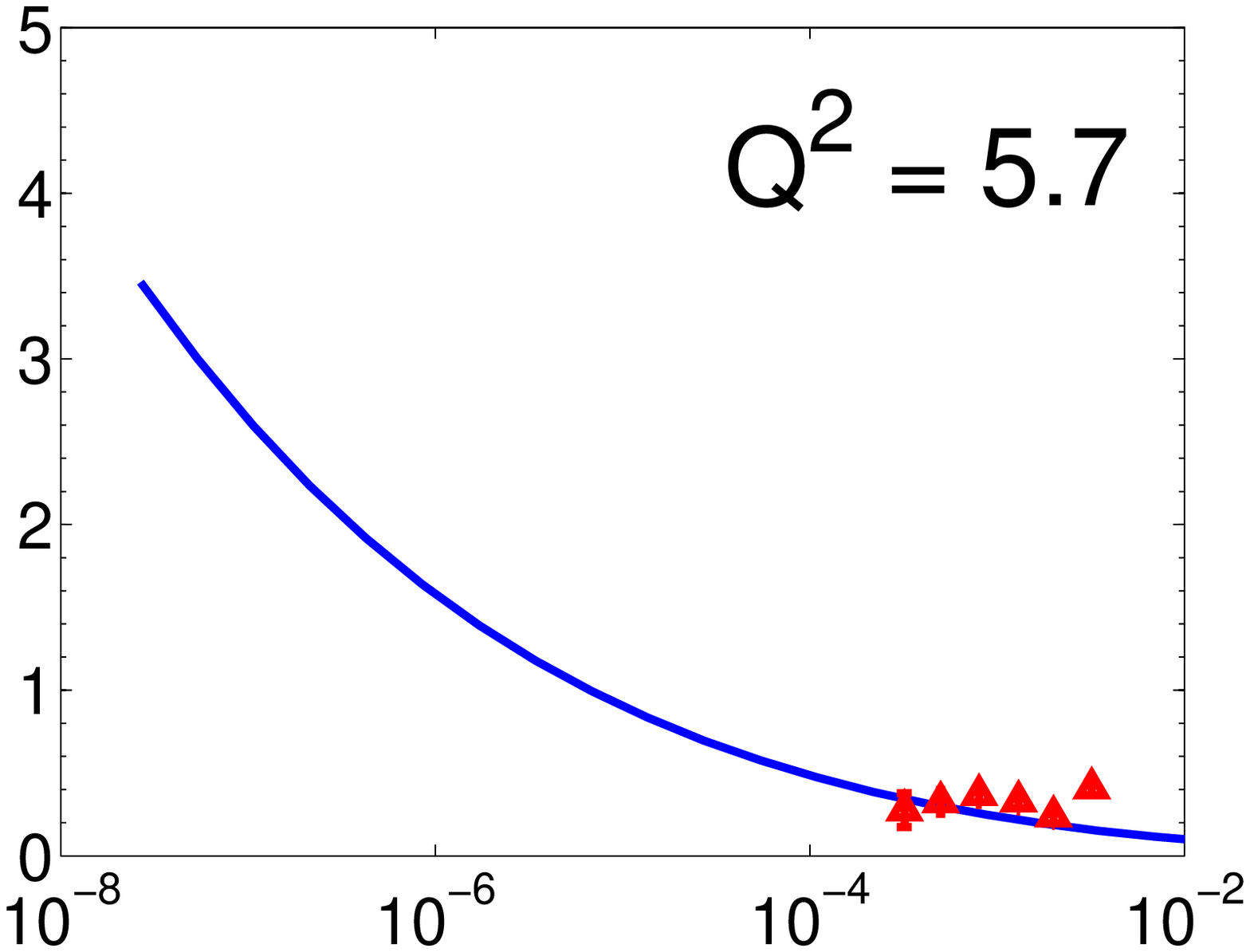,width=35mm, height=20mm}&
\epsfig{file=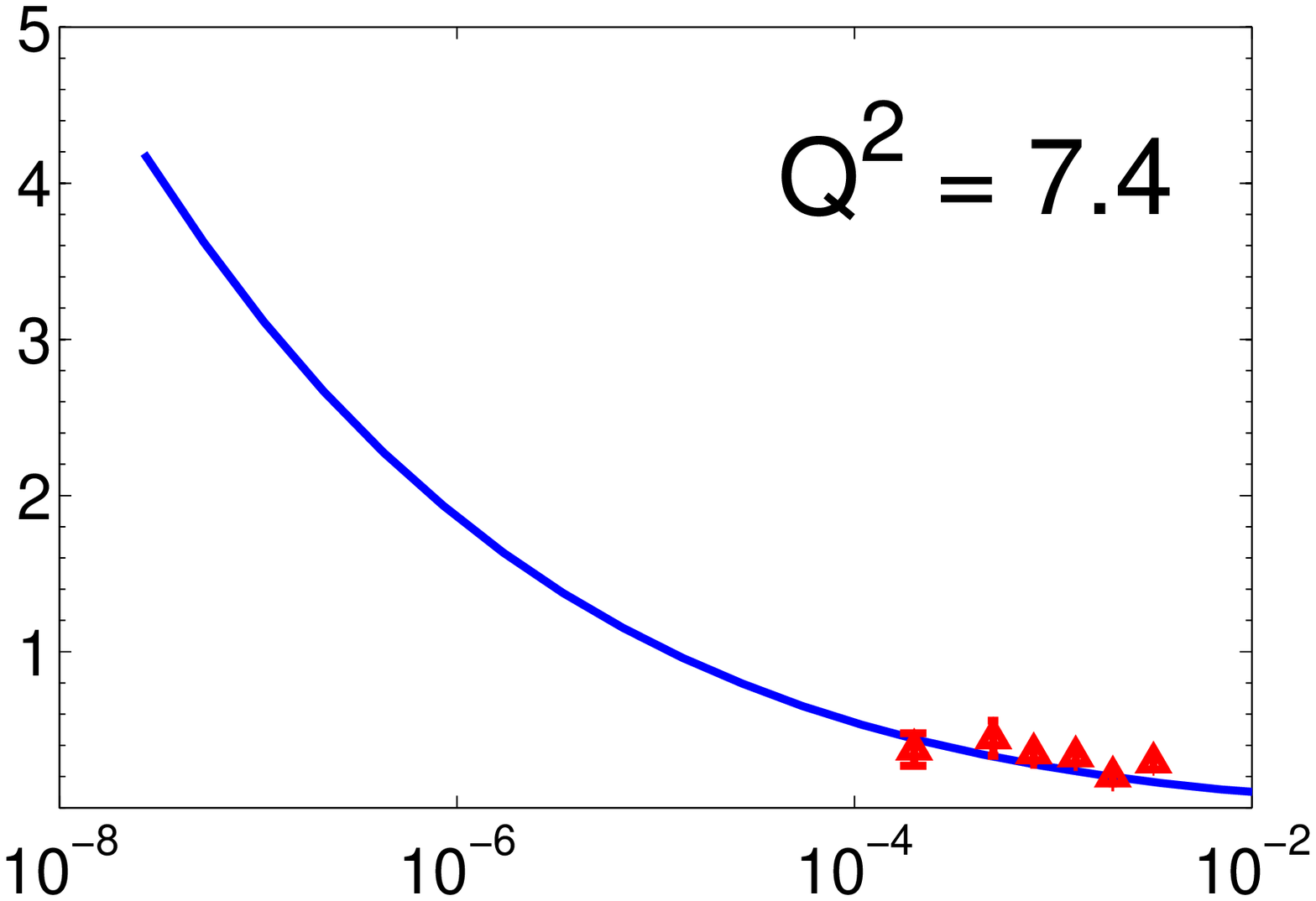,width=35mm, height=20mm}&
\epsfig{file=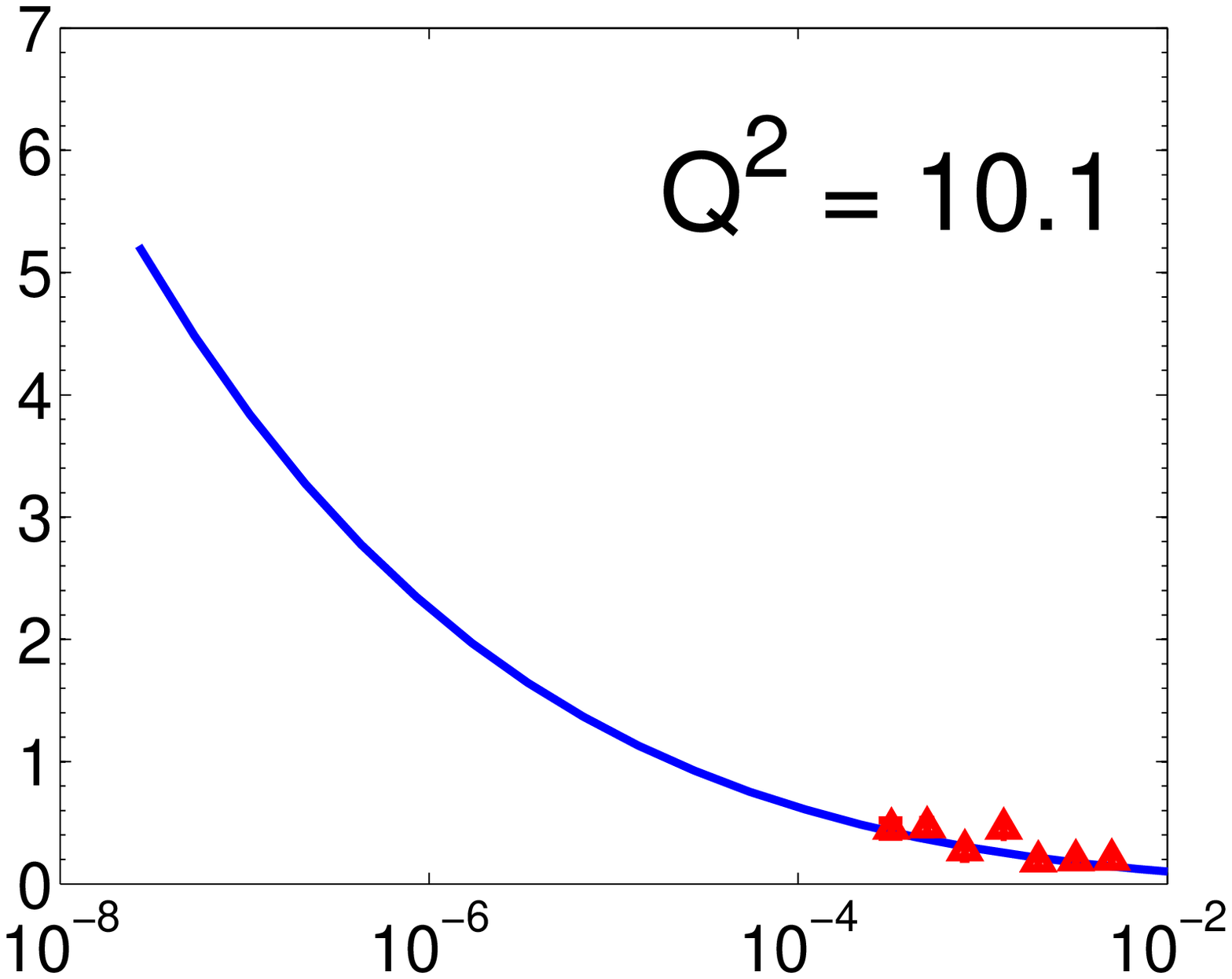,width=35mm, height=20mm}&
\epsfig{file=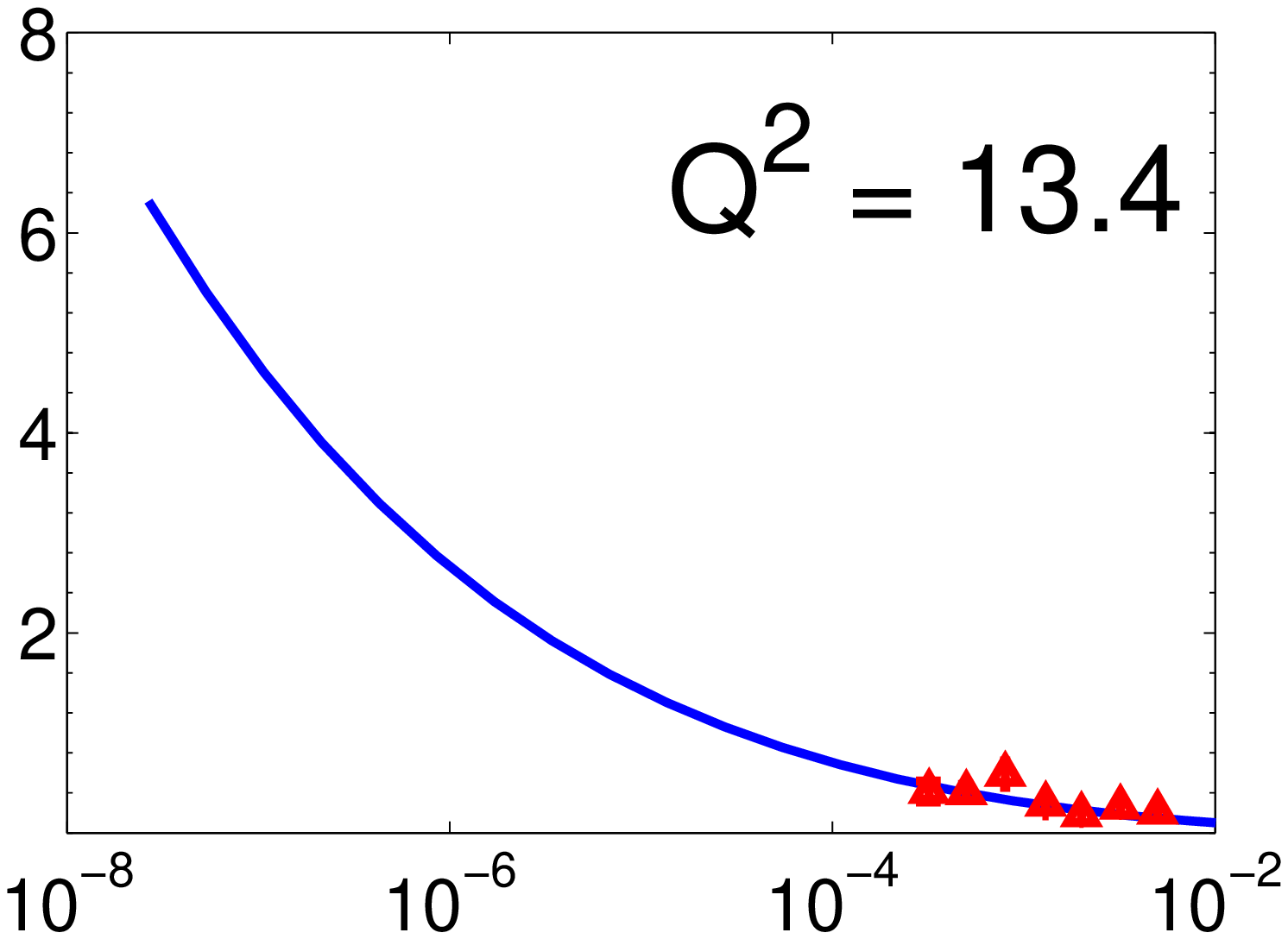,width=35mm, height=20mm}\\
\epsfig{file=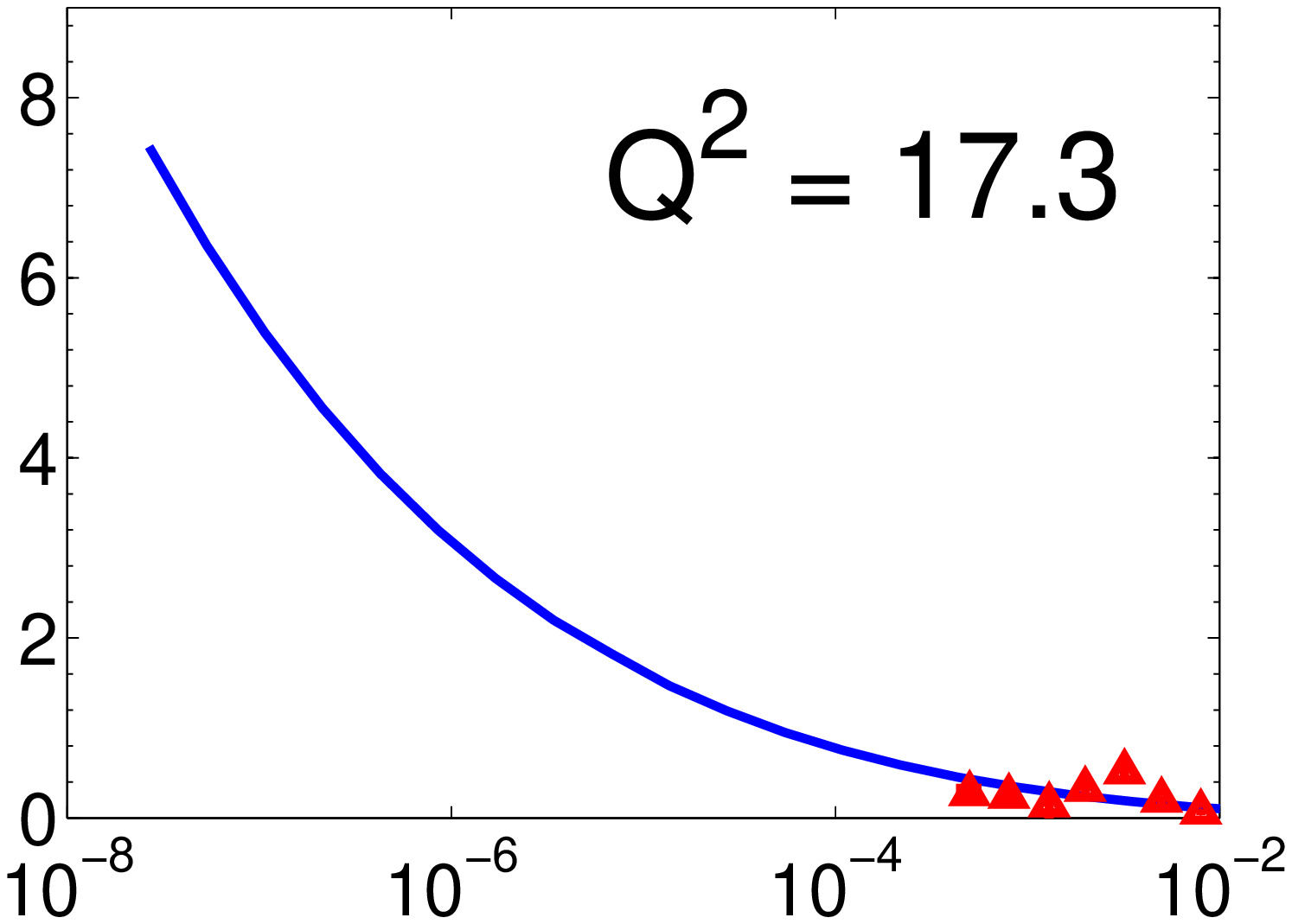,width=35mm, height=20mm}&
\epsfig{file=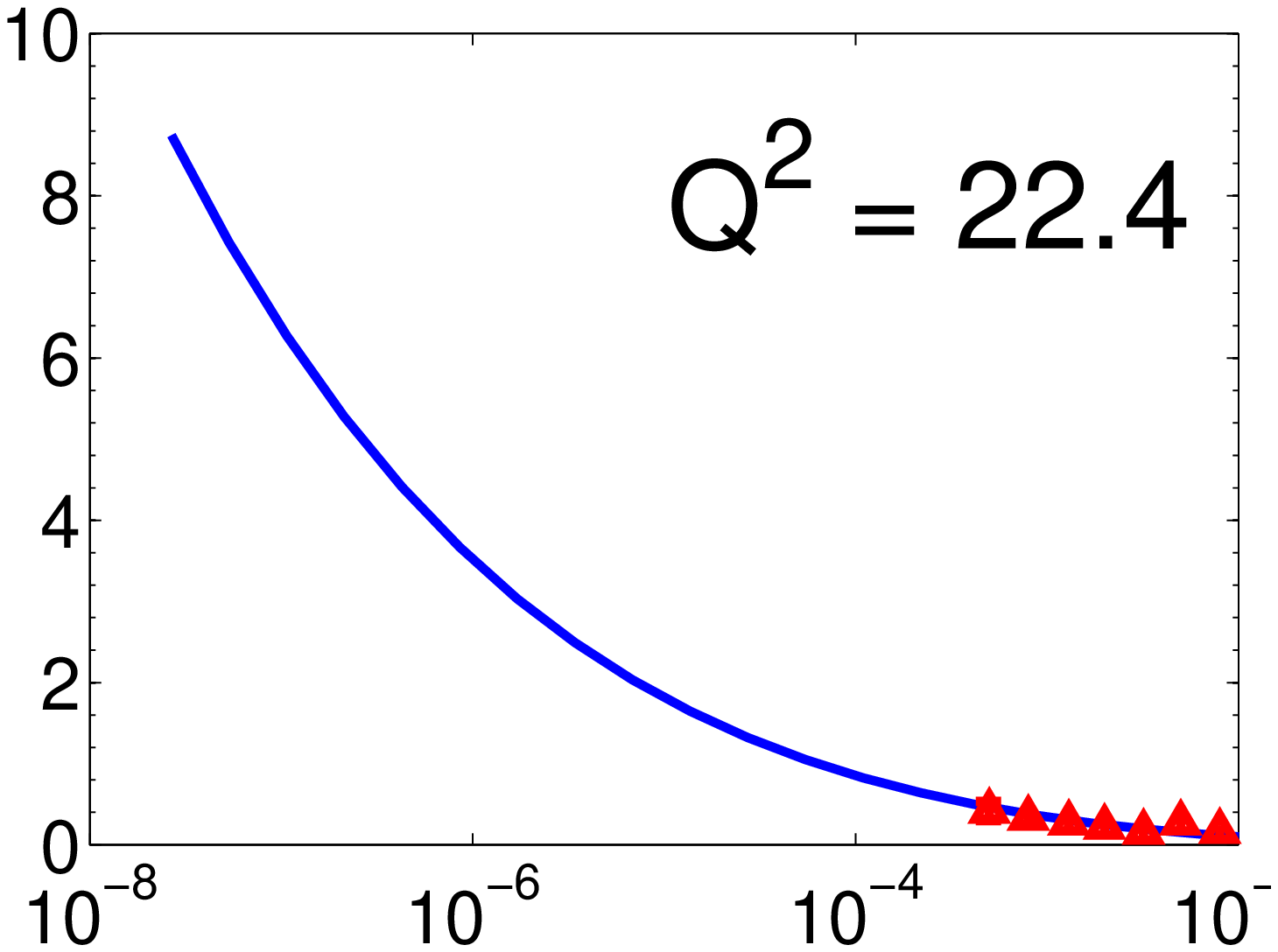,width=35mm, height=20mm}&
\epsfig{file=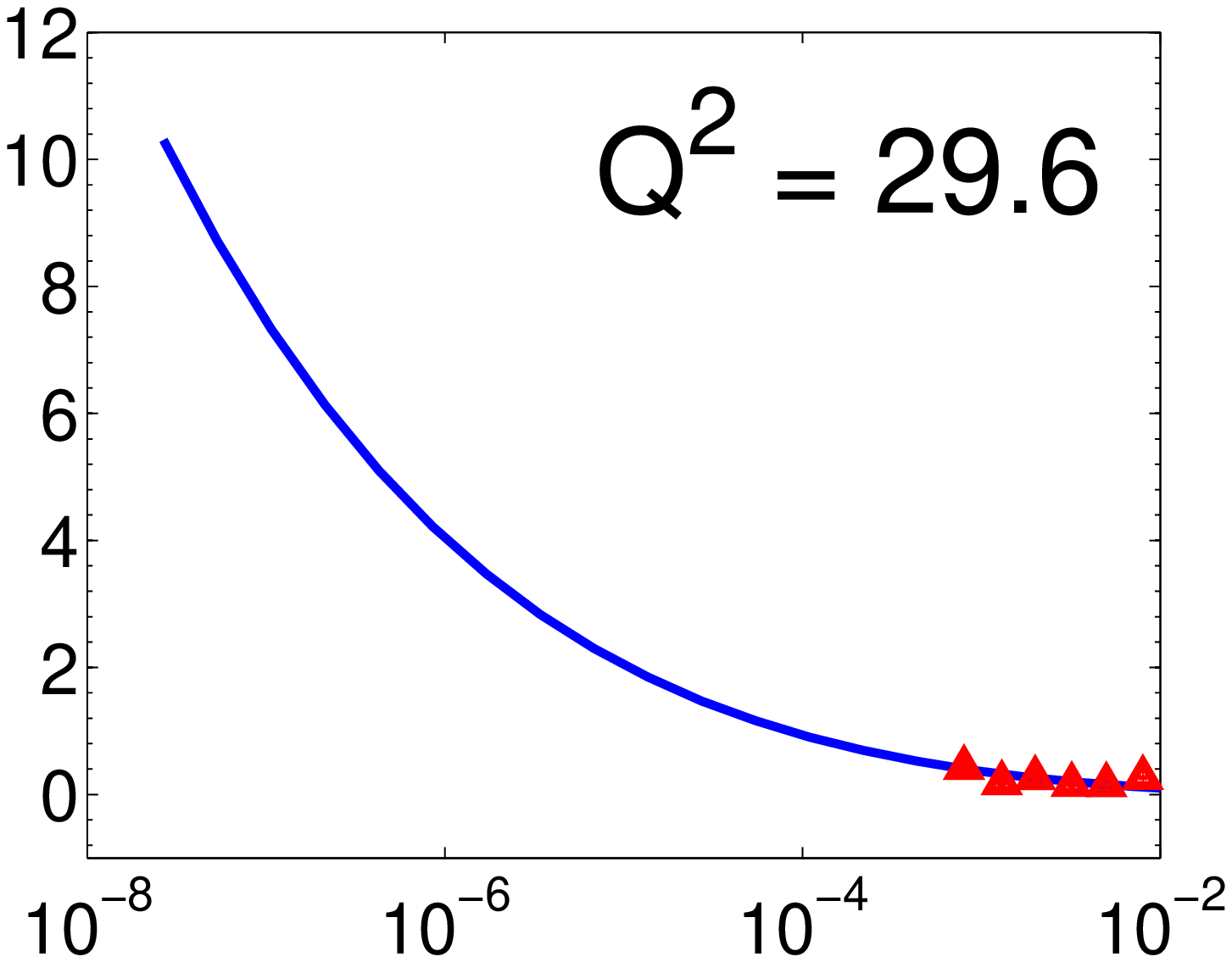,width=35mm, height=20mm}&
\epsfig{file=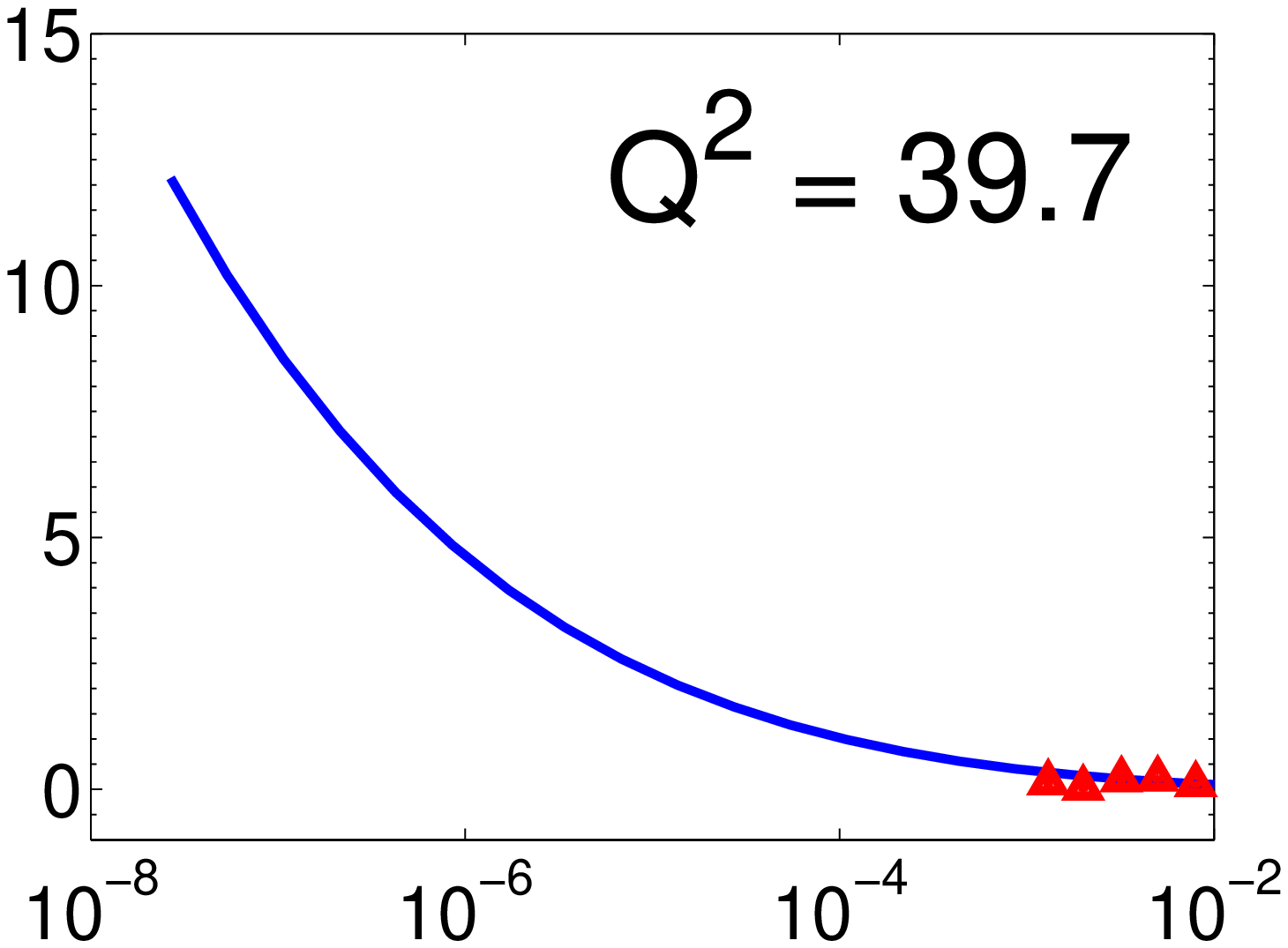,width=35mm, height=20mm}\\
\epsfig{file=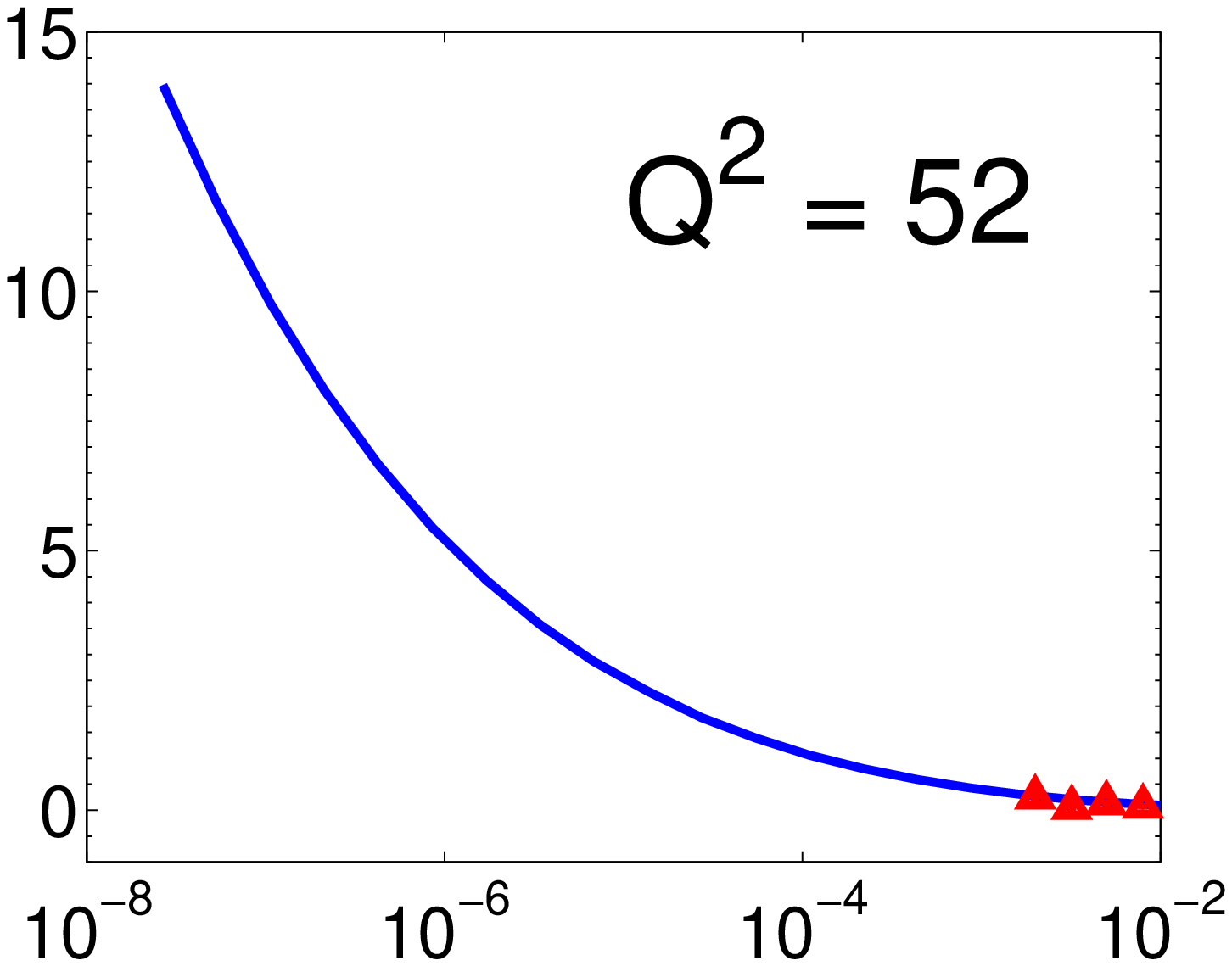,width=35mm, height=20mm}&
\epsfig{file=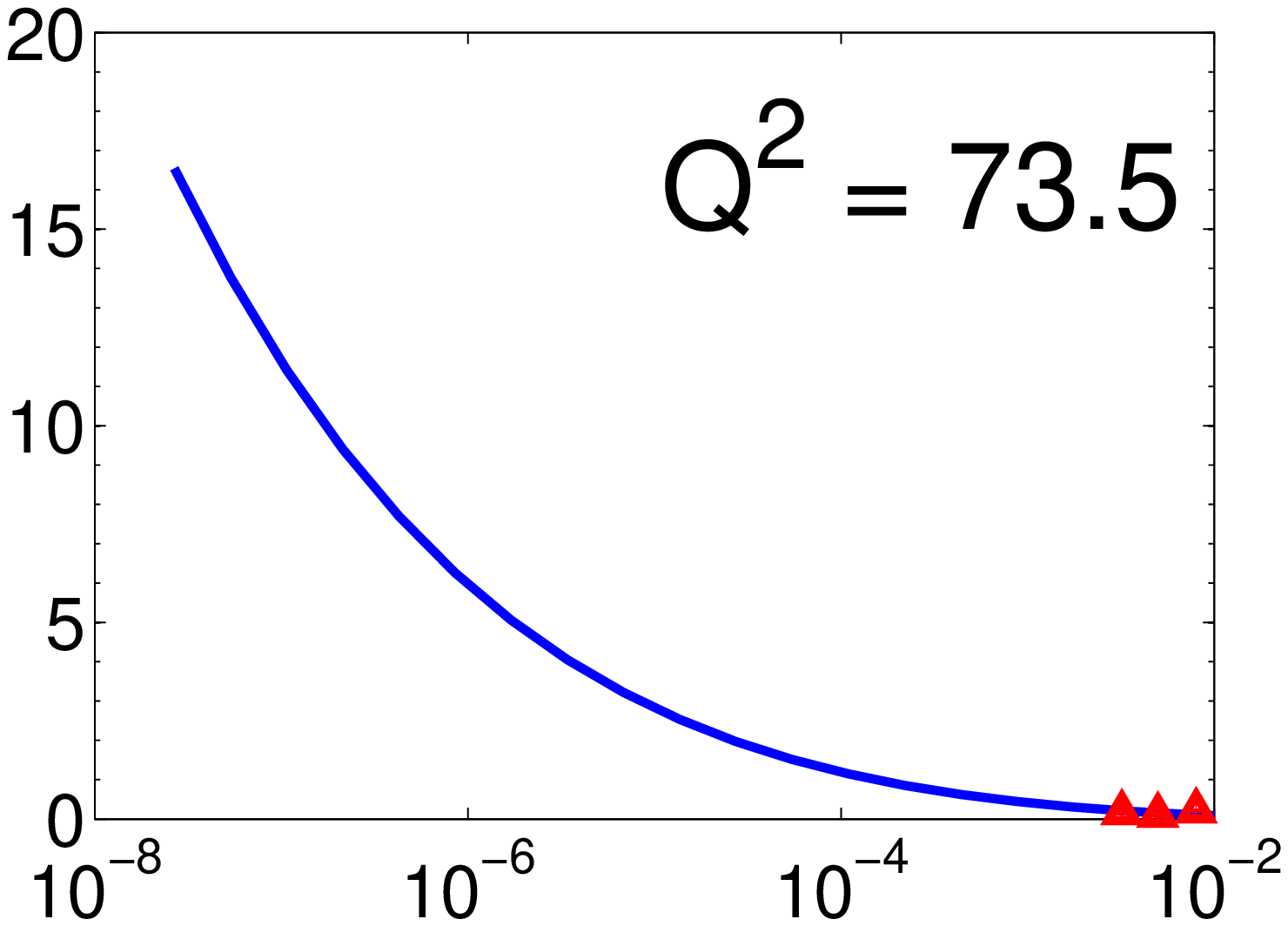,width=35mm, height=20mm}&
\epsfig{file=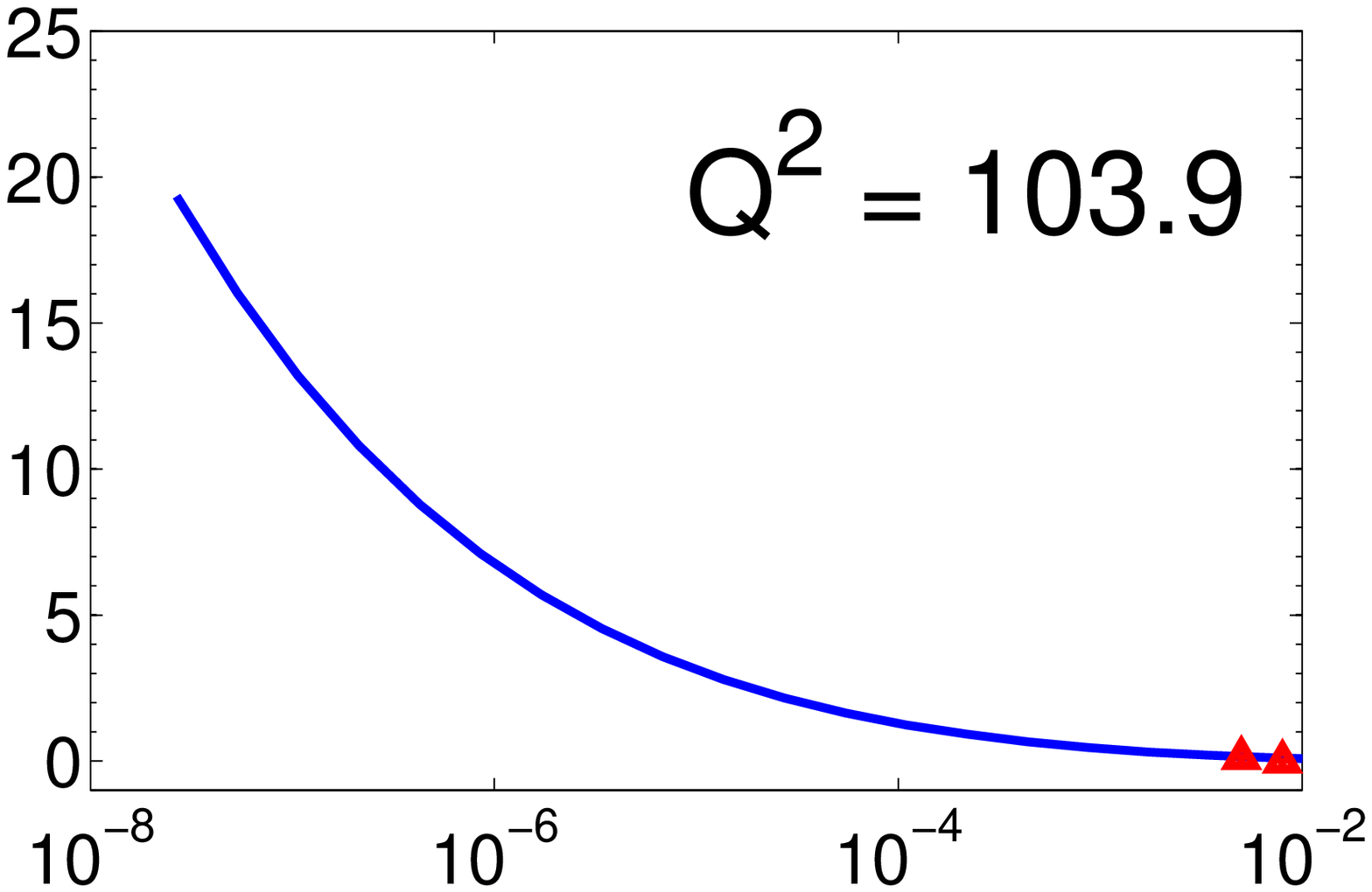,width=35mm, height=20mm}&
\end{tabular}{\begin{center}x\end{center}}\caption{\it Prediction of our model to logarithmic derivative
$\partial F_{2}/\partial(\ln Q^{2})$ as a function of Bjorken-$x$
for fixed values of $Q^{2}$.}
\end{figure}

\begin{figure}[htbp]
\centering
{\begin{rotate}{90}\vspace{2cm}$\partial F_{2}/\partial(\ln
Q^{2})$\end{rotate}}
\begin{tabular}{cccc cccc cccc ccc}
\epsfig{file=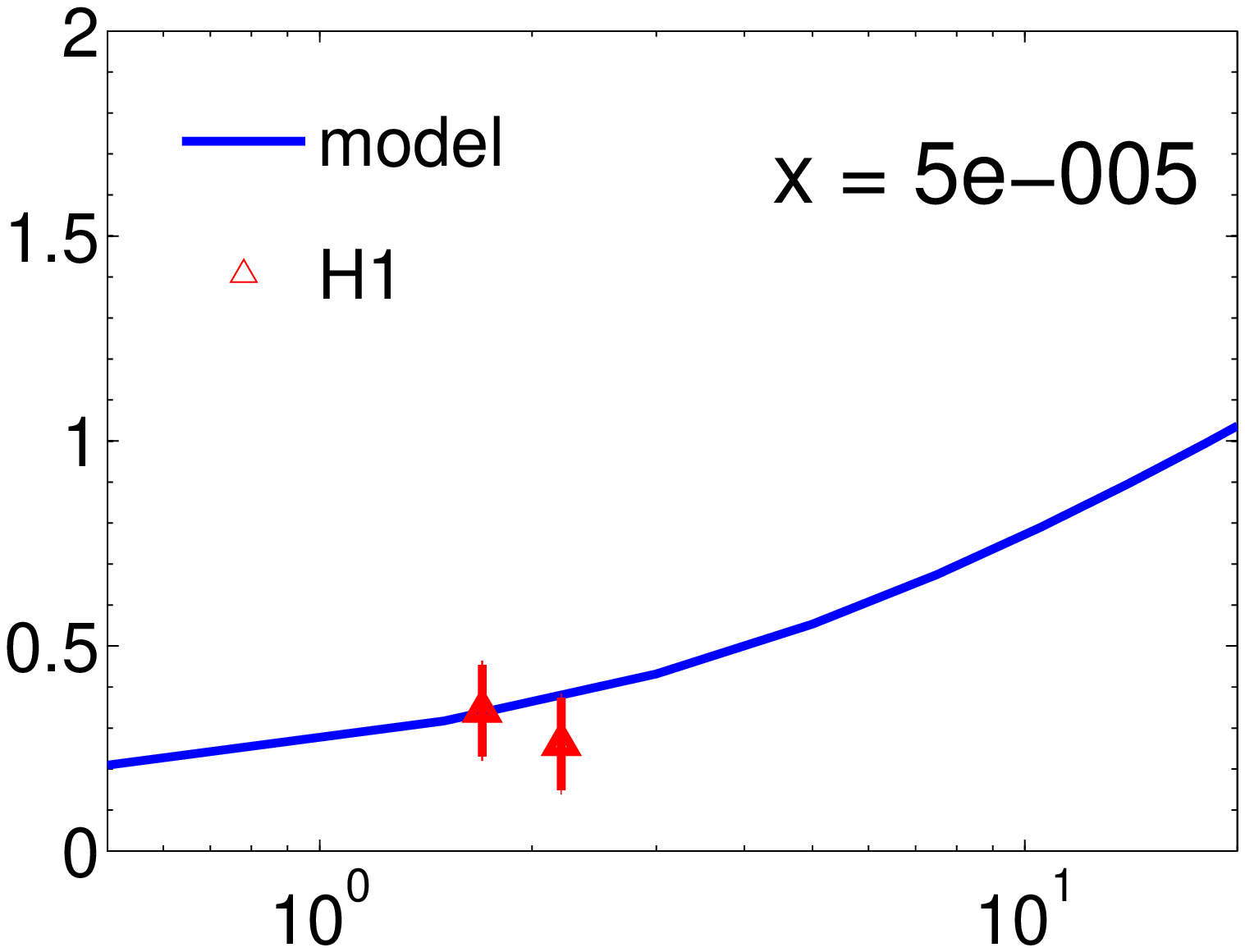,width=35mm, height=20mm}&
\epsfig{file=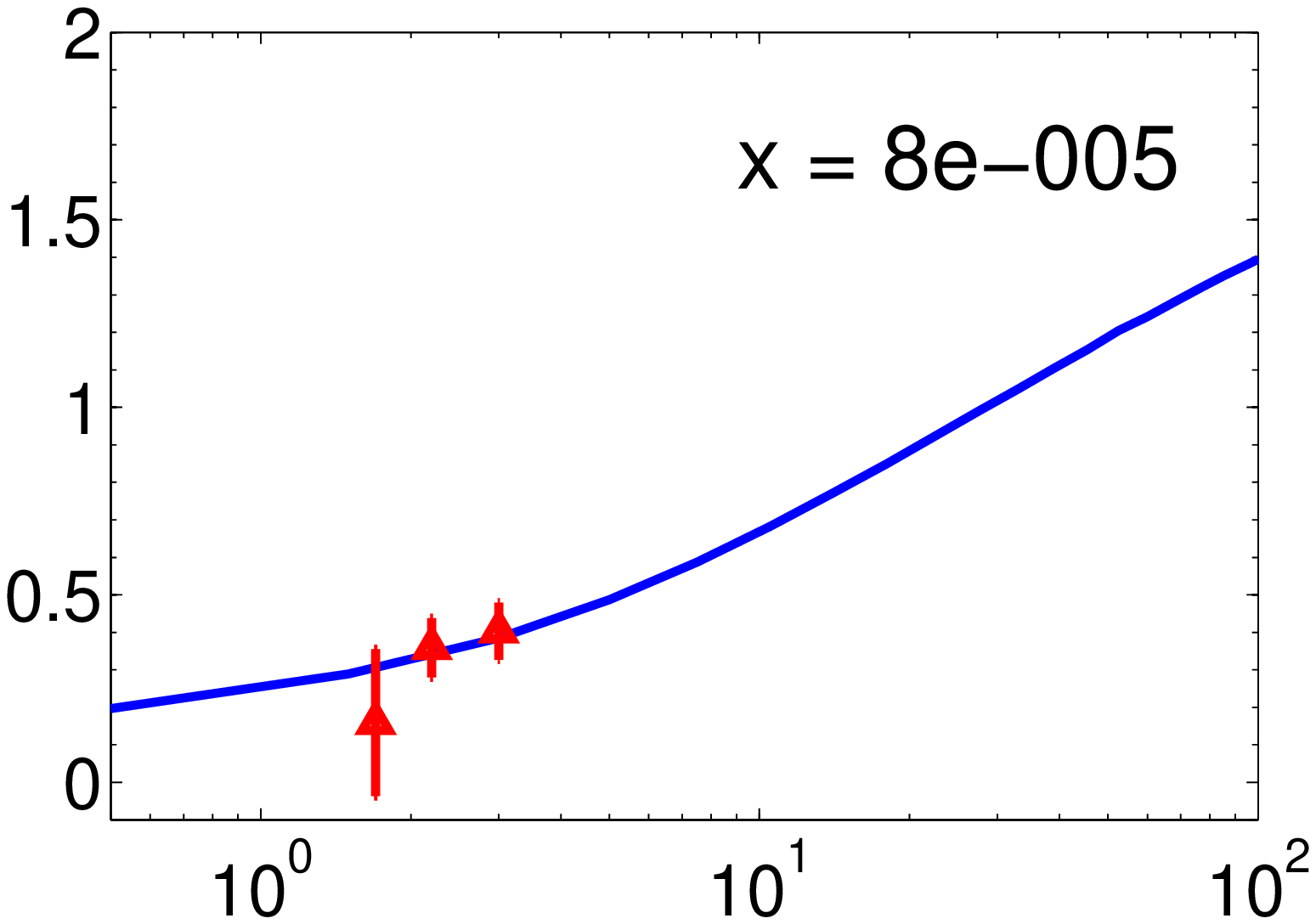,width=35mm, height=20mm}&
\epsfig{file=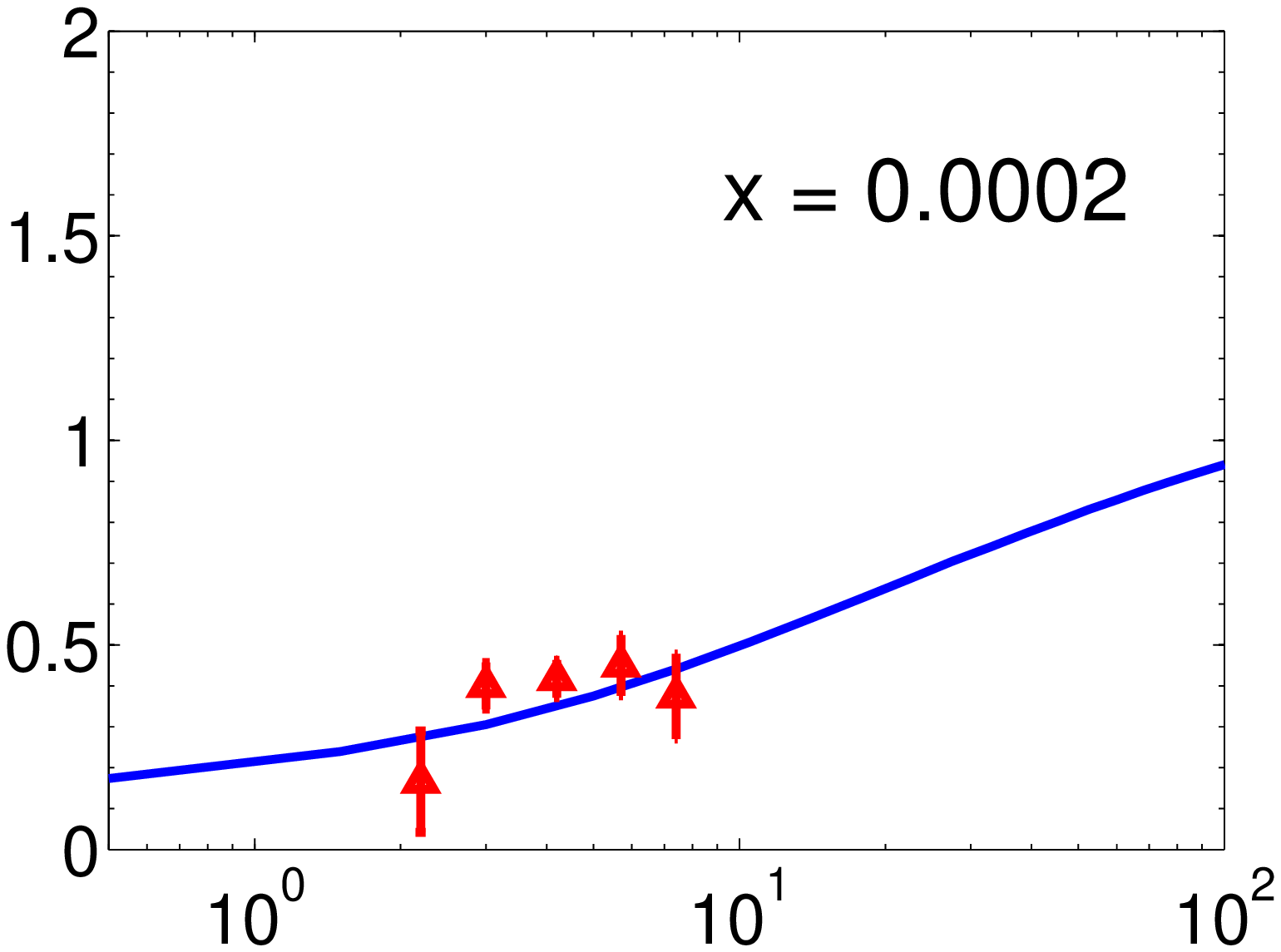,width=35mm, height=20mm}&
\epsfig{file=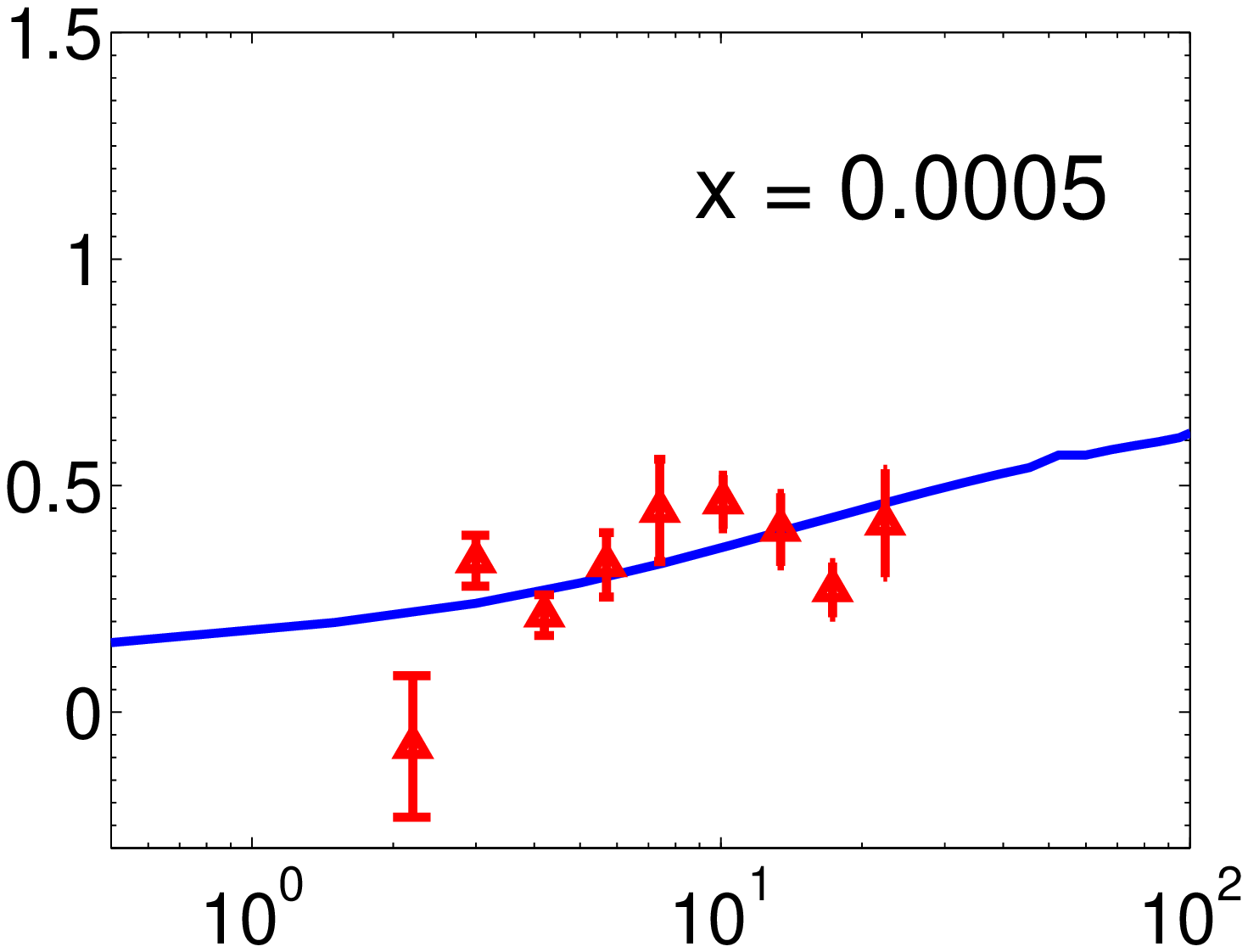,width=35mm, height=20mm}\\
\epsfig{file=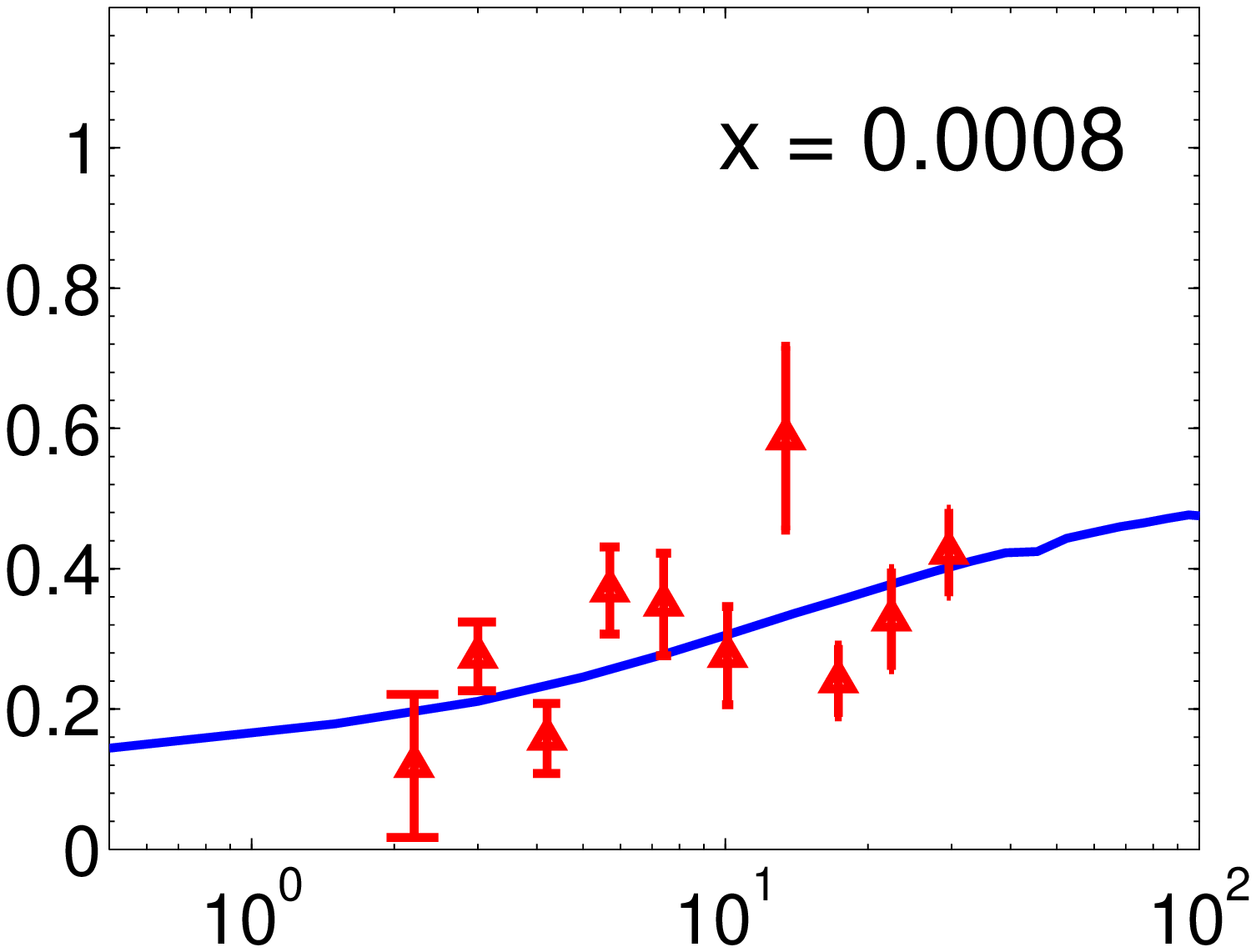,width=35mm, height=20mm}&
\epsfig{file=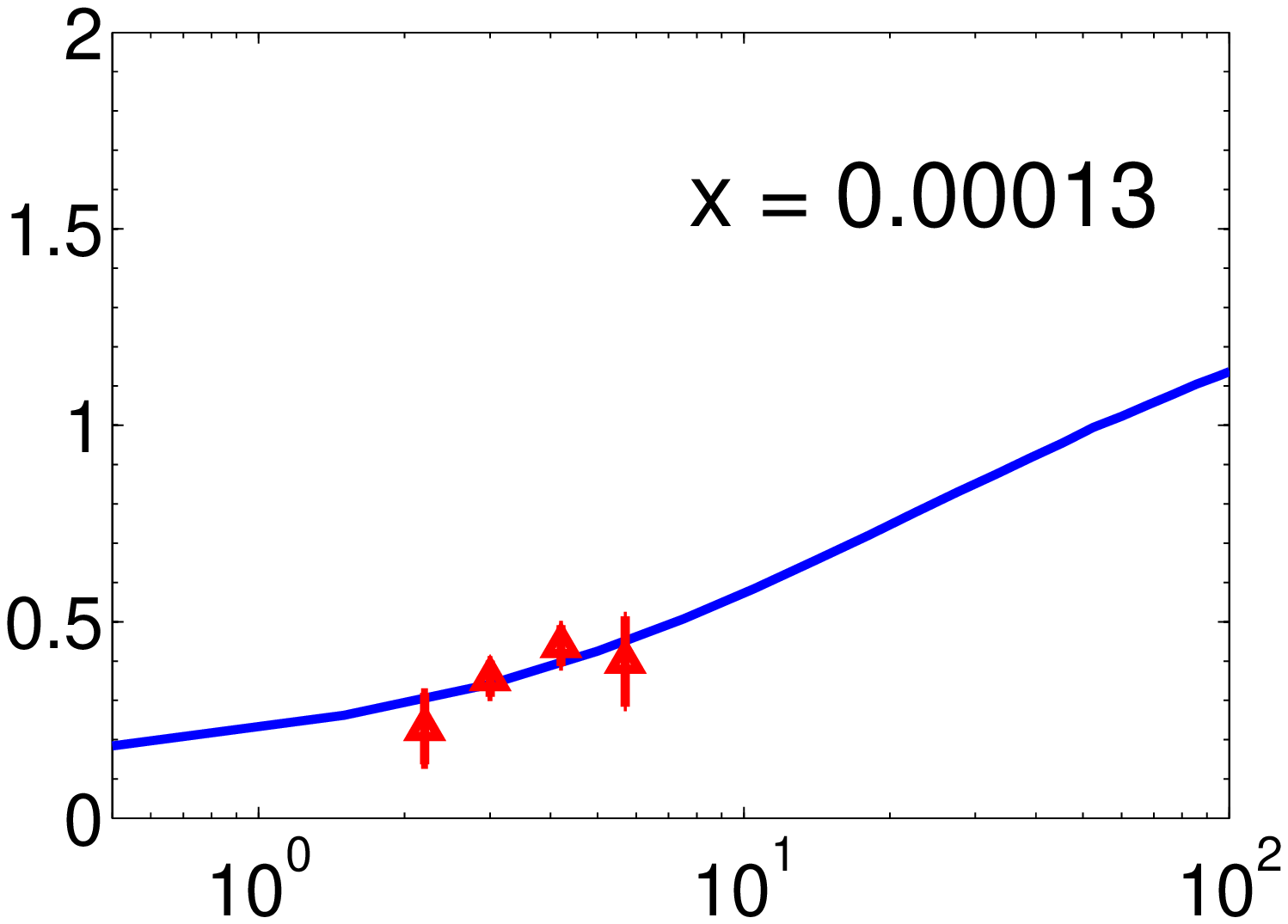,width=35mm, height=20mm}&
\epsfig{file=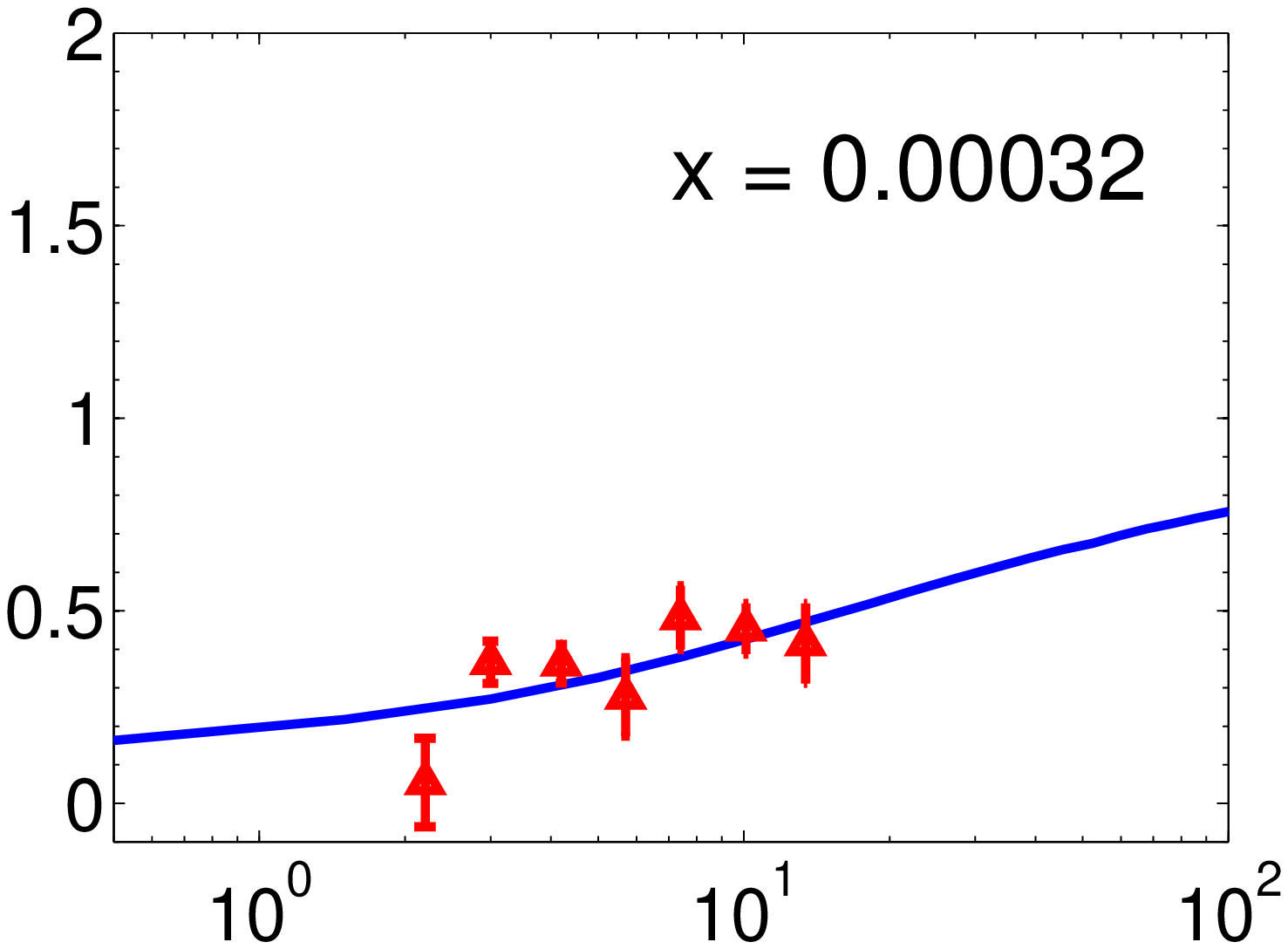,width=35mm, height=20mm}&
\epsfig{file=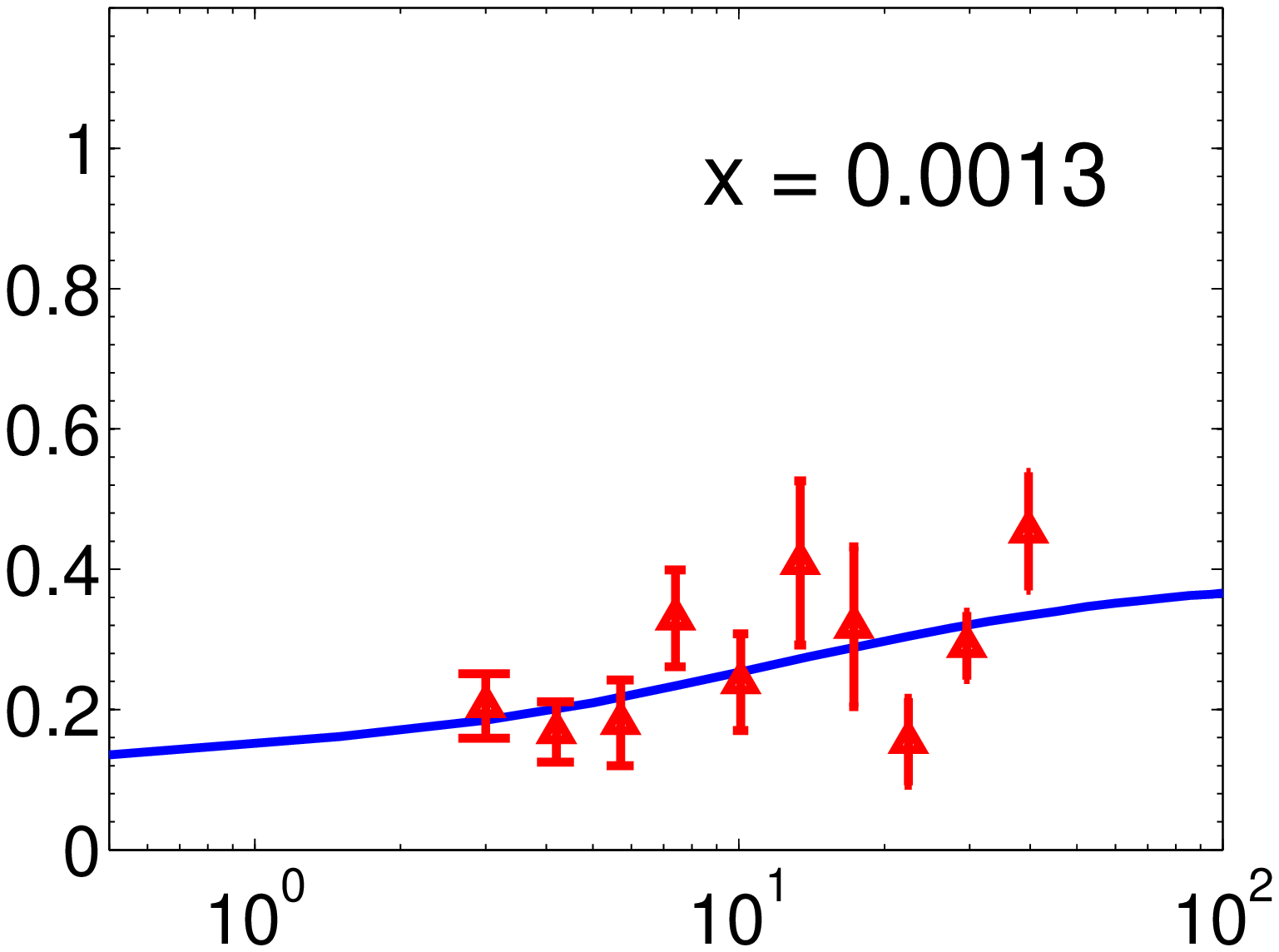,width=35mm, height=20mm}\\
\epsfig{file=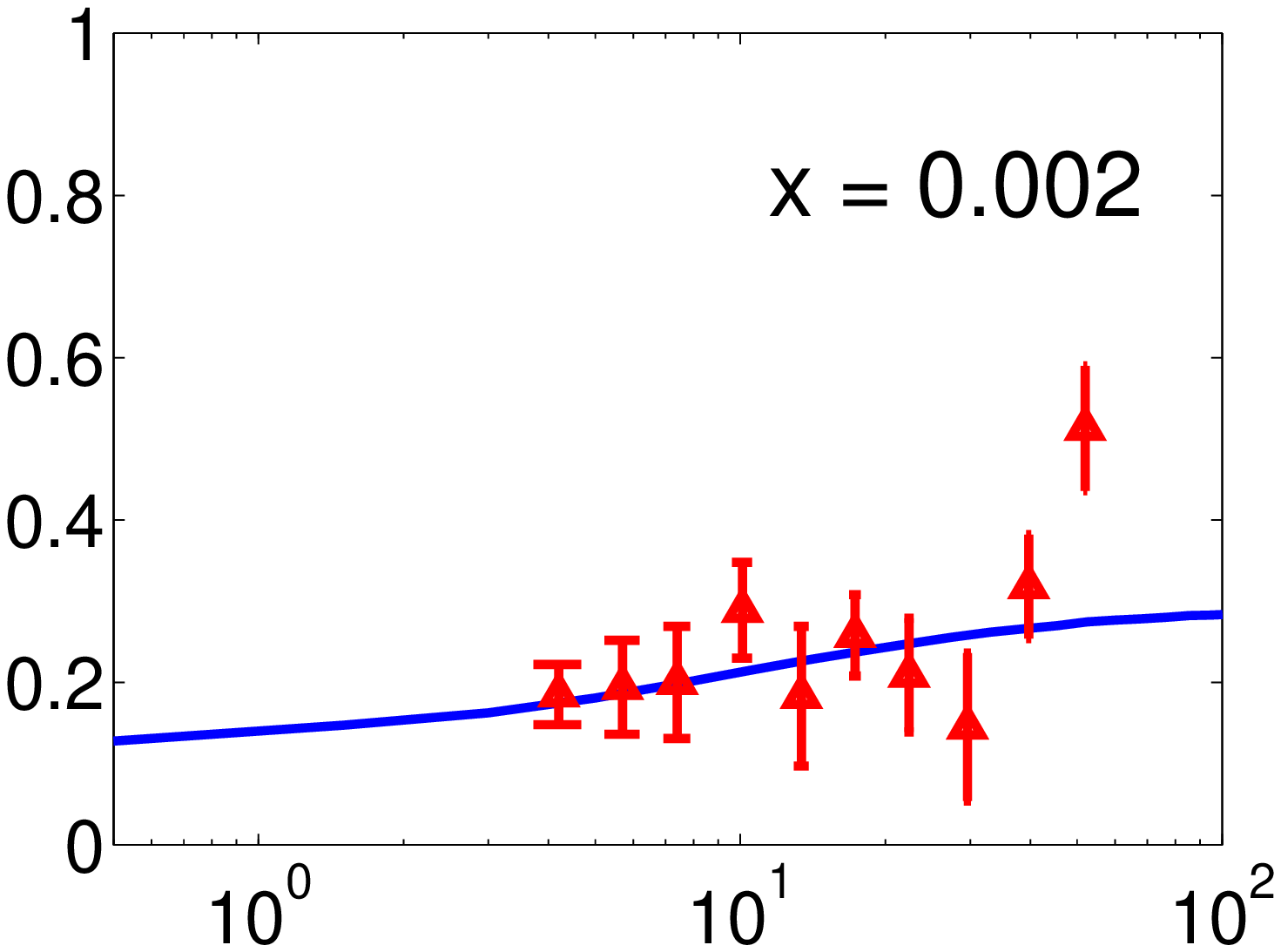,width=35mm, height=20mm}&
\epsfig{file=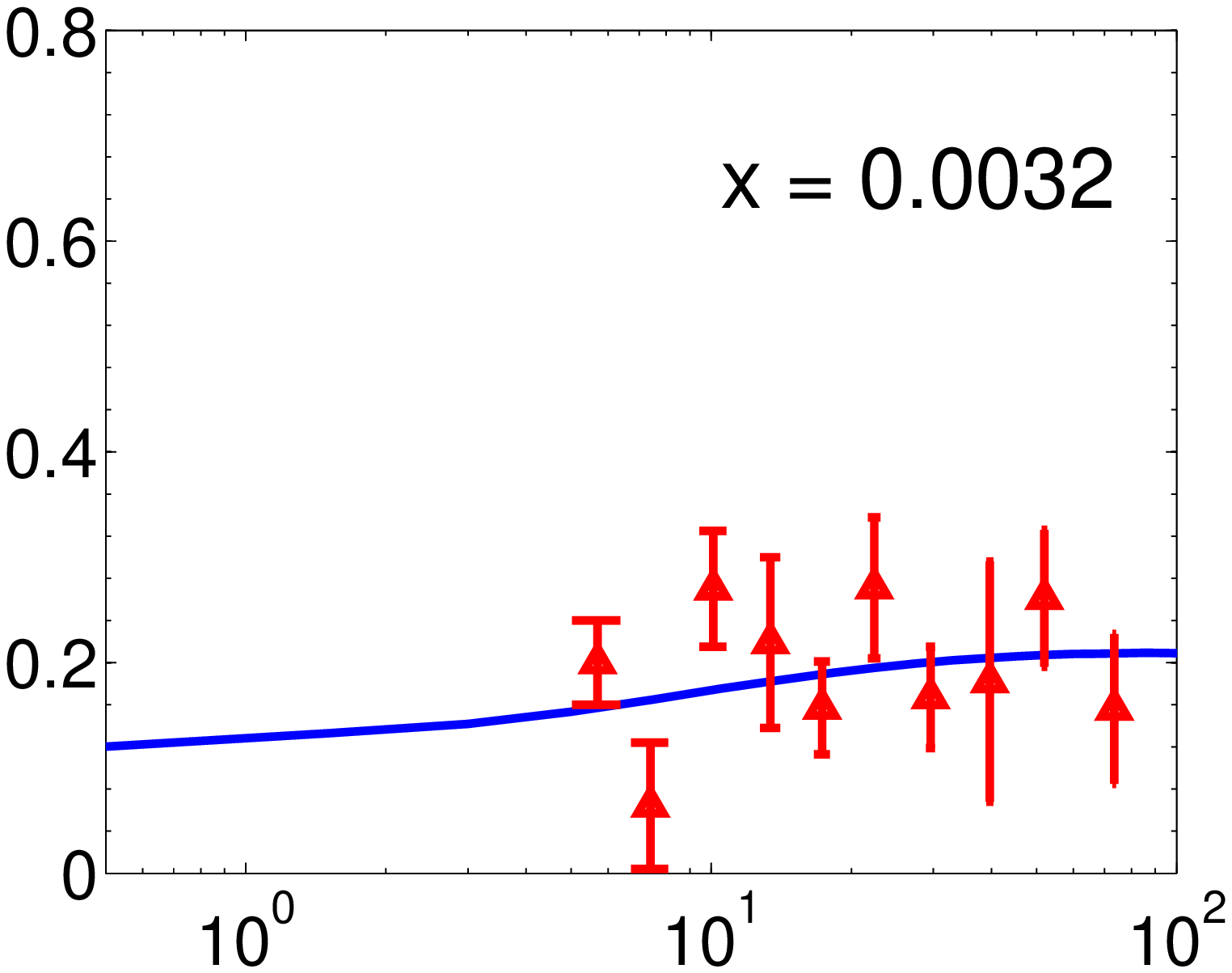,width=35mm, height=20mm}&
\epsfig{file=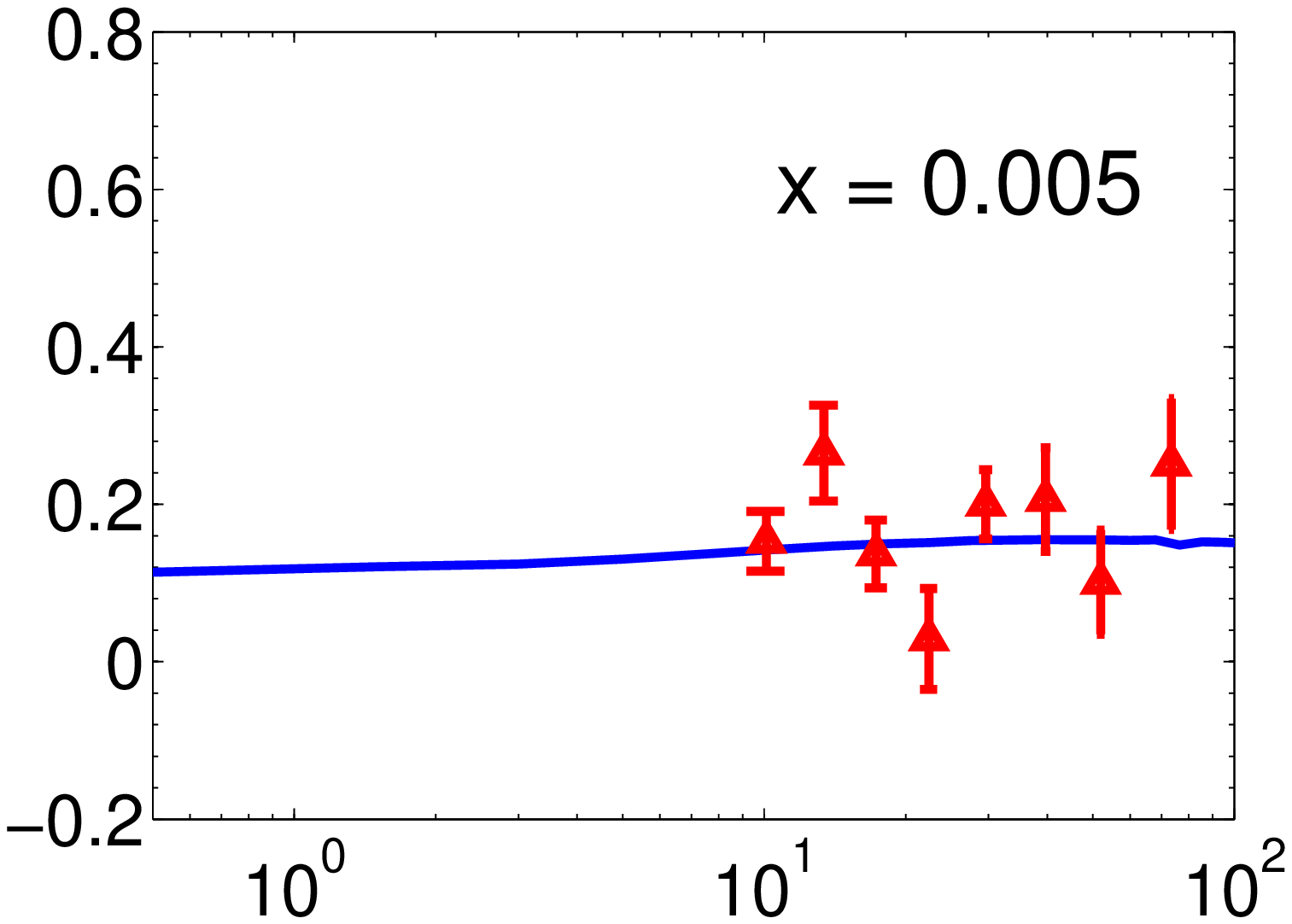,width=35mm, height=20mm}&
\epsfig{file=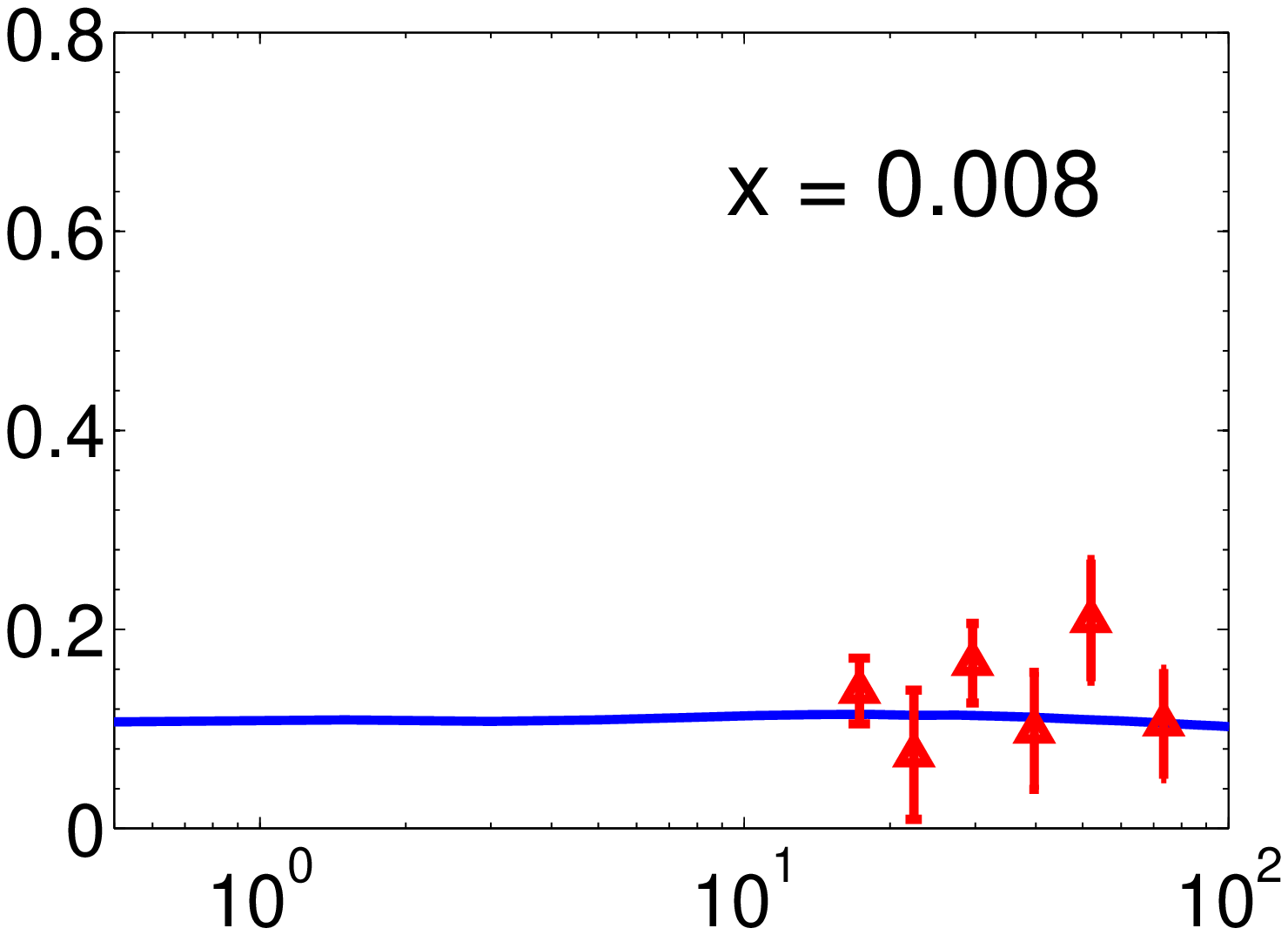,width=35mm, height=20mm}&
\end{tabular}{\begin{center}$Q^{2}$\;\;$(GeV^{2})$\end{center}}
\caption{\it Logarithmic derivative of $F_{2}$ as a function of
$Q^{2}$ for fixed values of Bjorken-$x$. Solid line is our model and
triangles correspond to \emph{H1 96-97} \cite{Adloff:2000qk}
respectively.}\label{dF2_dLnQ_2}\end{figure}

In this section, we want to check how our model is able to describe
the slope of the structure function, as a function of the photon
virtuality $Q^{2}$, for fixed values of Bjorken-$x$, and vice versa.
For this purpose, we calculate the logarithmic derivative of $F_{2}$
\beqn \lambda_{Q^{2}}\;\equiv\;\frac{\partial F_{2}}{\partial(\ln
Q^{2})}\eeqn Since the only dependence on the virtuality $Q^{2}$, is
in the photon wavefunction, the resulting expression for the
calculation has the following form

\beqn \frac{\partial F_{2}}{\partial(\ln
Q^{2})}\;=\;Q^{2}\frac{\partial F_{2}}{\partial
Q^{2}}\;=\;Q^{2}\frac{\partial}{\partial
Q^{2}}\left(\frac{Q^{2}}{4\pi^{2}\alpha}\sigma(x,Q^{2})\right)\;=\;F_{2}\;
+\;\frac{Q^{4}}{4\pi^{2}\alpha}\frac{\partial\sigma(x,Q^{2})}{\partial
Q^{2}}\eeqn and \beqn \frac{\partial\sigma(x,Q^{2})}{\partial
Q^{2}}\;=\;2\int_{0}^{\infty}d^{2}b\int_{0}^{\infty}d^{2}r\int_{0}^{1}dz\;
\frac{\partial|\Psi(r,z,Q^{2})|^{2}}{\partial Q^{2}}\,N(r,x,b)\eeqn
where $|\Psi(r,z,Q^{2})|^{2}$ and $N(r,x,b)$  are defined in
\eq{psi} and \eq{int_amp_fin}, respectively. The resulting plots are
presented in Figs. \ref{dF2_dLnQ_1},~\ref{dF2_dLnQ_2} for fixed
values of $Q^{2}$ and $x$ respectively. Note, that these sets of
experimental data, were not take into account in the fitting
procedure.

We can see, that predictions which are based on our model, fits well
with all experimental data on logarithmic derivative of $F_{2}$. We
enlarged the kinematic region towards very low $x$ values, to give
predictions for $\lambda_{Q^{2}}$ in the HERA and the LHC kinematic
region.

\subsection{Description of $\partial \ln F_{2}/\partial(\ln 1/x)$}

\begin{figure}[htbp]
\centering
{\begin{rotate}{90}$\partial\ln
F_{2}/\partial(\ln\;1/x)$\end{rotate}}
\begin{tabular}{c c c c c}
\epsfig{file=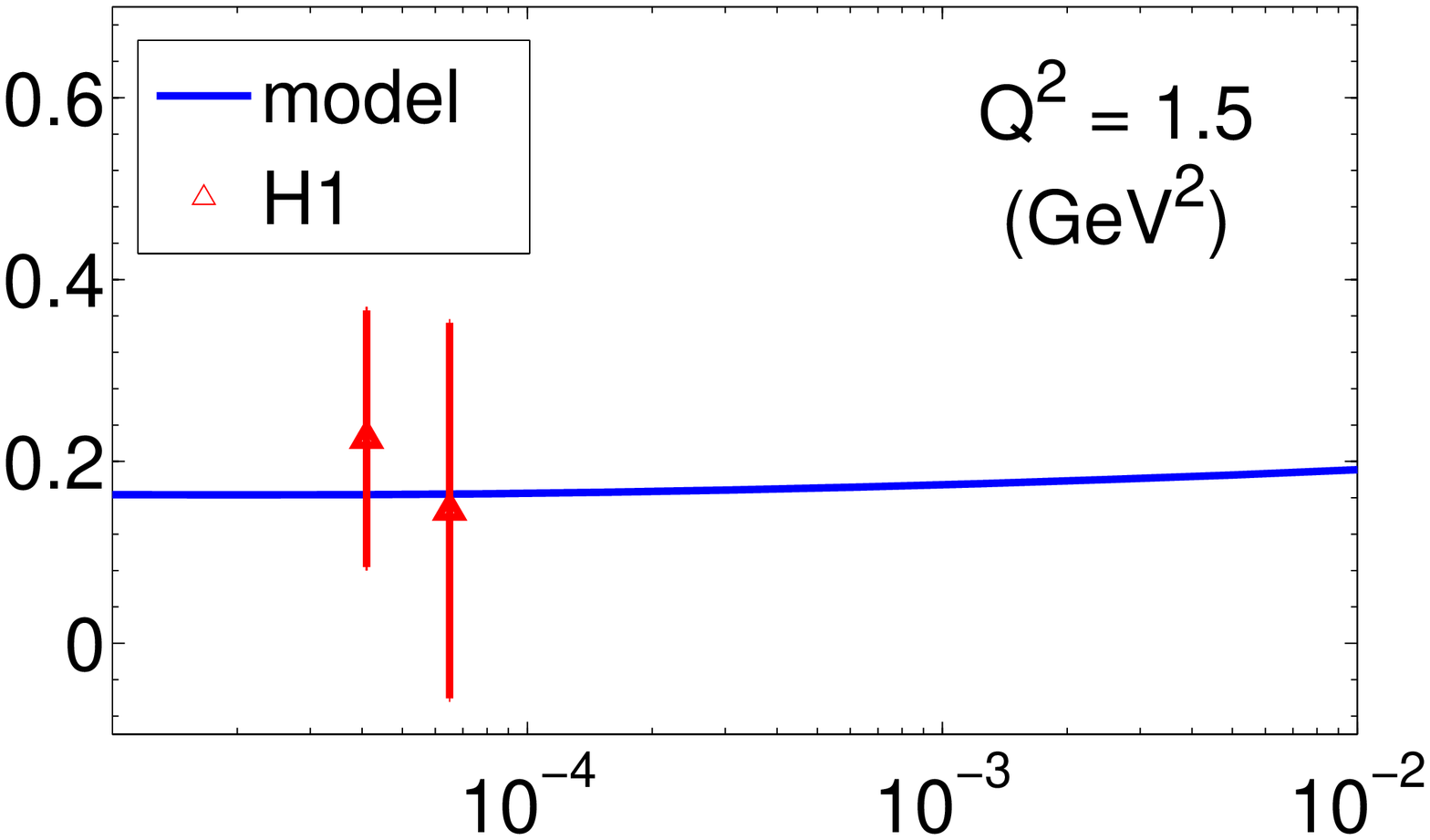,width=35mm, height=19mm} &
\epsfig{file=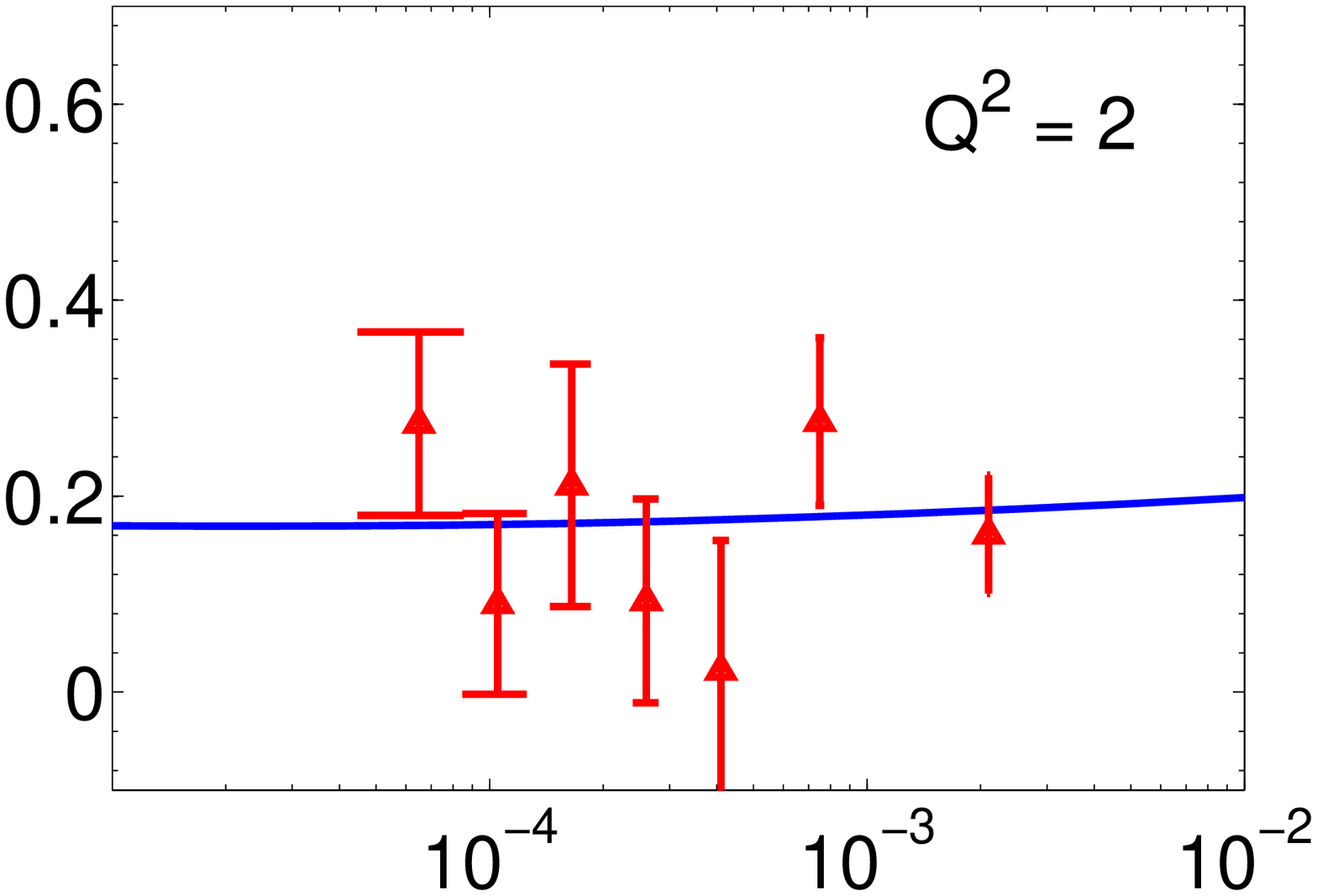,width=35mm,height=19mm} &
\epsfig{file=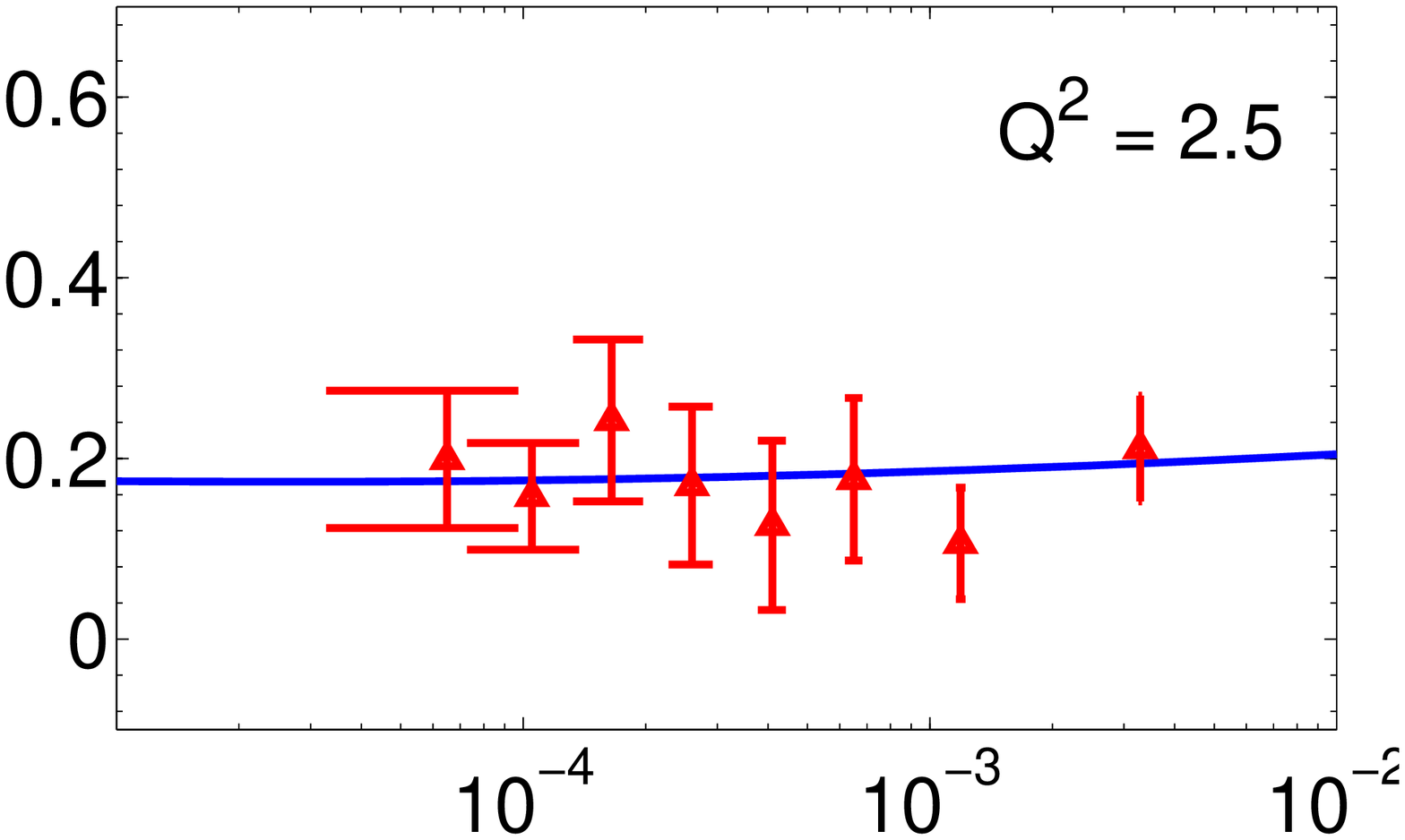,width=35mm,height=19mm} &
\epsfig{file=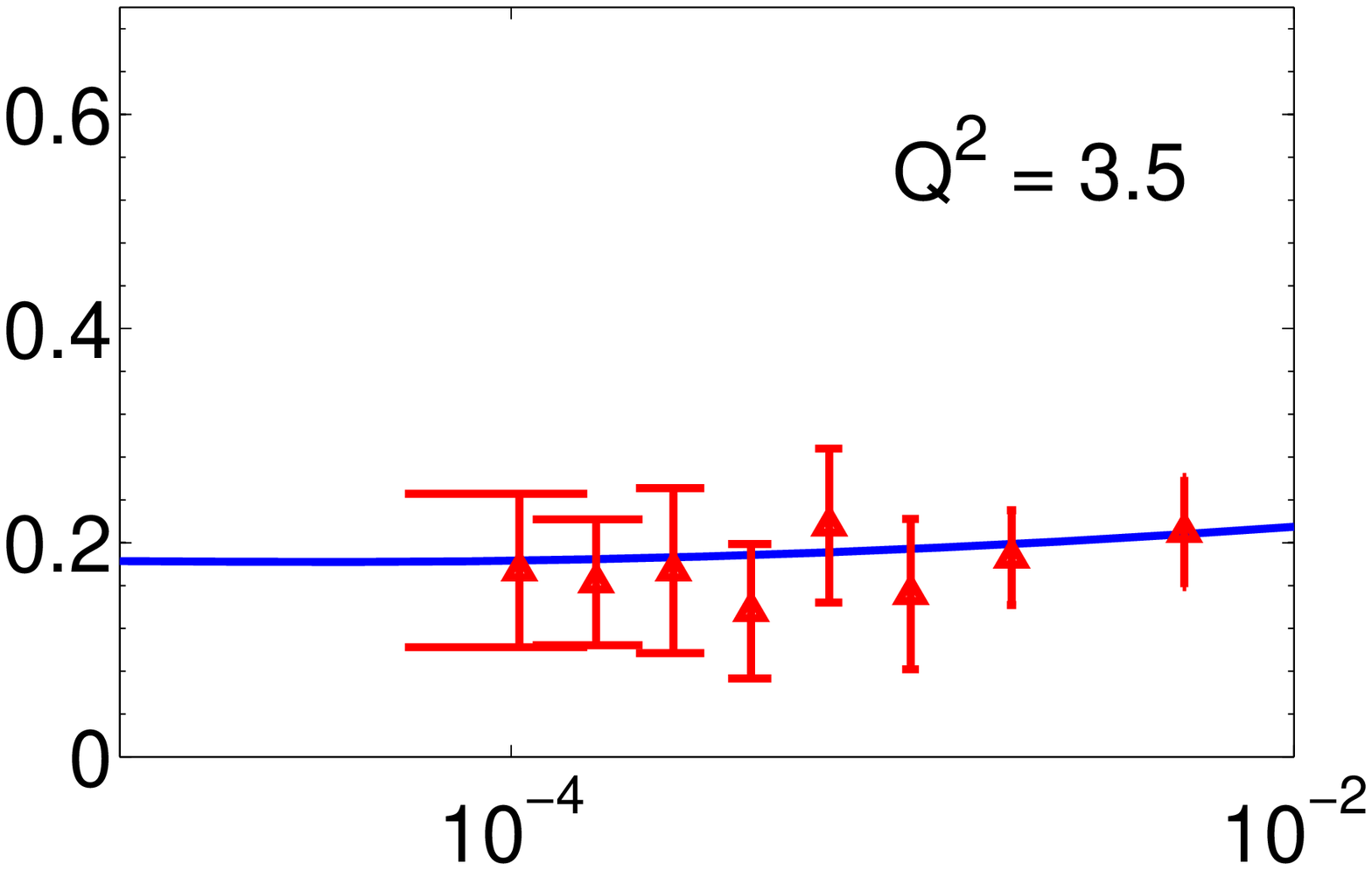,width=35mm, height=19mm} \\
\epsfig{file=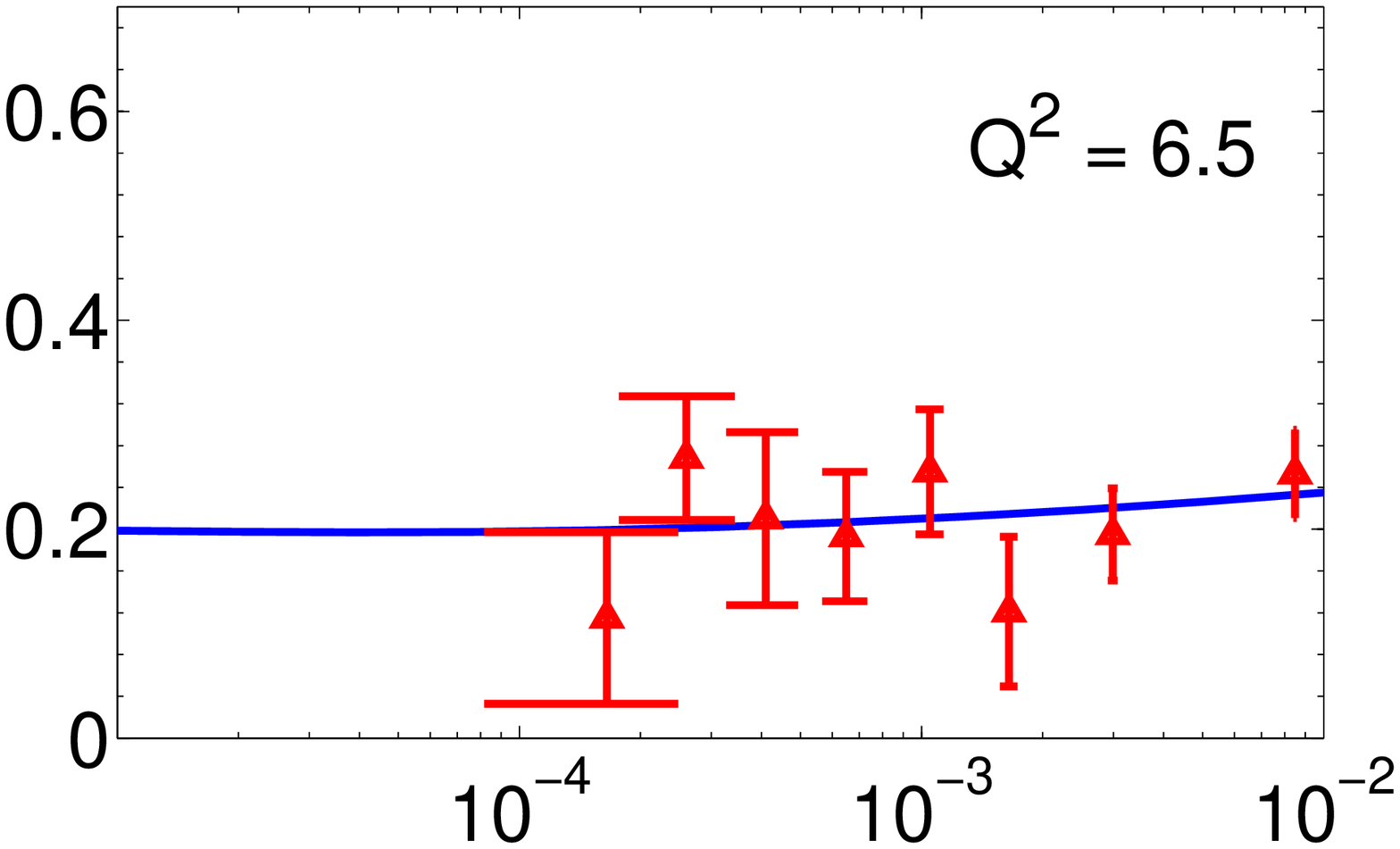,width=35mm, height=19mm}&
\epsfig{file=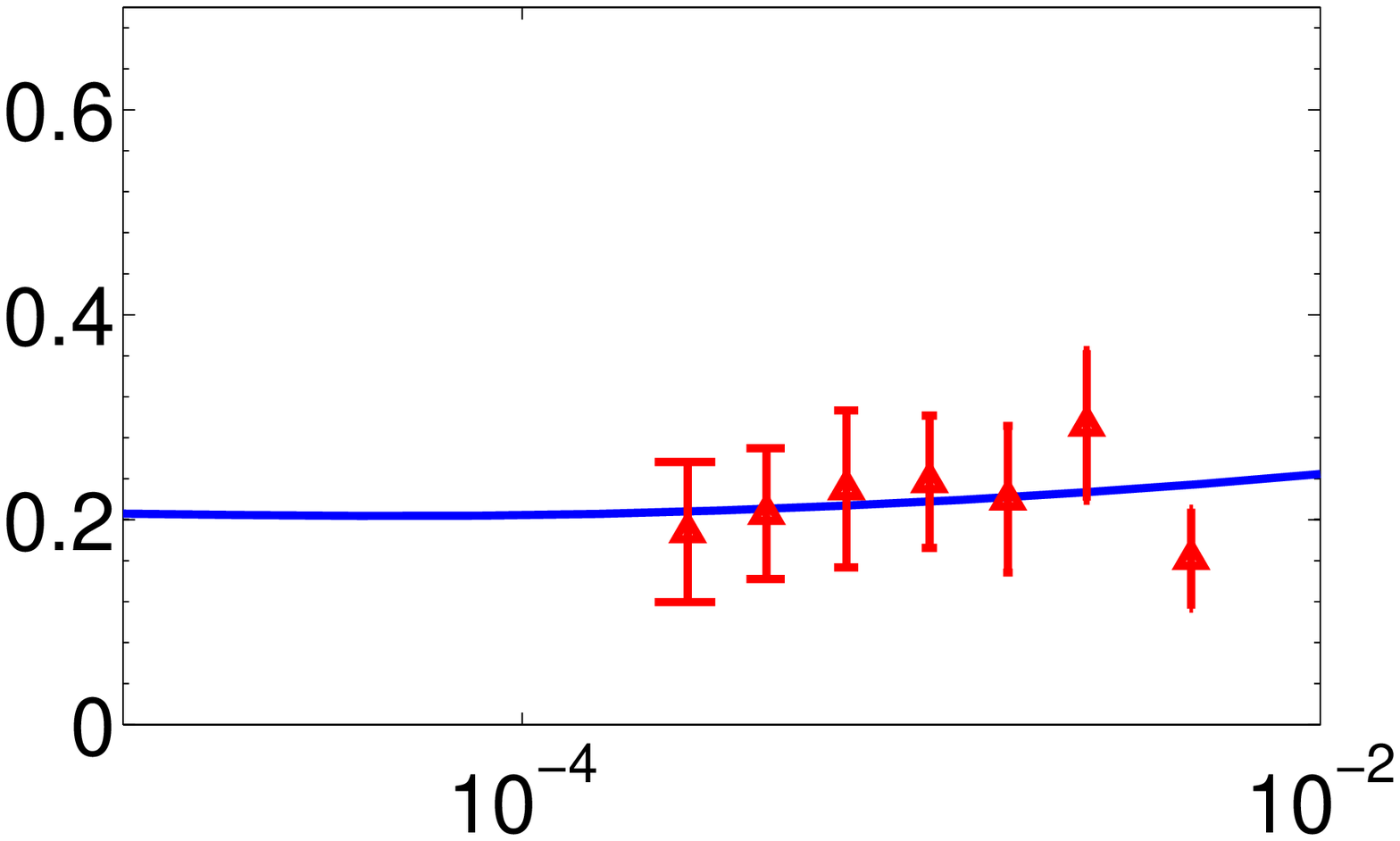,width=35mm,height=19mm} &
\epsfig{file=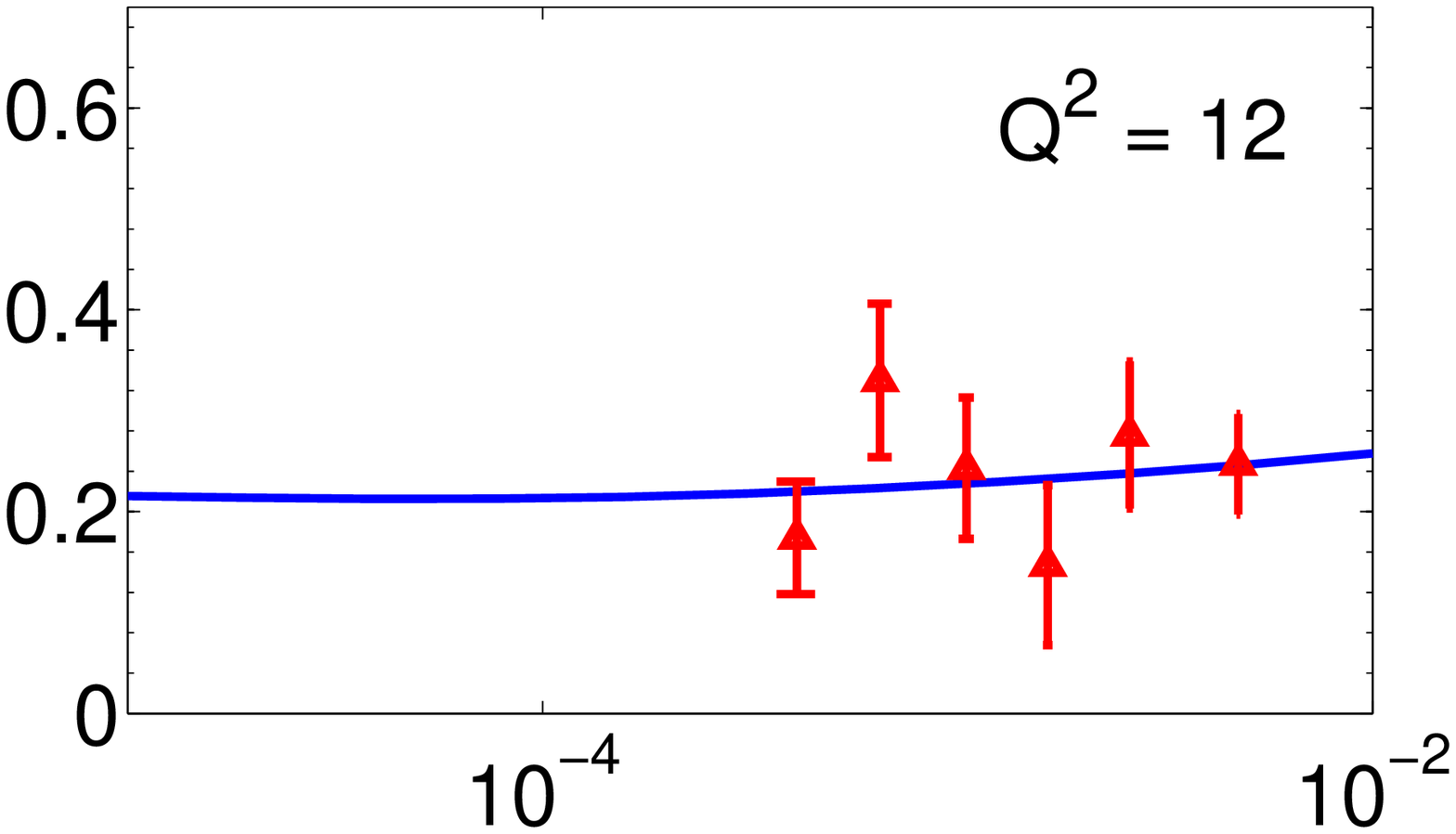,width=35mm,height=19mm} &
\epsfig{file=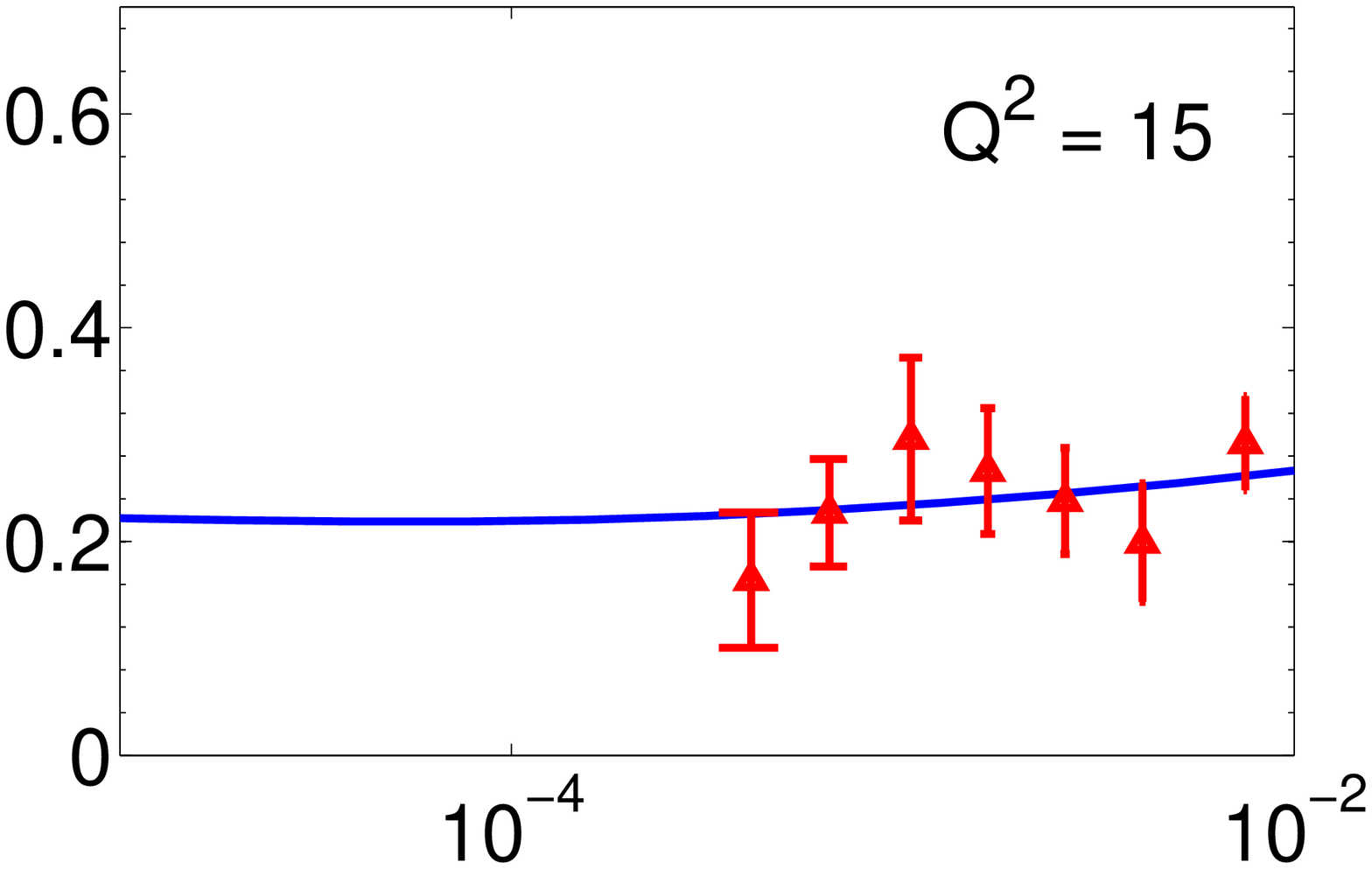,width=35mm, height=19mm}\\
\epsfig{file=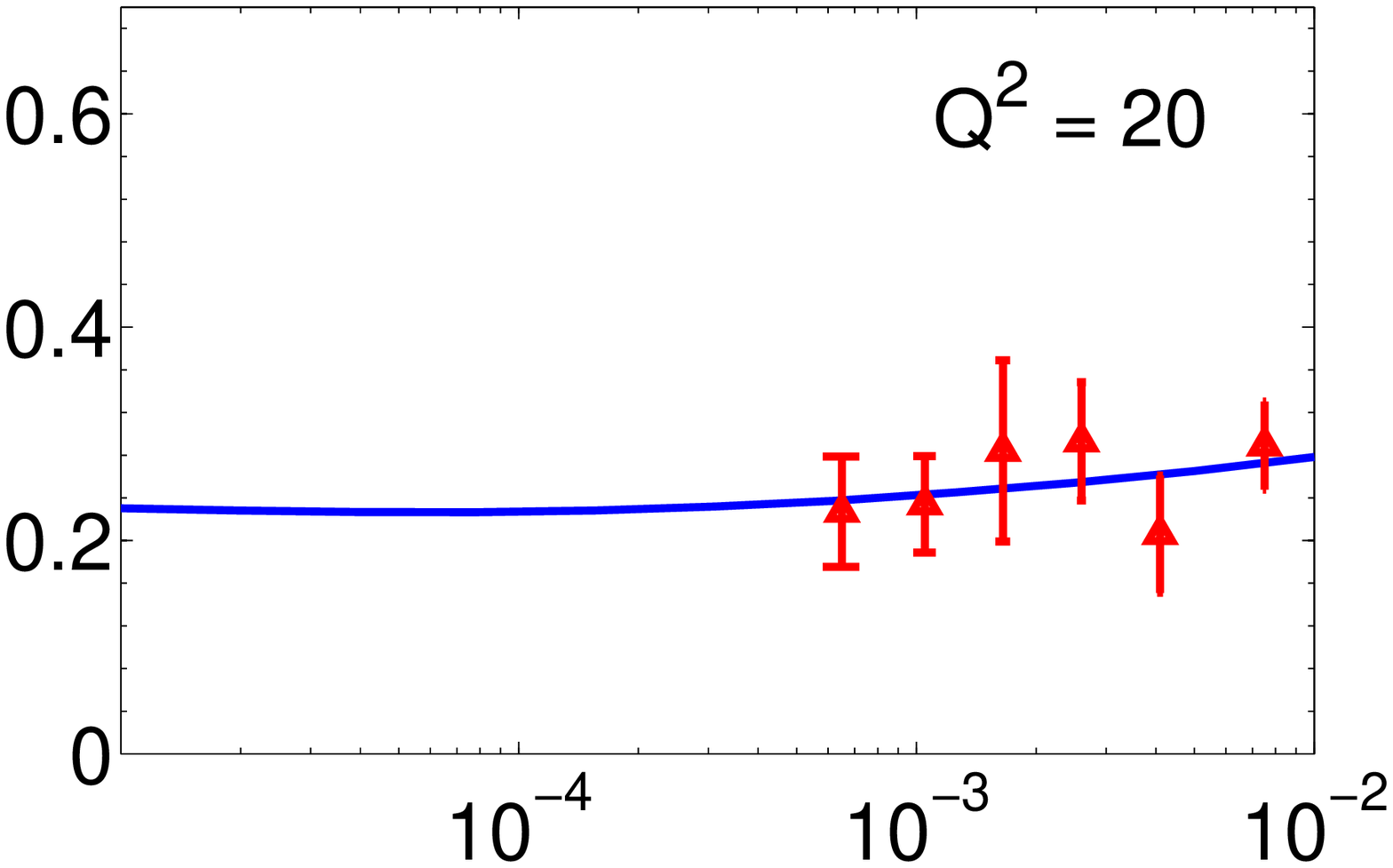,width=35mm, height=19mm} &
\epsfig{file=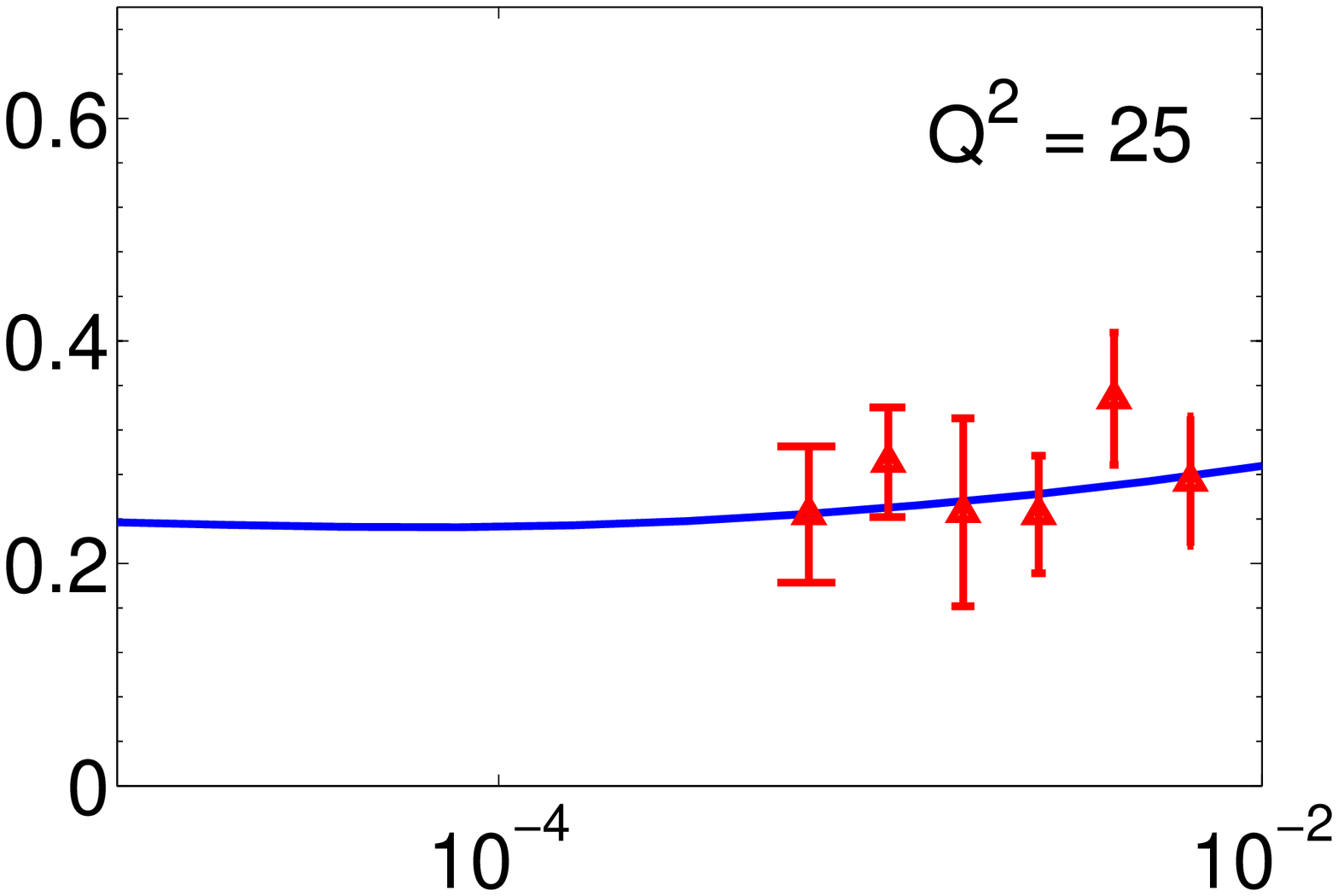,width=35mm,height=19mm} &
\epsfig{file=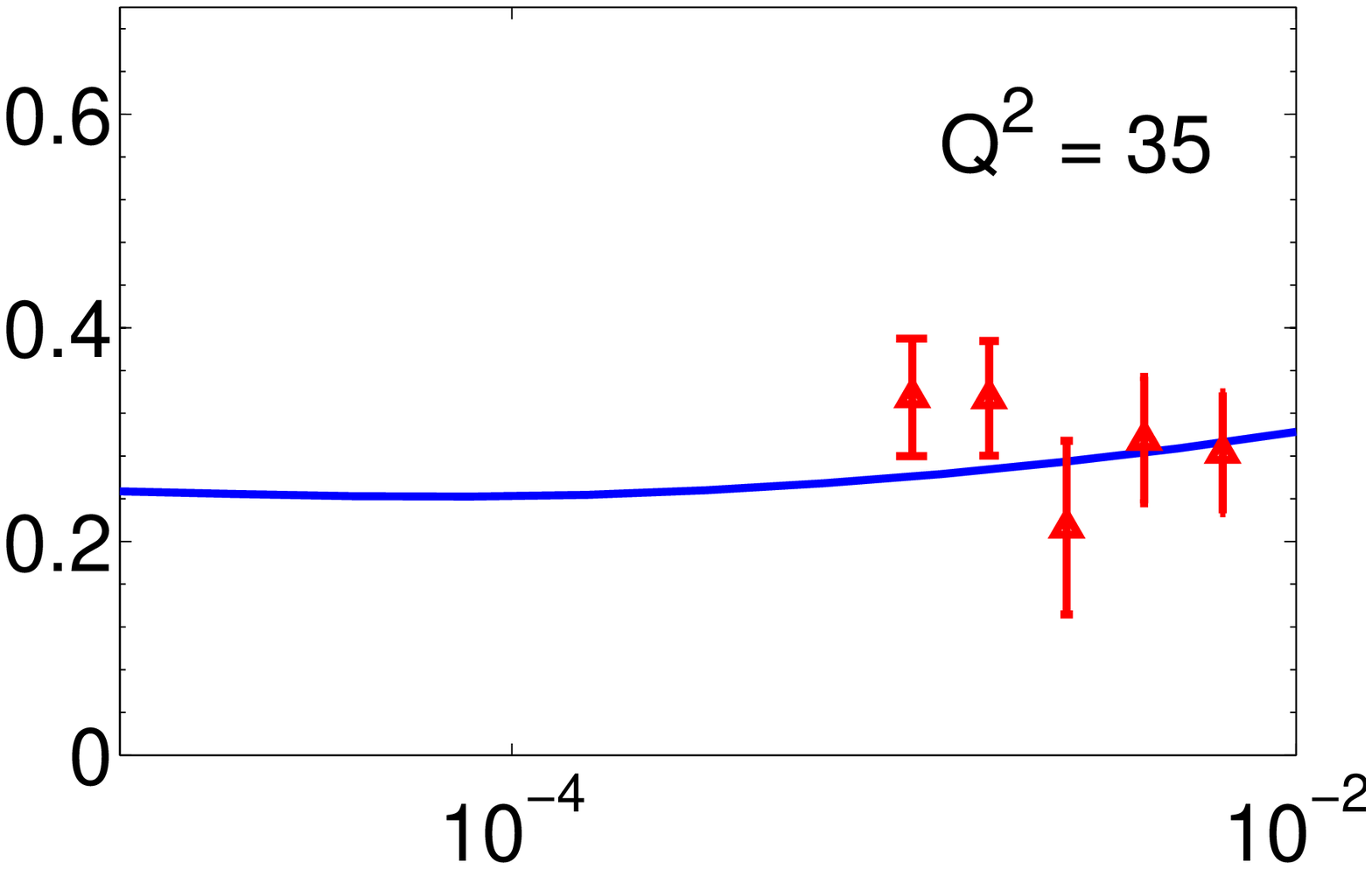,width=35mm,height=19mm} &
\epsfig{file=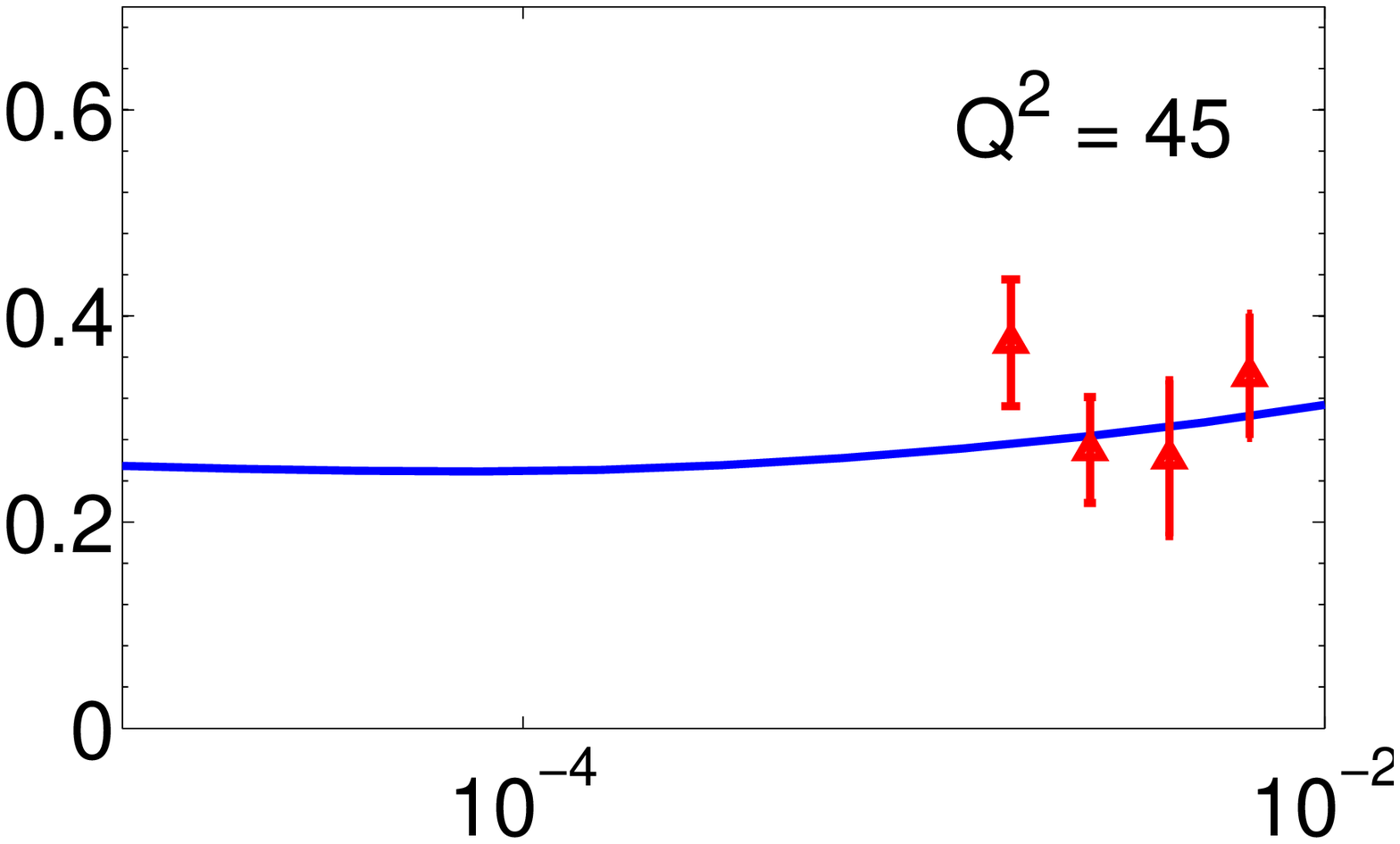,width=35mm, height=19mm}\\
\epsfig{file=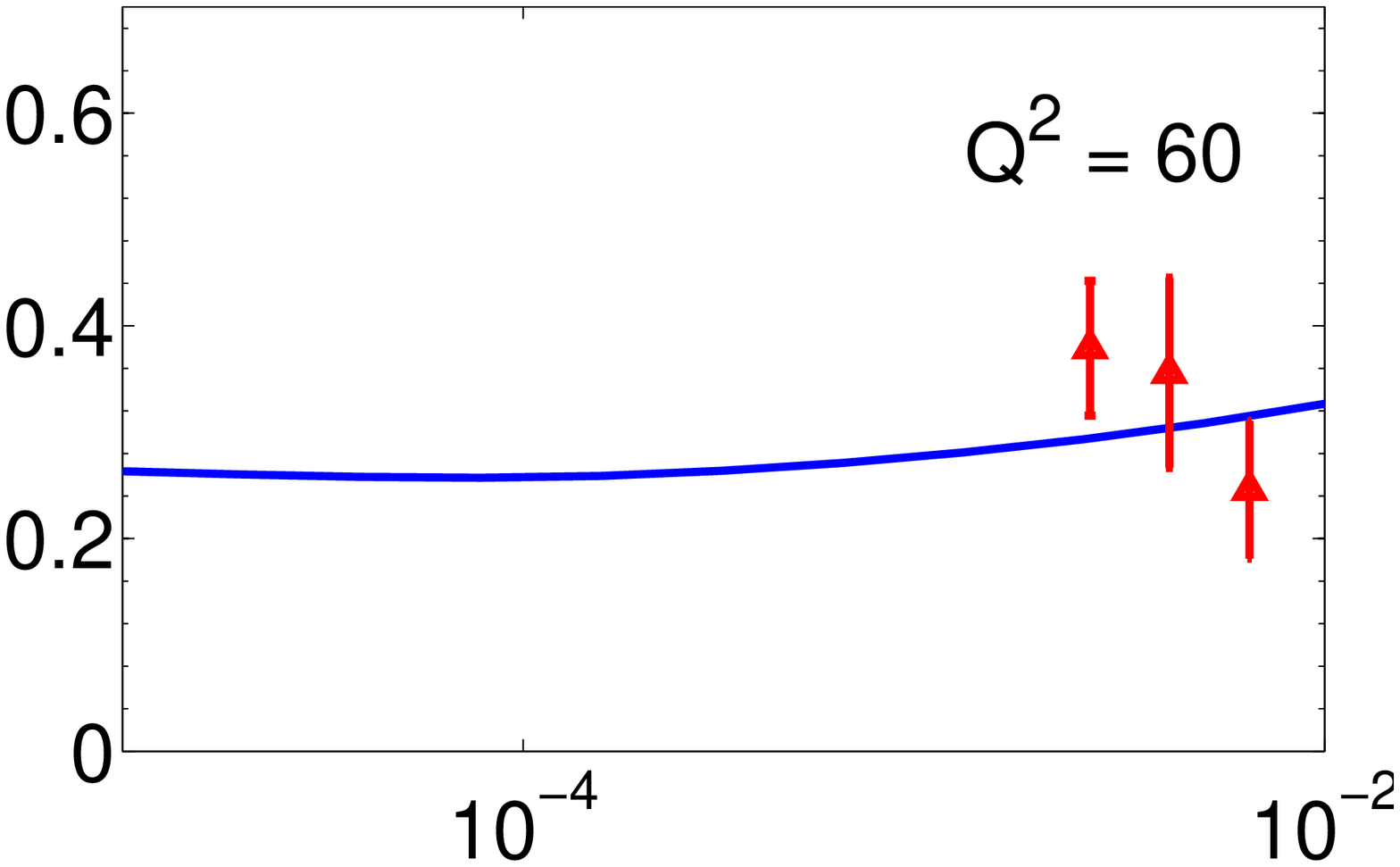,width=35mm, height=19mm} &
\epsfig{file=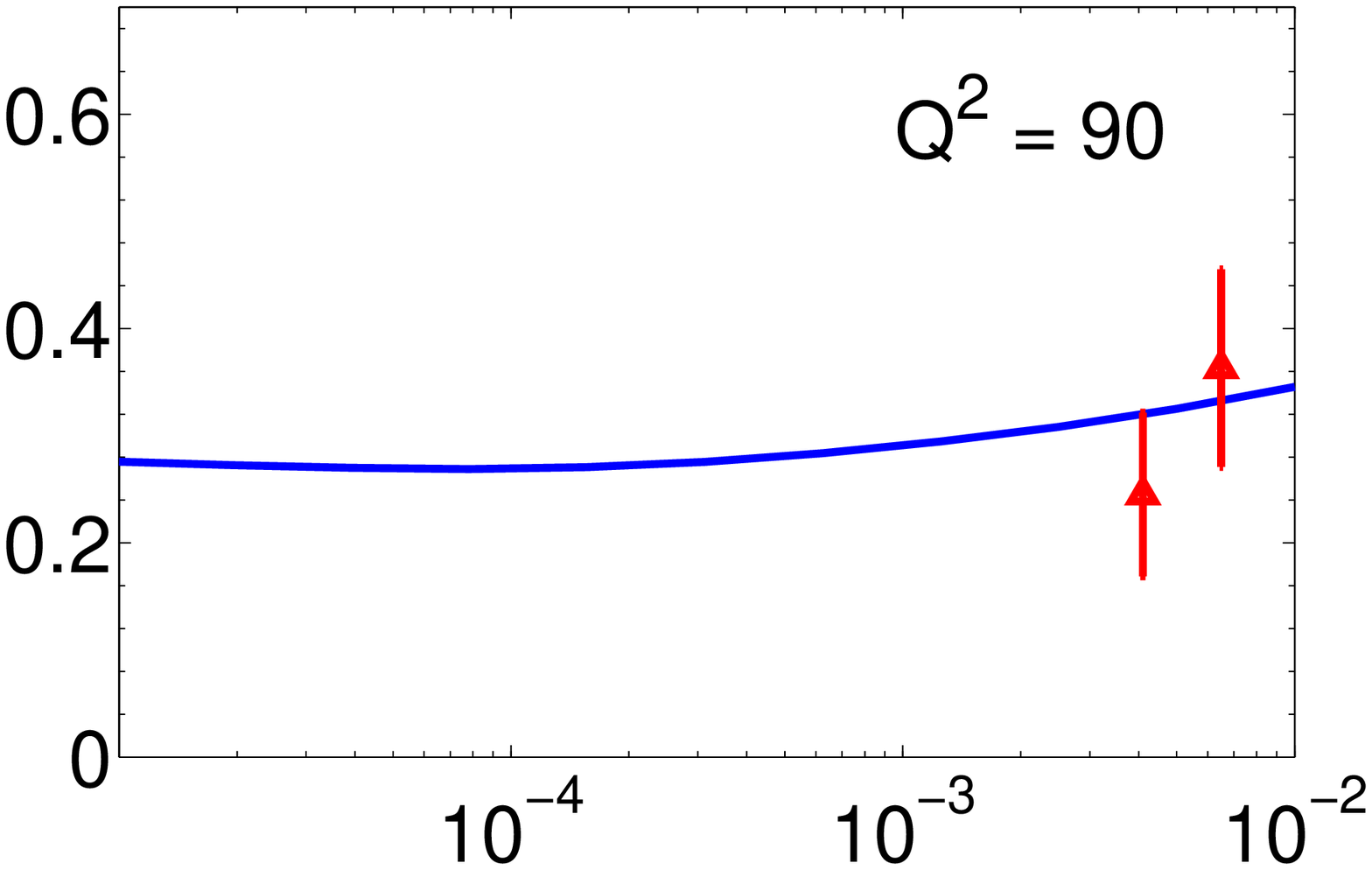,width=35mm,height=19mm} &
\epsfig{file=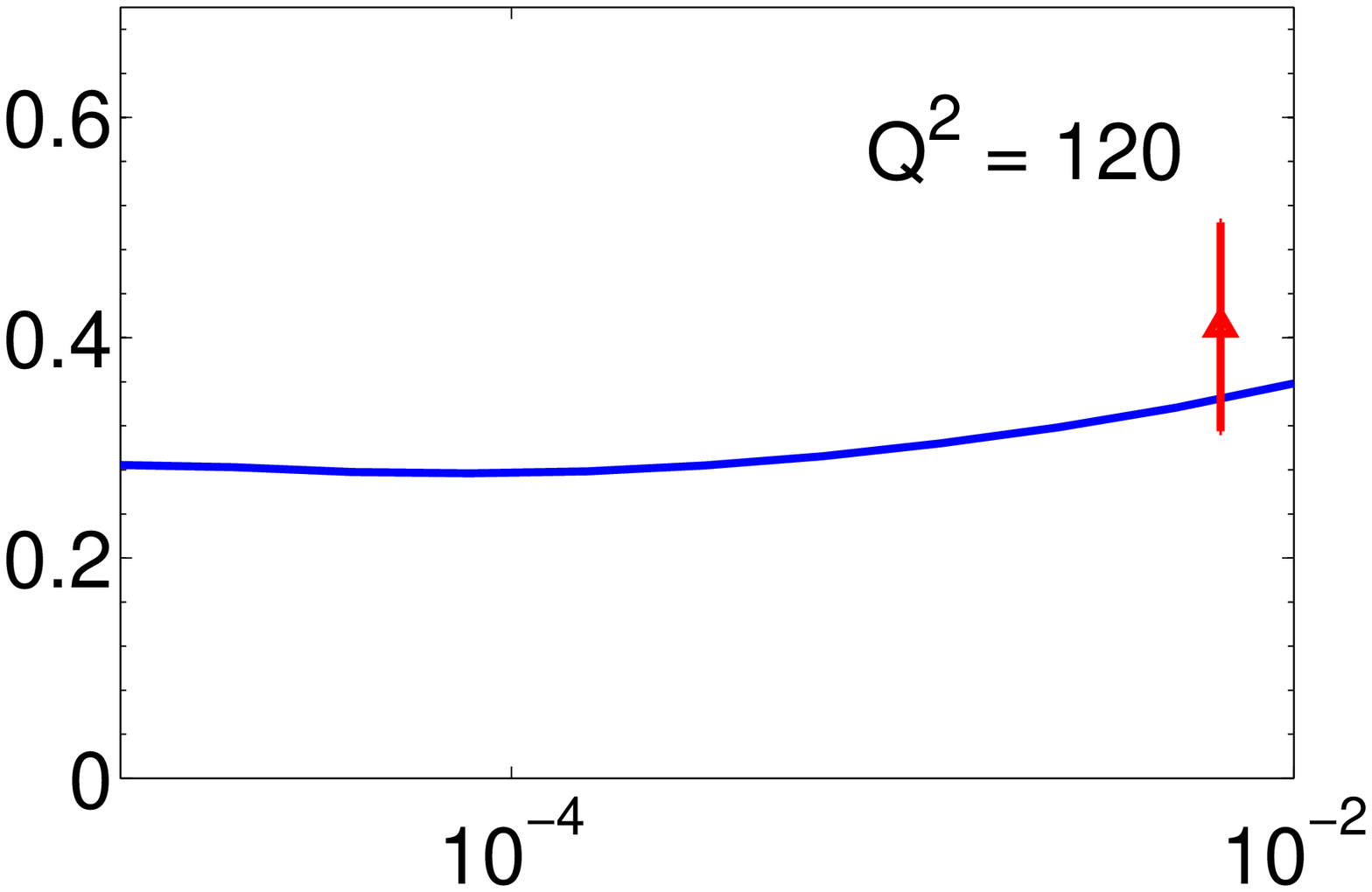,width=35mm,height=19mm}&
\epsfig{file=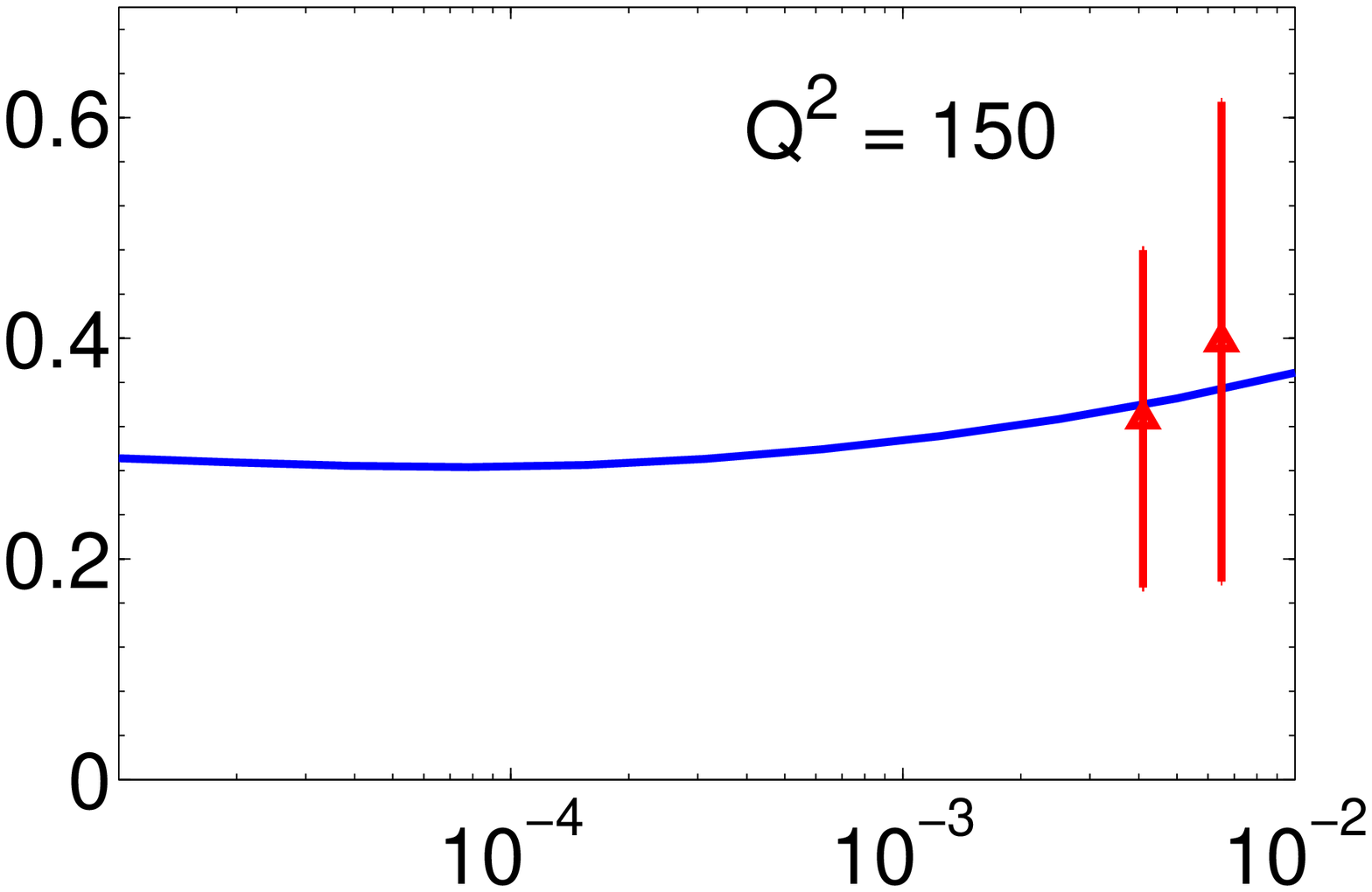,width=35mm, height=19mm}
\end{tabular}{\begin{center}x\end{center}}
\caption{\it The logarithmic derivative $\lambda_{x}\;=\;\partial\ln
F_{2}/\partial(\ln\;1/x)$ as a function $x$ for fixed values of
photon virtuality $Q^{2}$. Data taken from
\cite{Adloff:2001rw}.}\label{dLnF2_dLnY_1}
\end{figure}

\begin{figure}[htbp]
\centering
{\begin{rotate}{90}$\partial\ln
F_{2}/\partial(\ln\;1/x)$\end{rotate}}
\begin{tabular}{c c c c}
\epsfig{file=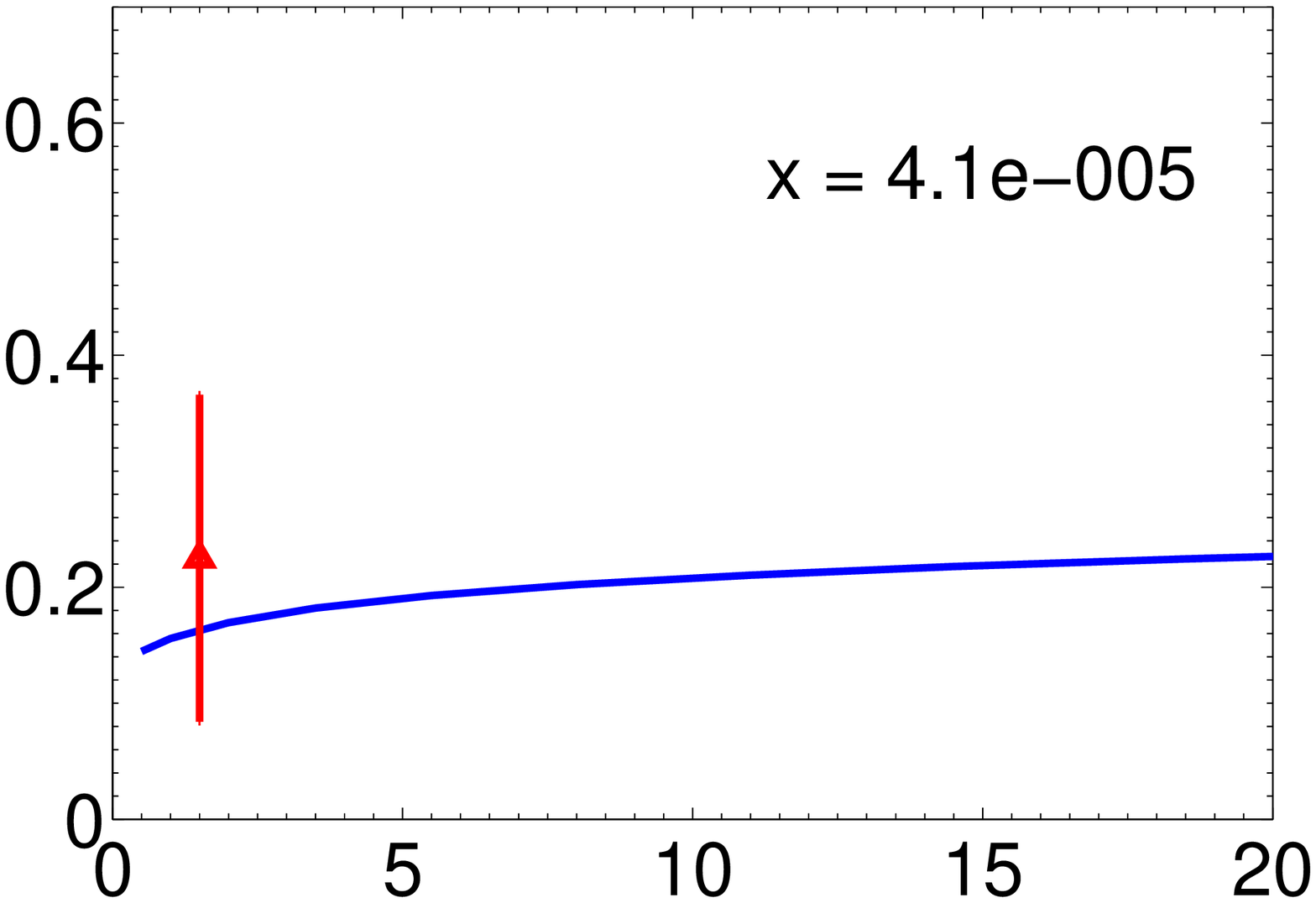,width=35mm, height=20mm} &
\epsfig{file=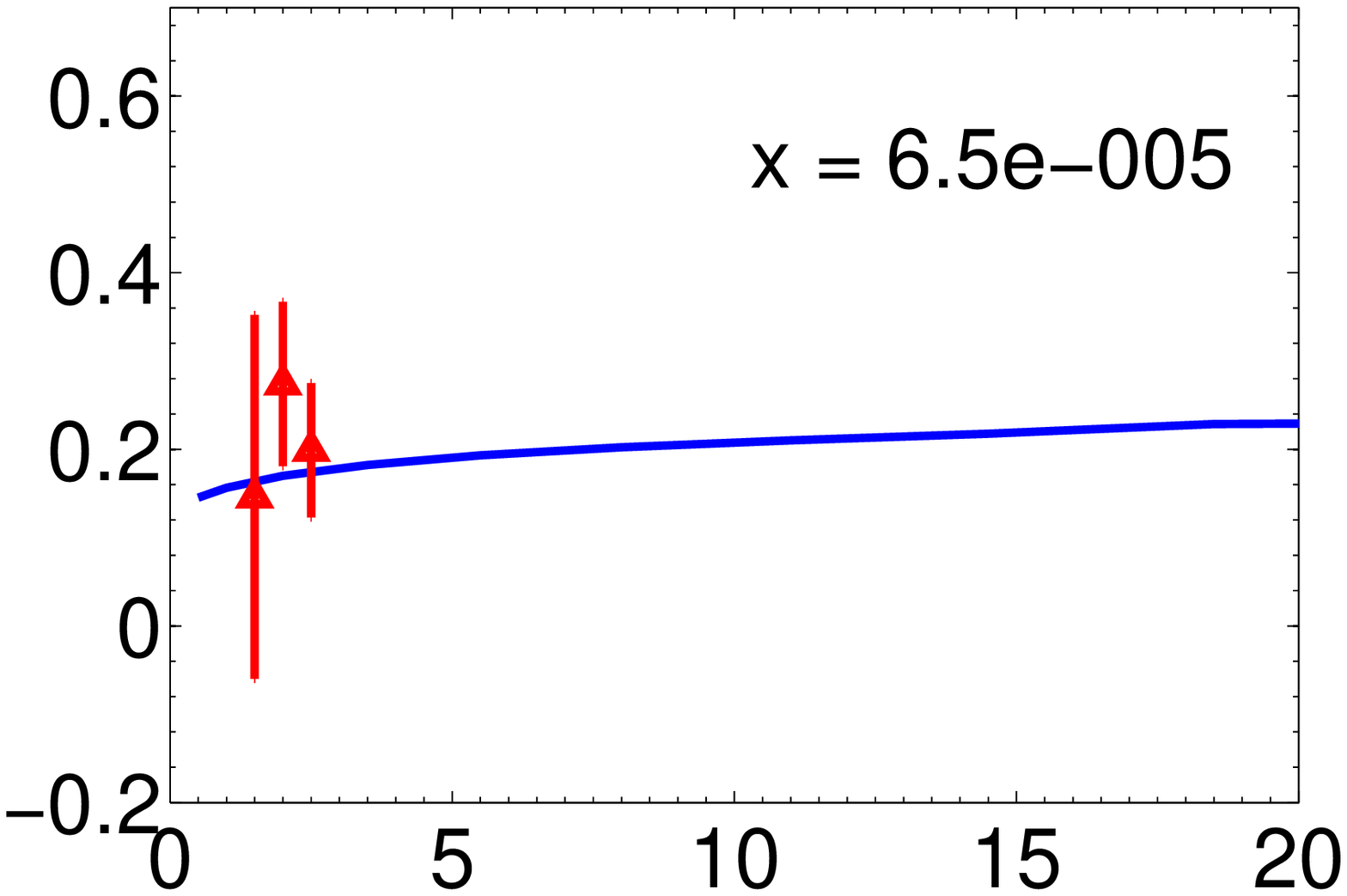,width=35mm,height=20mm} &
\epsfig{file=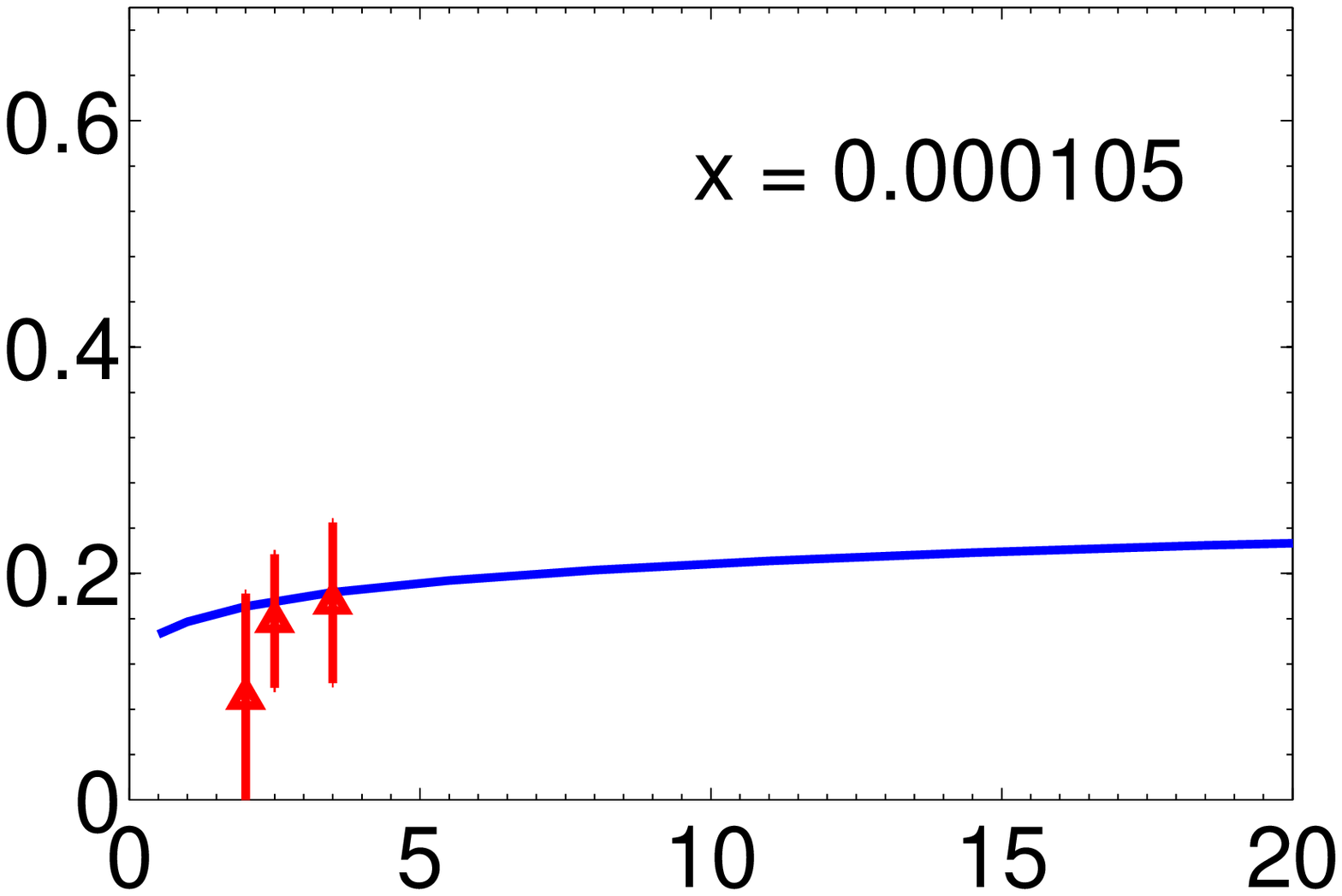,width=35mm, height=20mm} &
\epsfig{file=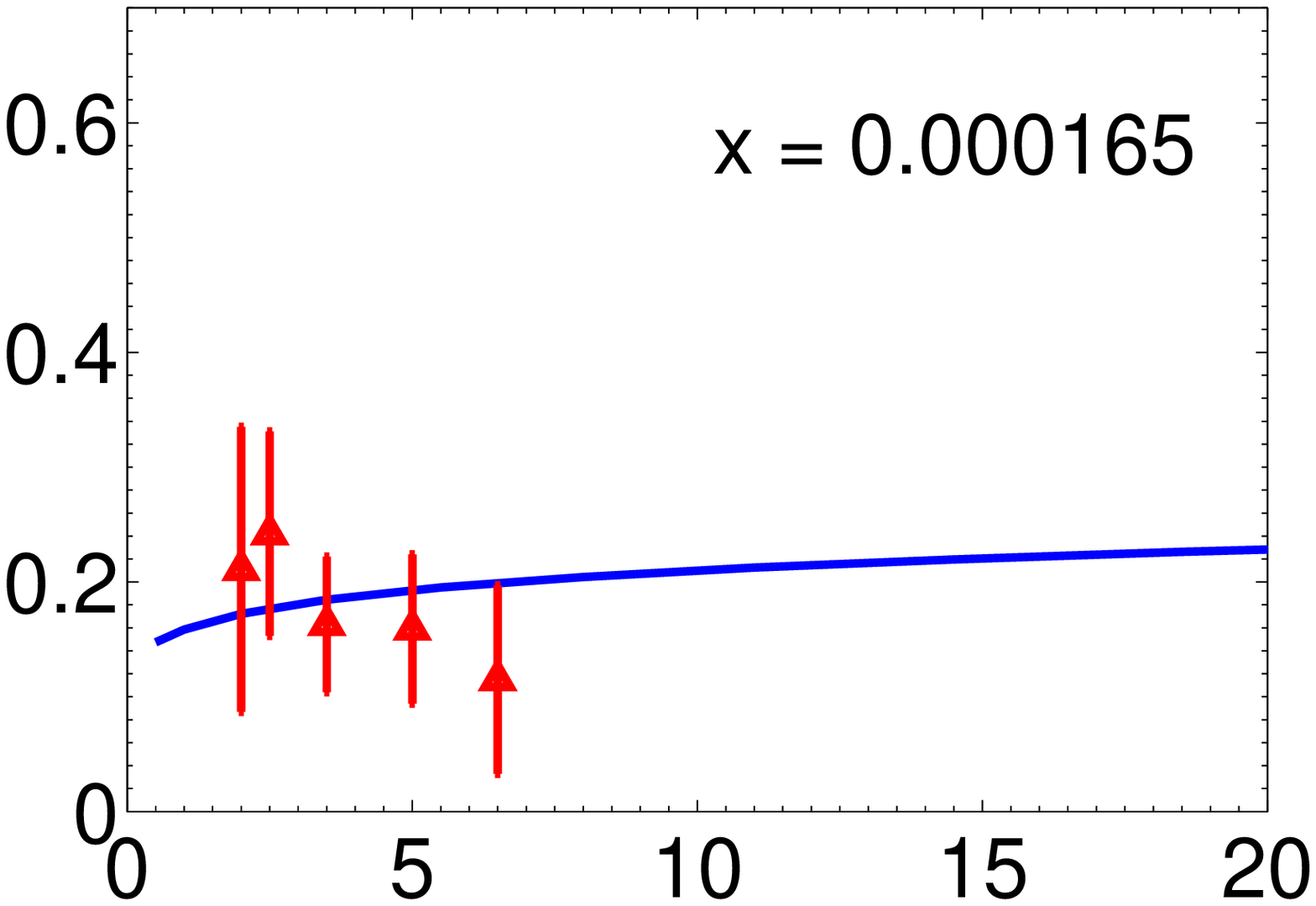,width=35mm,height=20mm} \\
\epsfig{file=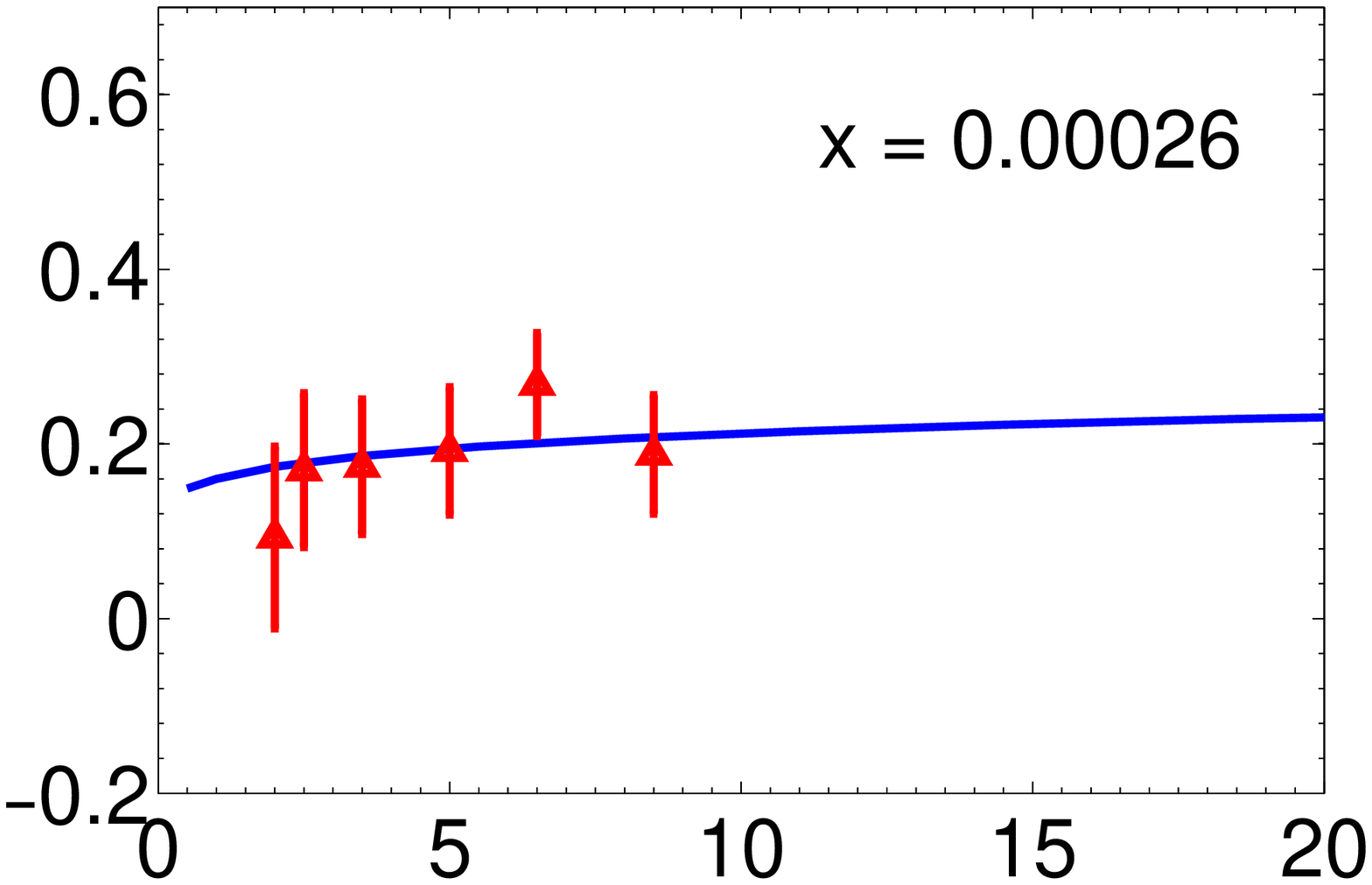,width=35mm, height=20mm} &
\epsfig{file=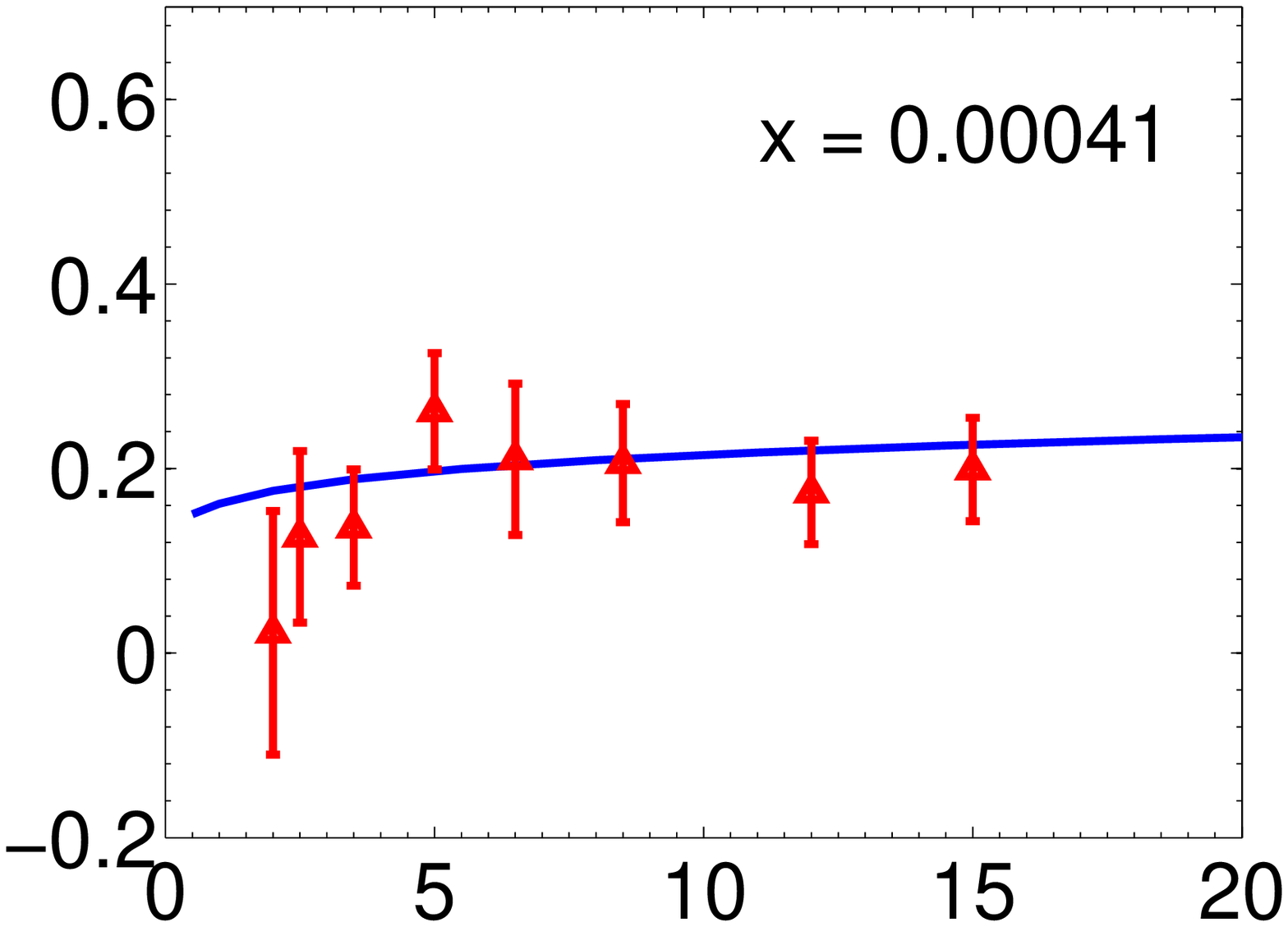,width=35mm,height=20mm} &
\epsfig{file=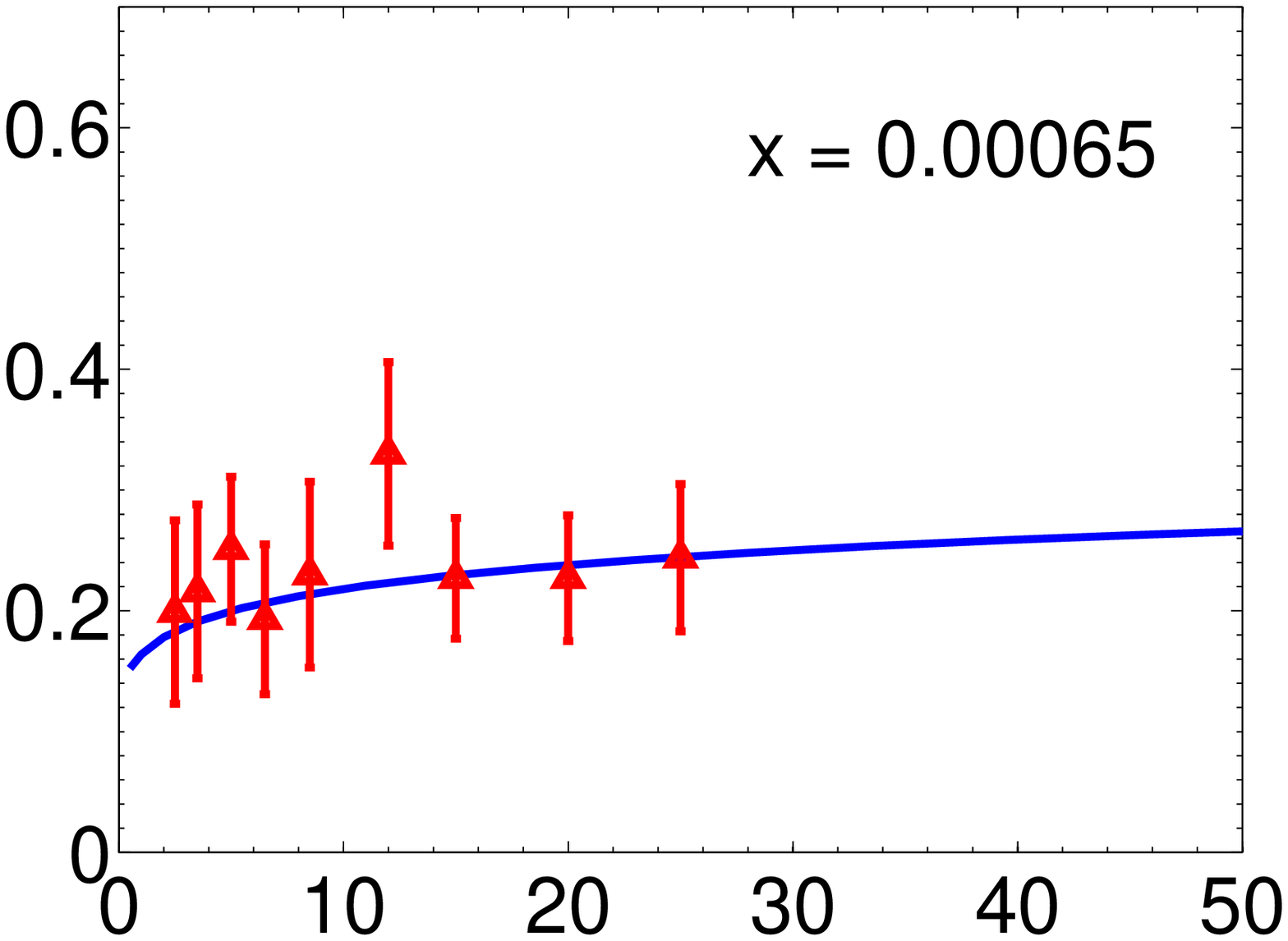,width=35mm, height=20mm} &
\epsfig{file=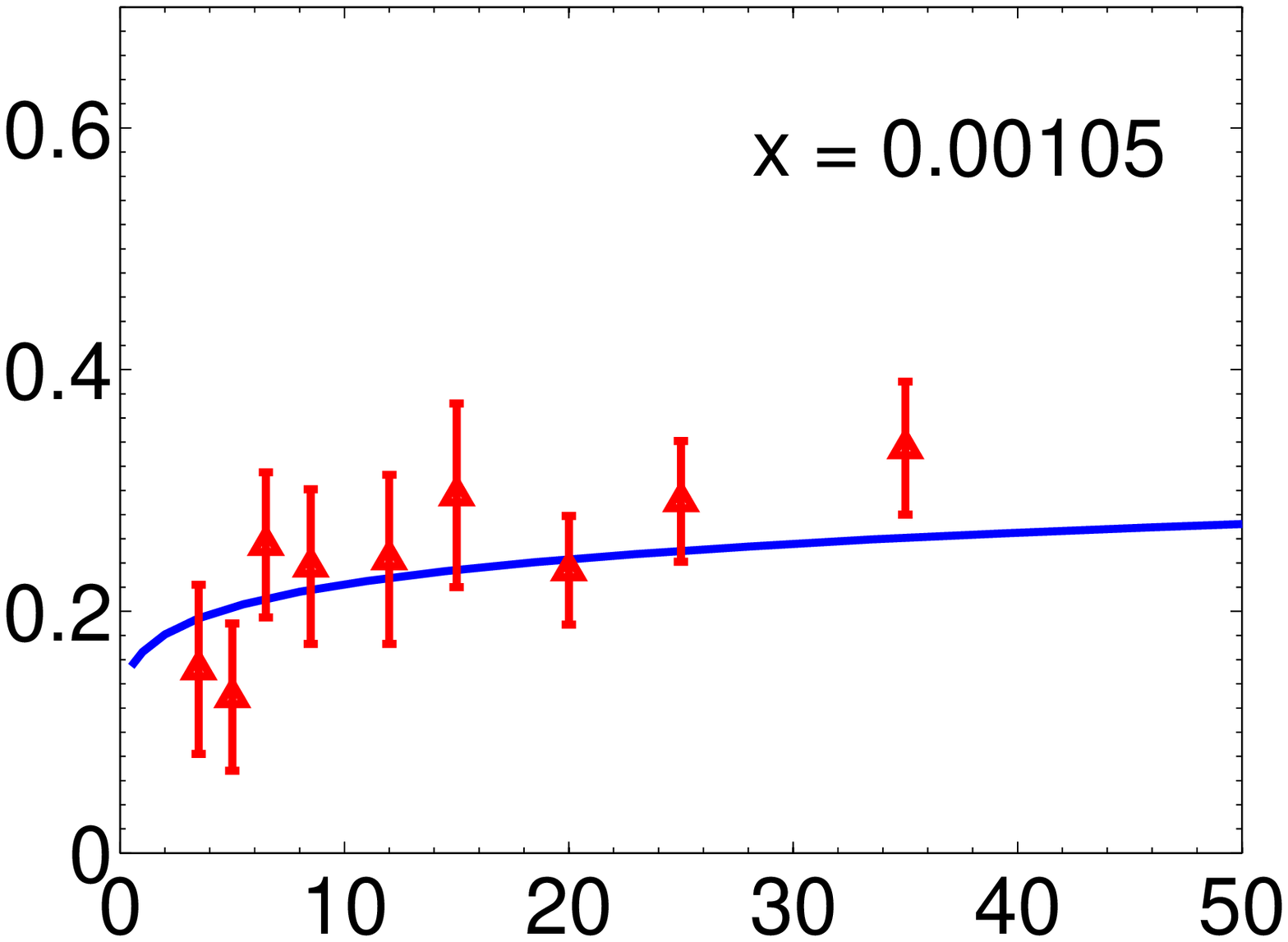,width=35mm,height=20mm} \\
\epsfig{file=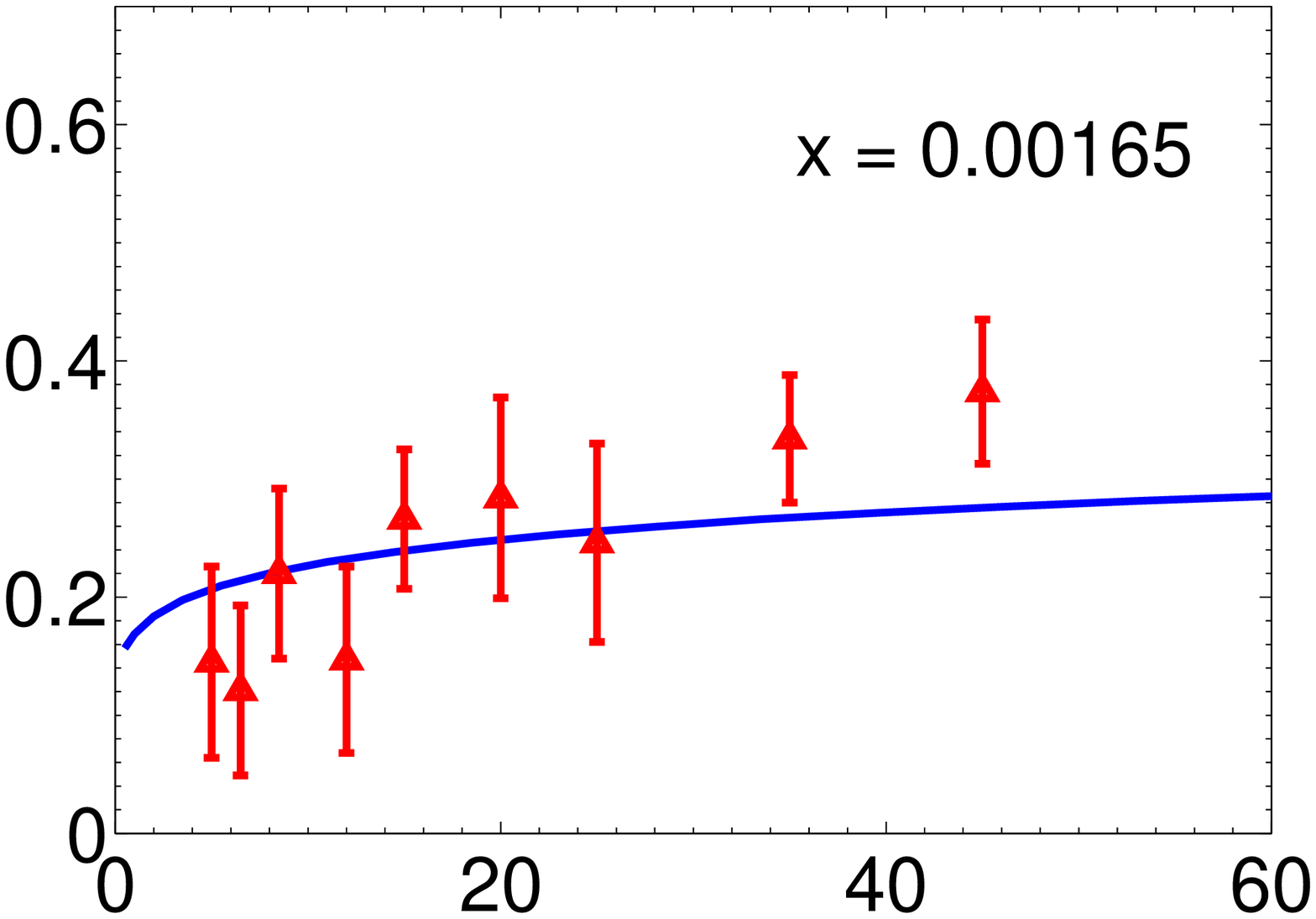,width=35mm, height=20mm} &
\epsfig{file=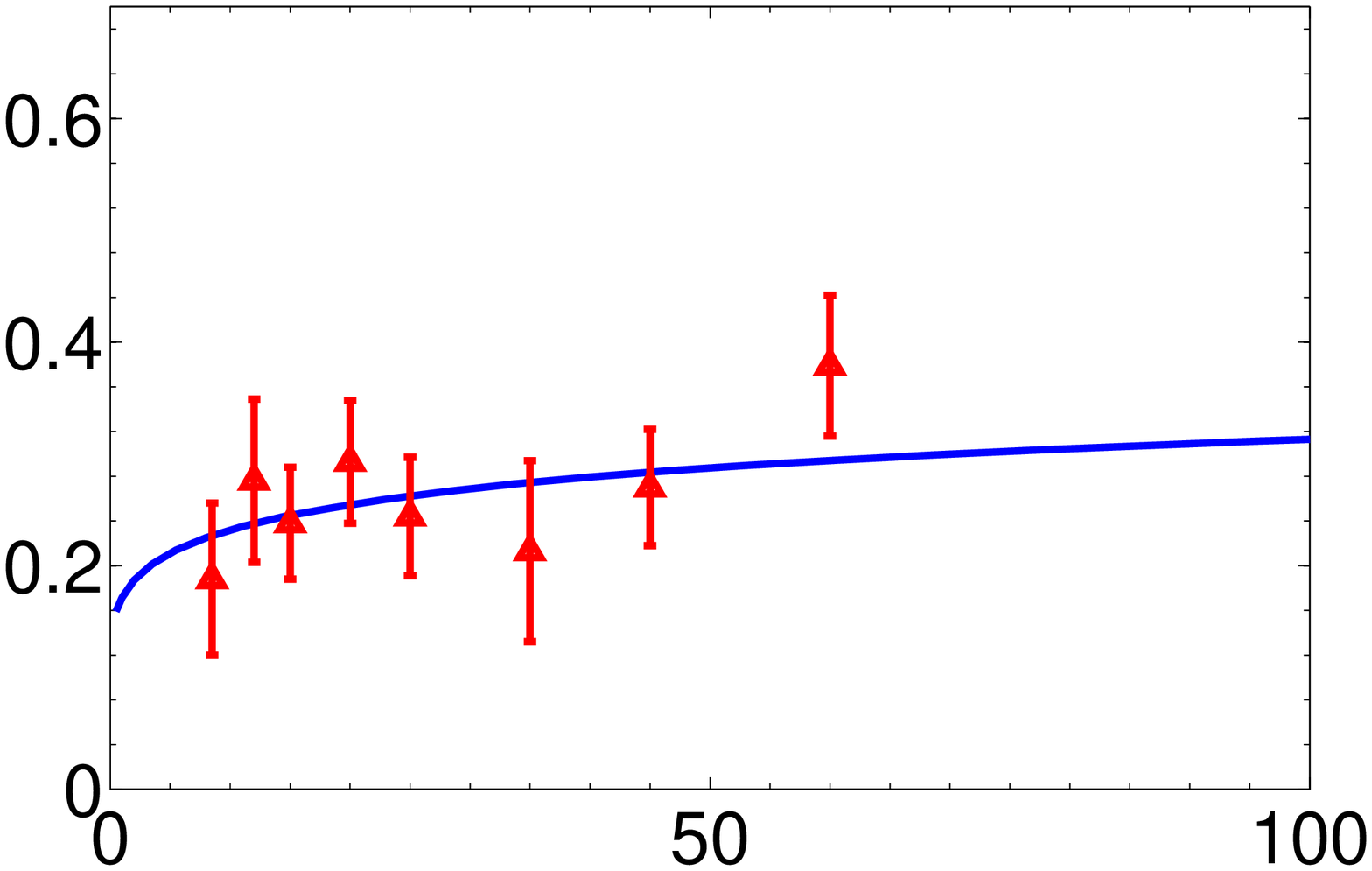,width=35mm,height=20mm} &
\epsfig{file=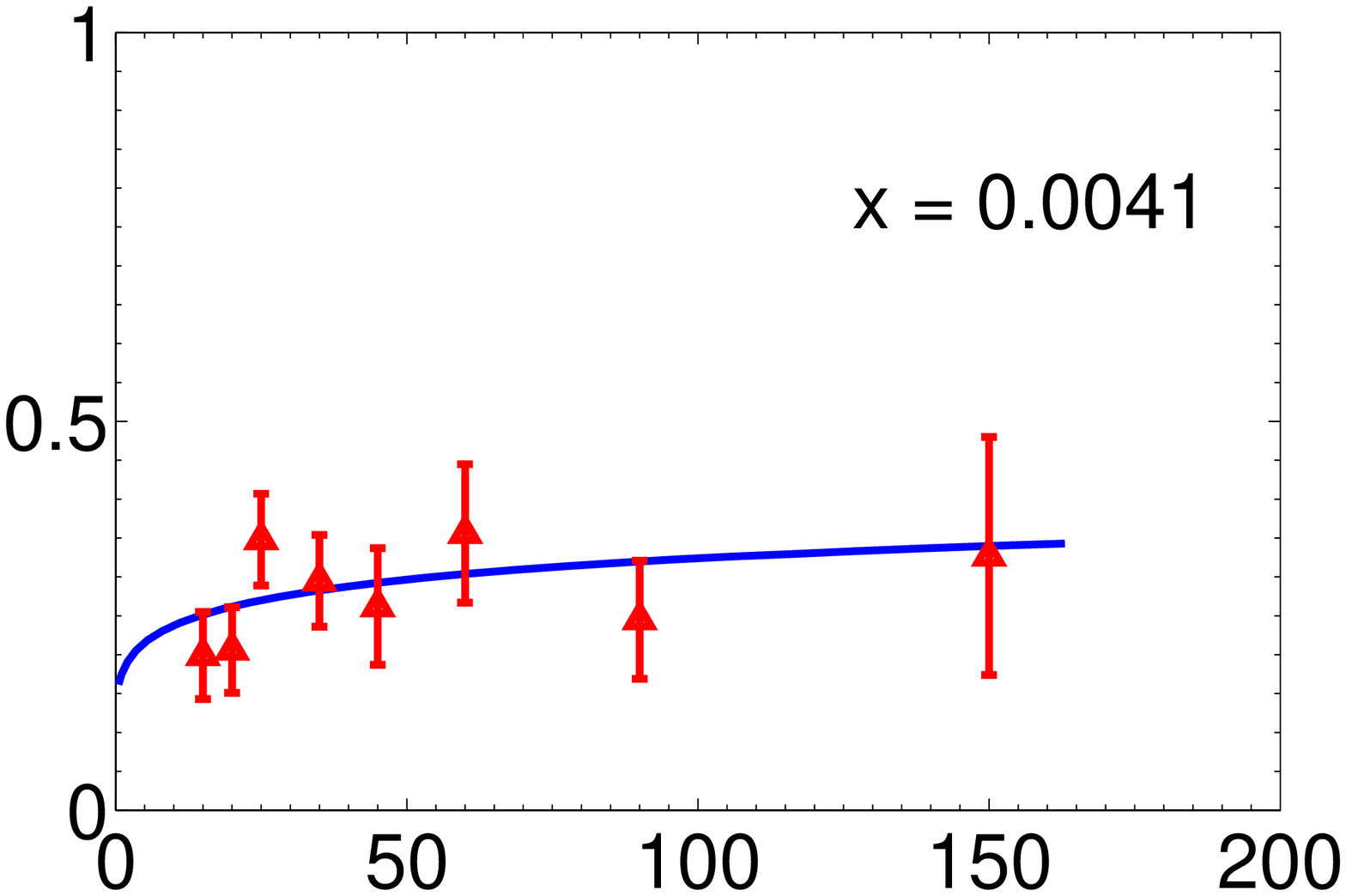,width=35mm, height=20mm} &
\epsfig{file=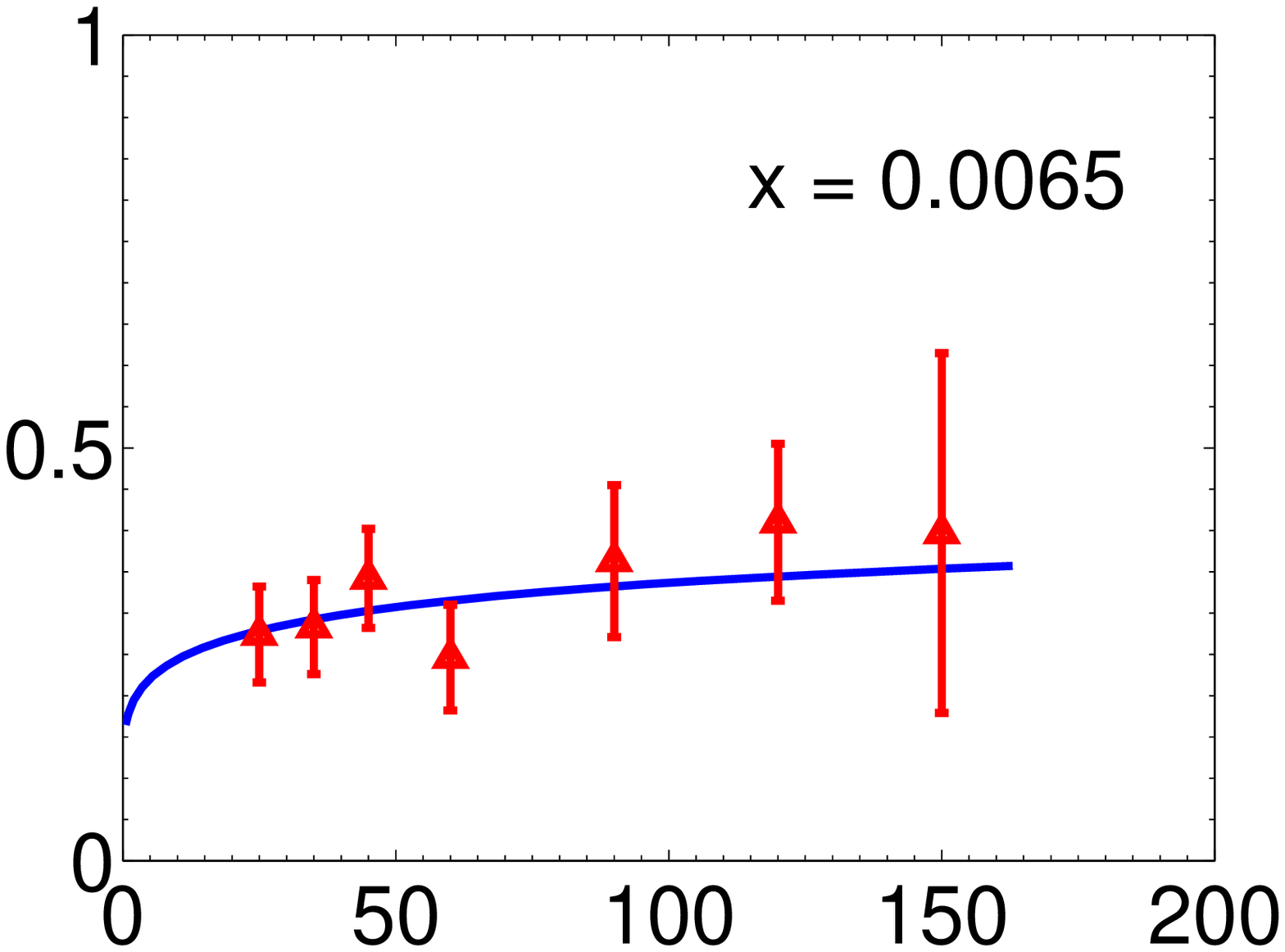,width=35mm,height=20mm} \\
\end{tabular}{\begin{center} $Q^{2}$\end{center}}\caption{\it The logarithmic
derivative $\lambda_{x}$ as a function of $Q^{2}$ for fixed values
of Bjorken-$x$.}\label{dLnF2_dLnY_2}
\end{figure}

In this subsection, we present our computation of
$\lambda_{x}\;\equiv\;\partial\ln F_{2}/\partial(\ln 1/x)$. A
comparison of our prediction with the H1 experimental data
\cite{Adloff:2001rw} is shown in Figs.
\ref{dLnF2_dLnY_1},~\ref{dLnF2_dLnY_2} for a fixed value of $Q^{2}$
and Bjorken-$x$ respectively.

\begin{figure}[htbp]
\centering
\centerline{\includegraphics[width=10cm,height=6cm]{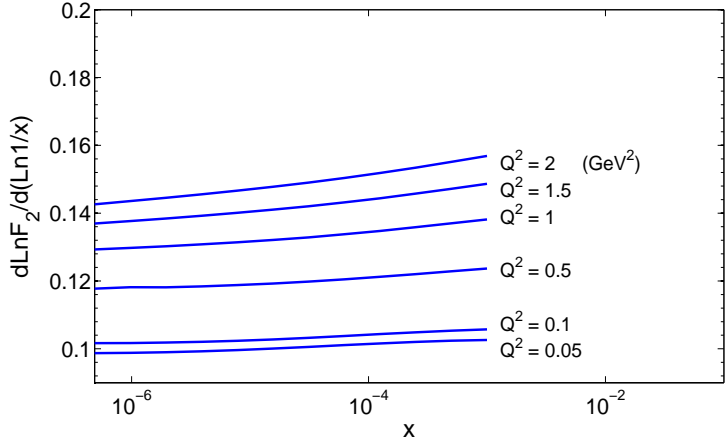}}
\caption{\it The prediction for logarithmic derivative
$\lambda_{x}\;=\;\partial\ln F_{2}/\partial(\ln\;1/x)$ plotted at
very low $x$ and $Q^{2}$.}\label{pred}
\end{figure}

We can see a good agrement with the experimental data. In order to
investigate the behavior of the slope $\lambda_{x}$, in the high
energy limit (very low $x$ and $Q^{2}$), we expand our predictions
to this region. The result is plotted in Fig. \ref{pred}. We want to
pointed out the fact, that our model predicts a behavior of the
structure function at the region of small photon virtualities
$Q^{2}<1\;GeV^{2}$ which is in good agreement with that obtained by
Donnachie and Landshoff \cite{Donnachie:1984xq} from the fit to
data.

\subsection{Prediction for $F_{L}$ at HERA}

\begin{figure}[htbp]
\begin{tabular}{c c c c}
\epsfig{file=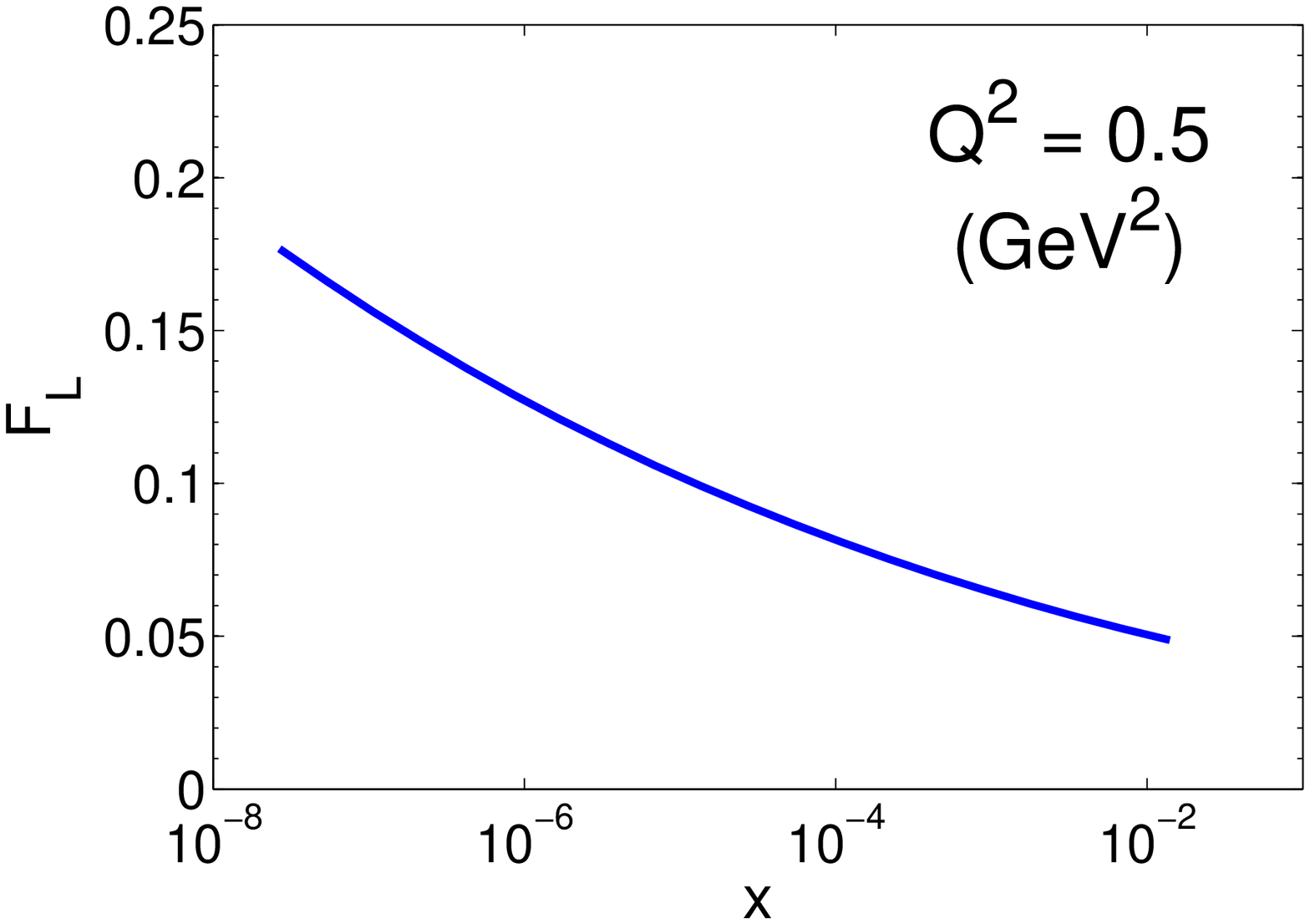,width=40mm, height=25mm} &
\epsfig{file=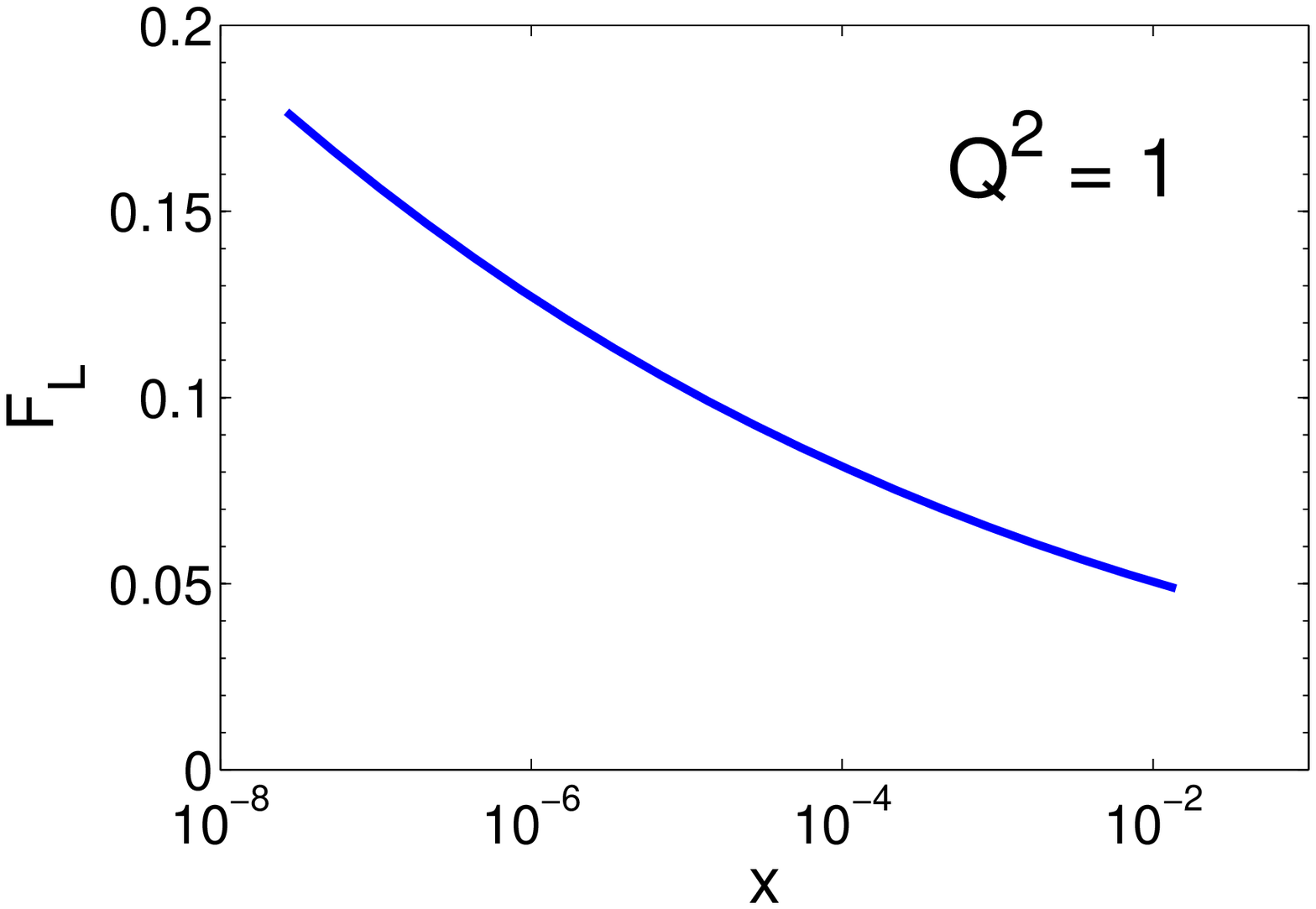,width=40mm,height=25mm} &
\epsfig{file=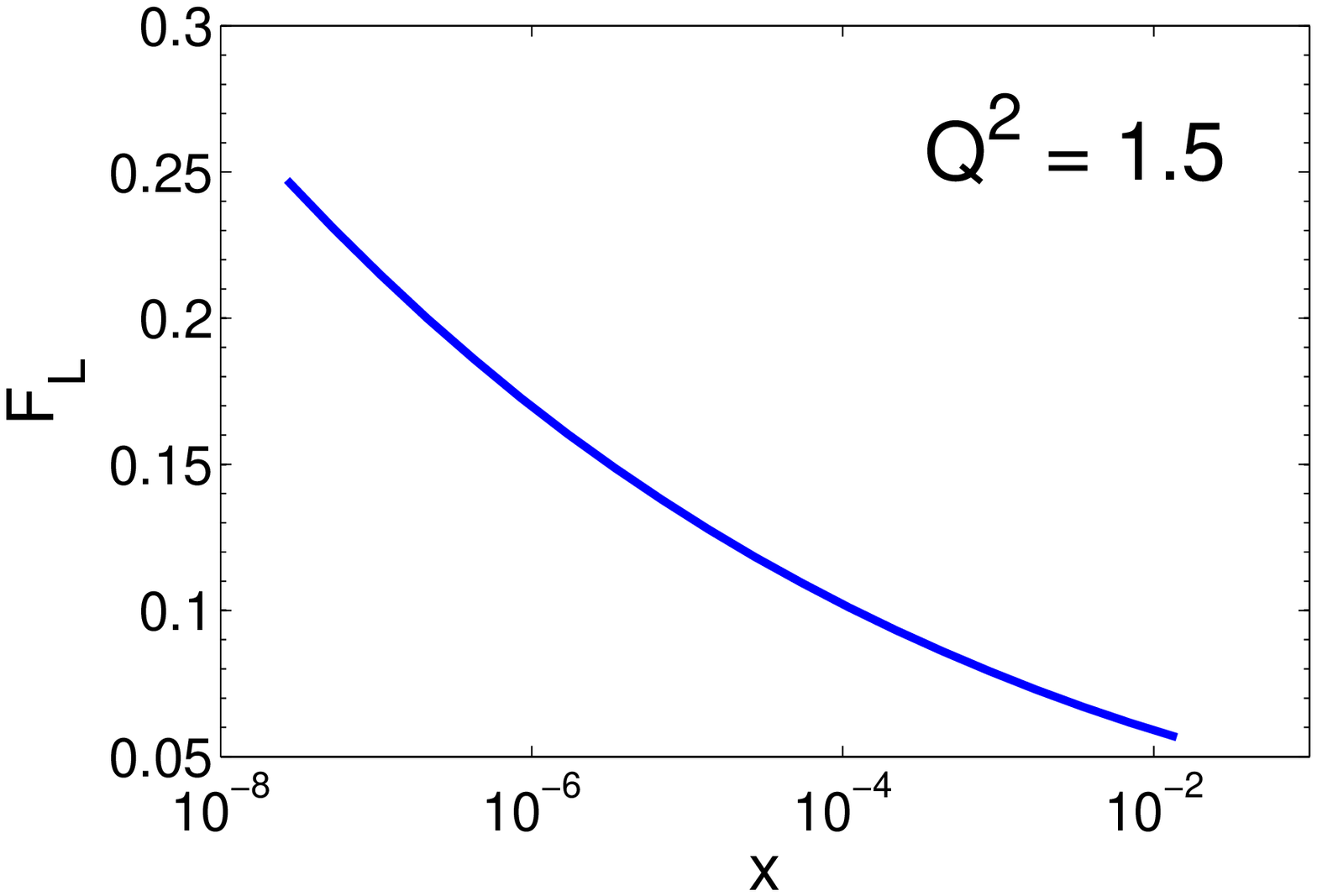,width=40mm, height=25mm} &
\epsfig{file=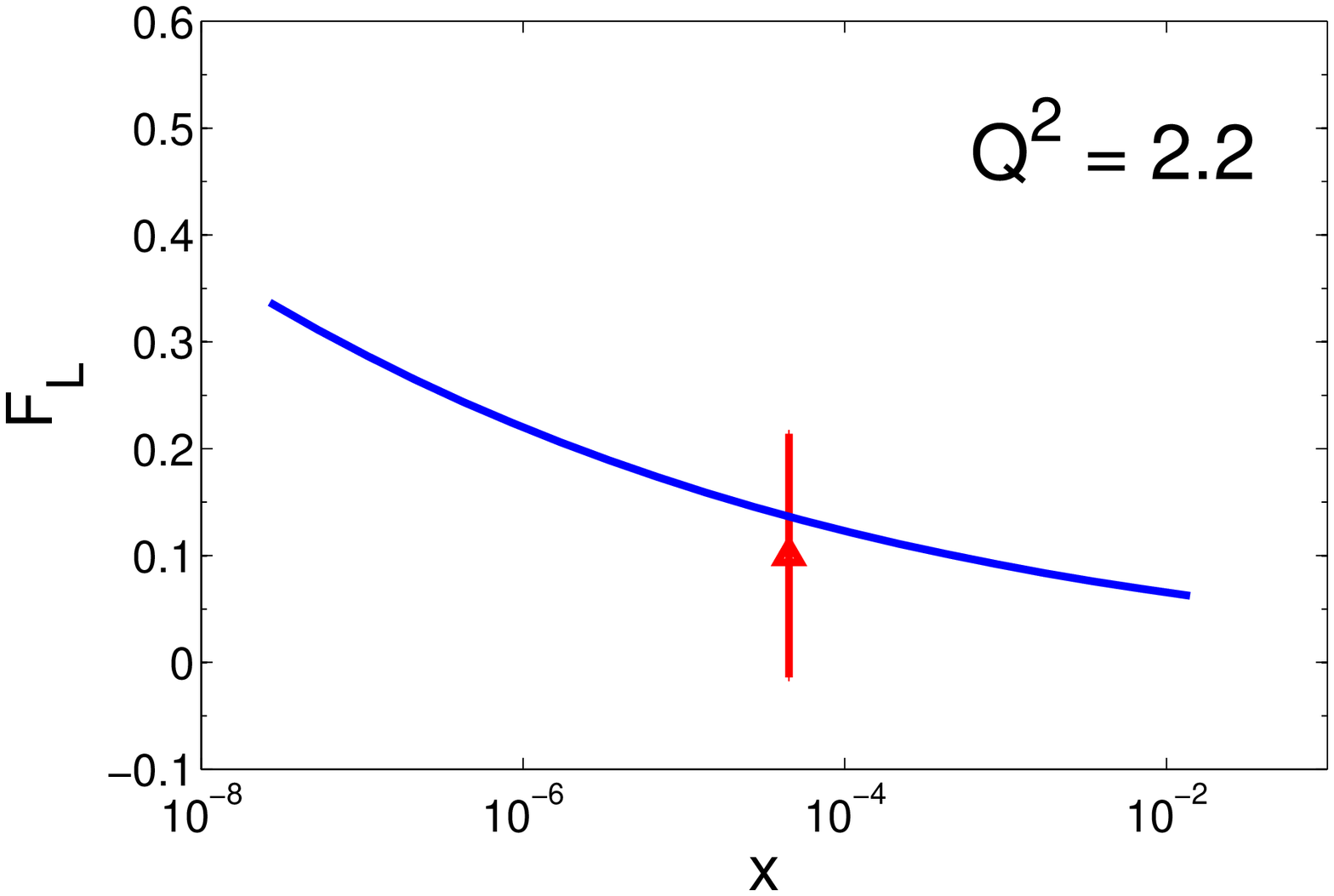,width=40mm,height=25mm} \\
\epsfig{file=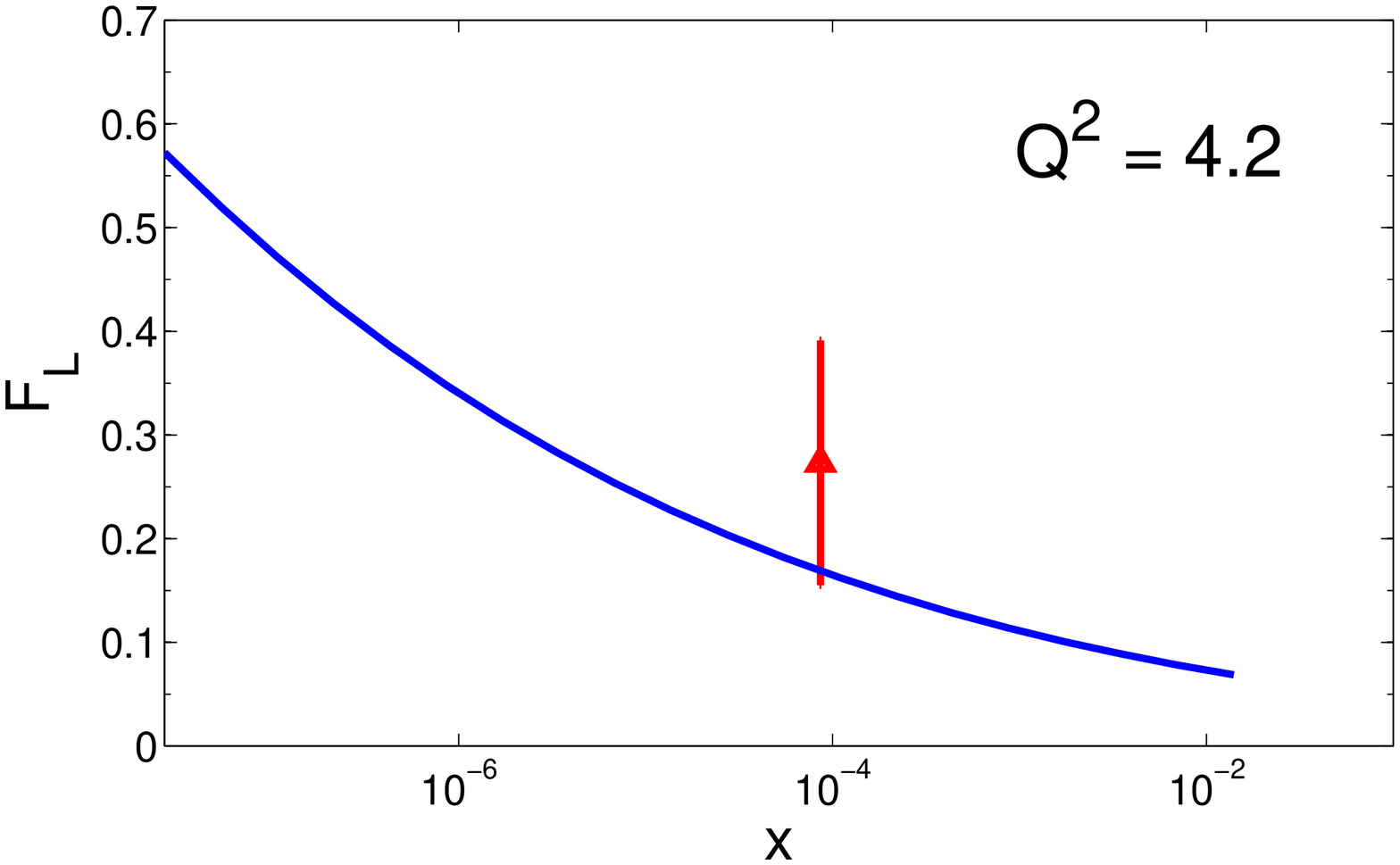,width=40mm, height=25mm} &
\epsfig{file=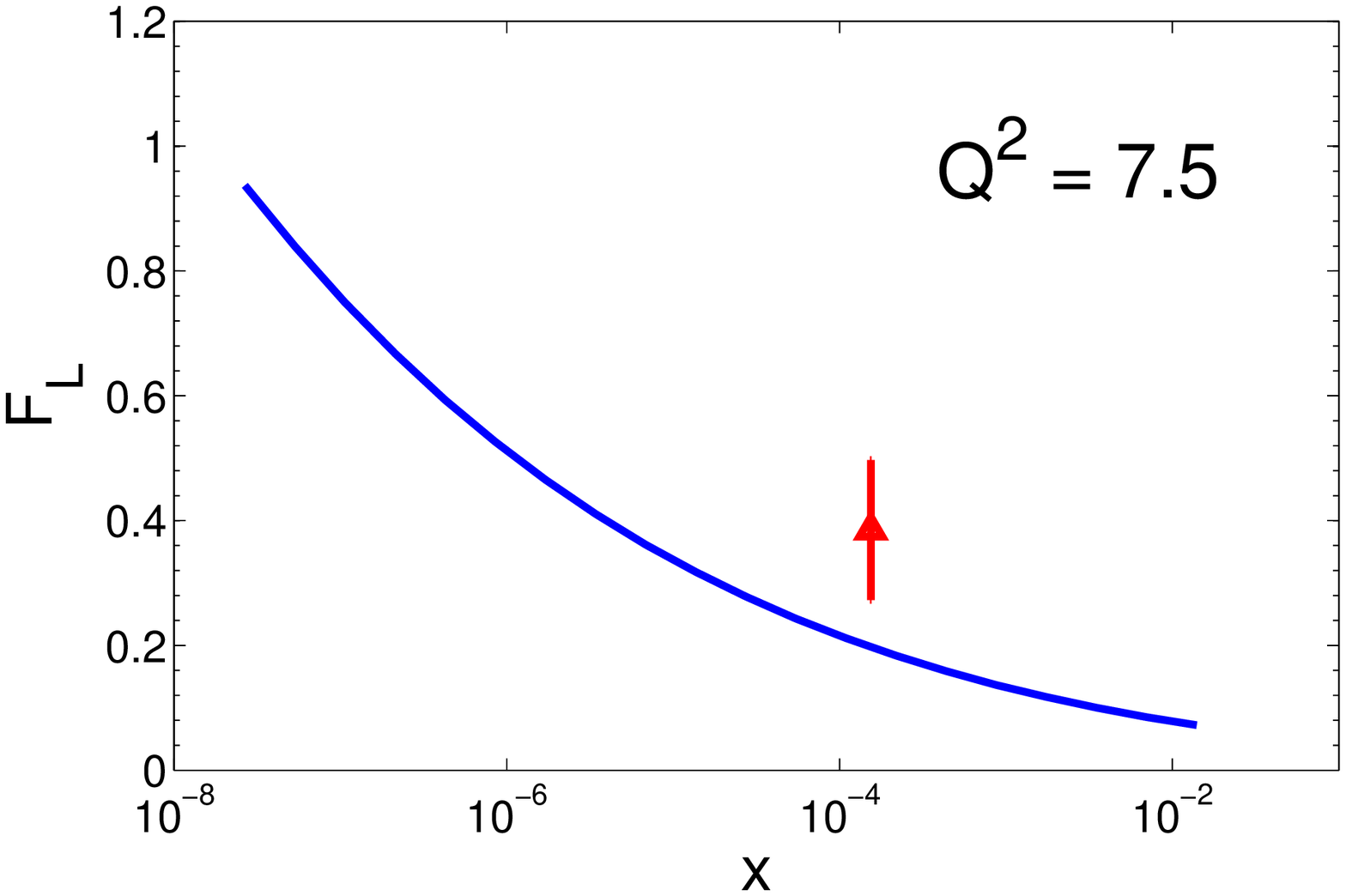,width=40mm,height=25mm} &
\epsfig{file=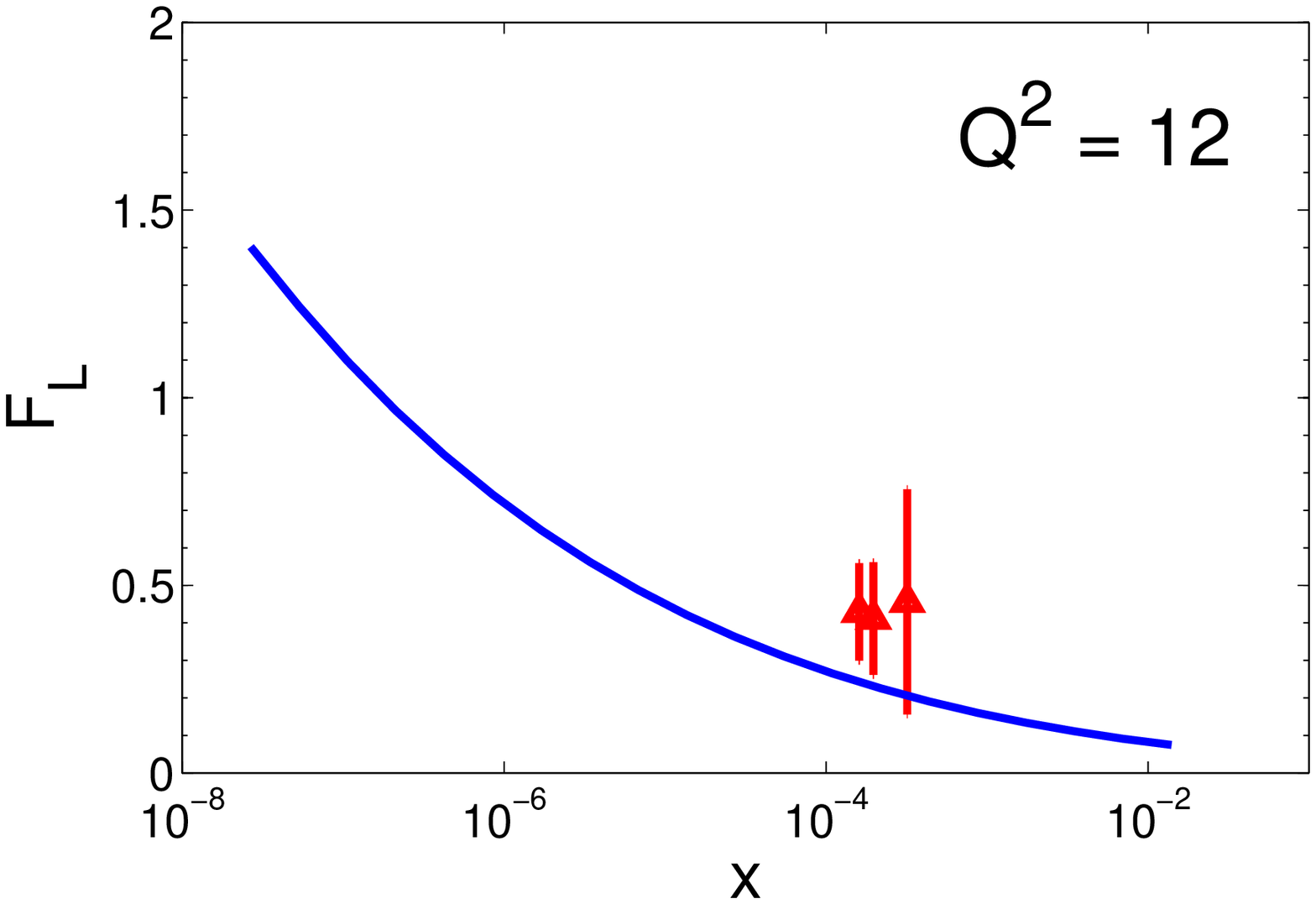,width=40mm, height=25mm} &
\epsfig{file=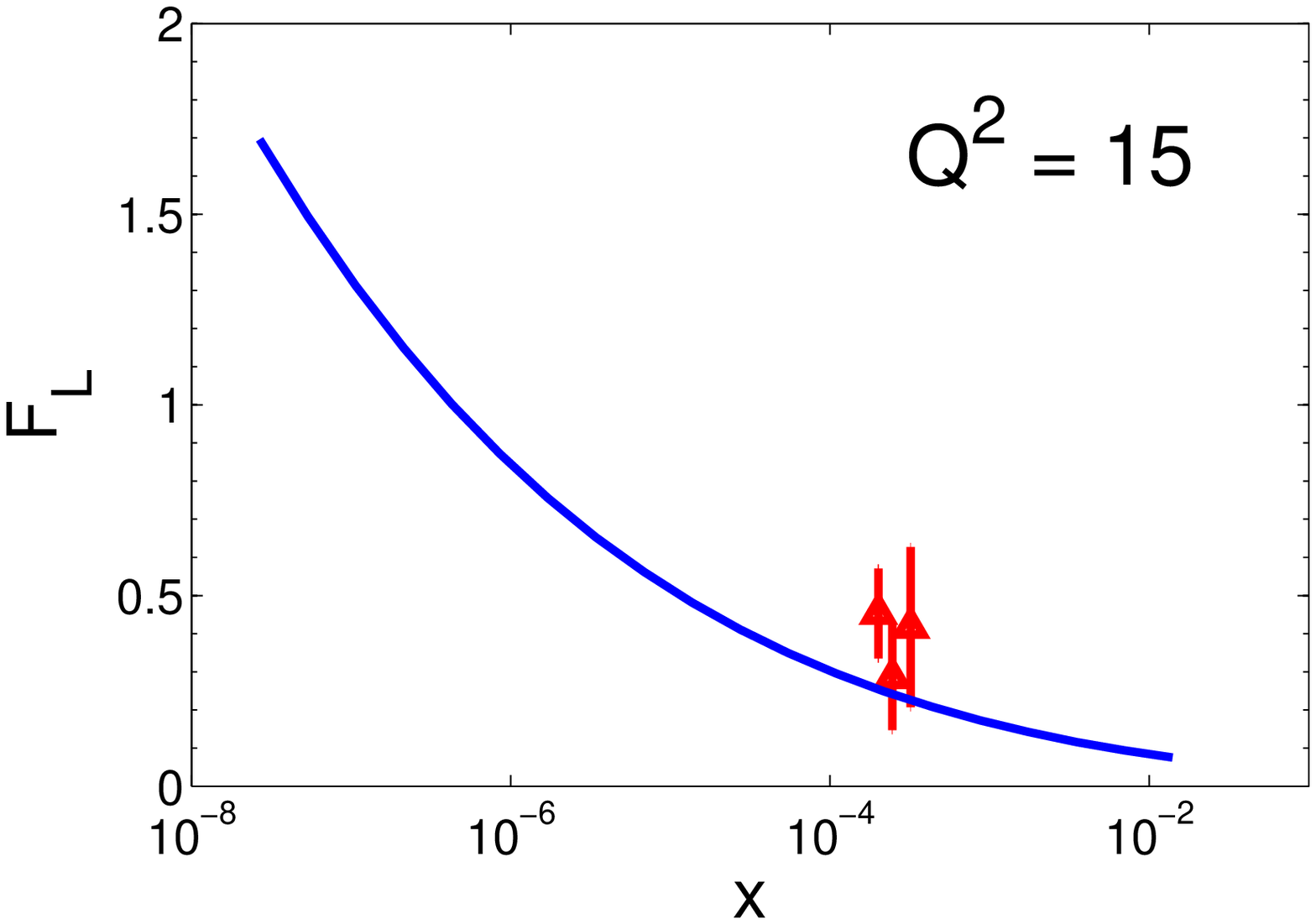,width=40mm,height=25mm} \\
\epsfig{file=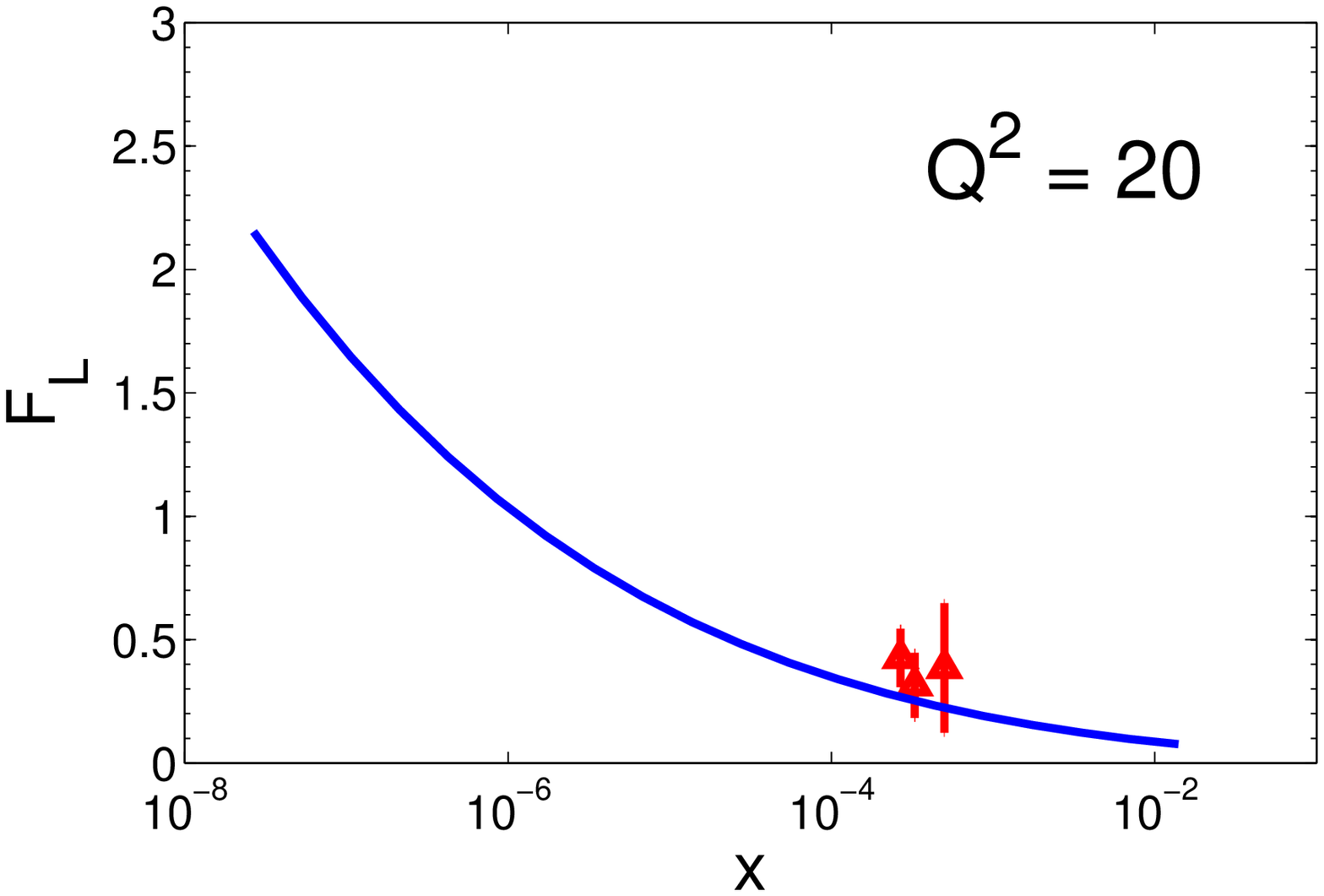,width=40mm, height=25mm} &
\epsfig{file=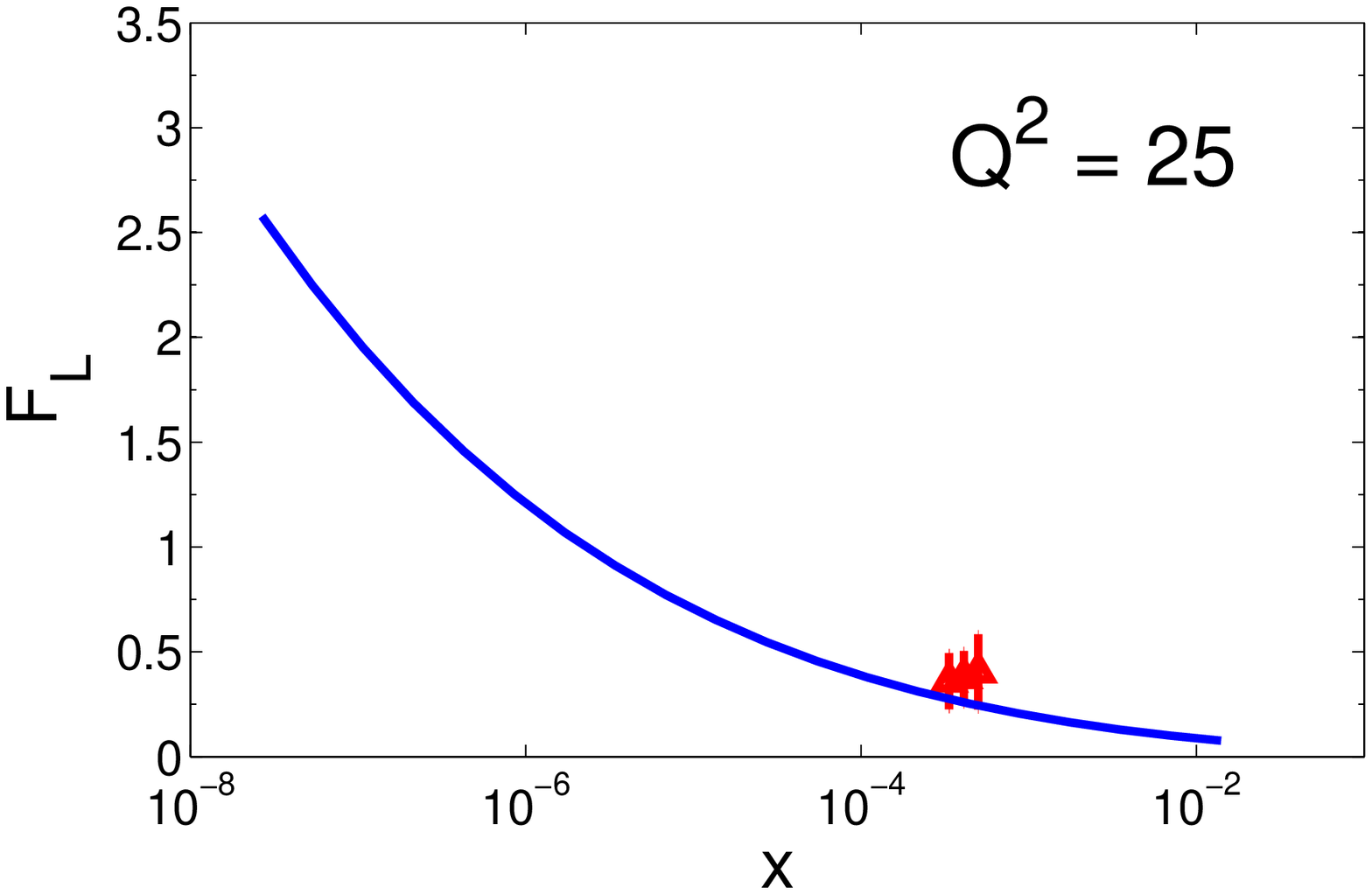,width=40mm,height=25mm} &
\epsfig{file=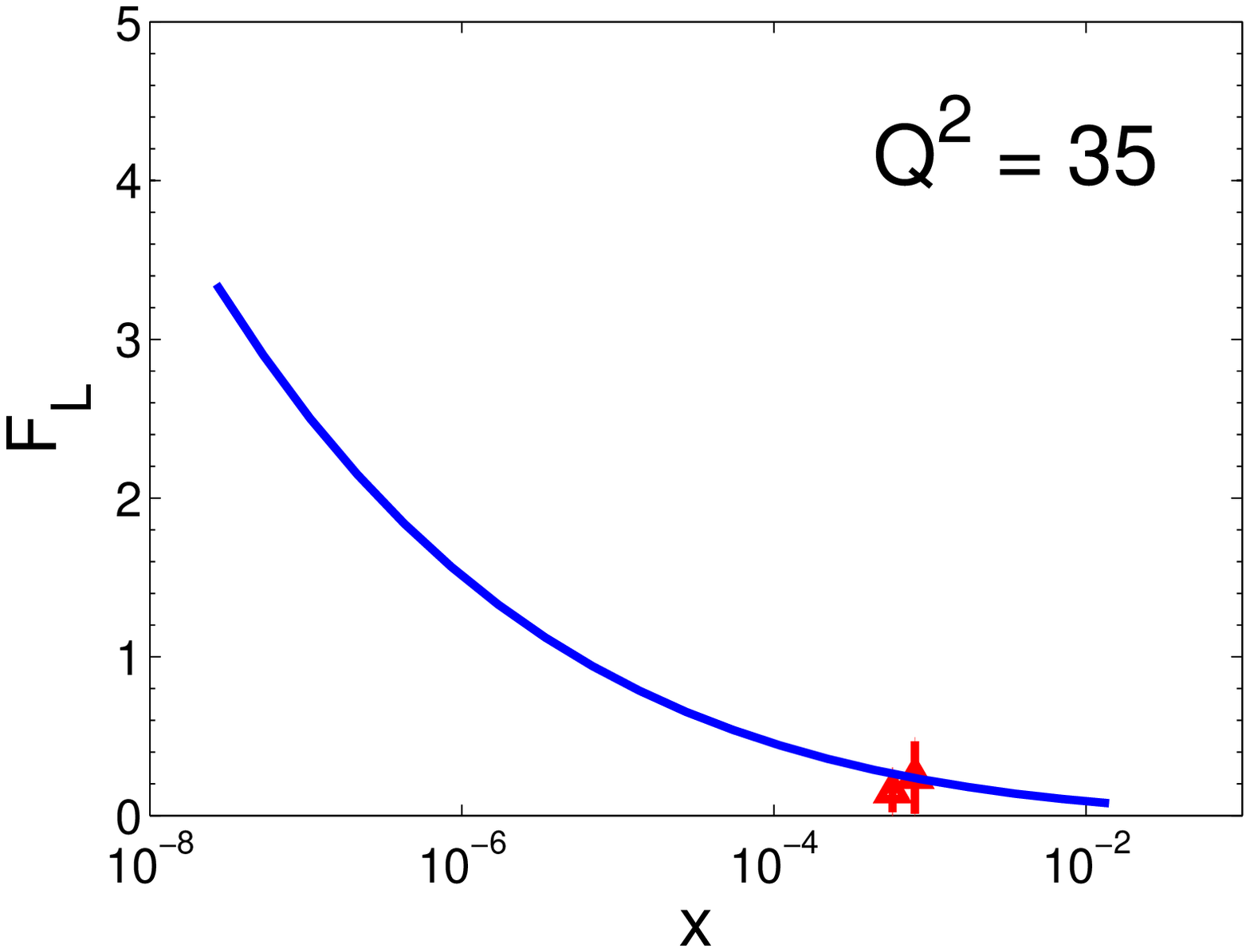,width=40mm, height=25mm} &
\epsfig{file=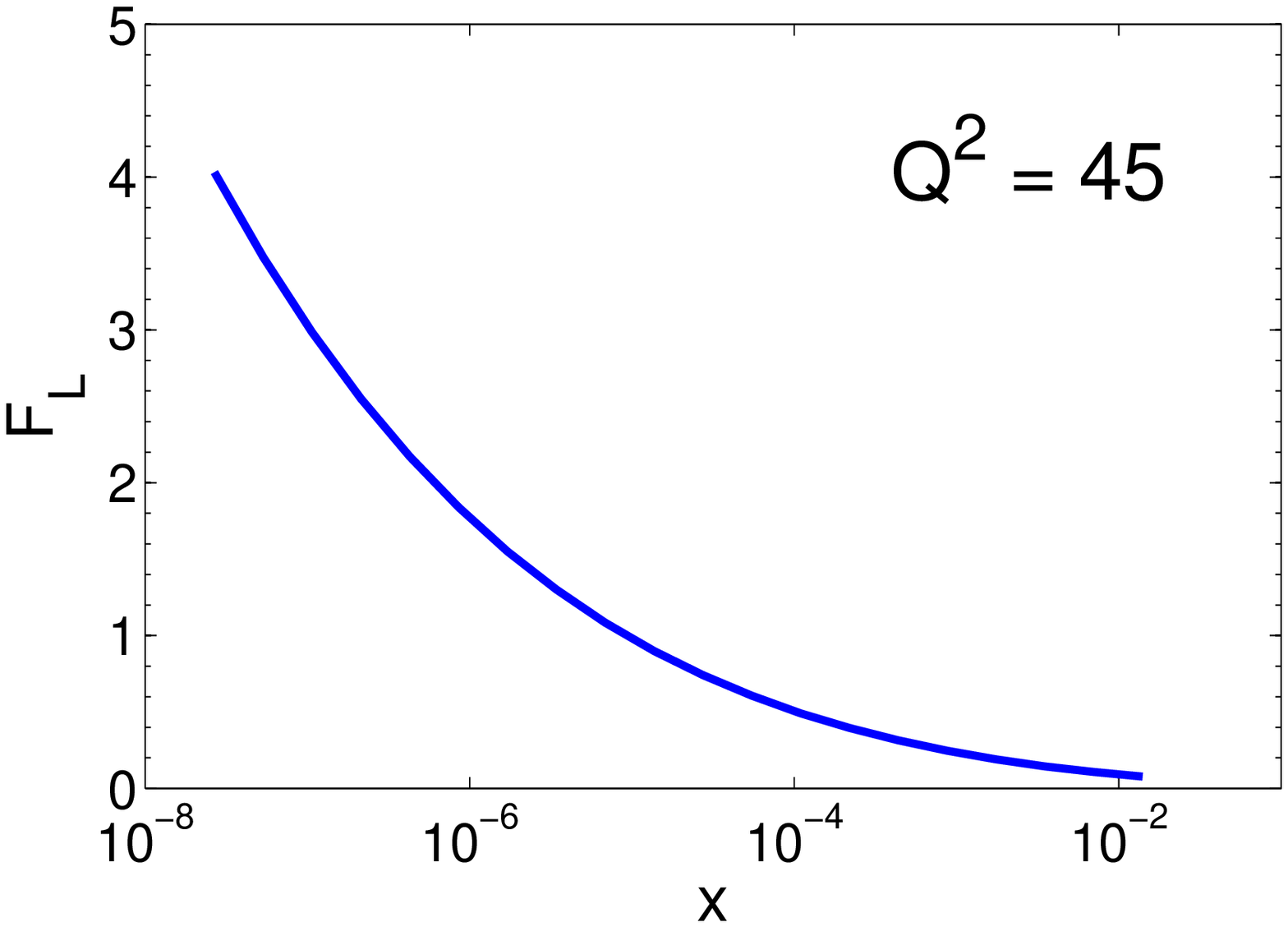,width=40mm,height=25mm} \\
\epsfig{file=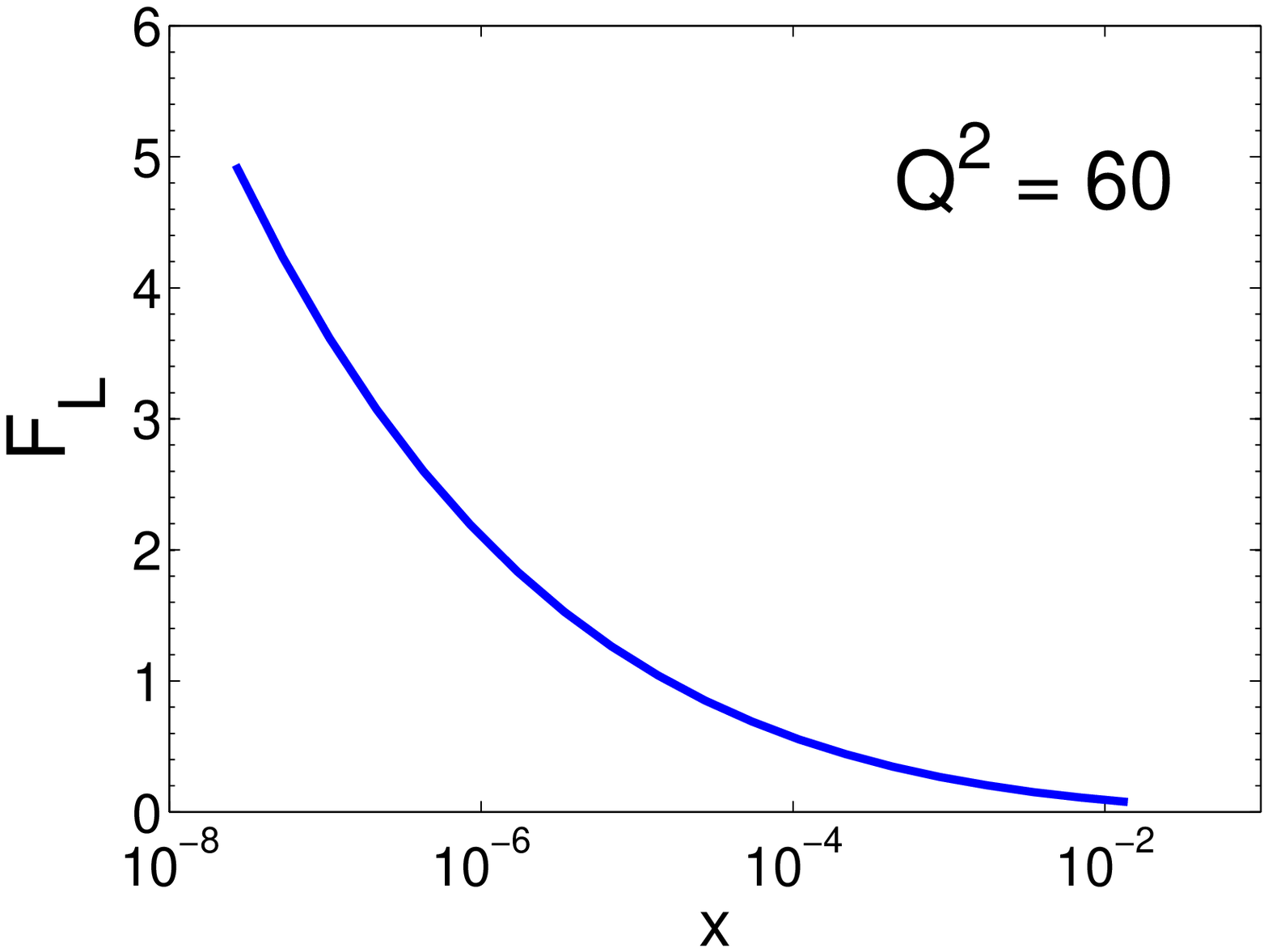,width=40mm, height=25mm} &
\epsfig{file=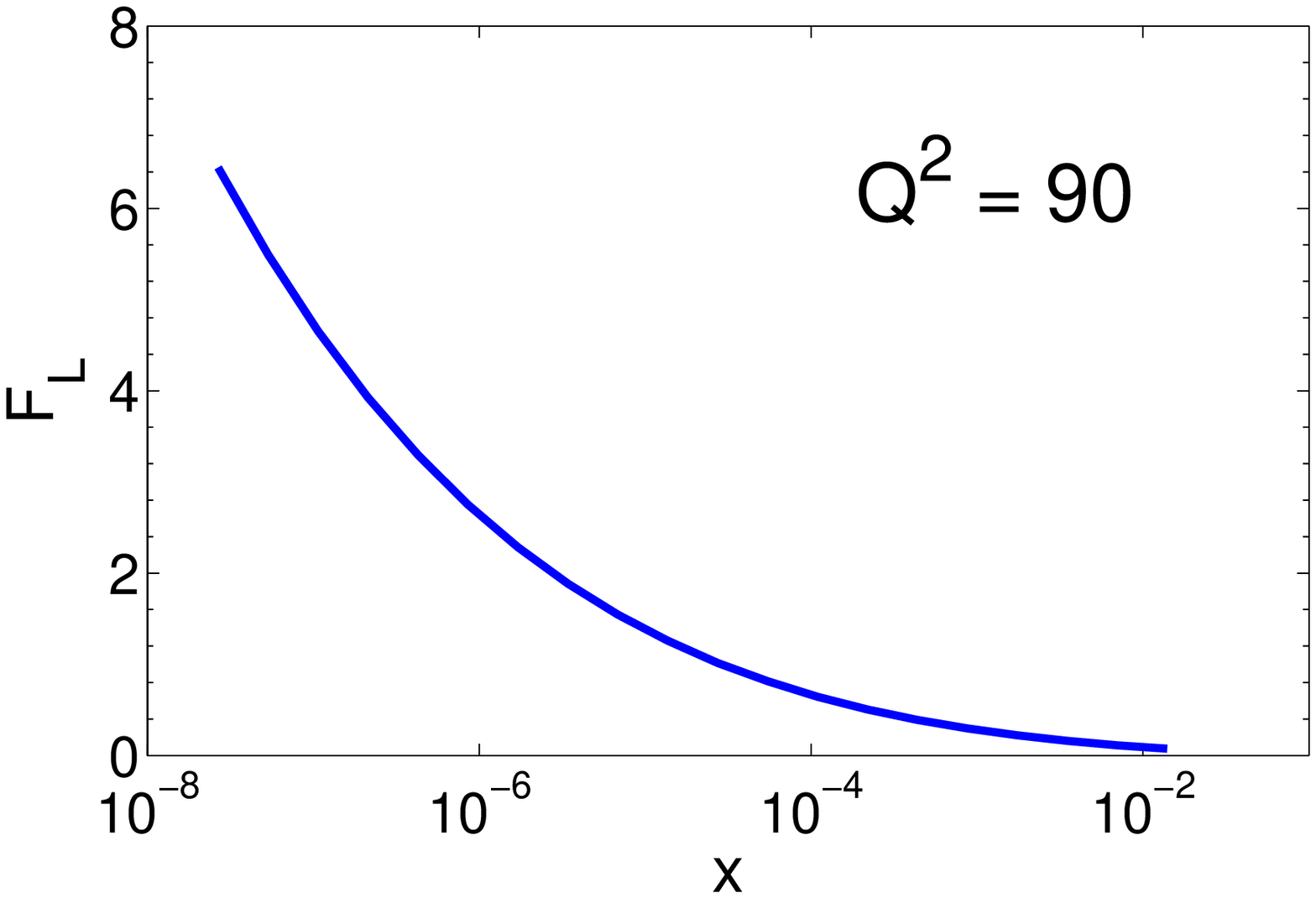,width=40mm,height=25mm} &
\epsfig{file=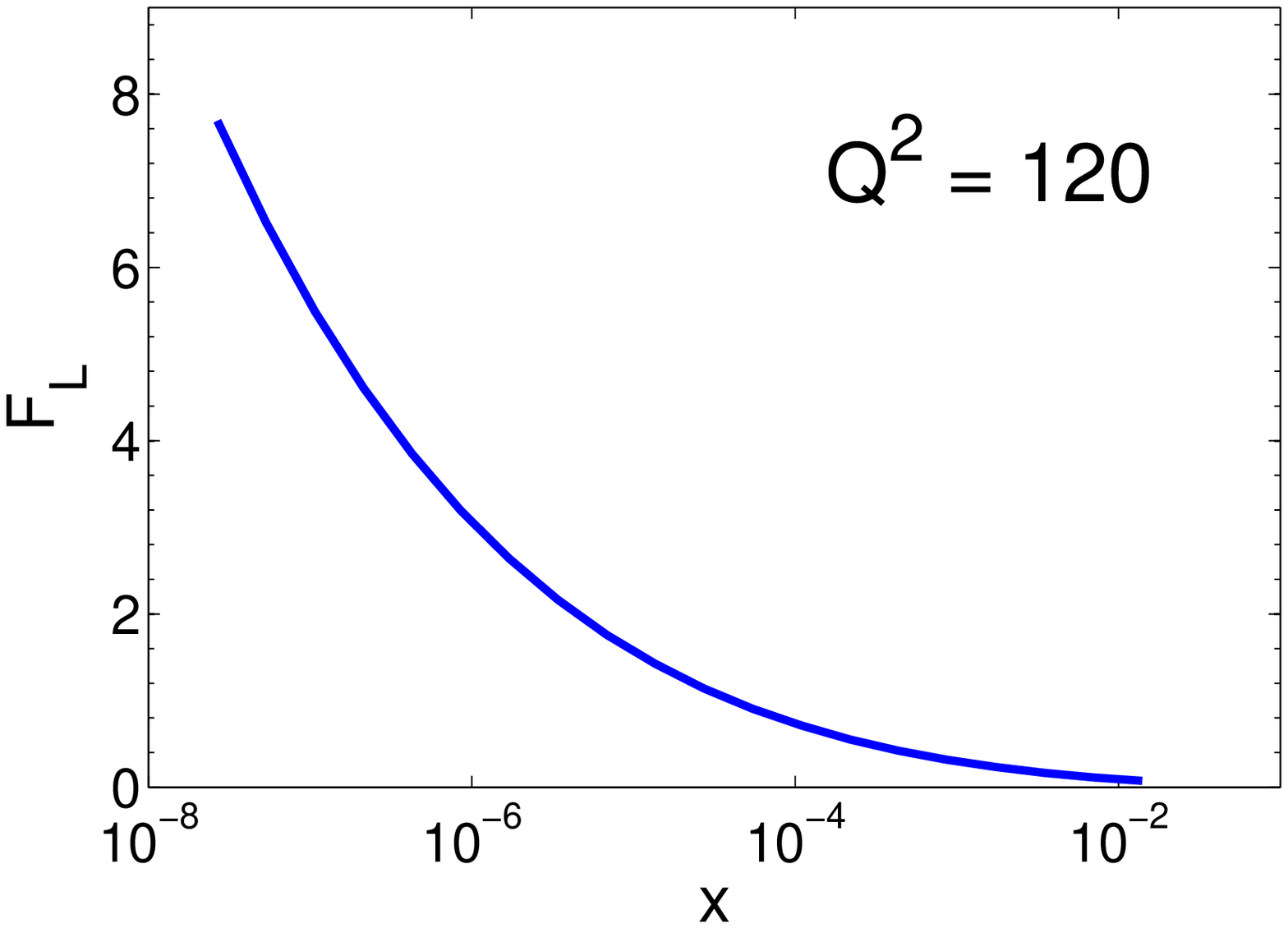,width=40mm, height=25mm} &
\epsfig{file=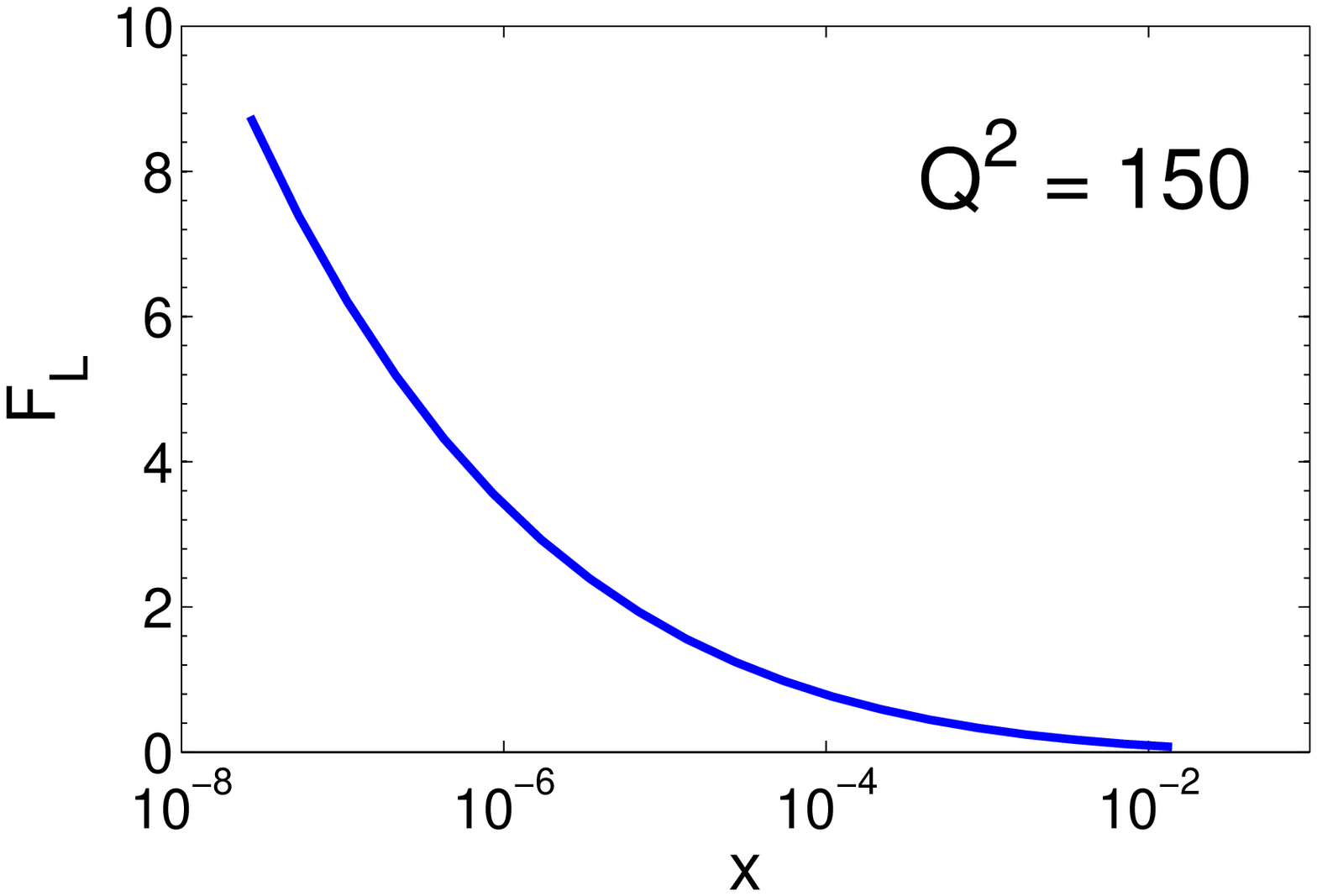,width=40mm,height=25mm}
\end{tabular}\caption{\it Prediction for the $F_{L}$
structure function at the HERA kinematics. The values of $Q^{2}$ are
given in $GeV^{2}$.}\label{F_L}
\end{figure}

Here, we want to present our predictions for the longitudinal part
of the $F_{2}$ structure function $F_{L}$. This longitudinal part,
originates from the scattering of the longitudinally polarized
virtual photon, off the proton target. The expression for the
calculation has the following form \beqn
F_{L}(x,Q^{2})\;=\;\frac{Q^{2}}{4\pi^{2}\alpha}2\int d^{2}b\int
d^{2}r \int dz |\Psi_{L}(r,z,Q^{2})|^{2}N(r,b,x)\eeqn where
$|\Psi_{L}(r,z,Q^{2})|^{2}$ corresponds to the longitudinal part of
the photon wavefunction squared \eq{psi}, and $N(r,b,x)$ is the
interaction amplitude \eq{int_amp_fin}. We used the relevant data on
$F_{L}$ from H1 collaboration \cite{Adloff:2000qk} in order to
estimate our calculations. The experimental values of the
longitudinal structure function, are not measured, rather they are
extracted from the total structure function $F_{2}$. The extraction
of the longitudinal structure function, is based on the reduced
cross section \eq{sig_r}, which depends on two proton structure
functions, $F_{2}(x,Q^{2})$ and $F_{L}(x,Q2)$
\beqn\label{sig_r}\sigma_{r}\;\equiv\;F_{2}(x,Q^{2})\;-\;\frac{y^{2}}{1
+ (1 + y)^{2}}F_{L}(x,Q2)\eeqn where is $y$ defined in
\eq{dis_kin_var}. From the reconstruction of the kinematical
variable, the desired data on $F_{L}$ were obtained. Our main idea,
is to predict the behavior of the longitudinal part of the structure
function, in the HERA kinematic region ($x<10^{-5}$), and to check
how it fits the existing extracted data. The resulting plots are
presented in Fig. \ref{F_L}

\subsection{Diffractive production in DIS}

\begin{figure}[htbp]
\centerline{\includegraphics[width=7cm,height=5cm]{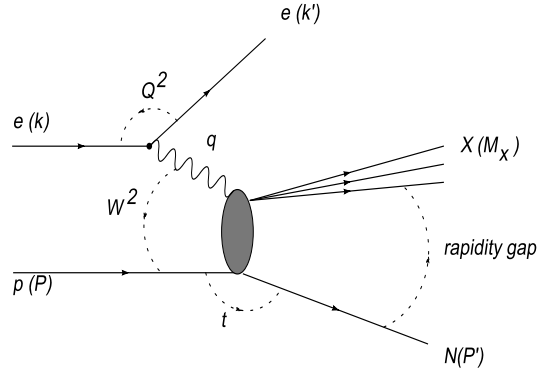}}
\caption{Kinematic variables of semi-inclusive reaction $ep
\rightarrow e N X$ diffractive dissociation}\label{diff_kin}
\end{figure}

Diffractive deep inelastic scattering is usually characterized by
two variables, the mass of the diffractive system $M^{2}_{X}$, and
the momentum transfer $t\,=\,(P - P^{'})^{2}$. These variables are
usually rewritten in terms of other, dimensionless variables as
\beqn\label{diff_var} x_{\Pom}\,=\,\frac{Q^{2} + M^{2} - t}{Q^{2} +
W^{2}} \eeqn which is the fractional energy-loss suffered by the
incident proton and \beqn \beta\,=\,\frac{Q^{2}}{Q^{2} + M^{2} - t}
\eeqn which corresponds to the momentum fraction, carried by a
struck parton. The pomeron, which carries longitudinal momentum
$x_{\Pom}$, is emitted by the proton, and subsequently undergoes
hard scattering satisfying \beqn x_{B}\,=\,x_{\Pom}\beta \eeqn The
diffractive dissociation process is depicted in Fig. \ref{diff_kin}

The important property of the wavefunction formalism, is the ability
to describe the diffractive scattering processes
\cite{Nikolaev:1991et}.  At small values of the diffractive mass
$M^{2}$, the elastic scattering of the $q\overline{q}$ pair,
dominates, and the corresponding diffractive structure function
reads as \beq\label{F2_diff}
F^{D}_{2}(x,Q^{2})_{T,L}\,=\,\frac{Q^{2}}{4\pi^{2}\alpha_{em}}\sum_{f
= 1}^{n_{f}}\int d^{2}b \int d^{2}r \int_{0}^{1} dz |\Psi_{T,L}(r,z;
m_{f},e_{f})|^{2}N^{2}(r,b,x)\eeq

\begin{figure}[htb]
\centerline{\epsfig{file=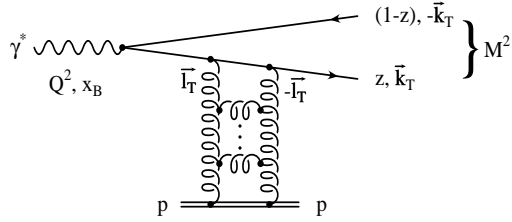,width=7cm, height=3cm}}
\caption{\sl $q\overline{q}$ contribution to diffractive DIS, where
$z$ represents the fraction of the energy of the photon that is
carried by a quark. $M$ is an invariant mass of diffractive system
\label{qq_prod}}
\end{figure}

However, at larger values of the mass $M^{2}$, the $q\overline{q}g$
contribution dominates, due to gluon production in the final
diffractive state. We take into account the three leading twist
terms \beq\label{diff_tree_contributions}
F^{D(3)}_{2}\,=\,F_{q\overline{q}}^{T} + F_{q\overline{q}}^{L} +
F_{q\overline{q}g}^{T} \eeq We introduce the last term in
\eq{diff_tree_contributions}, to describe high mass diffraction, and
as a simple approximation of the first "fan" diagram, in which the
emission of large numbers of gluons is taken into account. The
longitudinal part $F_{q\overline{q}g}^{L}$, has no leading logarithm
in $Q^{2}$, and can be neglected. The Feynman diagram for the
interaction of a quark-antiquark dipole, with the target proton via
two-gluon exchange, is shown in Fig.\ref{qq_prod}. The gluons couple
to the quark in all possible ways.

We follow the procedure proposed in Ref. \cite{Levin_open_mass,
Golec-Biernat:2001mm} and we obtain \beqn
F^{L}_{q\overline{q}}(x_{\Pom},\beta,Q^{2})\,&=&\,\frac{3Q^{4}}{\pi^3x_{\Pom}(1-\beta)}\int_{0}^{\infty}\,d^{2}b\,\int^{\frac{1}{4}M^{2}-m^{2}}_{0}
\frac{dk^{2}_{\perp}}{\sqrt{1-4\frac{m^{2}_{\perp}}{M^{2}}}}
\left(\frac{m^{2}_{\perp}}{M^{2}}\right)^{3}\phi^{2}_{0}(k,b,\beta,x_{\Pom})
\eeqn and the transverse $q\overline{q}$ contribution has the form
\beqn F^{T}_{q\overline{q}}(x_{\Pom},\beta,Q^{2})\,&=& \,\frac{3Q^{2}}{x_{\Pom}4\pi^{2}}\sum_{f}e^{2}_{f}\int_{0}^{\infty}\,d^{2}b\,\int^{\frac{1}{4}M^{2}-m^{2}}_{0} \frac{dk^{2}_{\perp}}{\sqrt{1-4\frac{m^{2}_{\perp}}{M^{2}}}}\left(\frac{m_{\perp}^{2}}{M^{2}}\right)\times\nonumber\\
&&\left\{\left(1-\frac{2m^{2}_{\perp}}{M^{2}}\right)\left(\frac{\beta
k^{2}+m^{2}_{f}}{1-\beta}\right)
\phi^{2}_{1}(k,b,\beta,x_{\Pom})\,+\,m^{2}_{f}\,\phi^{2}_{0}(k,b,\beta,x_{\Pom})\right\}
\eeqn where \beqn M^{2}\,=\,\frac{Q^{2}(1 - \beta)}{\beta}\;\;\;\;
\mbox{with} \;\;\;\; m^{2}_{\perp}\,=\,k^{2}_{\perp}\,+\,m^{2}_{f}
\eeqn and the "impact factor" $\phi_{i}(k,b,\beta, x_{\Pom})$
$(i=0,1,2)$ given by: \beq\label{phi_i} \phi_{i}(k,b,\beta,
x_{\Pom})\,=\,\int\,dr\,r\,K_{i}\left(\sqrt{\frac{k^{2}_{\perp}\beta+m^{2}_{f}}{1-\beta}}r\right)J_{i}(kr)N(x_{\Pom},r,b)
\eeq where $K_{i}$ and $J_{i}$ are Bessel functions. This impact
factor, represents the interaction between the produced dipole from
the virtual photon, and the target.

\begin{figure}[htb]
\centerline{\epsfig{file=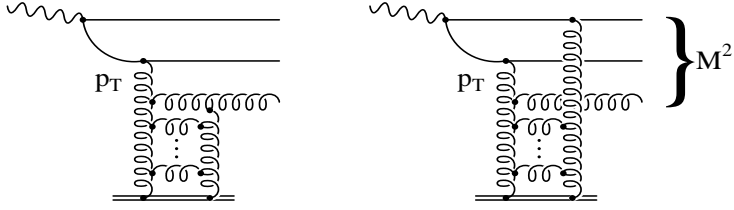,width=10cm, height=3cm}}
\caption{\sl $q\overline{q}g$ contribution with strong ordering with
the gluon. Gluon's transverse momentum is considered to be smaller
than the (anti)quark's one. \label{qq_interaction_prot}}
\end{figure}

The next contribution  $q\overline{q}g$, was calculated assuming the
strong ordering in the transverse momenta of the gluon and the
$q\overline{q}$ dipole, namely $k_{\perp g}\ll k_{\perp
q,\overline{q}}$. This assumption allows us to treat the
$q\overline{q}$, and $q\overline{q}g$, as an effective color dipole
in the transverse $r$ space. The corresponding diagram is plotted in
Fig. \ref{qq_interaction_prot}.

Thus, we obtain:
\beqn\label{F2qqg_T}F_{q\overline{q}g}^{T}(\beta,Q^{2},x_{\Pom})\,=\,\frac{81\beta}{16\pi^{2}x_{\Pom}}\sum_{f}e^{2}\frac{\alpha_{s}(Q^{2})}{2\pi}
\int^{1}_{\beta}
\frac{d\beta^{'}}{\beta^{'}}\left[\left(1-\frac{\beta}{\beta^{'}}\right)^{2}\,+\,\left(\frac{\beta}{\beta^{'}}\right)^{2}\right]\nonumber\\
\times\,\frac{\beta^{'}}{(1-\beta^{'})^{3}}\int_{0}^{\infty}\,d^{2}b\,\int
^{Q^{2}(1-\beta^{'})}_{0}dk^{2}_{\perp}\,\ln\left(\frac{Q^{2}(1-\beta^{'})}{m^{2}_{\perp}}\right)\,\phi^{2}_{2}(k,b,
\beta^{'},x_{\Pom}) \eeqn

Using the developed model (see section~\ref{model}), which
successfully fitted all the experimental data on DIS, we want to
describe diffractive DIS, using the same model and parameters which
were obtained from the fit. For this purpose, we used the latest
data on diffractive dissociation, from \emph{ZEUS} collaboration
\cite{Chekanov:2005vv}. The resulting plots, are shown in Figs.
\ref{diff_1},~\ref{diff_2}.

\begin{sidewaysfigure}[htbp]
\centering
\begin{tabular}{cccccc ccccc ccccc ccc}
$M_{X} = 1.2\,GeV$ & $M_{X} = 3\,GeV$ & $M_{X} = 6\,GeV$ & $M_{X} = 11\,GeV$ & $M_{X} = 20\,GeV$ & $M_{X} = 30\,GeV$\\
\begin{sideways}{\small $Q^{2}=2.7\,GeV^{2}$}\end{sideways}
\epsfig{file=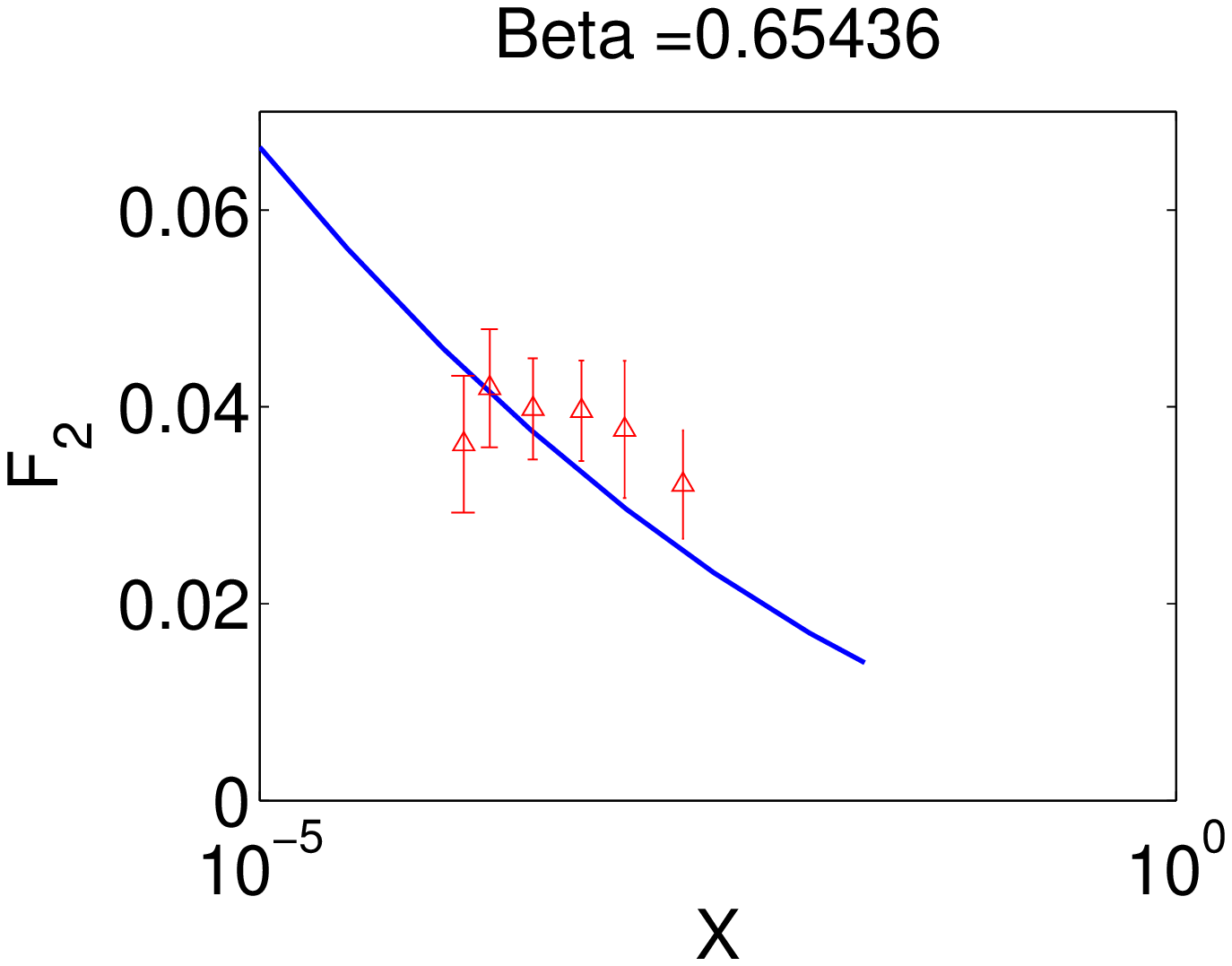,width=32mm, height=28mm}&
\epsfig{file=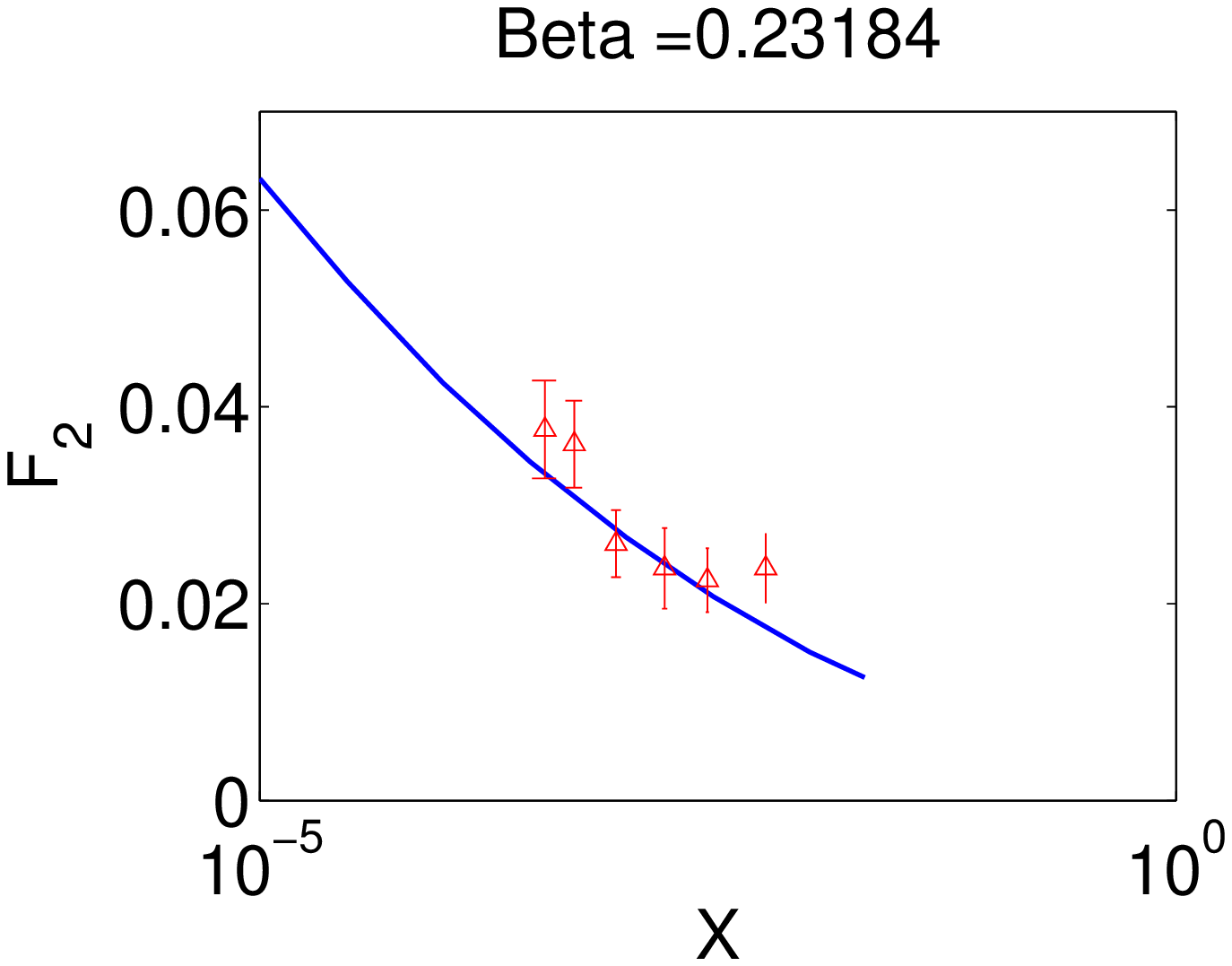,width=32mm, height=28mm}&
\epsfig{file=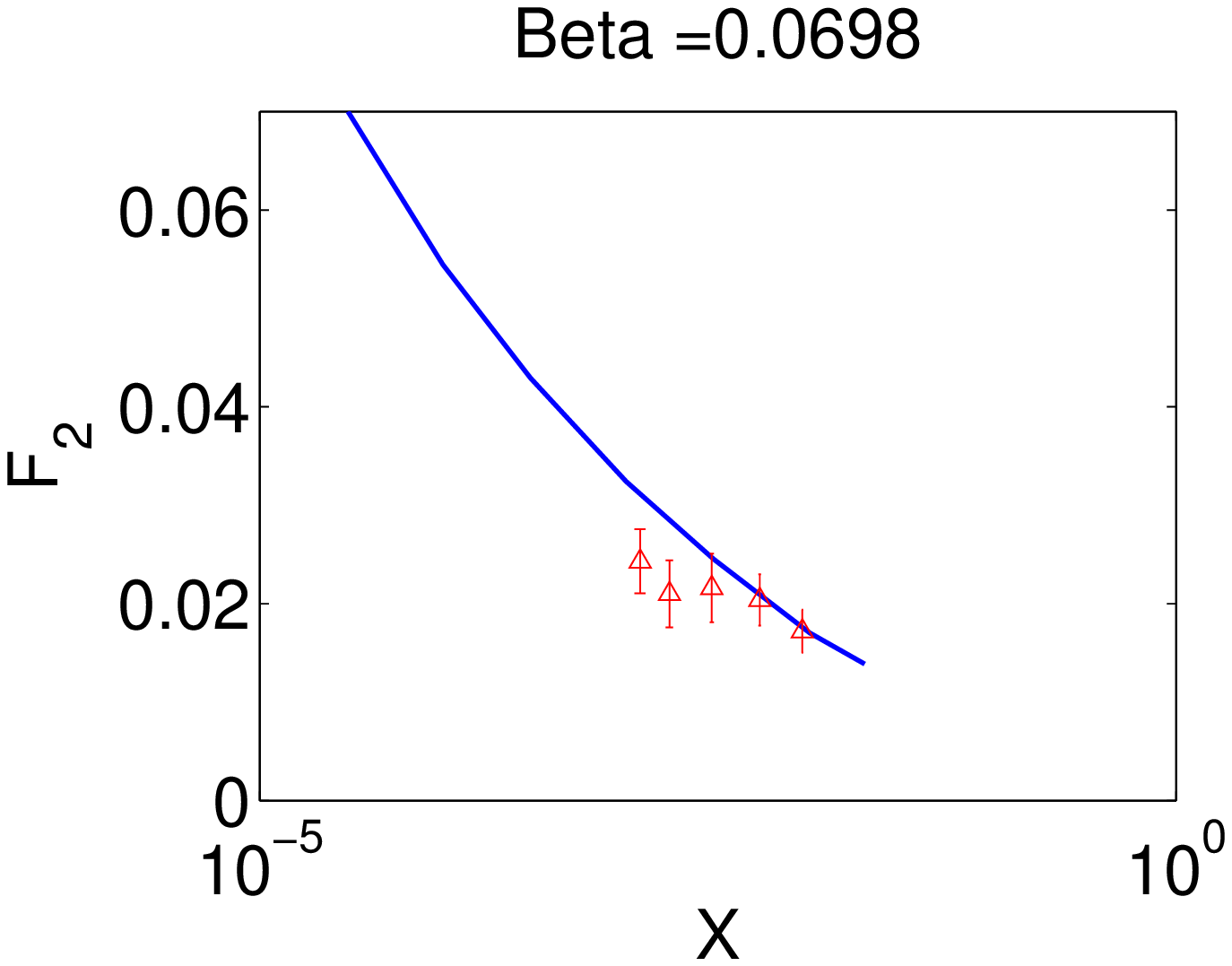,width=32mm, height=28mm}&
\epsfig{file=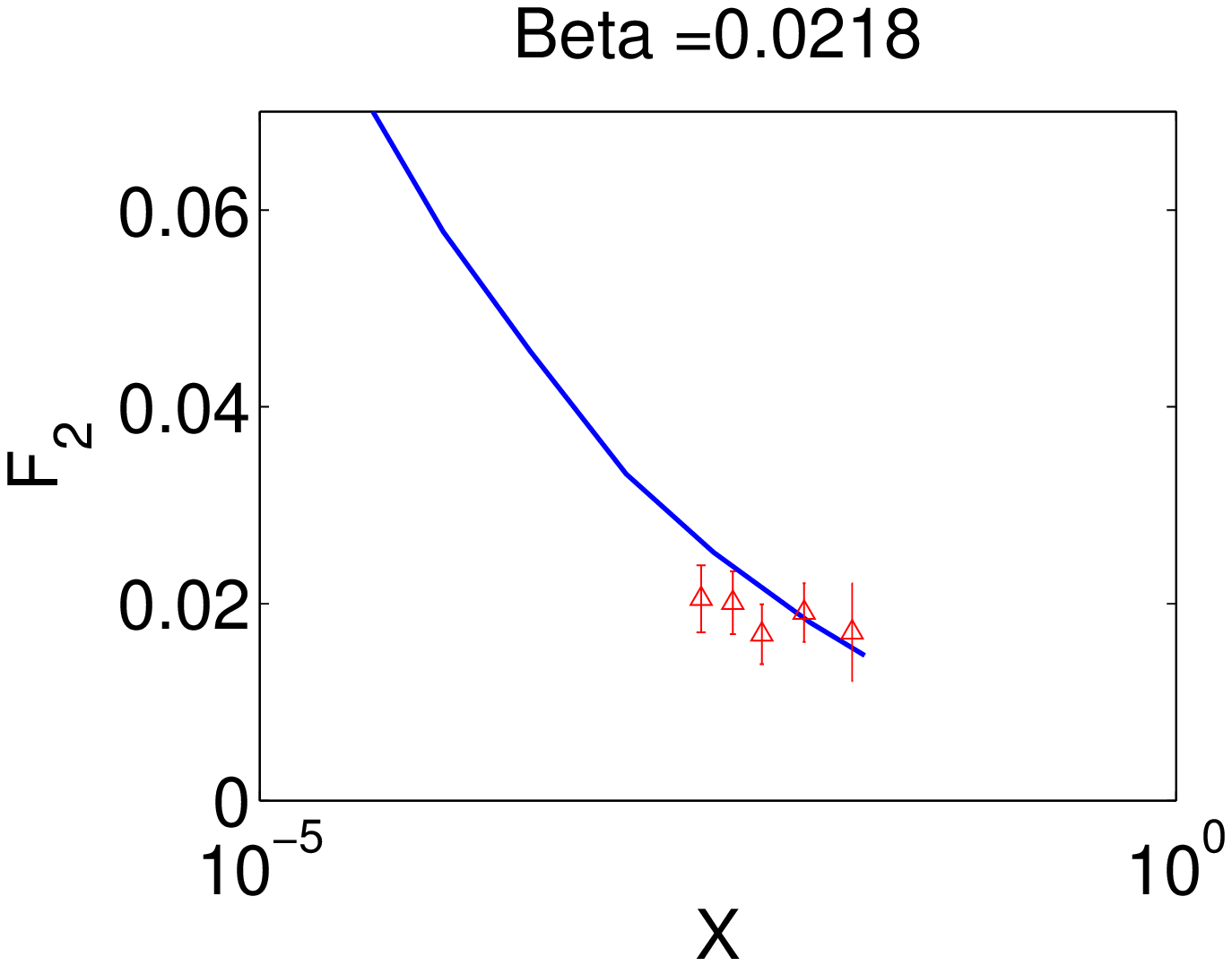,width=32mm, height=28mm}&
\epsfig{file=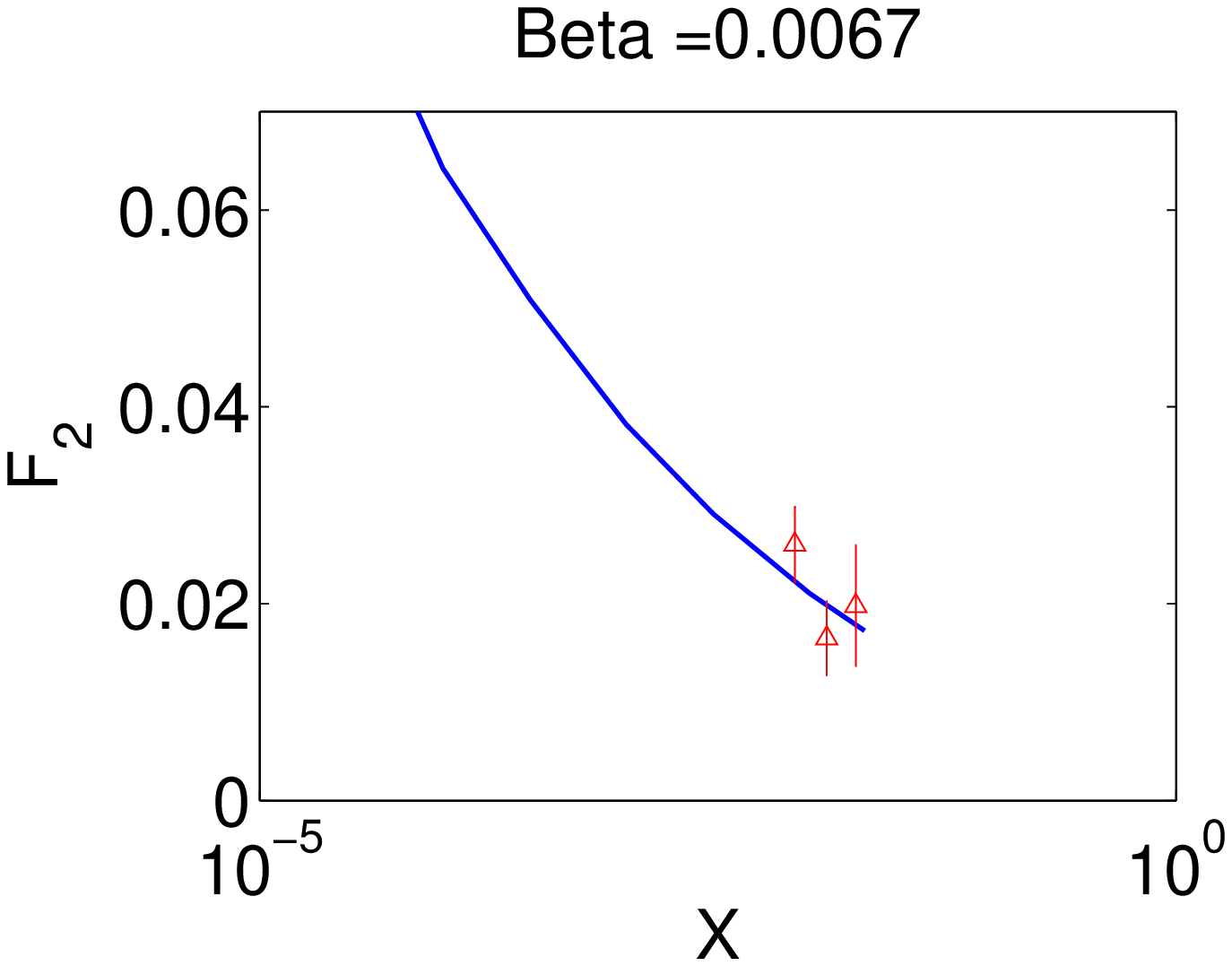,width=32mm, height=28mm}&
\epsfig{file=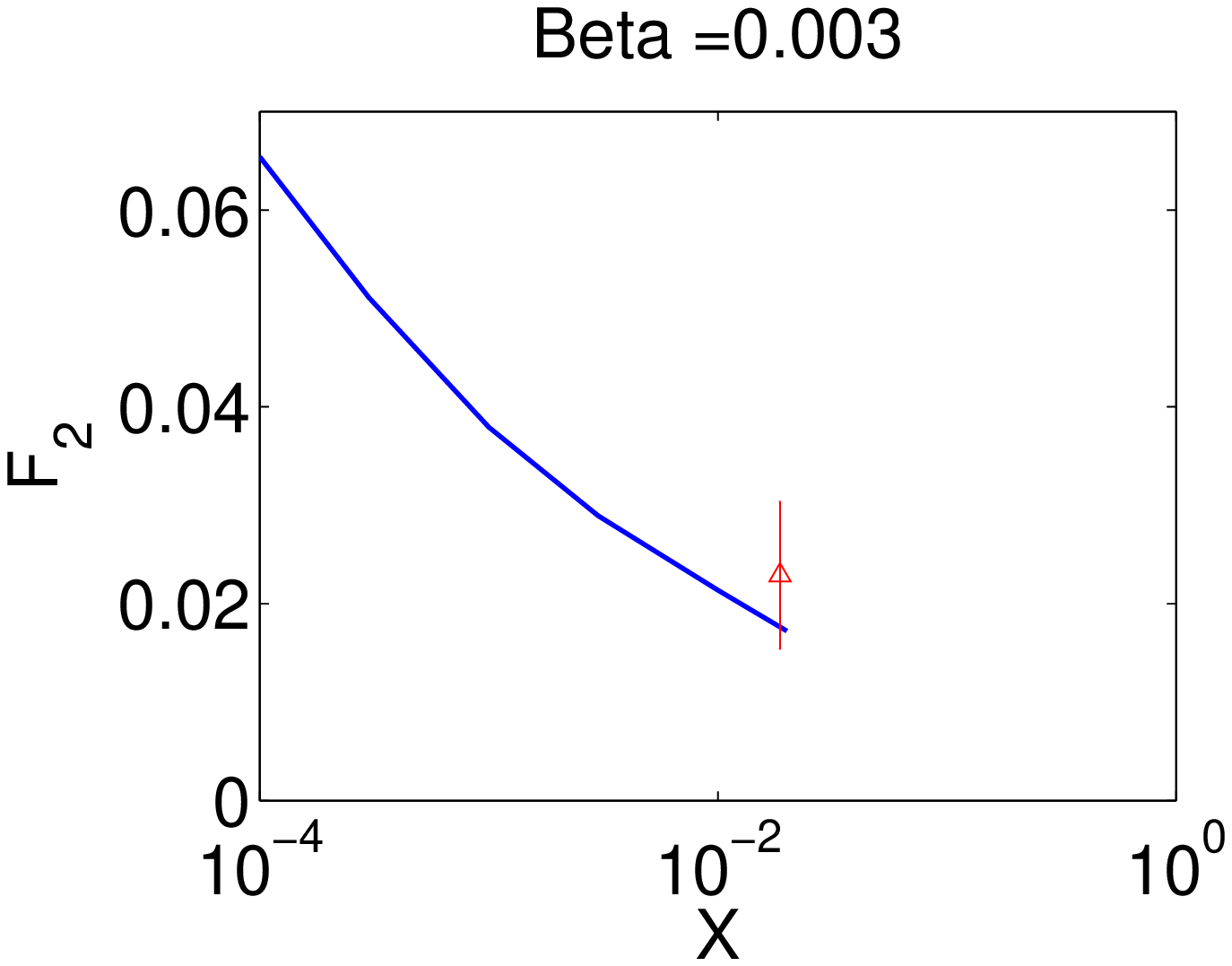,width=32mm, height=28mm}\\
\begin{sideways}{\small $4\,GeV^{2}$}\end{sideways}
\epsfig{file=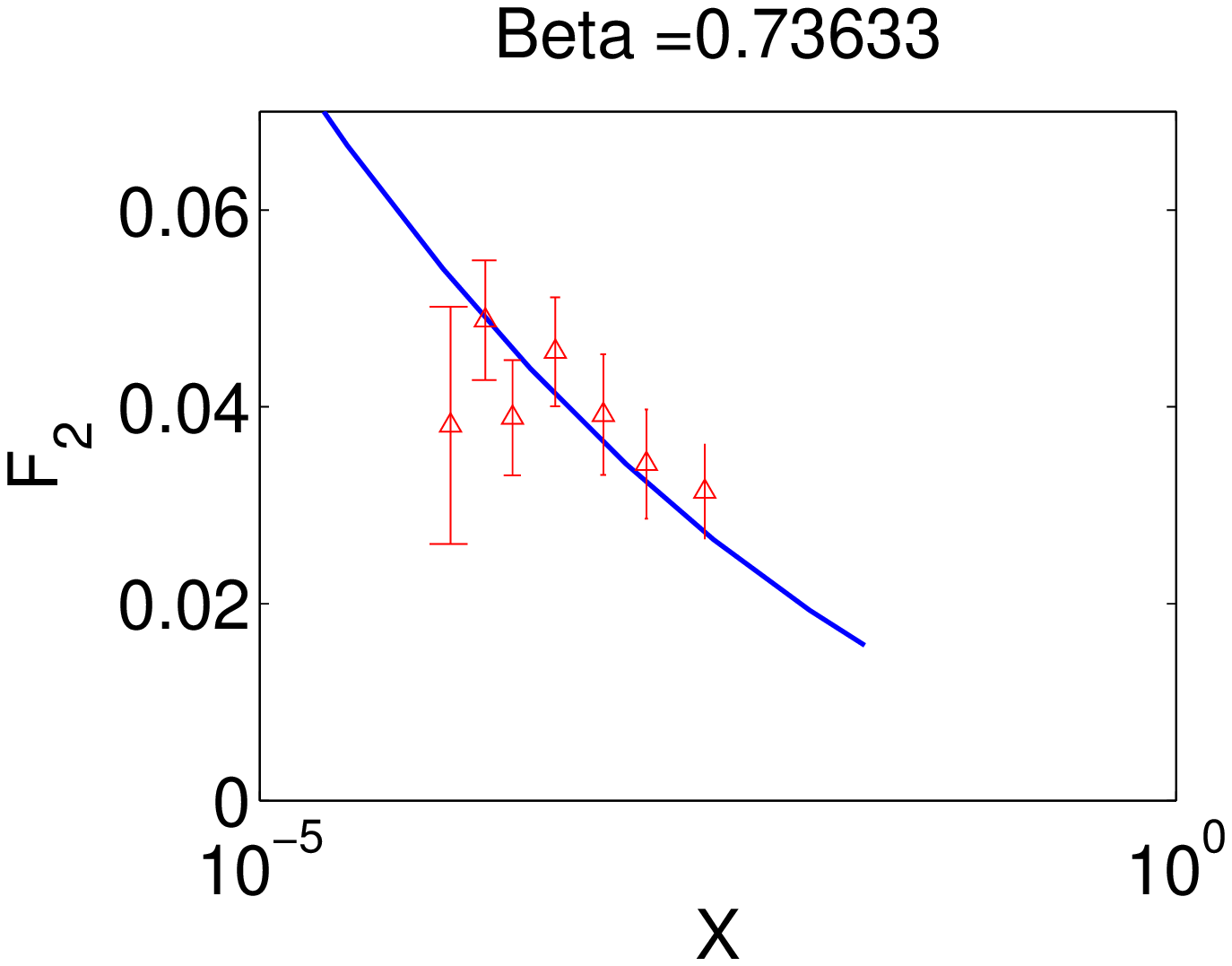,width=32mm, height=28mm}&
\epsfig{file=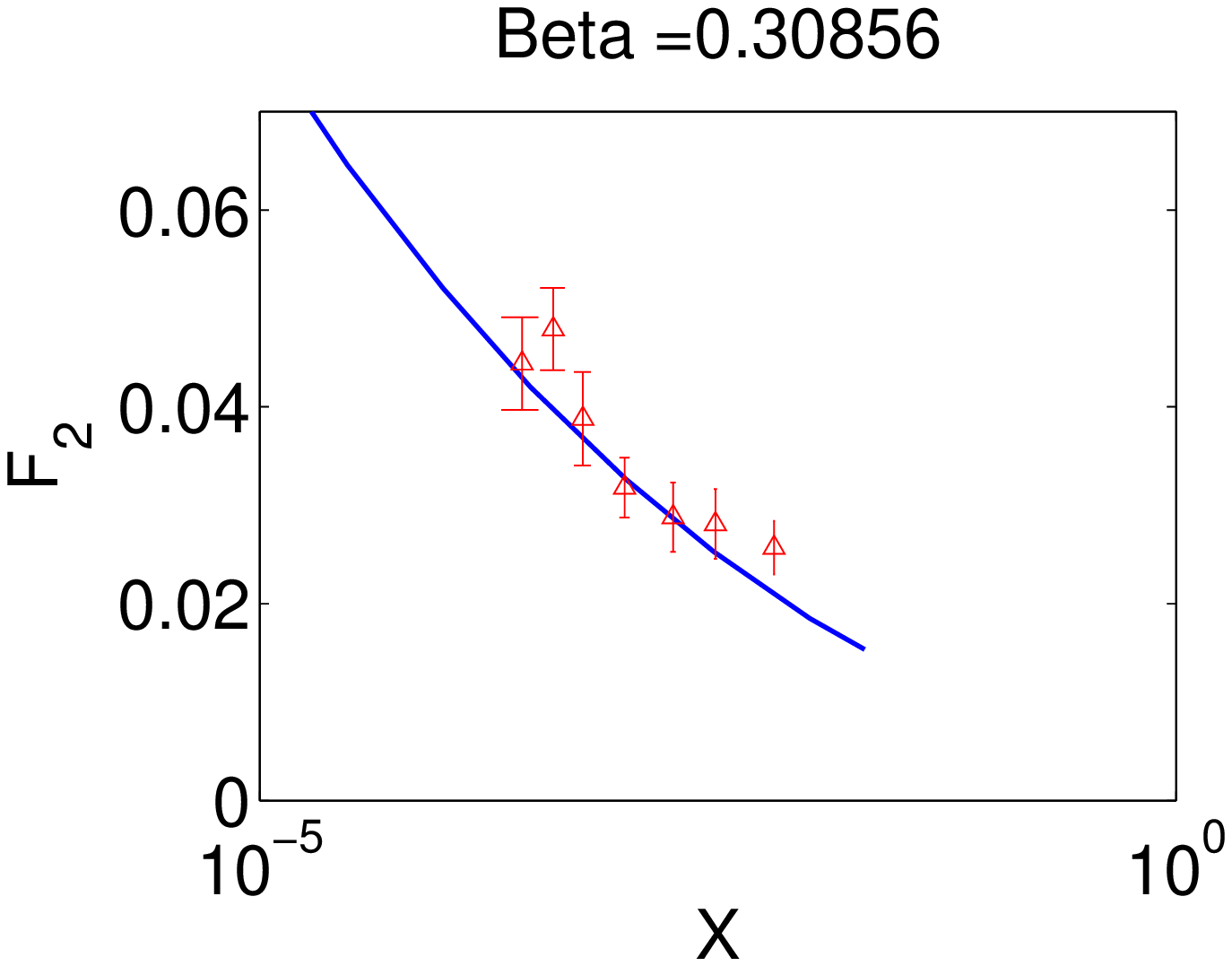,width=32mm, height=28mm}&
\epsfig{file=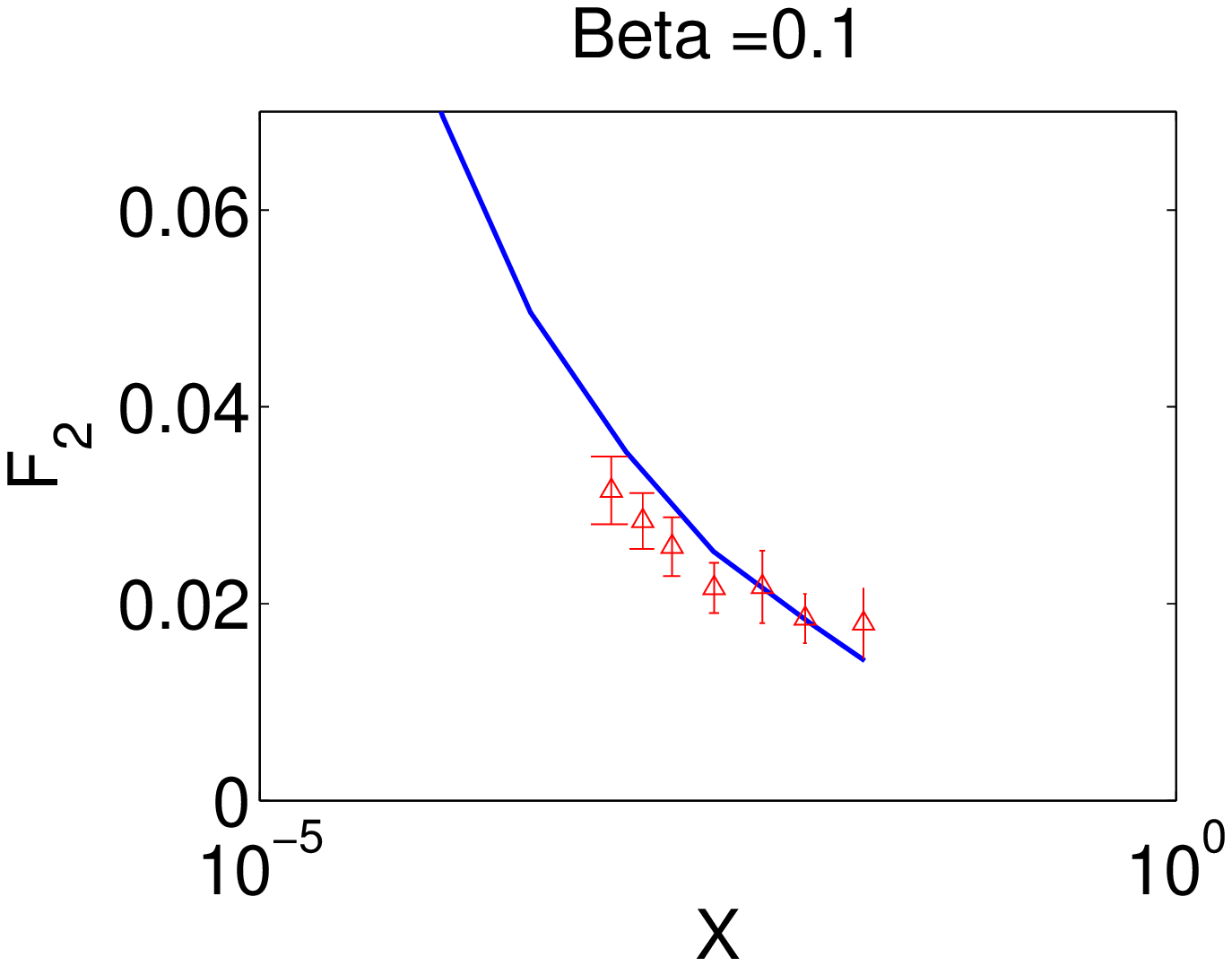,width=32mm, height=28mm}&
\epsfig{file=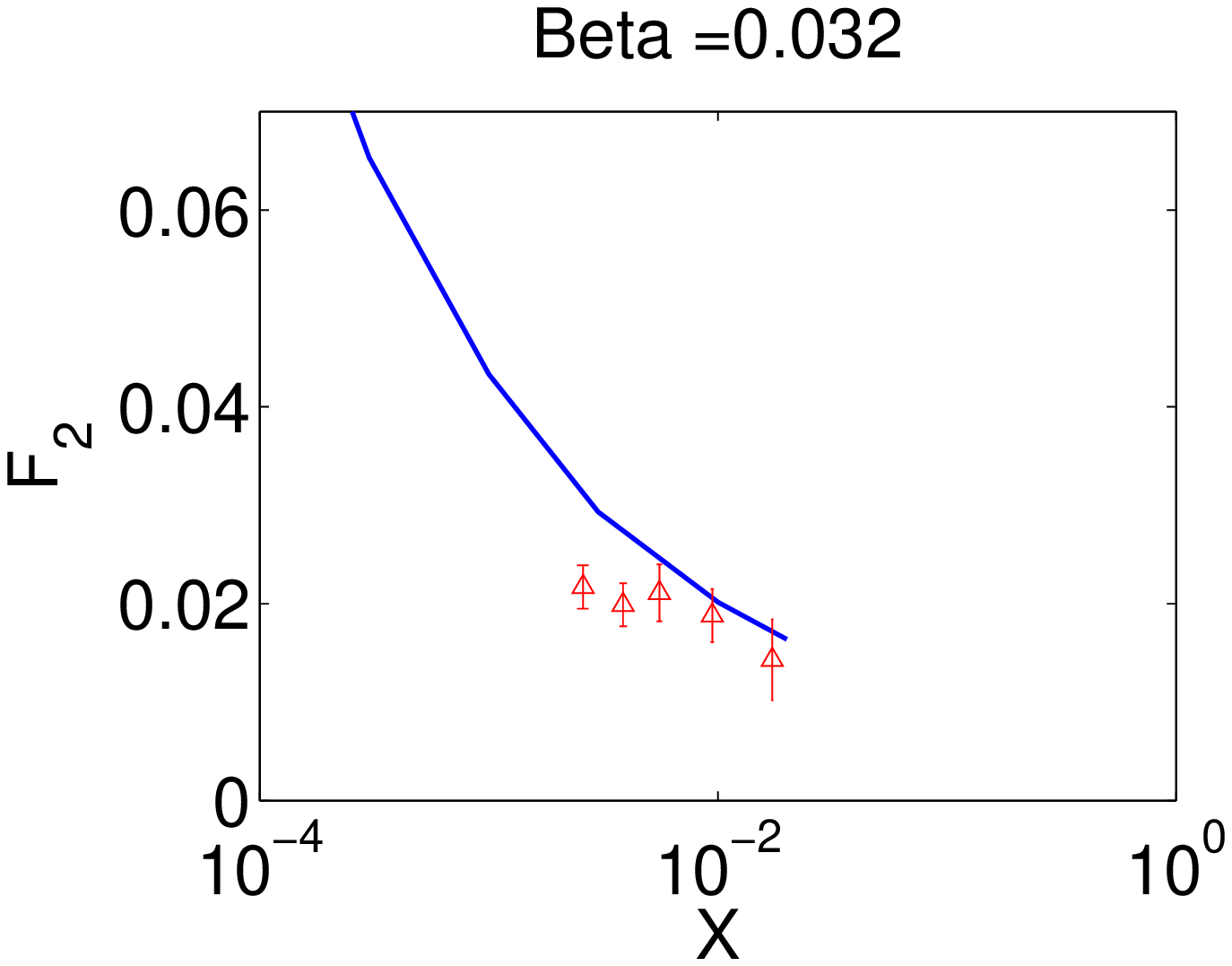,width=32mm, height=28mm}&
\epsfig{file=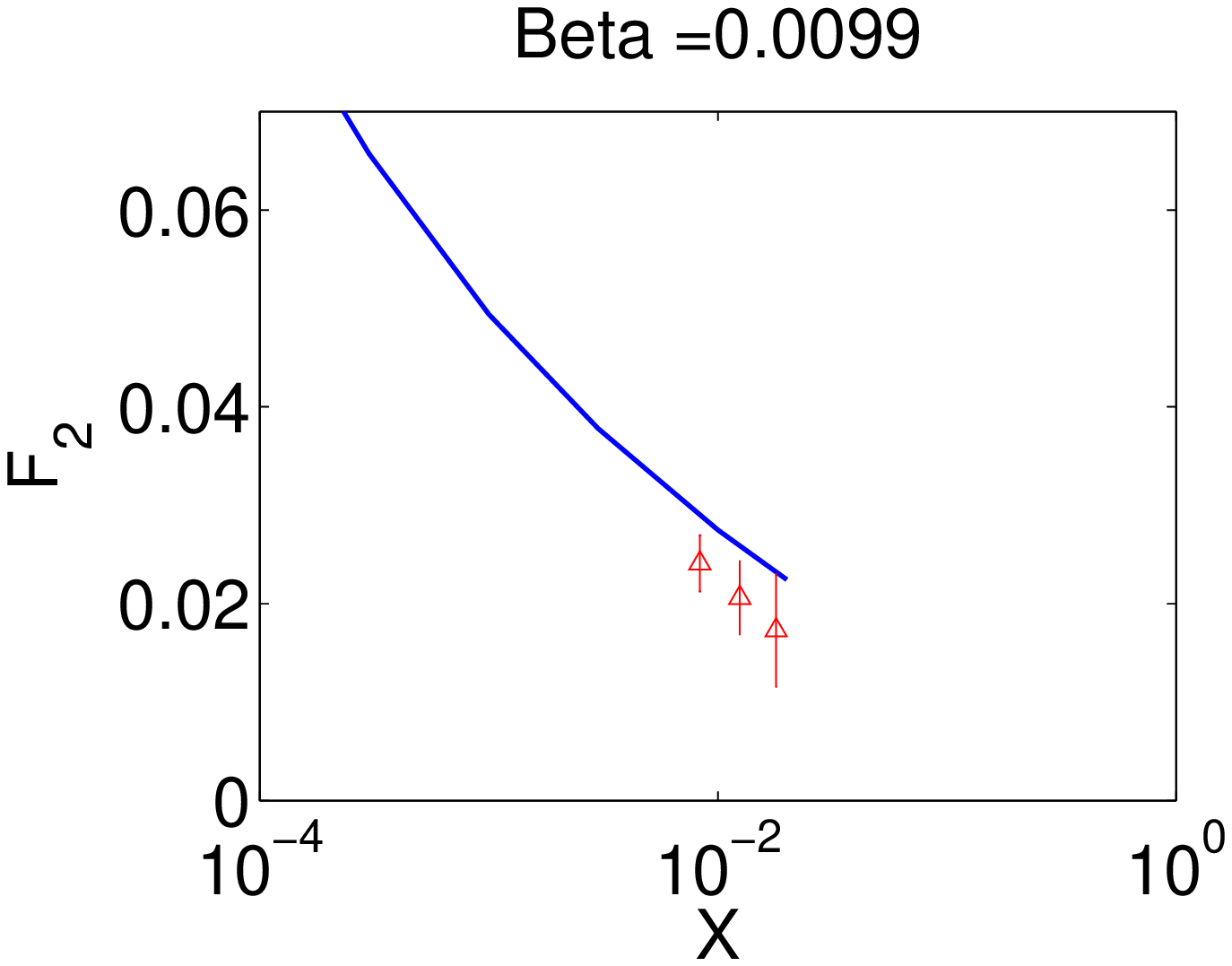,width=32mm, height=28mm}&
\epsfig{file=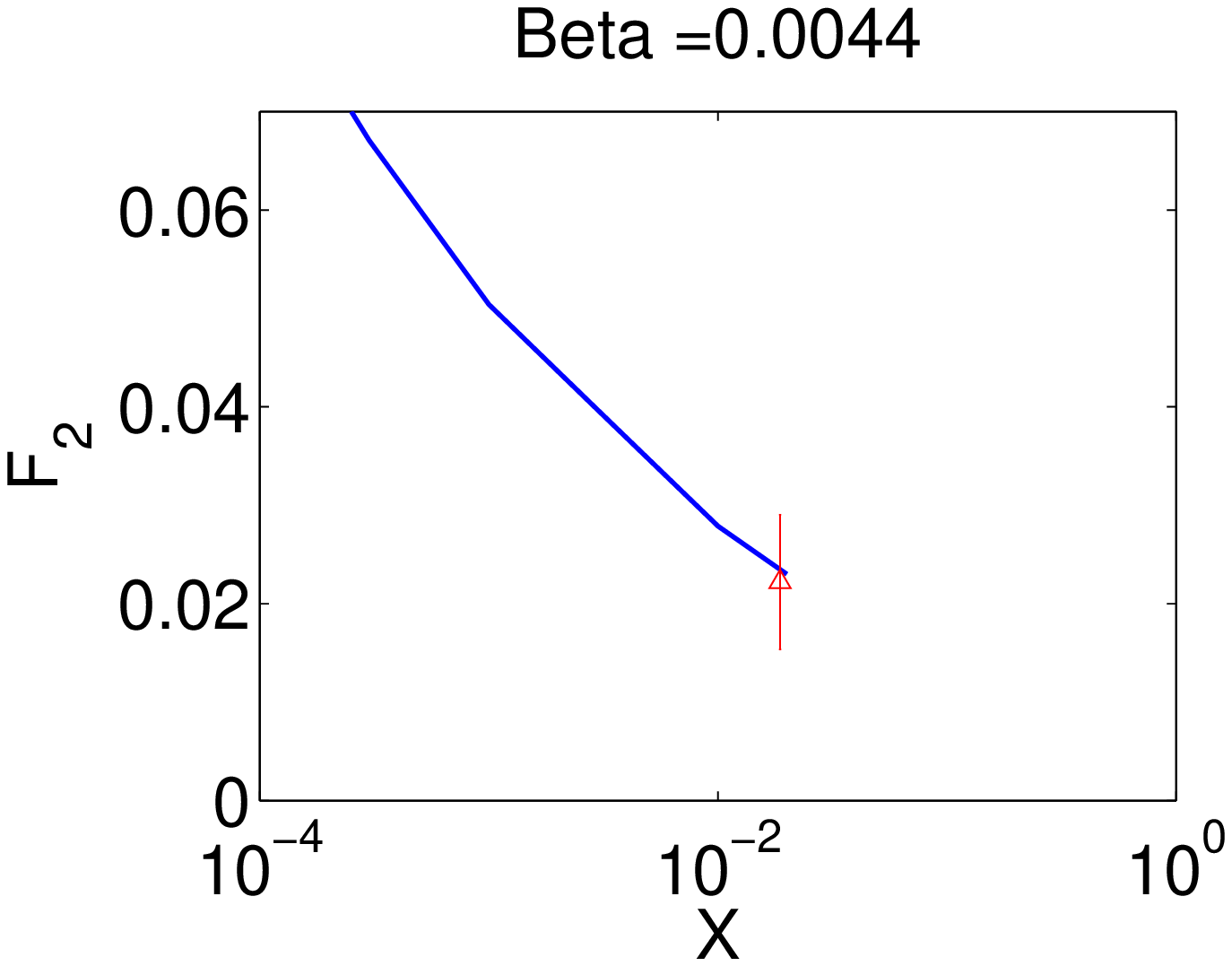,width=32mm, height=28mm}\\
\begin{sideways}{\small $6\,GeV^{2}$}\end{sideways}
\epsfig{file=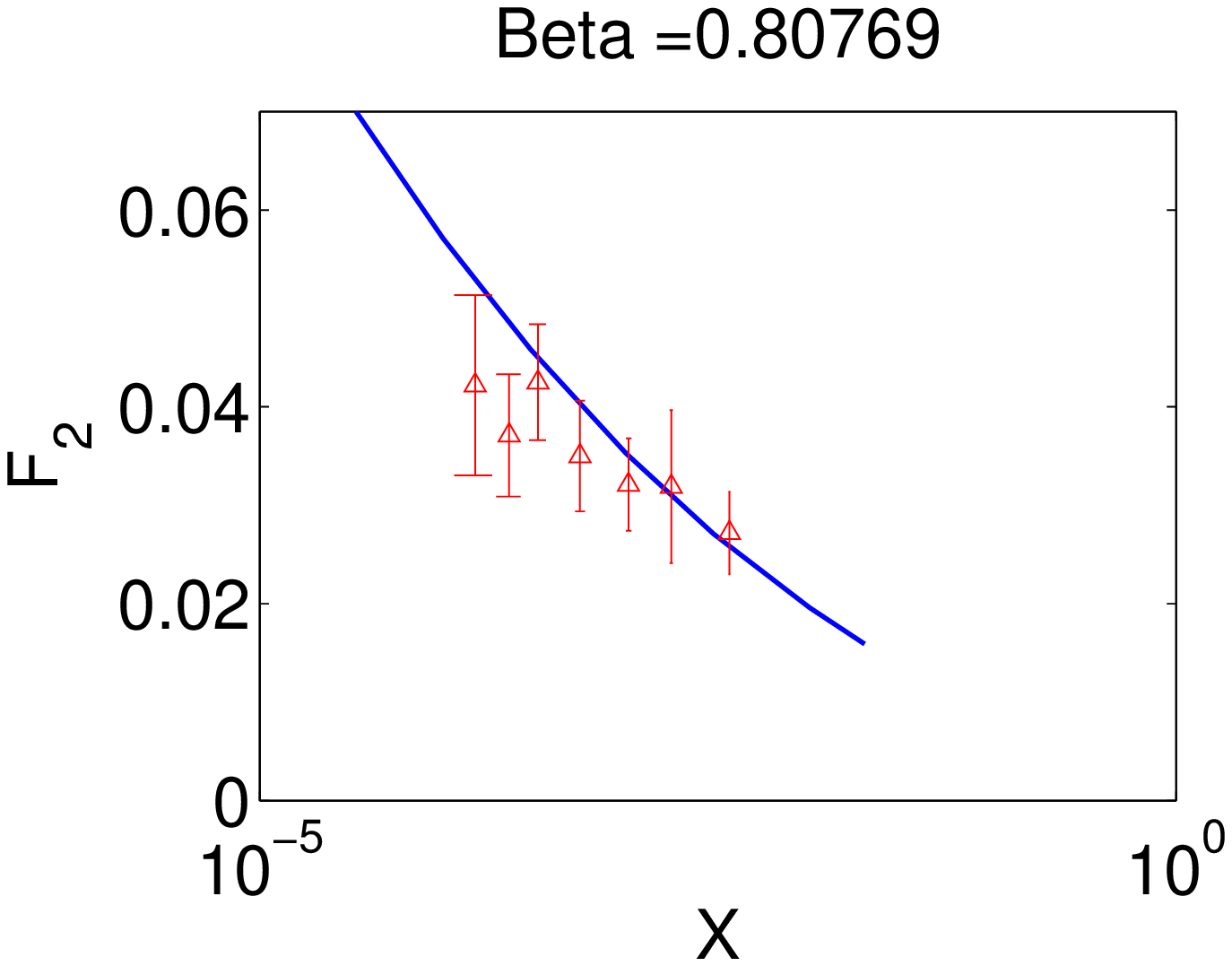,width=32mm, height=28mm}&
\epsfig{file=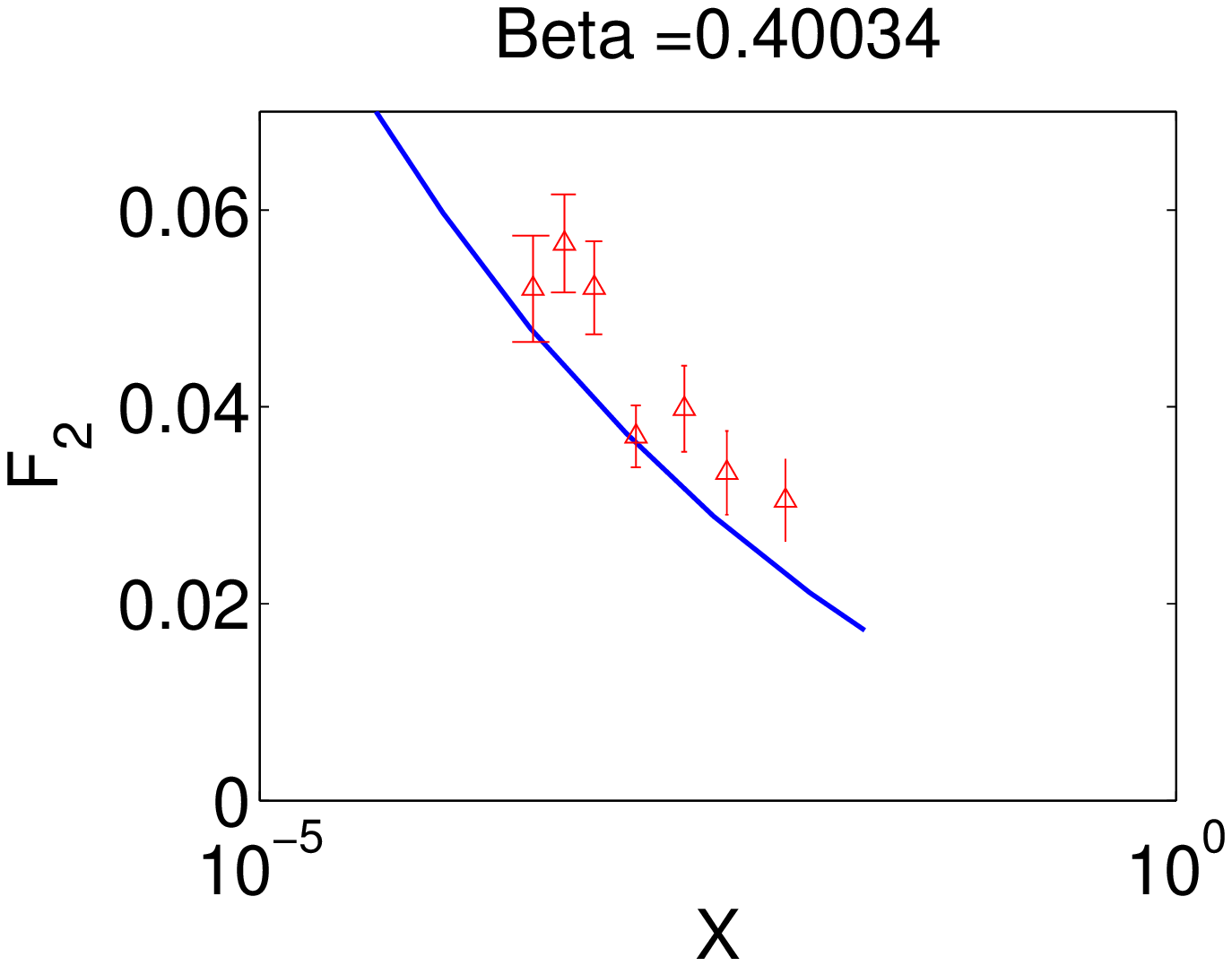,width=32mm, height=28mm}&
\epsfig{file=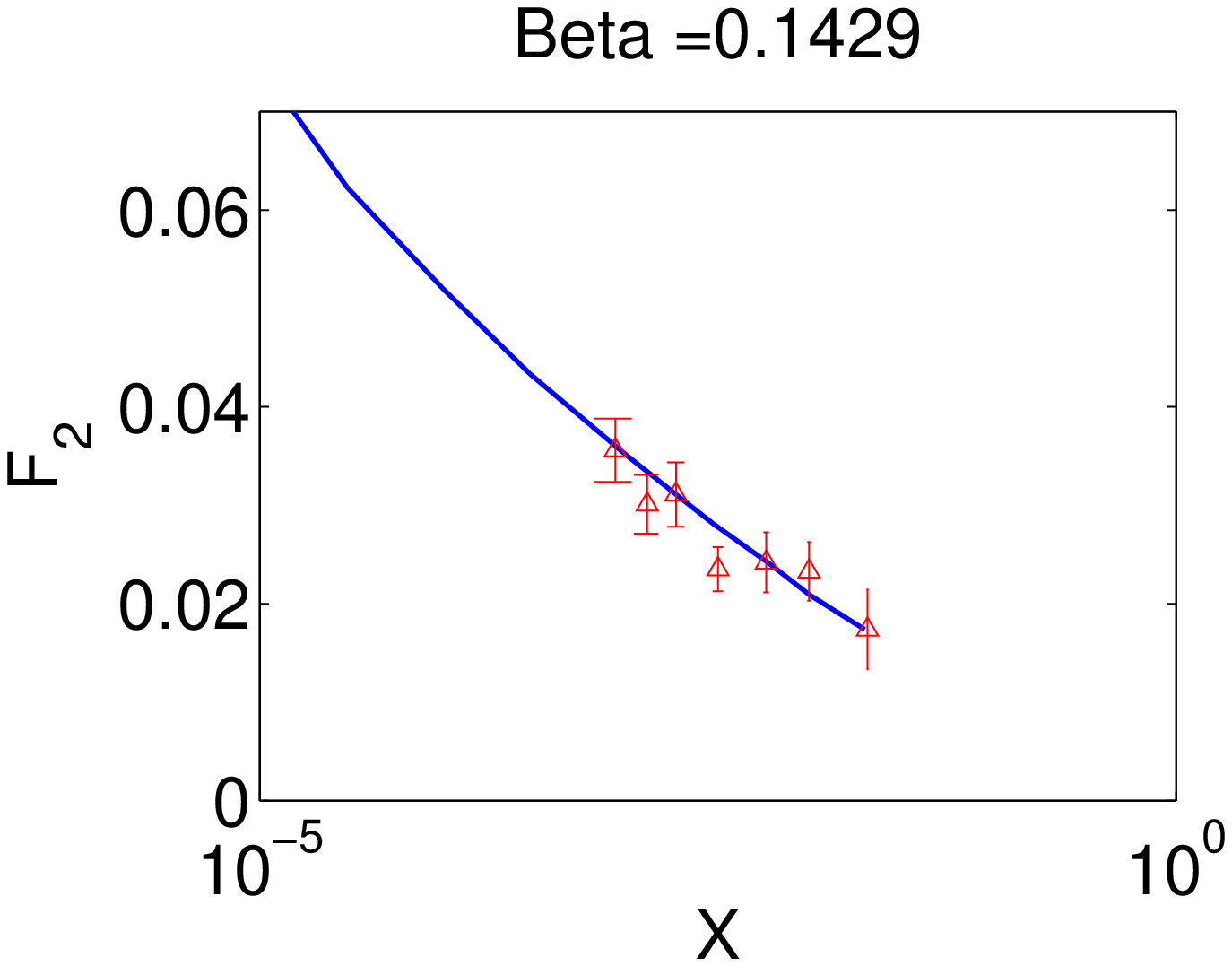,width=32mm, height=28mm}&
\epsfig{file=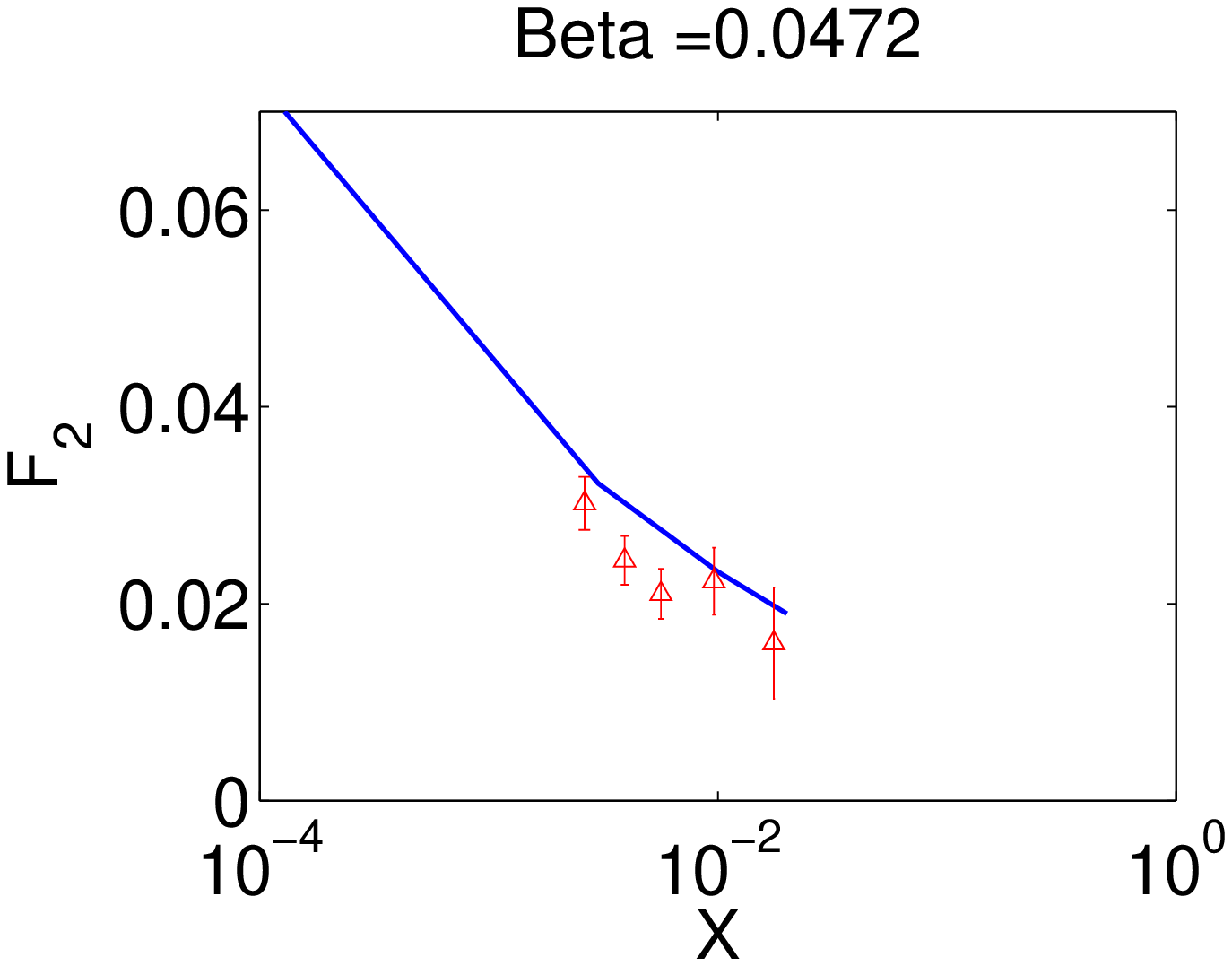,width=32mm, height=28mm}&
\epsfig{file=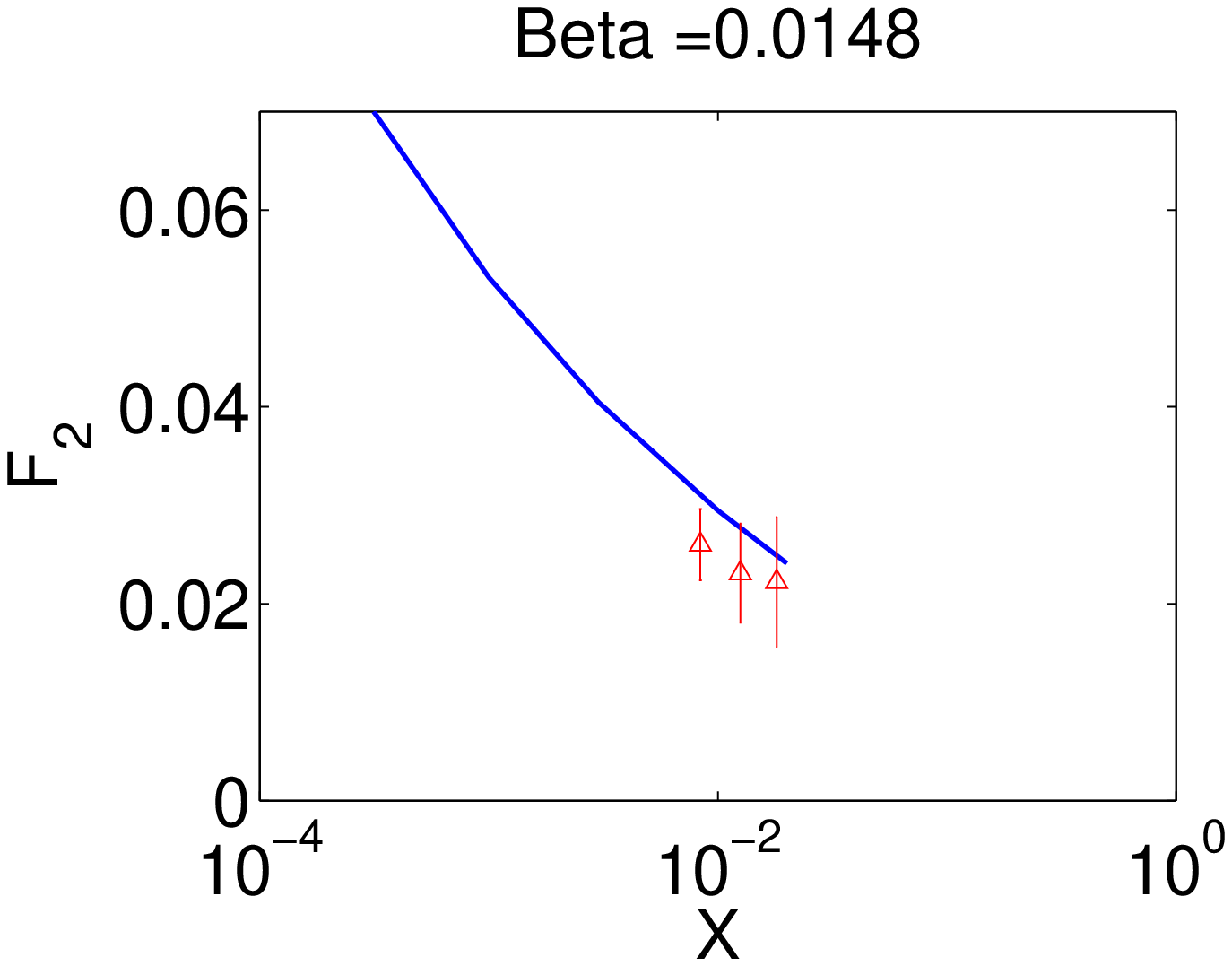,width=32mm, height=28mm}&
\epsfig{file=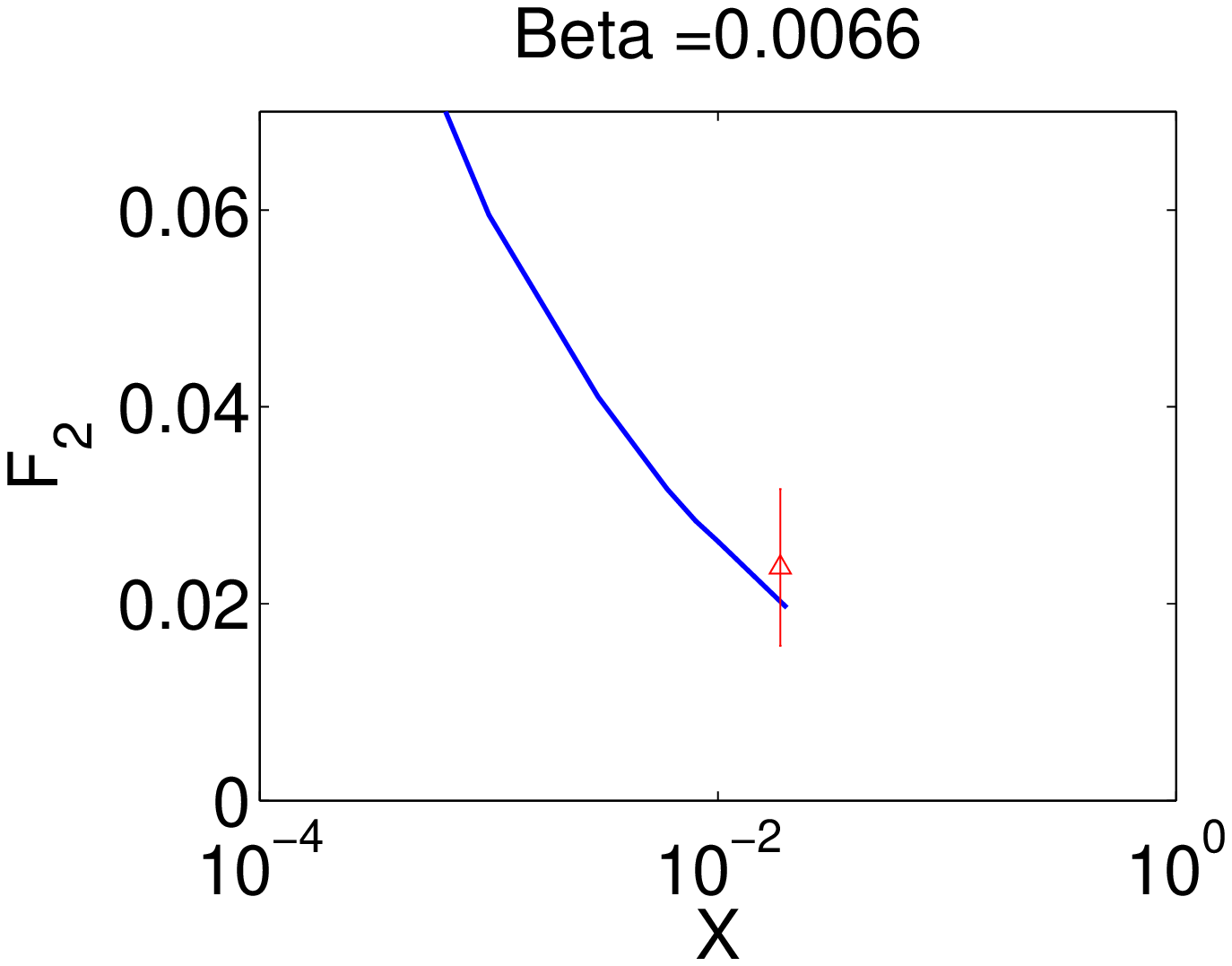,width=32mm, height=30mm}\\
\begin{sideways}{\small $8\,GeV^{2}$}\end{sideways}
\epsfig{file=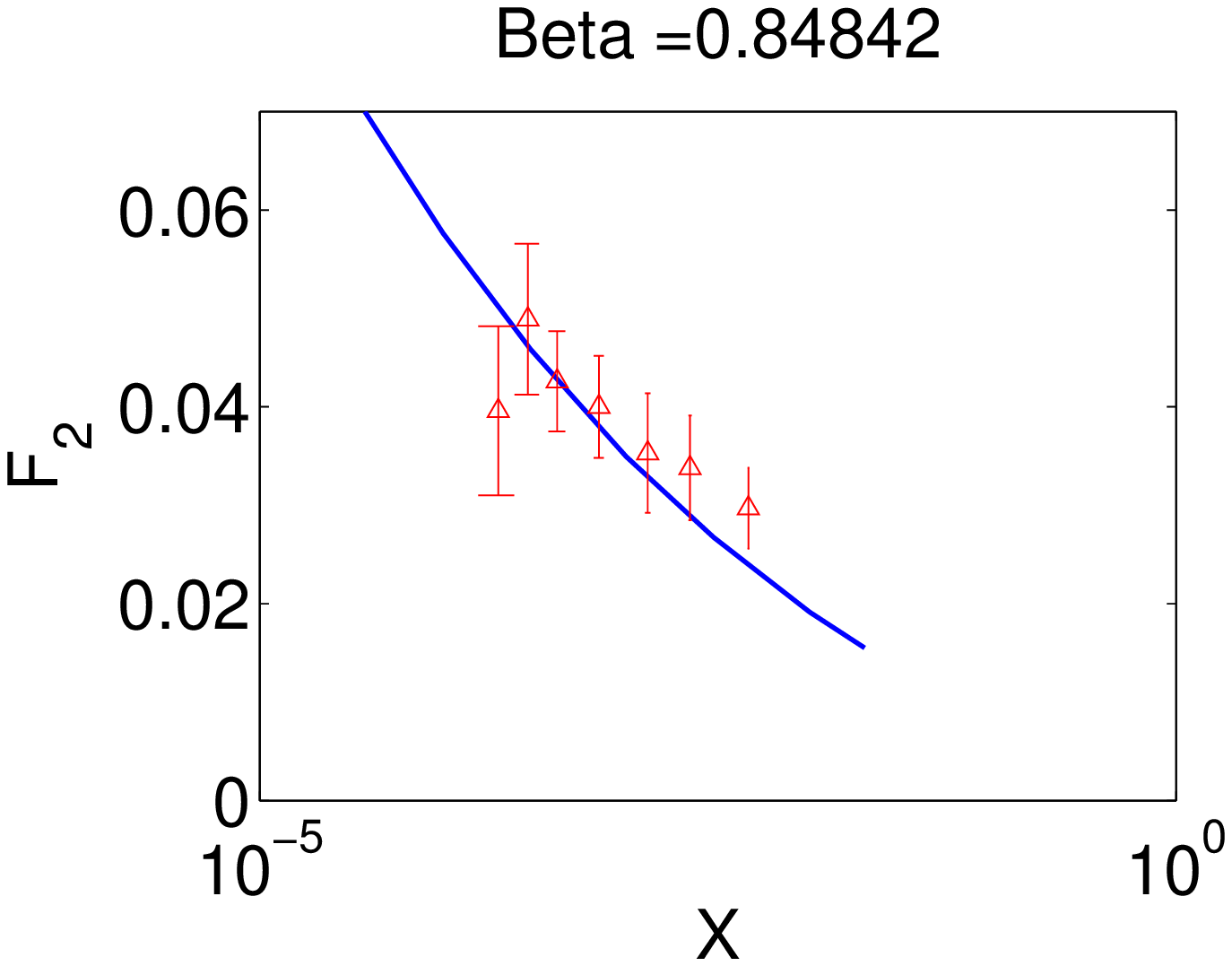,width=32mm, height=28mm}&
\epsfig{file=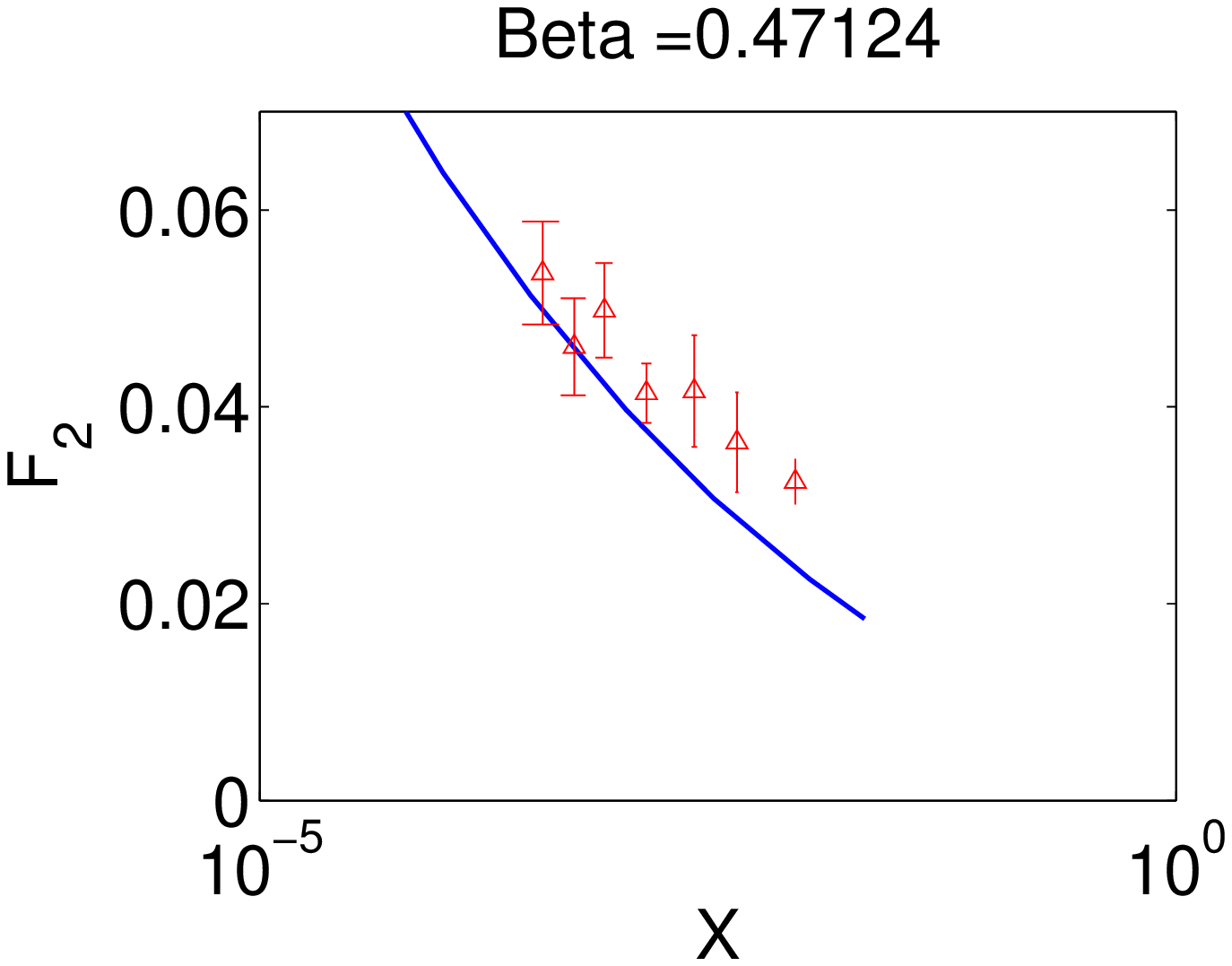,width=32mm, height=28mm}&
\epsfig{file=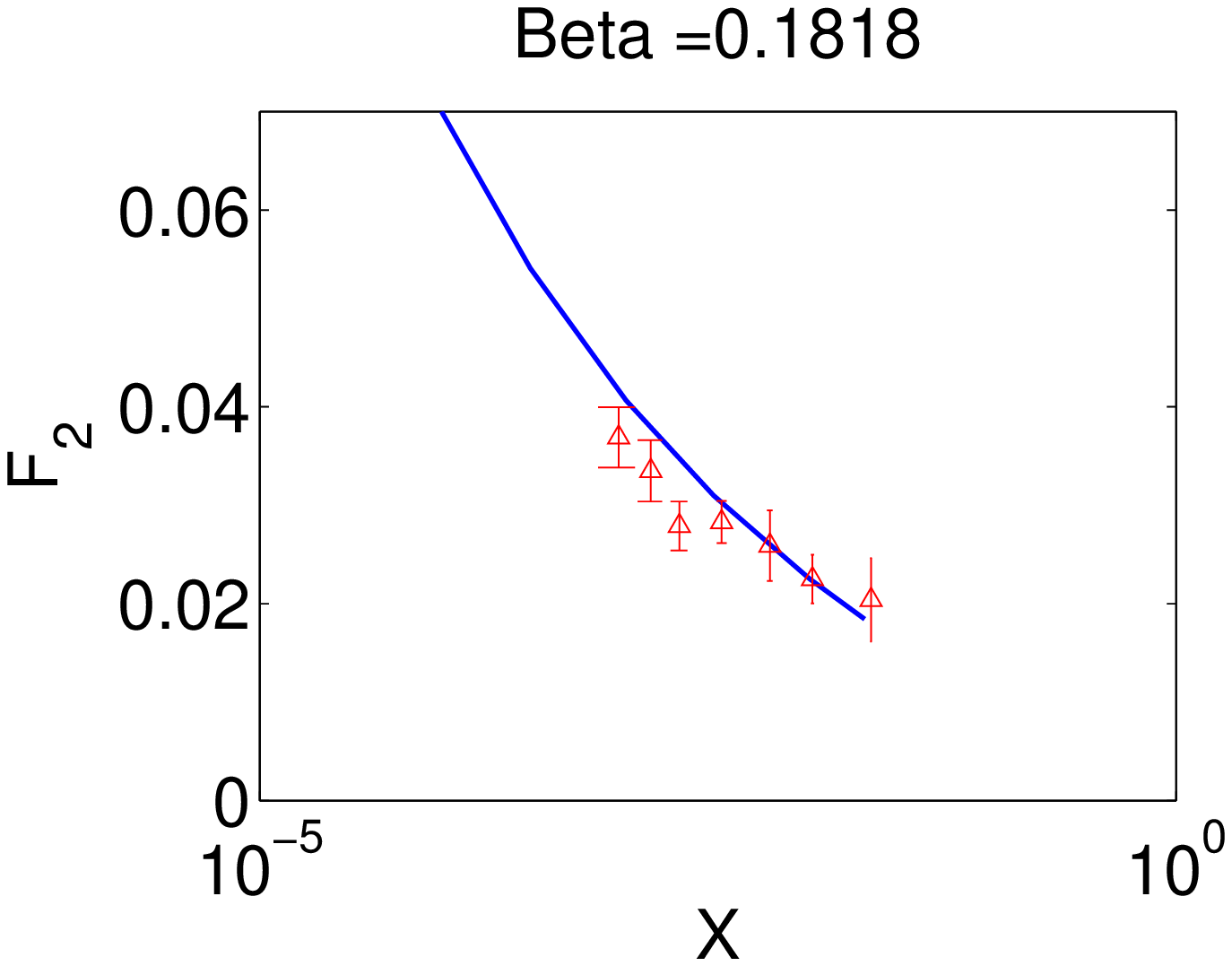,width=32mm, height=28mm}&
\epsfig{file=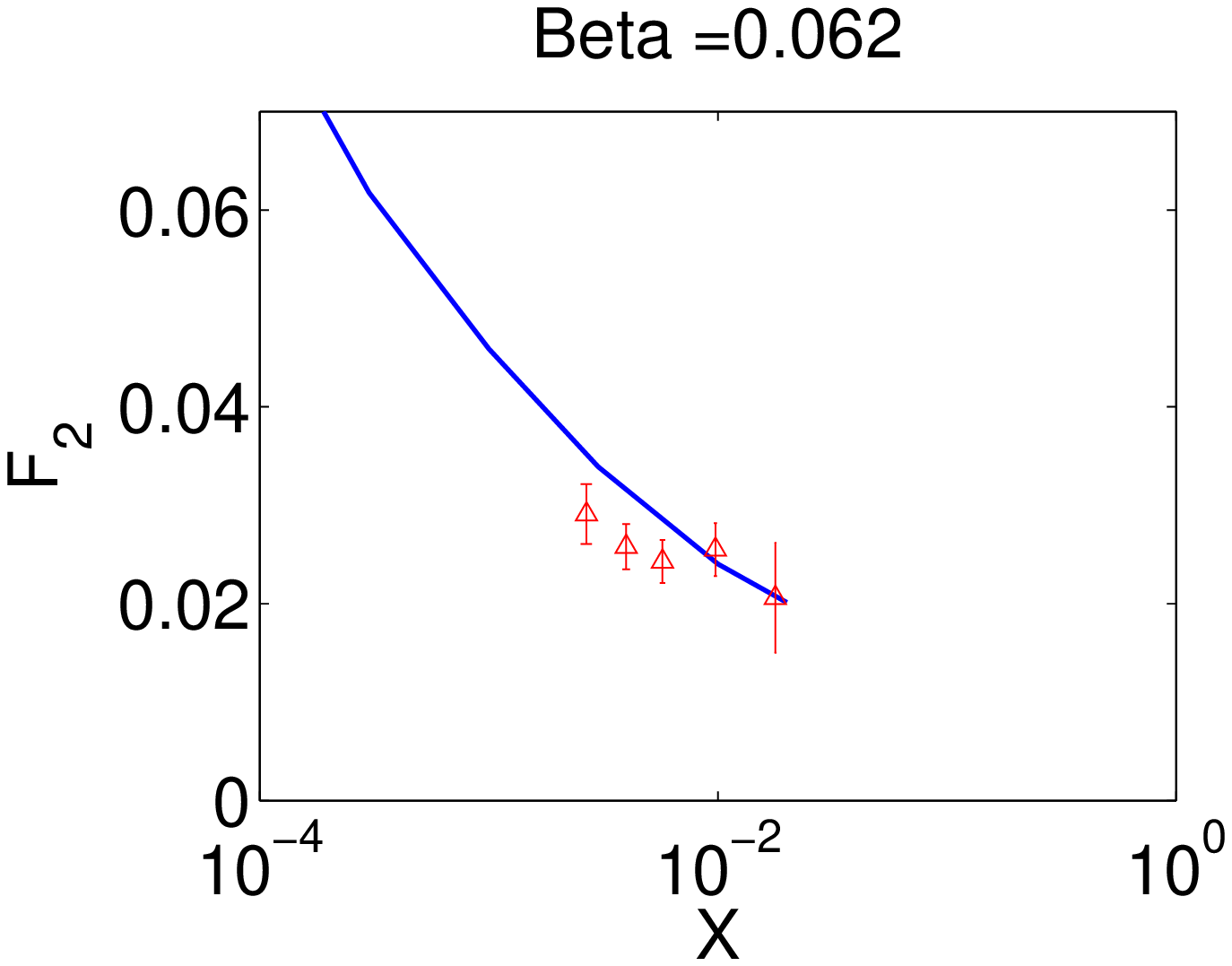,width=32mm, height=28mm}&
\epsfig{file=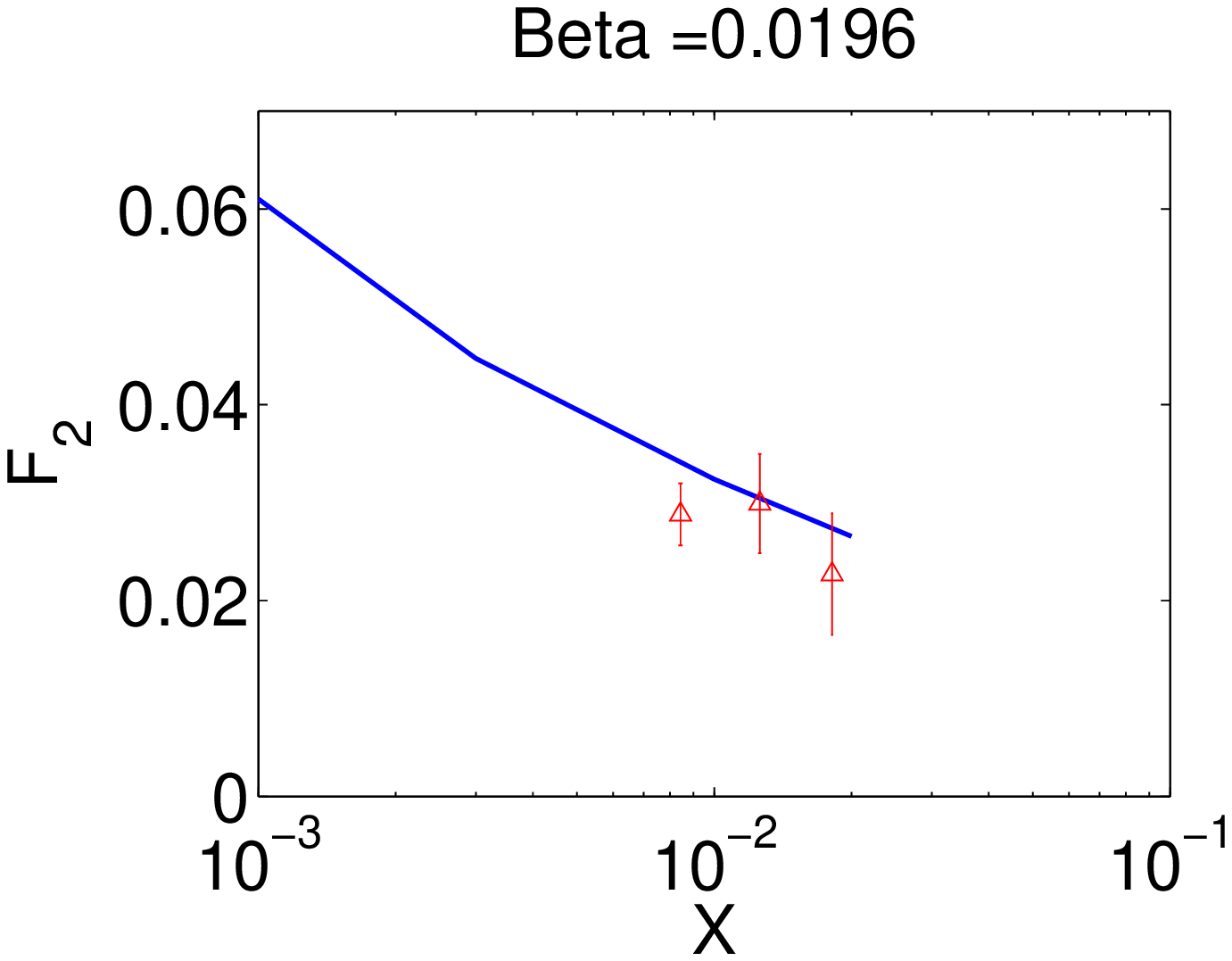,width=32mm, height=28mm}&
\epsfig{file=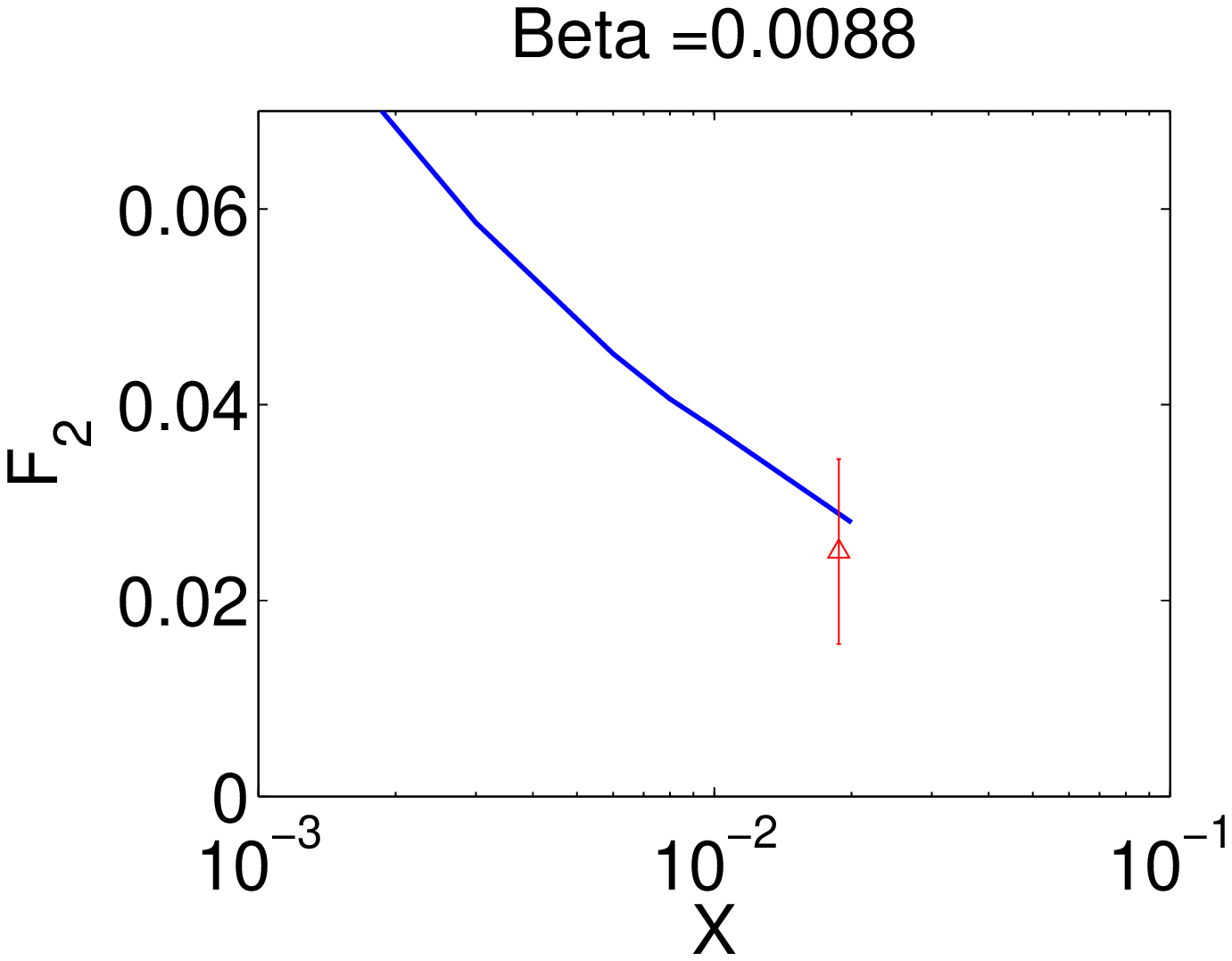,width=32mm, height=28mm}\\
\end{tabular}
\caption{\it Diffractive dissociation structure function
$F_{2}^{D(3)}(\beta,x_{\Pom},Q^{2})$ as a function of $x_{\Pom}$.
Comparison of the prediction of the model with ZEUS data
\cite{Chekanov:2005vv}.}\label{diff_1}
\end{sidewaysfigure}

\begin{sidewaysfigure}[htbp]
\centering
\begin{tabular}{cccccc ccccc ccccc ccc}
$M_{X} = 1.2\,GeV$ & $M_{X} = 3\,GeV$ & $M_{X} = 6\,GeV$ & $M_{X} = 11\,GeV$ & $M_{X} = 20\,GeV$ & $M_{X} = 30\,GeV$\\
\begin{sideways}{\small $14\,GeV^{2}$}\end{sideways}
\epsfig{file=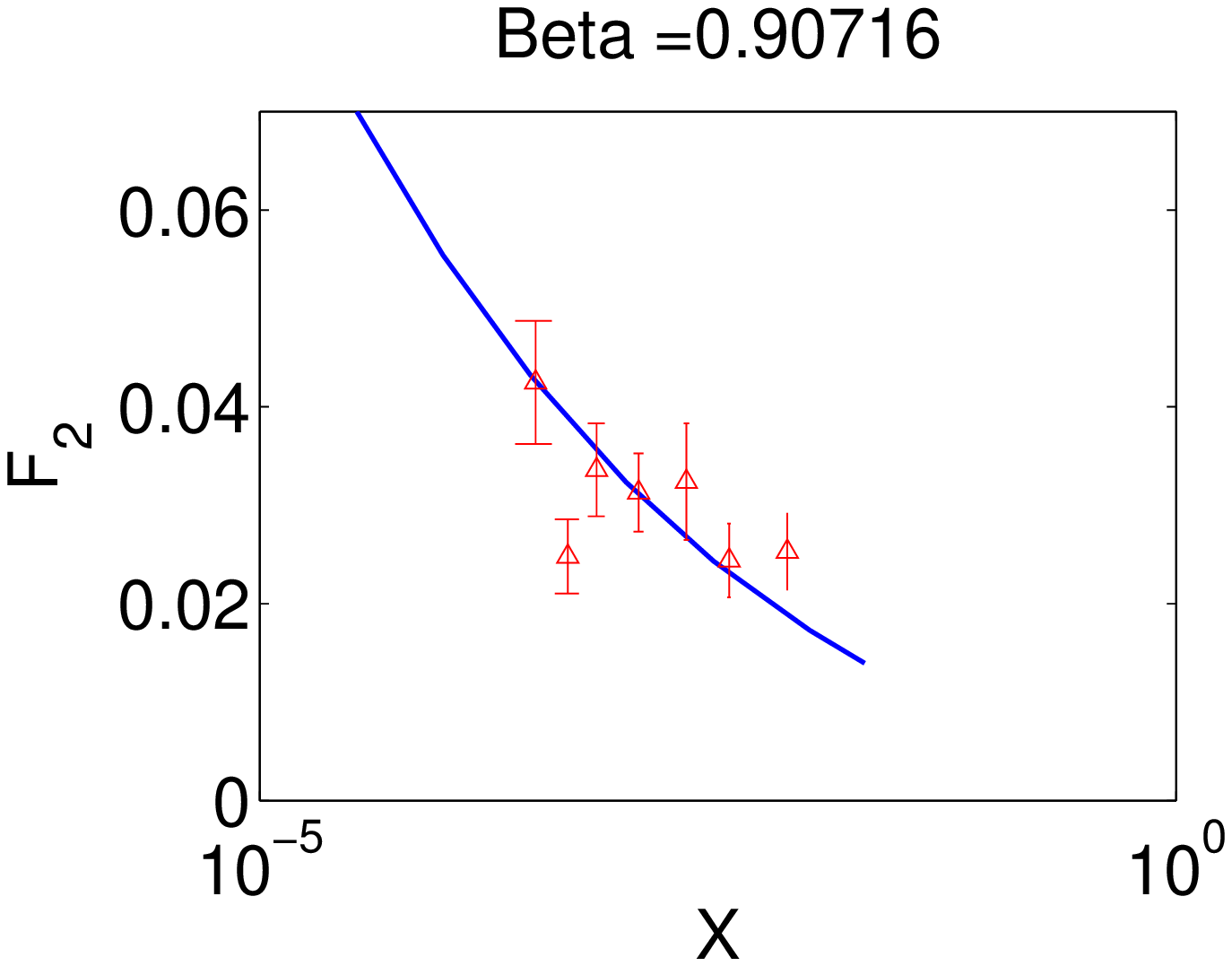,width=32mm, height=28mm}&
\epsfig{file=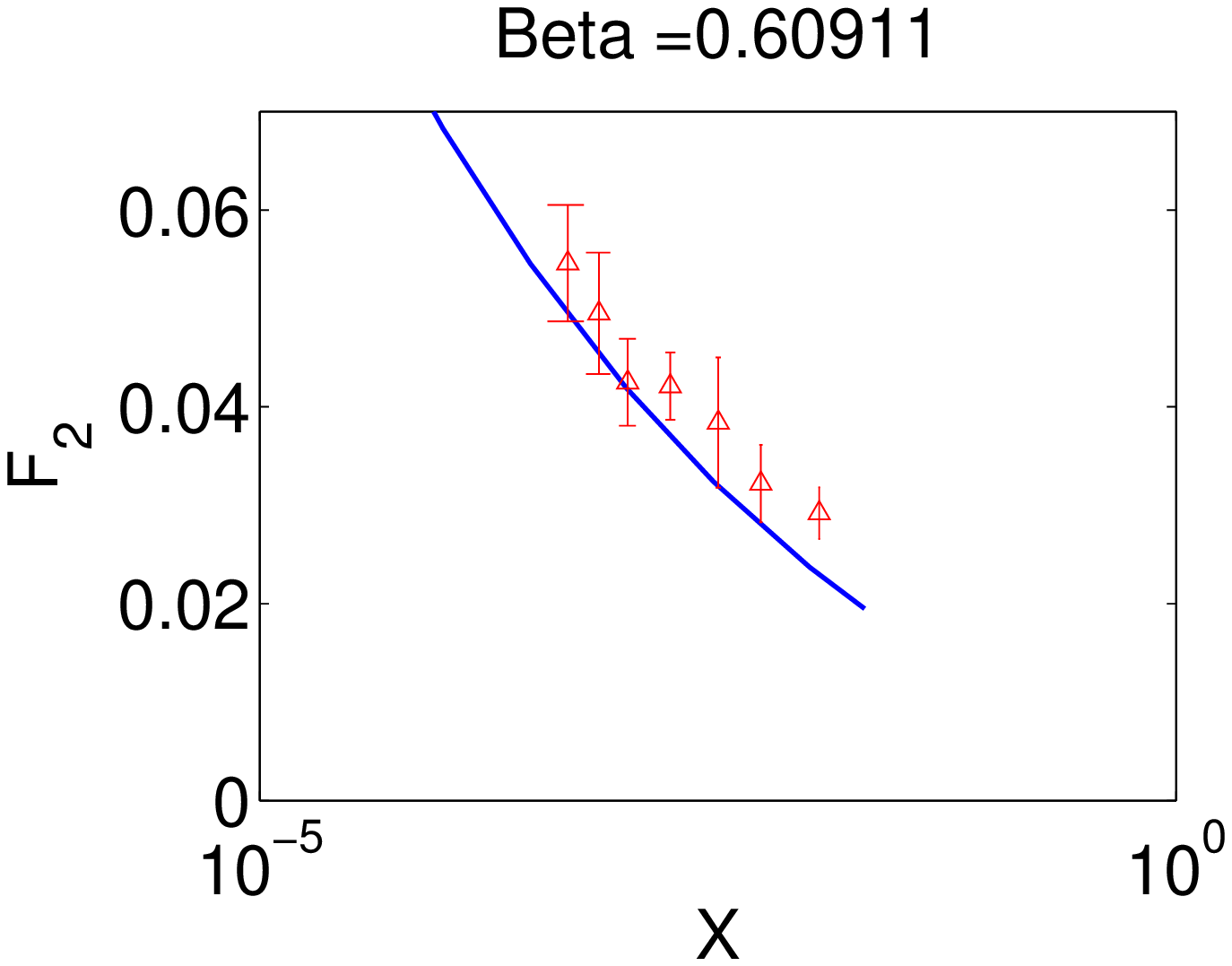,width=32mm, height=28mm}&
\epsfig{file=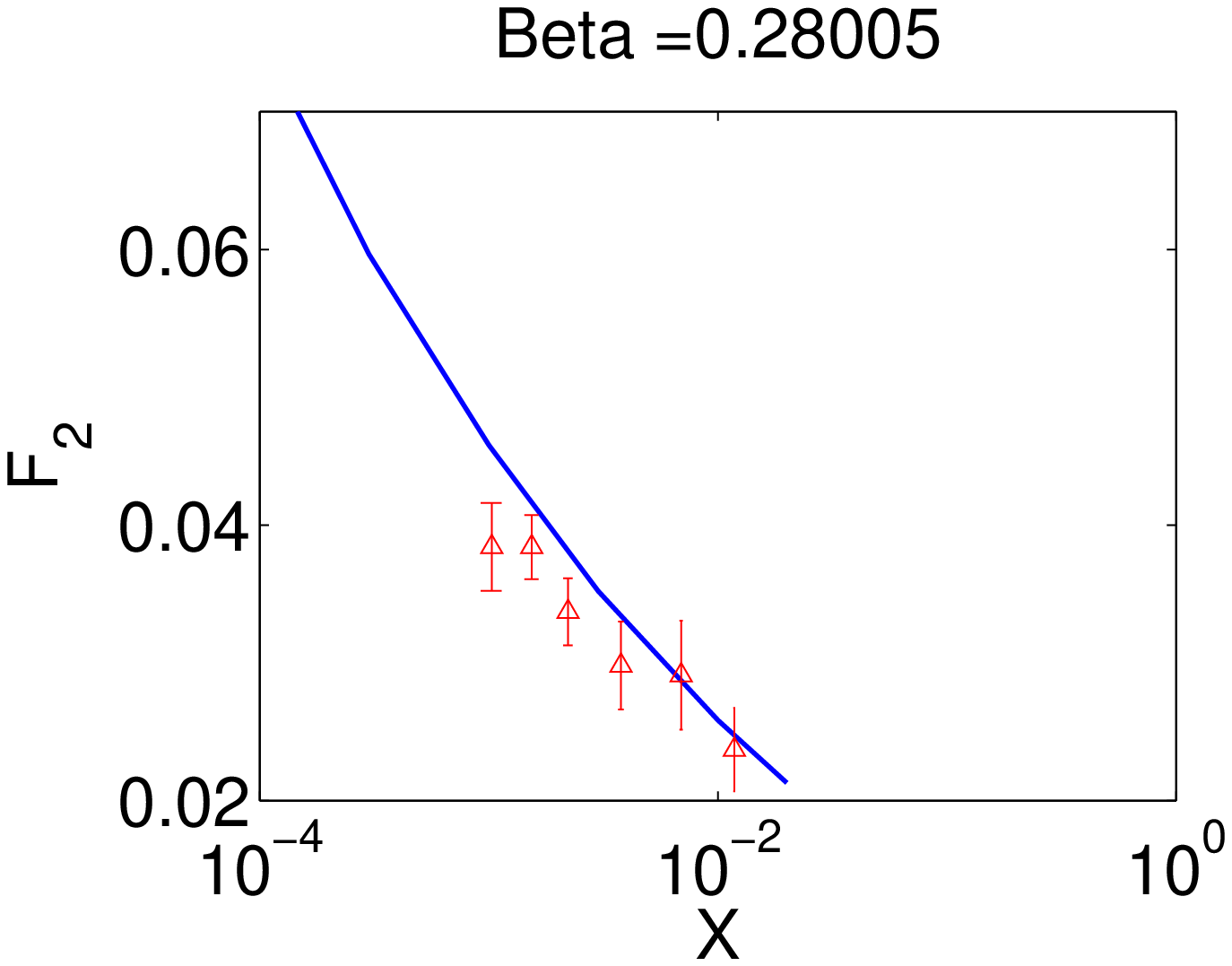,width=32mm, height=28mm}&
\epsfig{file=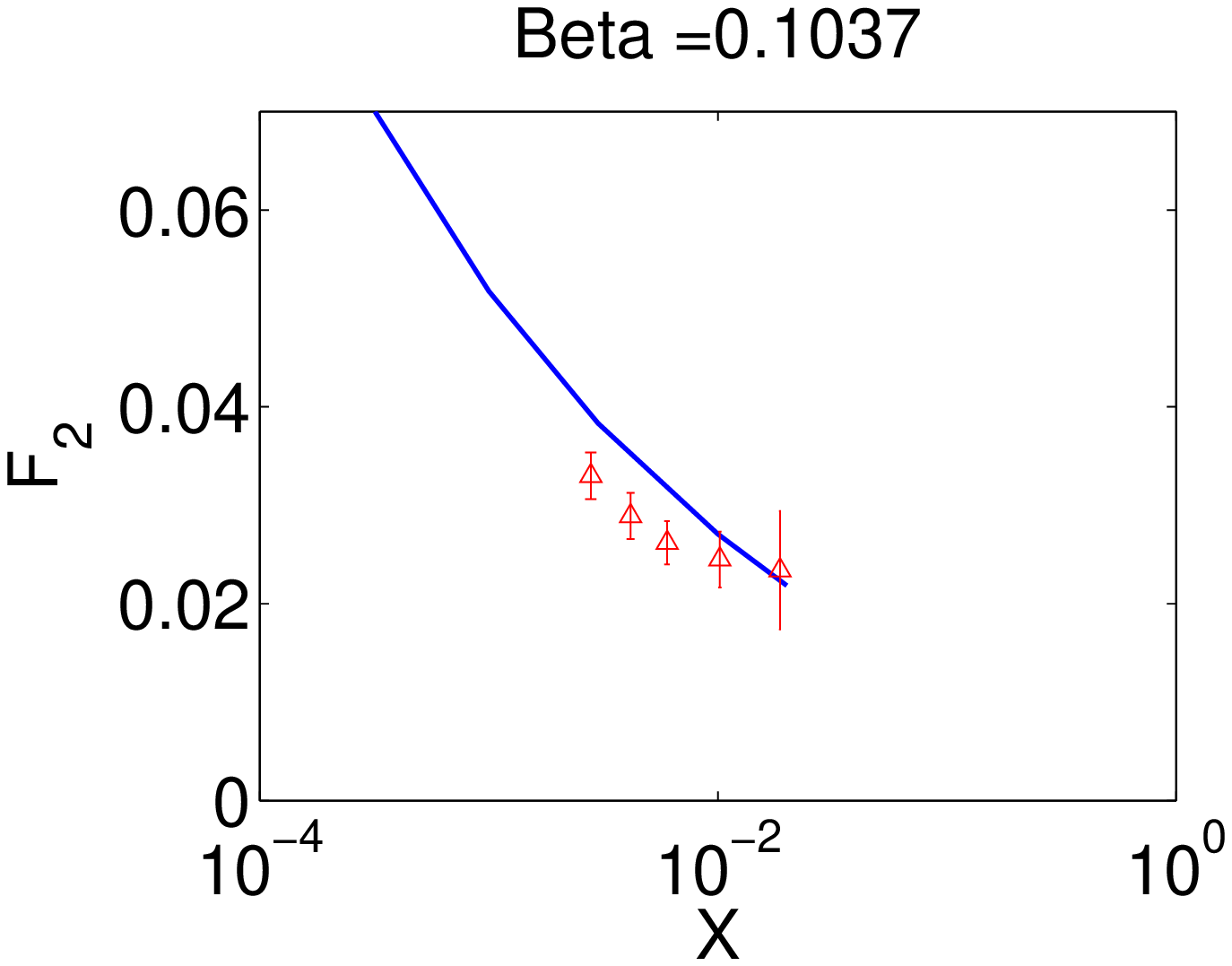,width=32mm, height=28mm}&
\epsfig{file=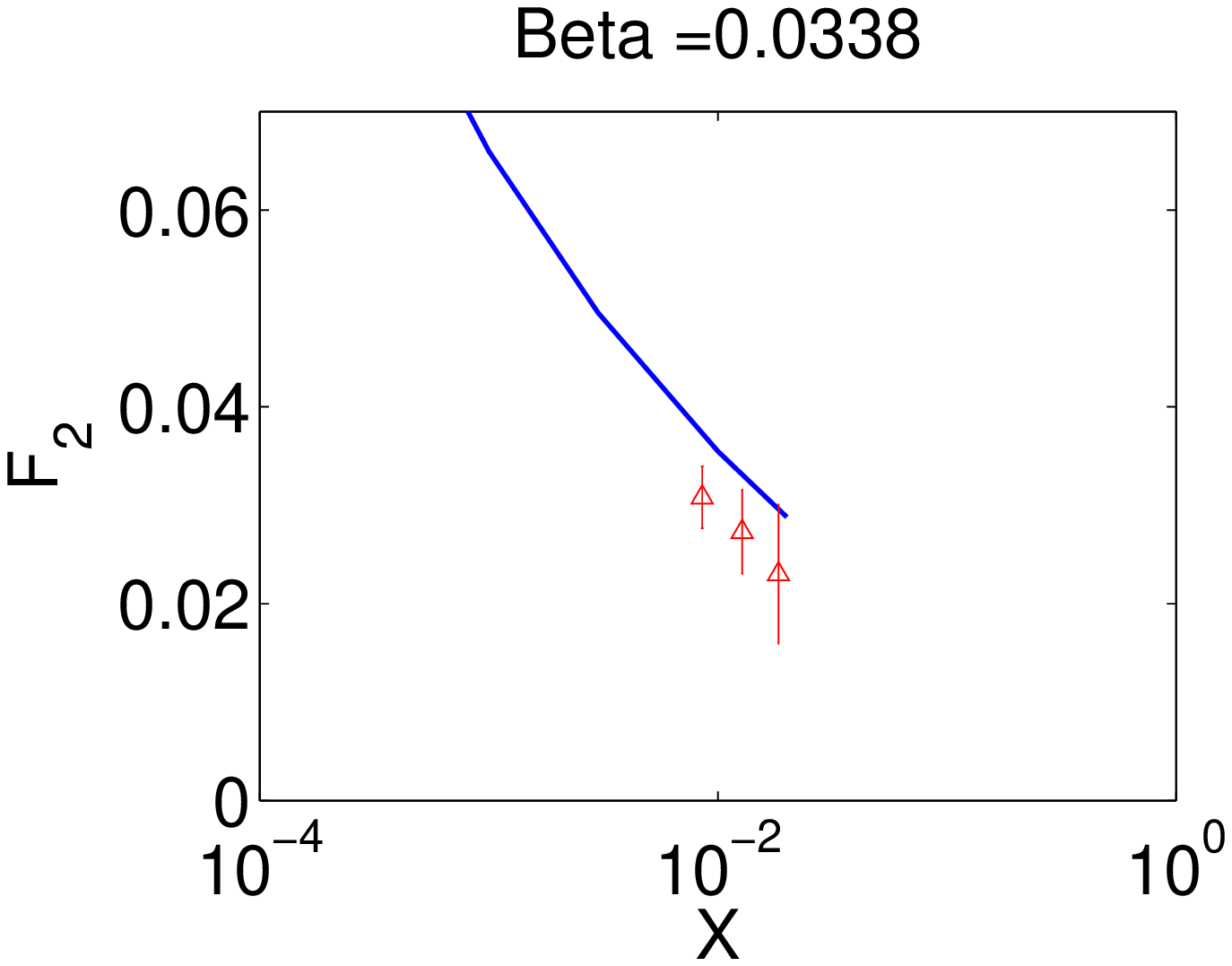,width=32mm, height=28mm}&
\epsfig{file=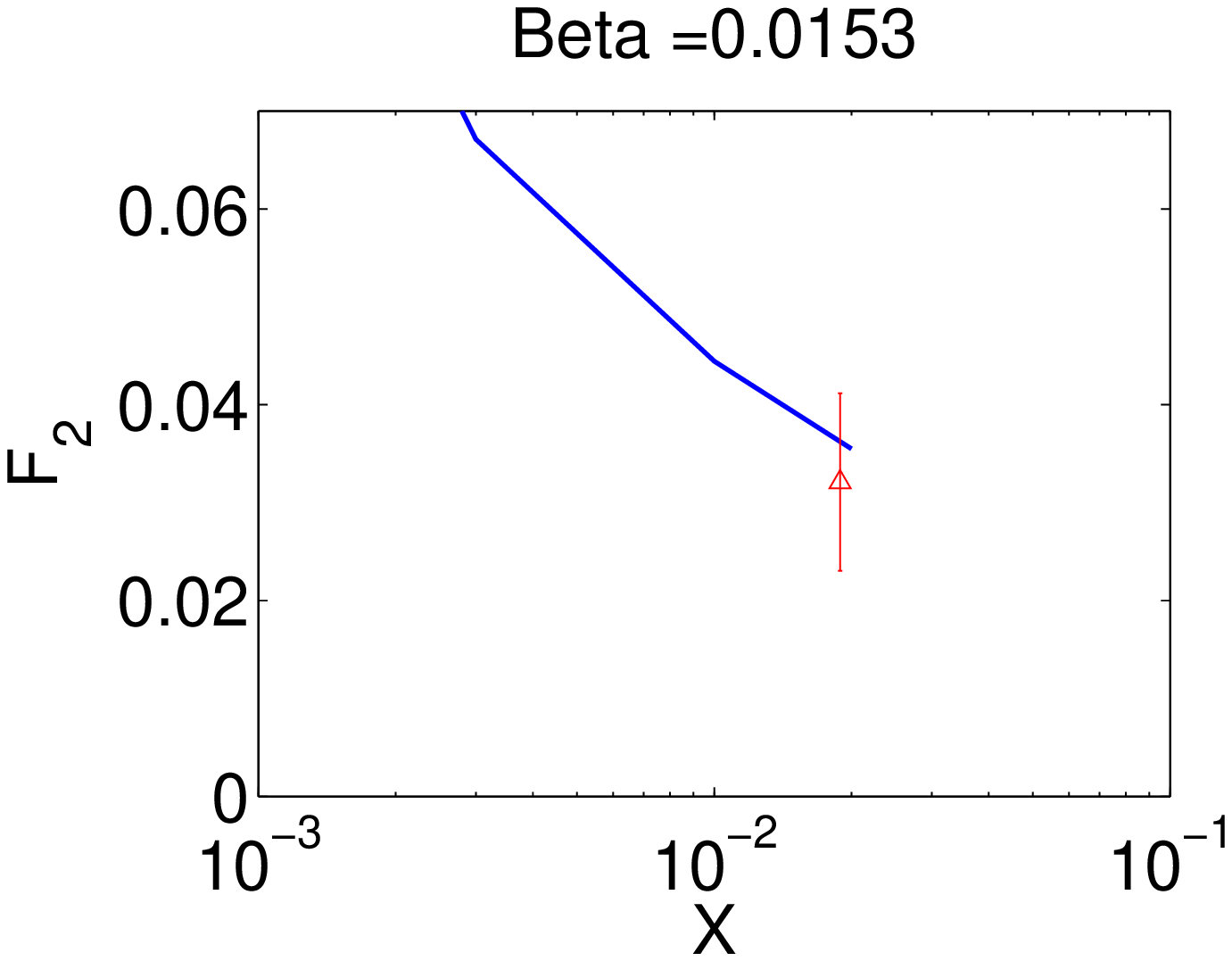,width=32mm, height=28mm}\\
\begin{sideways}{\small $27\,GeV^{2}$}\end{sideways}
\epsfig{file=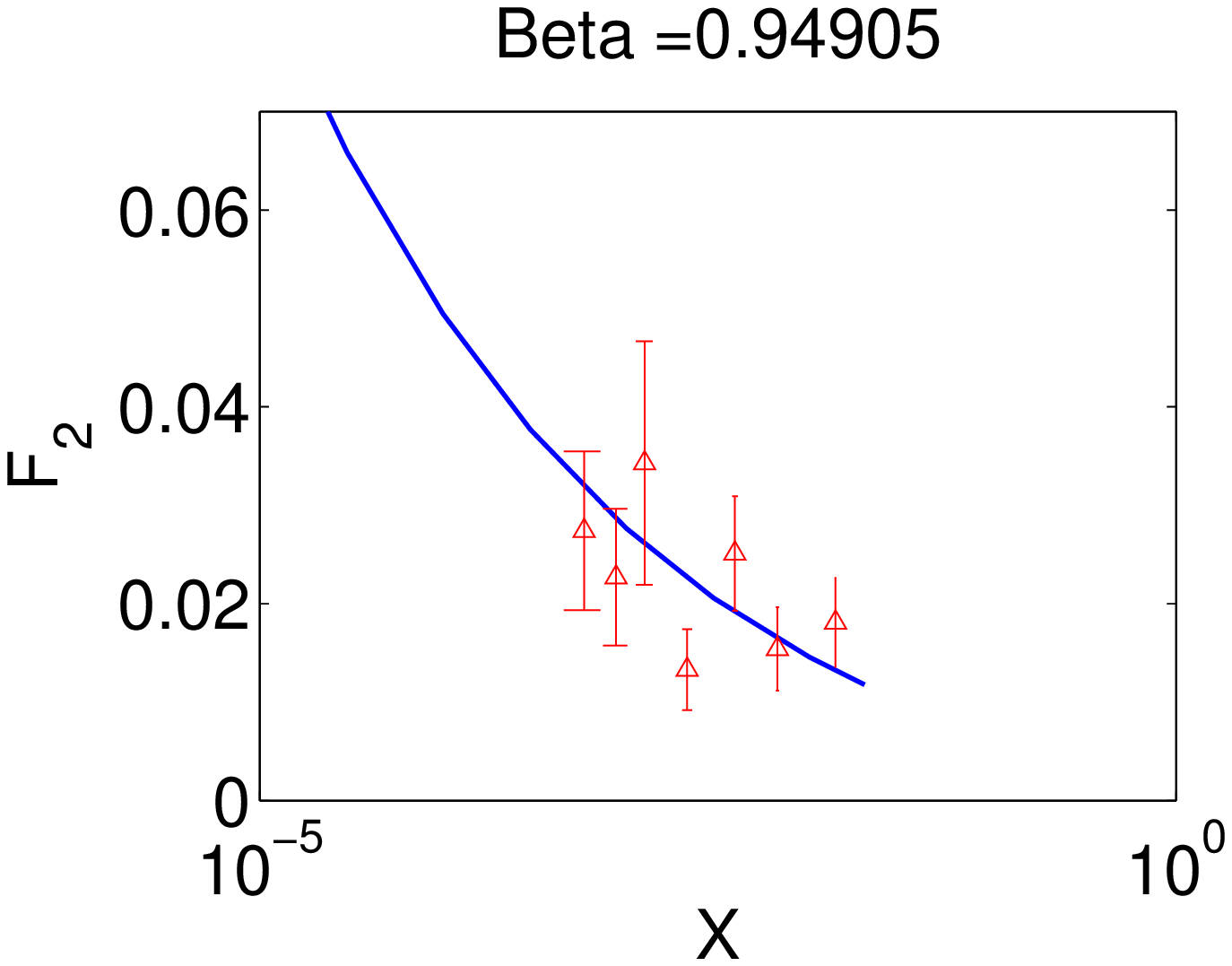,width=32mm, height=28mm}&
\epsfig{file=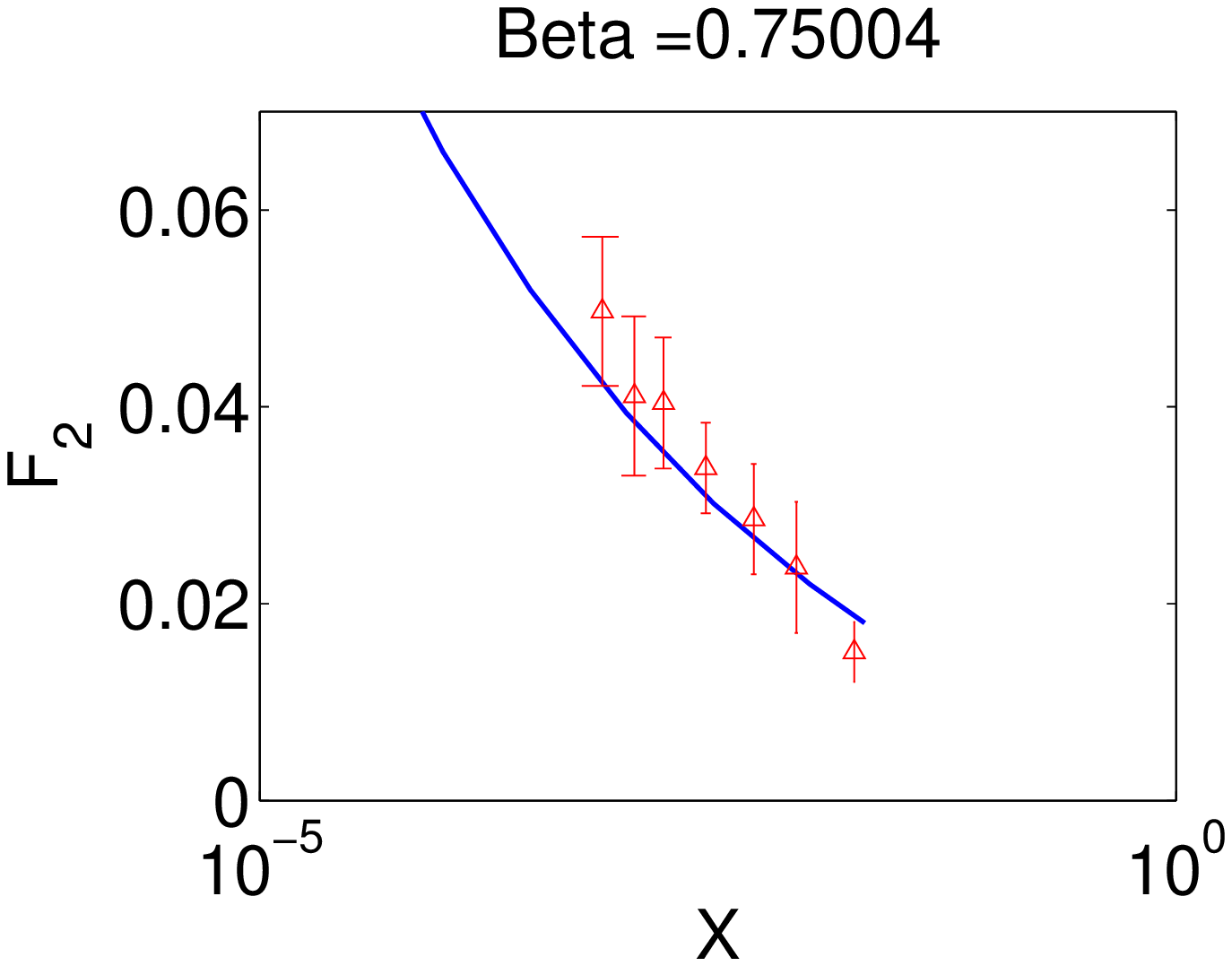,width=32mm, height=28mm}&
\epsfig{file=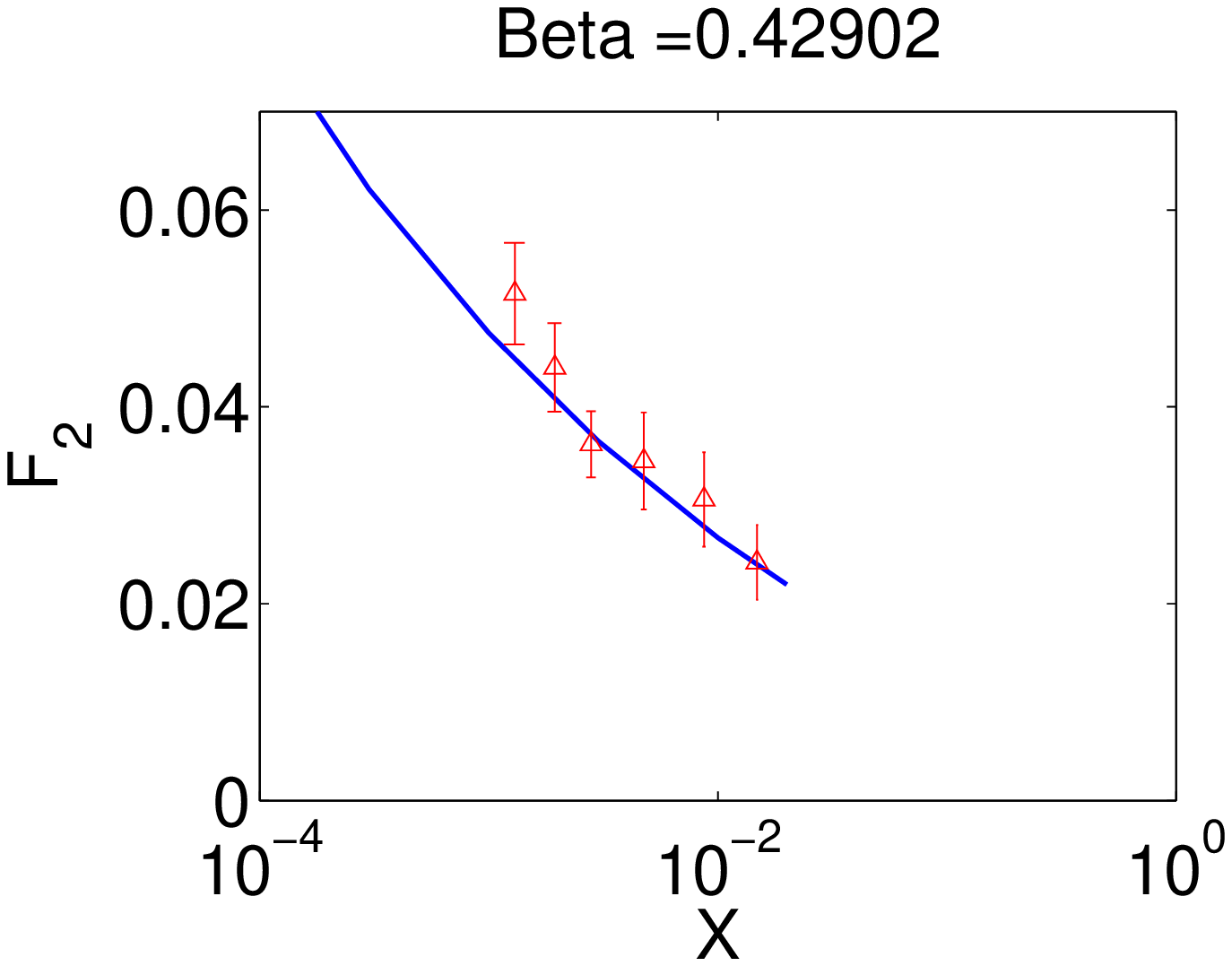,width=32mm, height=28mm}&
\epsfig{file=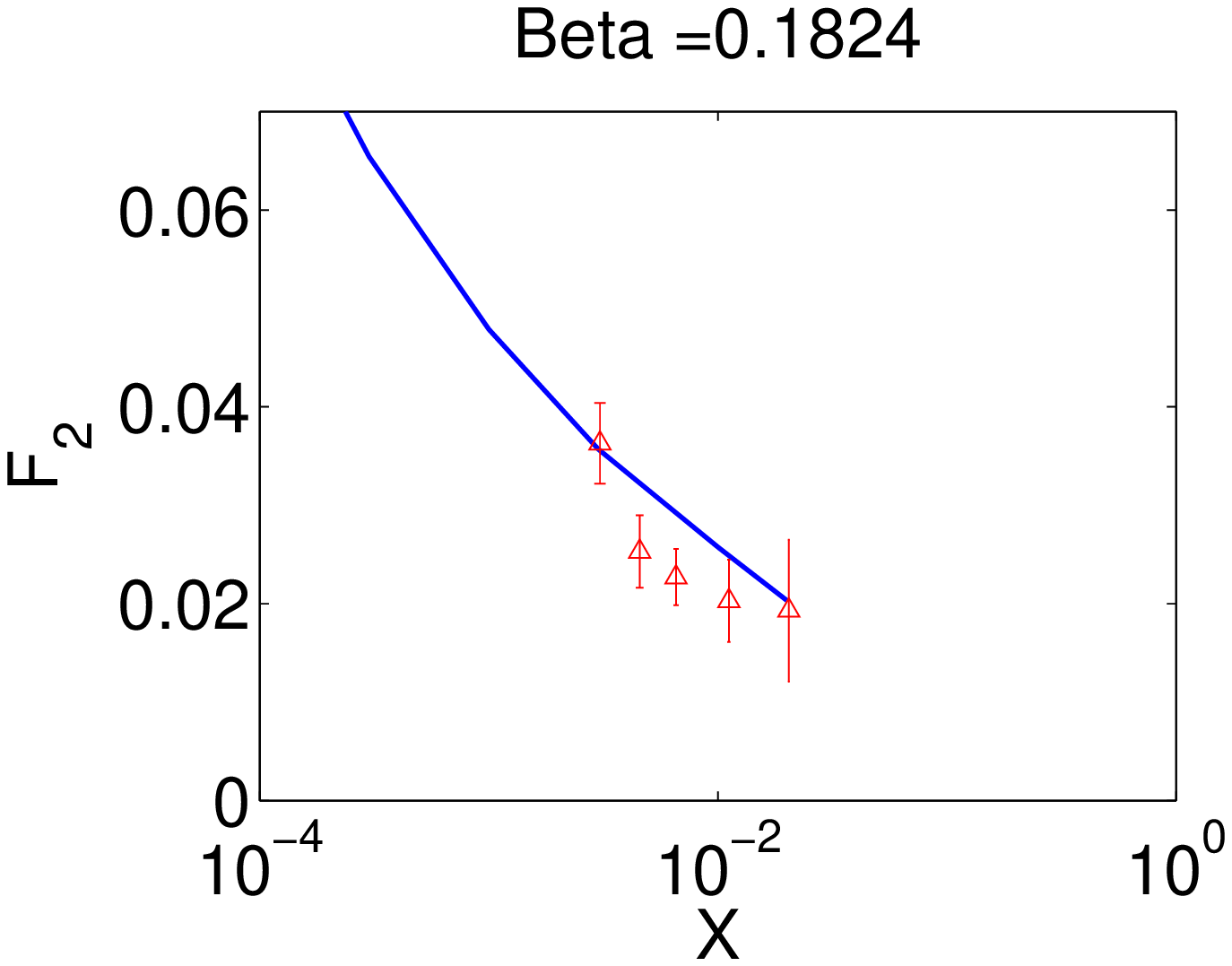,width=32mm, height=28mm}&
\epsfig{file=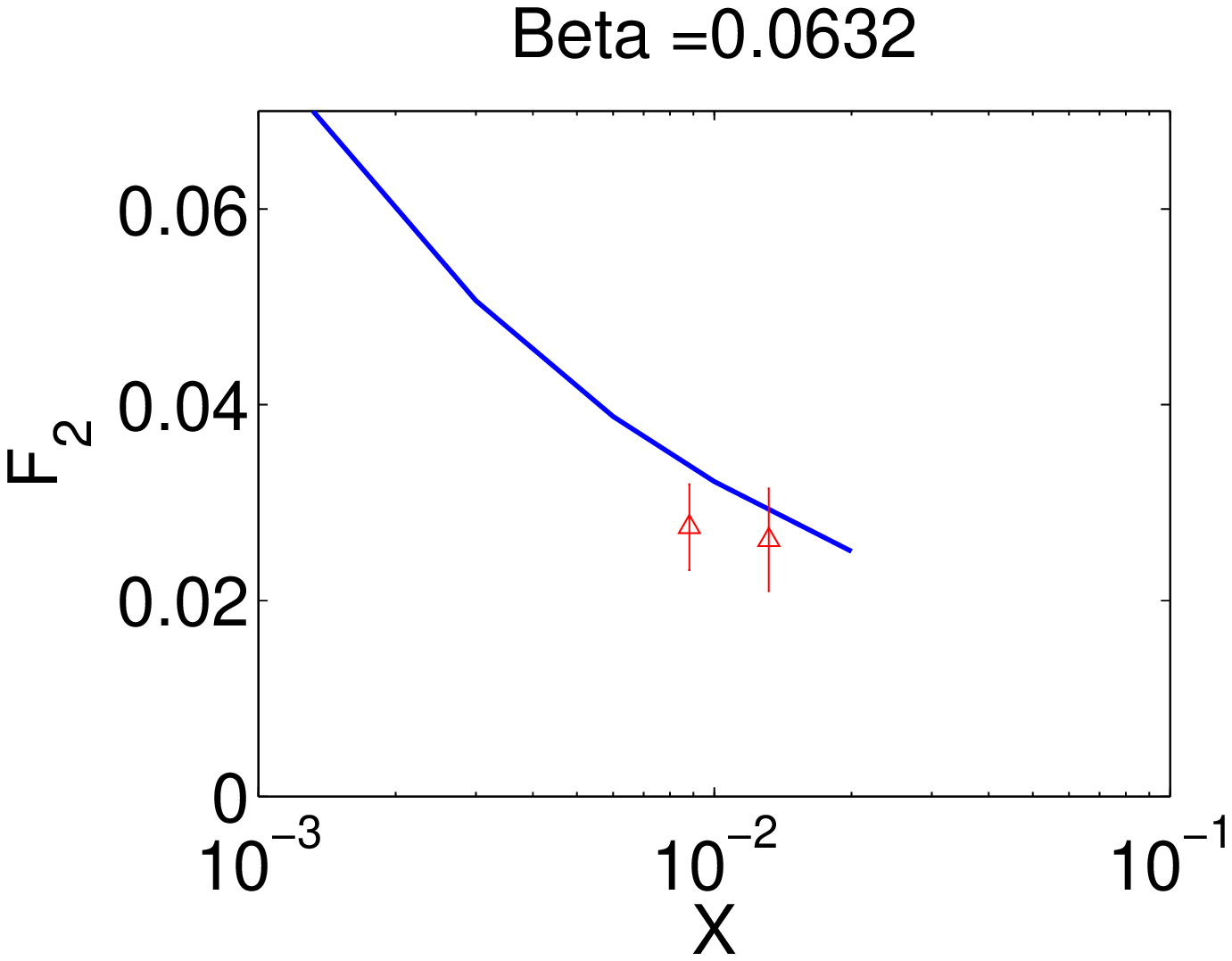,width=32mm, height=28mm}&
\epsfig{file=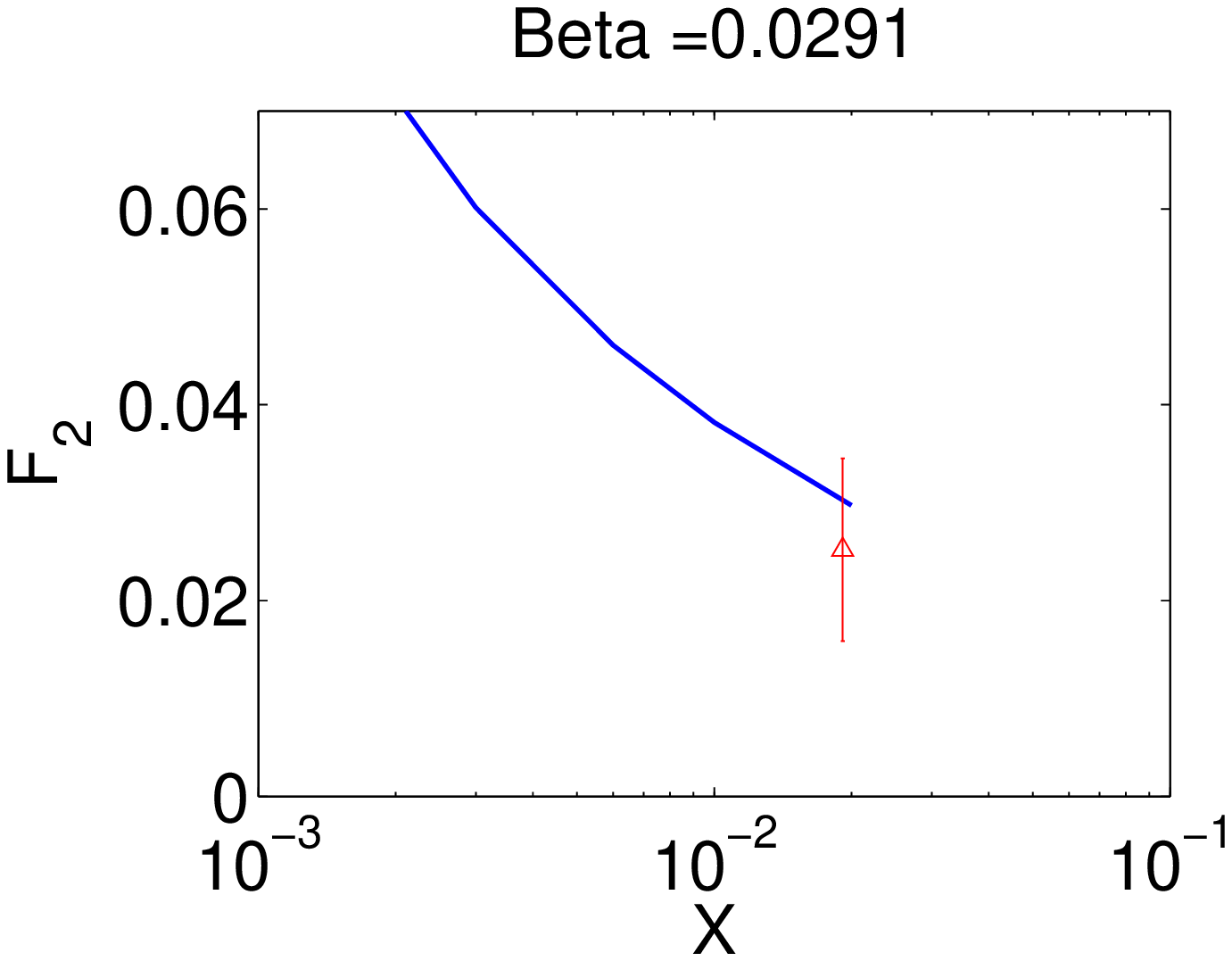,width=32mm, height=28mm}&\\
\begin{sideways}{\small $55\,GeV^{2}$}\end{sideways}
\epsfig{file=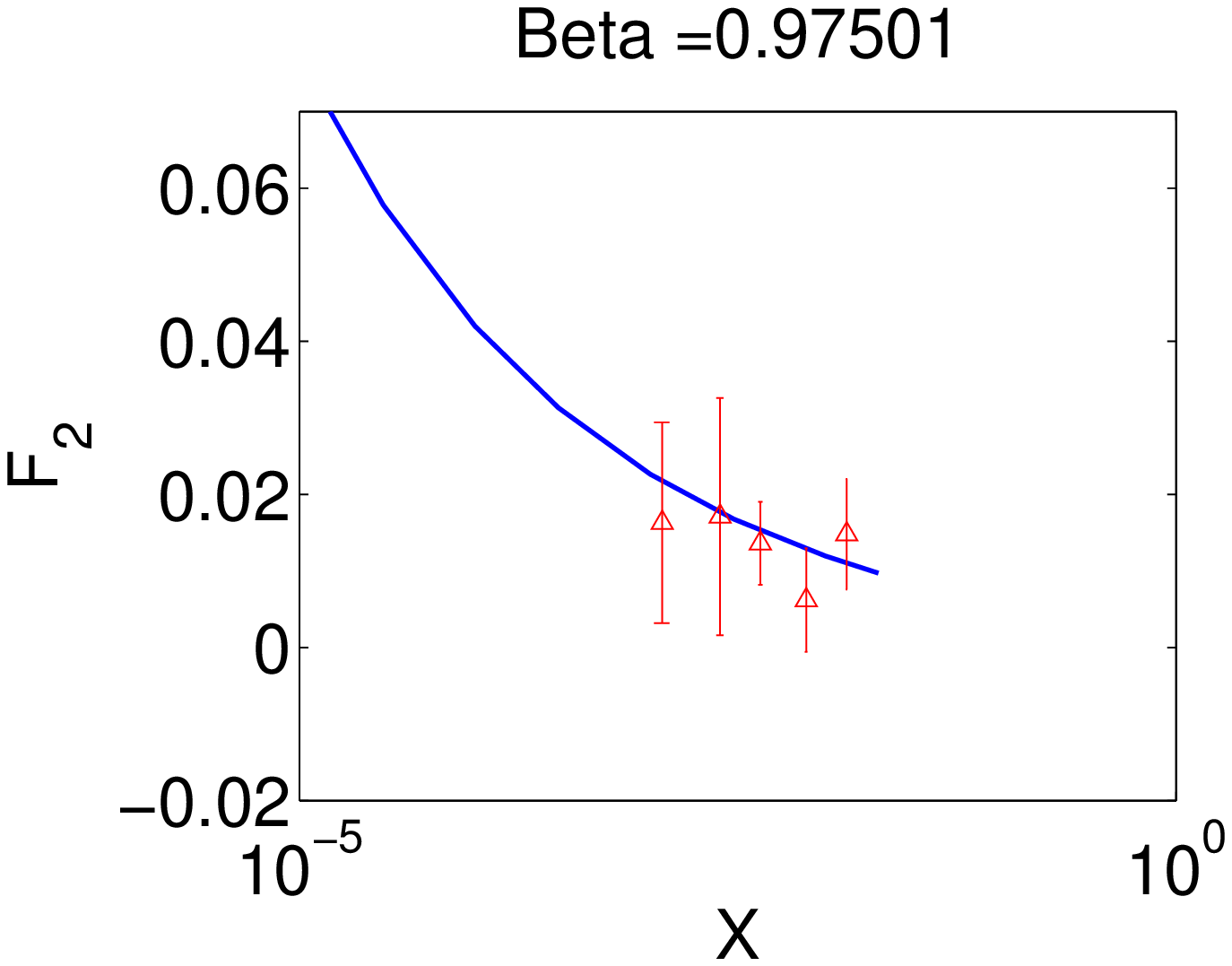,width=32mm, height=28mm}&
\epsfig{file=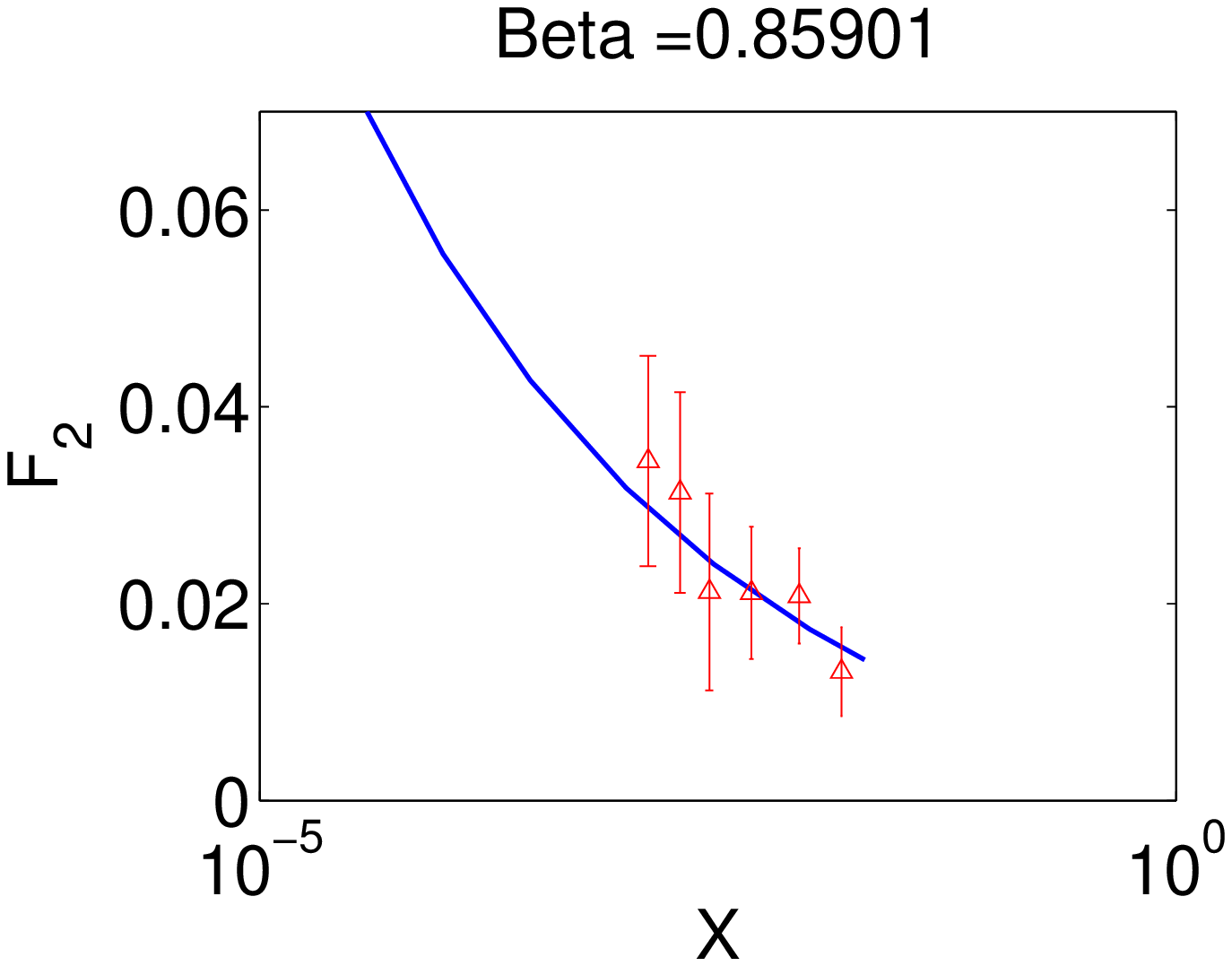,width=32mm, height=28mm}&
\epsfig{file=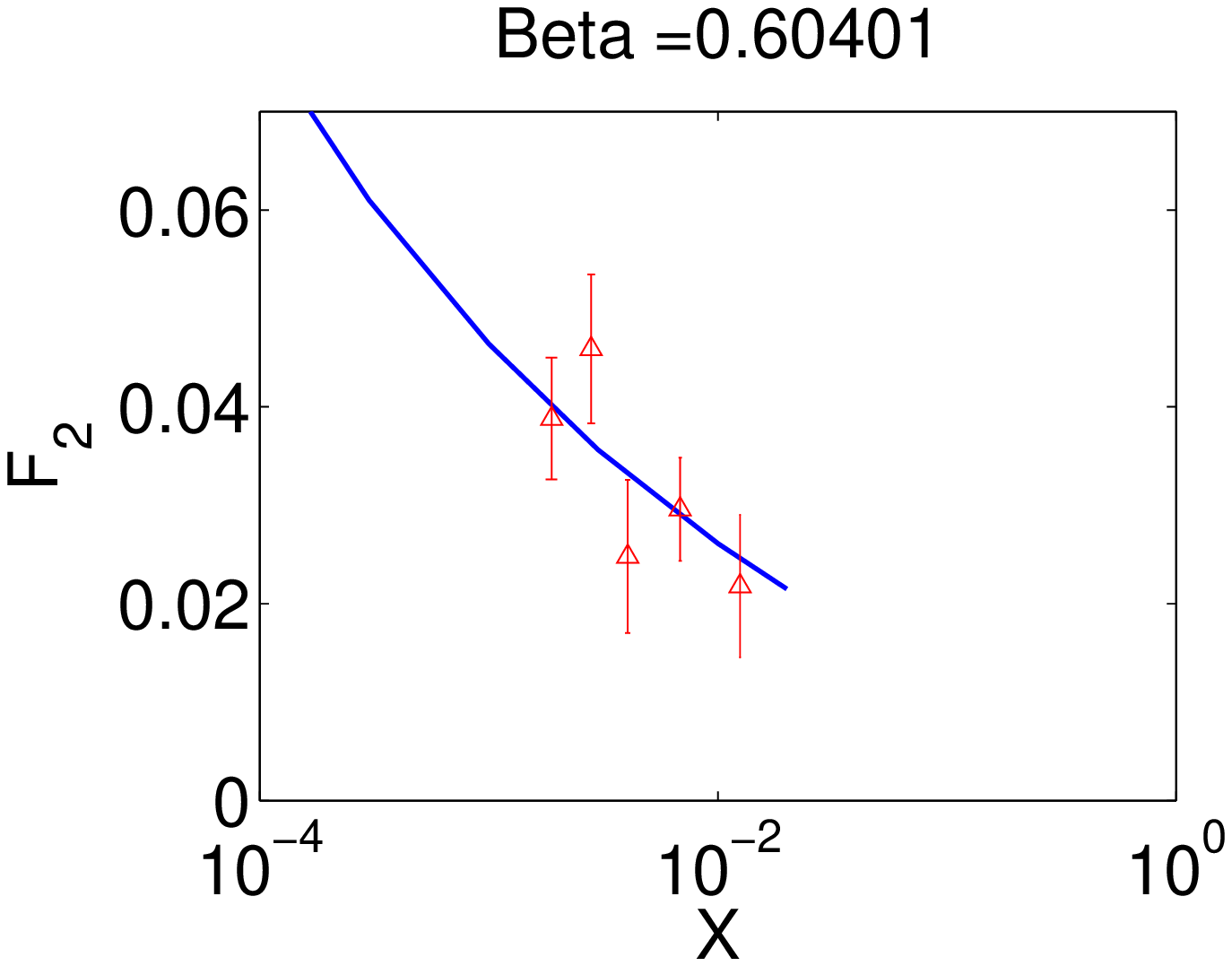,width=32mm, height=28mm}&
\epsfig{file=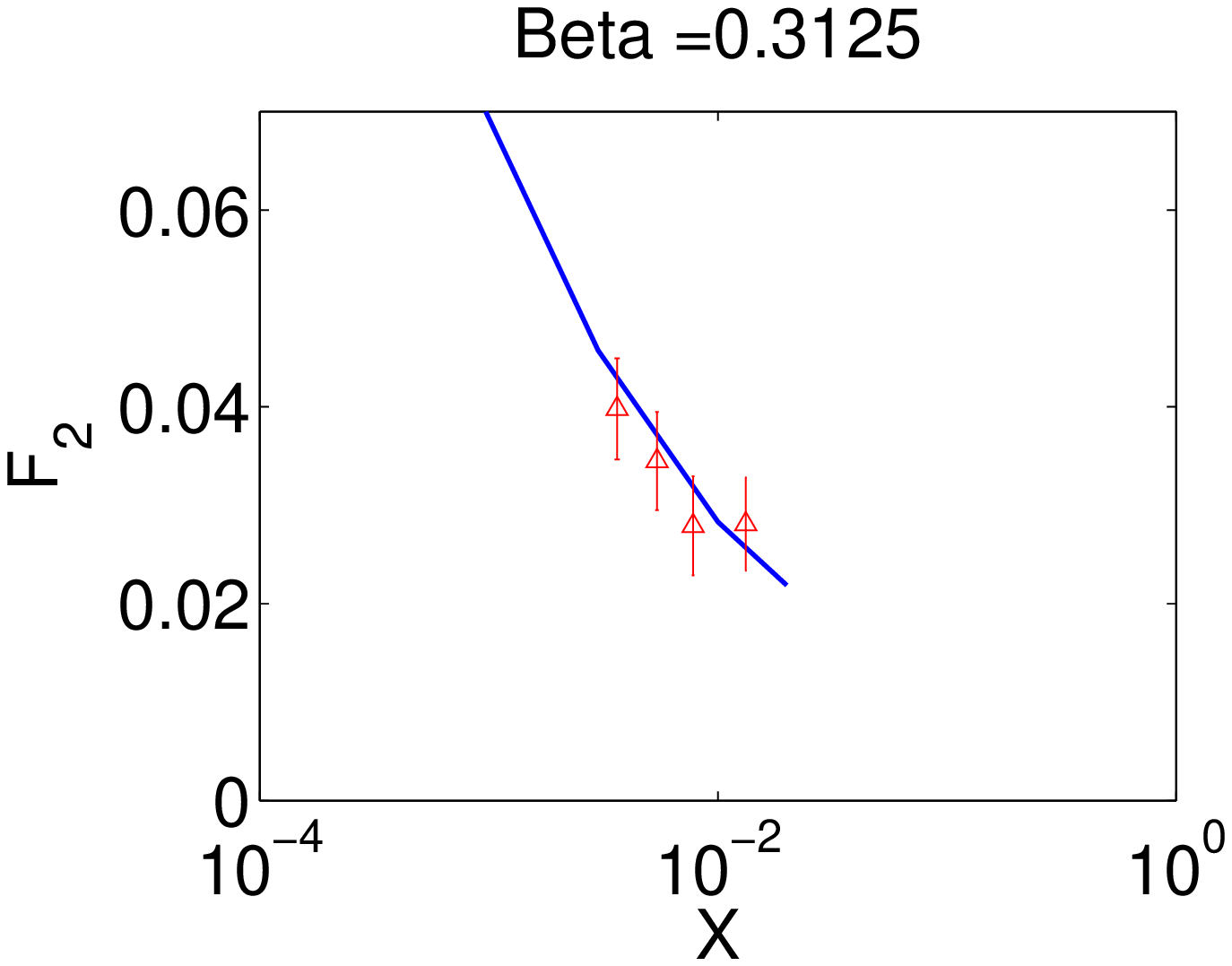,width=32mm, height=28mm}&
\epsfig{file=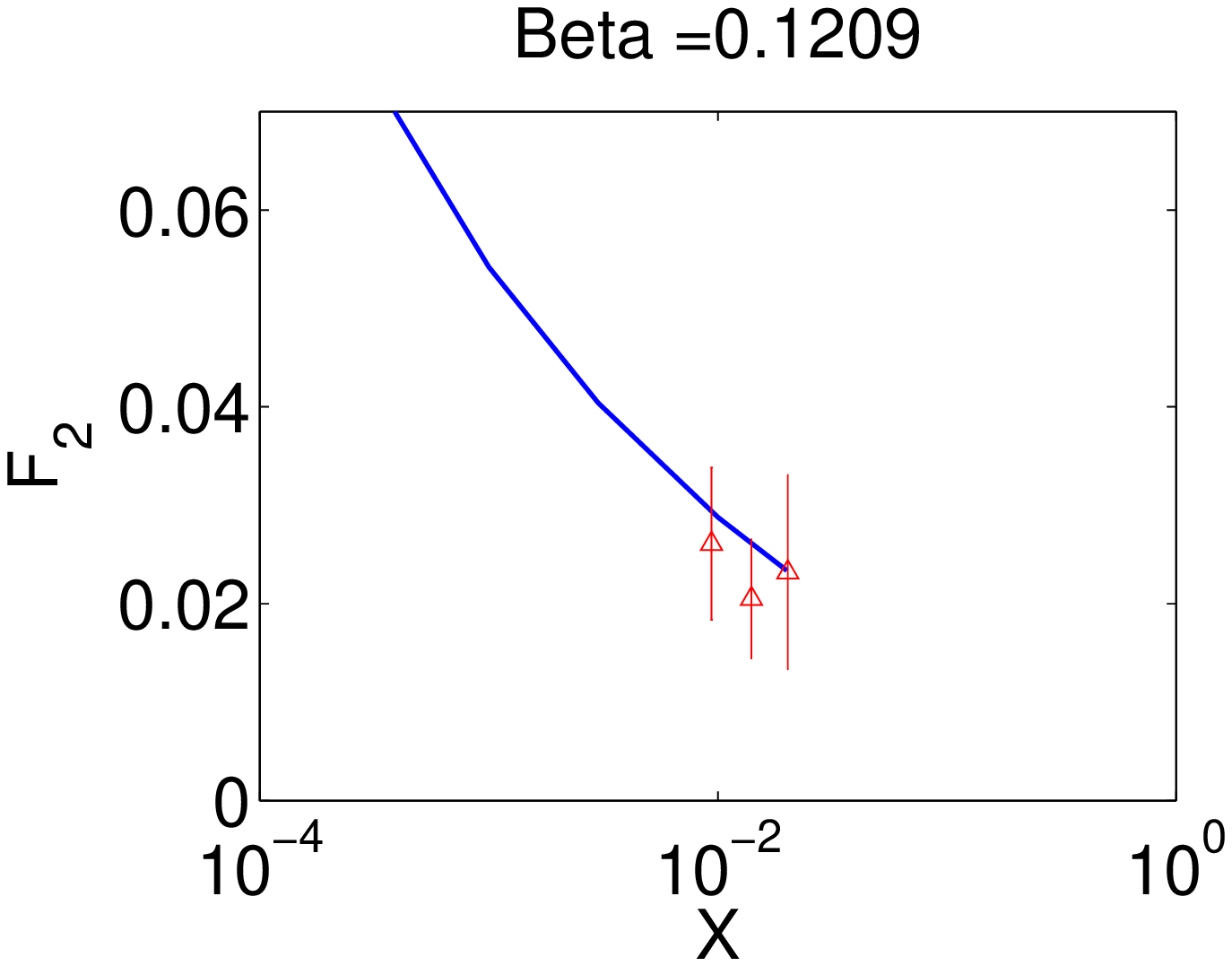,width=32mm, height=30mm}&\\
\end{tabular}\caption{\it (continued) Diffractive dissociation structure function
$F_{2}^{D(3)}(\beta,x_{\Pom},Q^{2})$ as a function of $x_{\Pom}$.
Comparison of the prediction of the model with ZEUS data
\cite{Chekanov:2005vv}.}\label{diff_2}
\end{sidewaysfigure}

\section{Summary}

Using the generating functional approach, we develop a new
saturation model, which describes well all the HERA data on deep
inelastic scattering and diffractive dissociation. We took into
account the QCD evolution, by evolving the initial gluon density
with the DGLAP evolution equation. The innovation used in this
current work, is the redefinition of Bjorken-$x$, in the saturation
region, since the transverse momentum of partons cannot be
neglected, and it is the saturation scale. We included in the
calculation of the proton structure function $F_{2}$, the
contribution from heavy quark production, and also investigated the
behavior of the $F_{2}^{c\overline{c}}$ structure function. The
resulting $\chi^{2}/d.o.f$ for the fit of DIS data, is very close to
one ($\chi^{2}/d.o.f\;=\;354/341$) and this reflects the fact that
our new model is able to give reliable predictions. This was
justified in the case of the description of diffractive dissociation
data, using the same parameters which were obtained from the fit to
DIS. Despite the fact that two completely different models are able
to describe well the experimental data, we can distinguish between
them by calculating the differential cross sections with various
multiplicities $d\sigma^{k}/d^{2}b$. From figures shown, we can
easily see the different behavior of these cross sections, as a
function of the impact parameter $b$. From the plot of the
saturation scale \ref{Qs_plot}, we can conclude, that up to the
scale $\thicksim\;$3 - 4 $GeV^{2}$, saturation is significant, and
the non-linear interaction term in the evolution equation, plays an
essential role. We believe that this paper, introduces an additional
argument for the saturation phenomenon. It shows that the saturation
models are able to describe all experimental data, including small
values of $Q^{2}$ and low $x$. The description does not depend on
the model assumptions  widely used instead of the solution to the
equation in the mean field approximation \cite{Gotsman:2004ra},
which turns out to be rather complicated.

\section{Acknowledgments}

I would like to express my deep appreciation to Eugene Levin for his
support in writing this paper. I am  also very grateful to Asher
Gotsman, Alex Prygarin, Jeremy Miller, and Alex Palatnik for
fruitful discussions on the subject. Special thanks to Erez Etzion,
Gideon Bella, Jonatan Ginzburg, Nir Amram and Eran Naftali for the
technical support and essential discussions on the experimental
background of the research. This research was supported in part  by
the Israel Science Foundation, founded by the Israeli Academy of
Science and Humanities and by BSF grant \# 20004019.

\end{document}